\documentstyle[12pt]{article}

\newcommand{\dual}[1]{{}^*\hspace{-1mm}#1}

\newcommand{\be}{\begin{equation}}
\newcommand{\ee}{\end{equation}}
\newcommand{\bea}{\begin{eqnarray}}
\newcommand{\eea}{\end{eqnarray}}
\newcommand{\beasn}{\begin{sneqnarray}}
\newcommand{\eeasn}{\end{sneqnarray}}
\newcommand{\bref}[1]{(\ref{#1})}
\newcommand{\ct}[1]{\cite{#1}}
\newcommand{\eps}{\epsilon}
\newcommand{\veps}{\varepsilon}

\newcommand{\der}[2]{\frac{\partial #1}{\partial #2}}
\newcommand{\lder}[2]{\frac{\partial_l#1}{\partial #2}}
\newcommand{\rder}[2]{\frac{\partial_r#1}{\partial #2}}
\newcommand{\gh}[1]{{\cal #1}}
\newcommand{\agh}[1]{\bar{\cal #1}}

\newcommand{\bra}[1]{\langle \, #1 \, |}
\newcommand{\ket}[1]{| \, #1 \, \rangle}
\newcommand{\bracket}[2]{\langle \, #1 \, | \, #1 \, \rangle}

\makeatletter

\def\slashit#1{\slash\mkern-10mu#1}

\def\mathoperat{\@ifnextchar [{\@mathoperat}{\@mathoperat[rm]}}
   \def\@mathoperat[#1]#2#3{\def#2{\mathop{\@nameuse{#1} #3{}}\nolimits}}
\def\restric#1#2{{\left. #1 \right|_{#2}}}
\def\dif{{\rm d}}
\def\deriv{\@ifnextchar[{\@deriv}{\@deriv[]}}
   \def\@deriv[#1]#2#3{\mathchoice%
{{\dif^{#1}#2\over\dif{#3}^{#1}}}{{\dif^{#1}#2/\dif{#3}^{#1}}}%
{{\dif^{#1}#2\over\dif{#3}^{#1}}}{{\dif^{#1}#2/\dif{#3}^{#1}}}}

\def\secteqno{\@addtoreset{equation}{section}%
\def\theequation{\thesection.\arabic{equation}}}
\def\endsecteqno{\def\theequation{\@ifundefined{chapter}%
{\arabic{equation}}{\thechapter.\arabic{equation}}}}
\newcounter{subequation}
\def\thesubequation{\alph{subequation}}
\def\sneqnarray{\stepcounter{equation}\let\@currentlabel=\theequation
\setcounter{subequation}{1}
\def\@eqnnum{{\rm (\theequation.\thesubequation)}}
\global\@eqcnt\z@\tabskip\@centering\let\\=\@eqncr\let\@@eqncr=\@@sneqncr
$$\halign to \displaywidth\bgroup\@eqnsel\hskip\@centering
 $\displaystyle\tabskip\z@{##}$&\global\@eqcnt\@ne
 \hskip 2\arraycolsep \hfil${##}$\hfil
 &\global\@eqcnt\tw@ \hskip 2\arraycolsep $\displaystyle\tabskip\z@{##}$\hfil
  \tabskip\@centering&\llap{##}\tabskip\z@\cr}
\def\endsneqnarray{\@@sneqncr\egroup $$\global\@ignoretrue}
\def\@@sneqncr{\let\@tempa\relax
   \ifcase\@eqcnt \def\@tempa{& & &}\or \def\@tempa{& &}
   \else \def\@tempa{&}\fi
     \@tempa \if@eqnsw\@eqnnum\stepcounter{subequation}\fi
     \global\@eqnswtrue\global\@eqcnt\z@\cr}
\def\nobiblabels{\def\@lbibitem[##1]##2{\@bibitem{##2}}}
\makeatother

\begin{document}

\begin{titlepage}

\begin{flushright}

CCNY-HEP-94/03\\
KUL-TF-94/12\\
UB-ECM-PF 94/15\\
UTTG-11-94\\
hep-th/9412228\\
May 1994\\

\end{flushright}

\vskip 2.0mm

\begin{center}

{\LARGE\bf  Antibracket, Antifields\\
\vskip 3.0mm
     and Gauge-Theory Quantization}\\

\vskip 6.mm

{ Joaquim Gomis$^{*1}$, Jordi Par\'{\i}s$^{\sharp 2}$
and Stuart Samuel$^{\dagger 3}$}

\vskip 0.4cm

{\small
$^*$
{\it{Theory Group, Department of Physics}}\\
{\it{The University of Texas at Austin}}\\
{\it{RLM\,5208, Austin, Texas}}\\
{\it{and}}\\
{\it{Departament d'Estructura i Constituents de la
           Mat\`eria}}\\
{\it{Facultat de F\'{\i}sica, Universitat de Barcelona}}\\
{\it{Diagonal 647, E-08028 Barcelona}}\\
{\it{Catalonia}}\\
[0.4cm]

$^\sharp$
{\it{Instituut voor Theoretische Fysica}}\\
{\it{Katholieke Universiteit Leuven}}\\
{\it{Celestijnenlaan 200D}}\\
{\it{B-3001 Leuven, Belgium}}\\
[0.4cm]

$^\dagger$
{\it{Department of Physics}}\\
{\it{City College of New York}}\\
{\it{138th St and Convent Avenue}}\\
{\it{New York, New York 10031 U.S.A.}}\\
[1.0cm]

}
\normalsize

\end{center}

\textwidth 6.5truein

\hrule width 5.cm

{\small
\noindent $^1$
Permanent address: Dept.\ d'Estructura i
Constituents de la Mat\`{e}ria, U.\ Barcelona.\\
E-mail: gomis@rita.ecm.ub.es\\
\noindent $^2$
Wetenschappelijk Medewerker, I.\,I.\,K.\,W., Belgium.\\
E-mail: jordi=paris\%tf\%fys@cc3.kuleuven.ac.be\\
\noindent $^3$
E-mail: samuel@scisun.sci.ccny.cuny.edu
}

\normalsize

\vfill
\eject

\textwidth 6truein

\pagestyle{empty}

{\ }
\vskip 1.5cm

\begin{center}

{\bf Abstract}

\end{center}

\begin{quote}

\hspace{\parindent}
{}\ \ \ The antibracket formalism for gauge theories,
at both the classical and quantum level,
is reviewed.
Gauge transformations and the
associated gauge structure are analyzed
in detail.
The basic concepts involved
in the antibracket formalism
are elucidated.
Gauge-fixing,
quantum effects,
and anomalies
within the field-antifield formalism
are developed.
The concepts, issues and constructions
are illustrated using eight gauge-theory models.

\end{quote}

\vfill
\eject

\pagestyle{empty}

\tableofcontents

\vfill
\eject

\pagestyle{empty}

{\ }
\vskip 4.5cm

\begin{center}

{This work is dedicated to Joseph and Marie,\\ }
{to Pilar,\\ }
{and to the memory of Pere and Francesca.}

\end{center}

\end{titlepage}

\setcounter{page}{2}

\secteqno

\pagestyle{myheadings}
\markright{{\it J.\,Gomis, J.\,Par\'{\i}s and S.\,Samuel} ---
     Antibracket, Antifields and $\dots $}

\section{Introduction}
\label{s:i}

\hspace{\parindent}
The known fundamental interactions of nature
are all governed by gauge theories.
The presence of a gauge symmetry
indicates that a theory
is formulated in a redundant way,
in which certain local degrees of freedom
do not enter the dynamics.
Conversely,
when there are degrees of freedom,
which do not enter the lagrangian,
a theory possesses local invariances.
Although one can in principle eliminate
the gauge degrees of freedom,
there are reasons for not doing so.
These reasons include manifest covariance,
locality of interactions,
and calculational convenience.

The first example of a gauge theory
was electrodynamics.
Electric and magnetic forces are generated
via the exchange of photons.
Being particles of spin $1$,
photons involve a vector field, $A^\mu$.
However, not all four components
of the electromagnetic potential $A^\mu$
enter dynamically.
Two degrees of freedom correspond to
the two possible physical polarizations of the photon.
The longitudinal degree of freedom
plays a role in interactions
via virtual exchanges of photons.
The remaining gauge degree of freedom
does not enter the theory.
Consequently,
electromagnetism is described by a gauge theory.
When it was realized that
the weak interactions could be unified
with electromagnetism
in an $SU(2) \times U(1)$ gauge theory
\ct{glashow61a,weinberg67a,salam68a}
and that this theory is renormalizable
\ct{thooft71a,thooft71b},
the importance of non-abelian gauge theories
\ct{ym54a}
grew enormously.
The strong interactions are also governed by
an SU(3) non-abelian gauge theory.
The fourth fundamental force
is gravity.
It is based on Einstein's general theory of relativity
and uses general coordinate invariance.
When formulated in terms of a metric
or any other convenient fields,
gravity also possesses gauge symmetries.

The quantization of gauge theories
is not always straightforward.
In the abelian case,
relevant for electromagnetism,
the procedure is well understood.
In contrast,
quantization of a non-abelian theory
and its renormalization
is more complicated.
Quantization generally involves
the introduction of ghost fields.
Typically, a gauge-fixing procedure
is used to render dynamical
all degrees of freedom.
Ghost fields are used
to compensate for the effects
of the gauge degrees of freedom
\ct{feynman63a},
so that unitarity is preserved.
In electrodynamics in the linear gauges,
ghosts decouple and can be ignored.
In non-abelian gauge theories,
convenient gauges
generically involve interacting ghosts.
A major step in understanding these issues
was the Faddeev-Popov quantization procedure
\ct{fp67a,dewitt67a},
which relied heavily on the functional-integral
approach to quantization
\ct{fh65a,al73a,iz80a}.
{}From this viewpoint,
the presence of ghost fields
is understood as a ``measure effect''.
In dividing out the volume of gauge transformations
in function space,
a Jacobian measure factor arises.
This factor is produced naturally
by introducing quadratic terms in the lagrangian
for ghosts and then integrating them out.
It was realized at a later stage
that the gauge-fixed action
retains a nilpotent, odd, global symmetry
involving transformations of both fields and ghosts.
This Becchi-Rouet-Stora-Tyutin (BRST) symmetry
\ct{brs74a,tyutin75a}
is what remains of the original gauge invariance.
In fact, for closed theories,
the transformation law for the original fields
is like a gauge transformation with gauge parameters
replaced by ghost fields.
In general,
this produces nonlinear transformation laws.
The relations among correlation functions
derived from BRST symmetry
involve the insertions of the BRST variation of fields.
These facts require the use of composite operators
and it is convenient to introduce
sources for these transformations.
The Ward identities
\ct{ward50a}
associated with the BRST invariance
treated in this way
are the Slavnov-Taylor identities
\ct{slavnov72a,taylor71a}.
The Slavnov-Taylor identities
and BRST symmetry
have played an important role
in quantization, renormalization, unitarity,
and other aspects
of gauge theories.

Ghosts fields have been useful
throughout the development of
covariant gauge-field-theory quantization
\ct{ko78a,ko79a,ku82a,on82a,ab83a}.
It is desirable to have a formulation of gauge theories
that introduces them from the outset
and
that automatically incorporates BRST symmetry
\ct{baulieu85a}.
The field-antifield formulation has these features
\ct{brs74a,zinnjustin75a,bv81a,bv83a,bv83b,bv84a}.
It relies on BRST symmetry as fundamental principle
and uses sources to deal with it
\ct{brs74a,tyutin75a,zinnjustin75a}.
It encompasses
previous ideas and developments
for quantizing gauge systems
and extends them to
more complicated situations
(open algebras, reducible systems, etc.)
\ct{fn76a,fnf76a,kallosh78a,stn78a,wh79a}.
In 1975, J. Zinn-Justin,
in his study of the renormalization of
Yang-Mills theories
\ct{zinnjustin75a},
introduced the above-mentioned sources
for BRST transformations
and a symplectic structure $( \ , \ )$
(actually denoted $*$ by him)
in the space of fields and sources,
He expressed
the Slavnov-Taylor identities
in the compact form $( \Gamma , \Gamma ) = 0$,
where $\Gamma$,
the generating functional
of the one-particle-irreducible diagrams,
is known as the effective action
(see also \ct{lee76a}).
These ideas were developed further
by B.\ L.\ Voronov and I.\ V.\ Tyutin in
\ct{vt82a,vt82b}
and by I.\ A.\ Batalin and G.\ A.\ Vilkovisky
in refs.\ct{bv81a,bv83a,bv83b,bv84a,bv85a}.
These authors generalized
the role of $( \ , \ )$ and of the sources
for BRST transformations
and called them
the antibracket and antifields respectively.
Due to their contributions,
this quantization procedure is
often referred to
as the Batalin-Vilkovisky formalism.

The antibracket formalism
gained popularity among string theorists,
when it was applied to the open bosonic string field theory
\ct{bochicchio87a,thorn87a}.
It has also proven
quite useful for the closed string field theory
and for topological field theories.
Only within the last few years
has it been applied to more general aspects
of quantum field theory.

In some sense,
the BRST approach,
which was driven, in part,
by renormalization considerations,
and the field-antifield formalism,
which was motivated by classical considerations
such as gauge structure,
are not so different.
When sources are introduced for BRST transformations,
the BRST approach resembles the field-antifield one.
Antifields, then, have a simple intepretation:
They are the sources
for BRST transformations.
In this sense,
the field-antifield formalism
is a general method for dealing with gauge theories
within the context of standard field theory.

The general structure of the antibracket formalism
is as follows.
One introduces an antifield
for each field and ghost,
thereby doubling the total number of original fields.
The antibracket $( \ , \ )$ is
an odd non-degenerate symplectic form
on the space of fields and antifields.
The original classical action $S_0$
is extended to a new action $S$,
in an essentially unique way,
to arrive at a theory
with manifest BRST symmetry.
One equation,
the master equation $( S , S ) = 0$,
reproduces in a compact way
the gauge structure of the original theory
governed by $S_0$.
Although the master equation resembles
the Zinn-Justin equation,
the content of the two is different
since $S$ is a functional of quantum fields and antifields
and $\Gamma$ is a functional of classical fields.

The antibracket formalism currently appears
to be the most powerful method for quantizing
a gauge theory.
Beyond tree level, order $\hbar$ terms
usually need to be added to the action,
thereby leading to a quantum action $W$.
These counterterms are expected
to render finite loop contributions,
after a suitable regularization procedure
has been introduced.
The master equation must be appropriately generalized
to the so-called quantum master equation.
It involves a potentially singular operator $\Delta$.
The regularization procedure and counterterms
should also render $\Delta$
and its action on $W$ well-defined.
Violations of the quantum master equation
are equivalent to gauge anomalies
\ct{tnp90a}.
To calculate correlation functions
and scattering amplitudes
in perturbation theory,
a gauge-fixing procedure is selected.
This procedure eliminates antifields
in terms of functionals of fields.
When appropriately implemented,
propagators exist,
and the usual Feynman graph methods can be used.
In addition,
for the study of symmetry properties,
renormalization and anomalies,
a modified version of the
gauge-fixing procedure is available
which keeps antifields.
In short,
the antibracket formalism
has manifest gauge invariance or BRST symmetry,
provides the extra fields needed
for covariant quantization,
permits a perturbative expansion
of the quantum theory,
and allows the study of
quantum corrections
to the symmetry structure of the theory.

The field-antifield formalism can treat systems
that cannot be handled
by Faddeev-Popov functional integration approach.
This is particularly clear for theories
in which quartic ghost interactions arise
\ct{kallosh78a,wh79a}.
Faddeev-Popov quantization leads to an action
bilinear in ghost fields,
and fails
for the case of open algebras.
An open algebra occurs
when the commutator of two gauge transformations
produces a term proportional to
the equations of motion
and not just another gauge transformation
\ct{wh79a,bv84a}.
In other words,
the gauge algebra closes only on-shell.
Such algebras occur in gravity
\ct{fv75b}
and supergravity
\ct{fnf76a,kallosh78a,wh79a,vannieuwenhuizen81a}
theories.
The ordinary Faddeev-Popov procedure
also does not work for reducible theories.
In reducible theories, the gauge generators
are all not independent
\ct{cs74a,kr74a,ad79a,bhnw79a,%
siegel80a,townsend80a,vasiliev80a,dts81a,thierrymieg90a}.
Some modifications of the procedure
have been developed by introducing ghosts for ghosts
\ct{siegel80a,kimura81a}.
However, these modifications
\ct{siegel80a,hko81a,kimura81a,fs88a}
do not work for the general reducible theory.
Even for Yang-Mills theories,
the Faddeev-Popov procedure can fail,
if one considers exotic gauge-fixing procedures
for which ``extraghosts'' appear
\ct{kallosh78a,nielsen78a,nielsen81a}.
The field-antifield formalism is sufficiently general
to encompass previously known lagrangian
approaches to the quantization of gauge theories.

Perhaps the most attractive feature
of the field-antifield formalism
is its imitation of a hamiltonian Poisson structure
in a covariant way.
In some instances,
the hamiltonian approach
to quantization has the advantage
of being manifestly unitary.
However, it is necessarily non-covariant
since the time variable
is treated in a manner different
from the space variables.
In addition, the gauge invariances usually must be fixed
at the outset.
In compensation for this,
one needs to impose
constraints on the Hilbert space of states.
In the field-antifield approach,
the antibracket plays the role of the Poisson bracket.
As a consequence,
hamiltonian concepts,
such as canonical transformations,
can be formulated and used
\ct{vlt82a,vt82a,vt82b,bv84a,fh90a,tnp90a}.
At the same time,
manifest covariance and BRST invariance are maintained.
Since the antibracket formalism proceeds via
the functional integral,
the powerful techniques of functional integration
are available.

A non-trivial aspect of the field-antifield approach
is the construction of the quantum action $W$.
When loop effects are ignored,
$W \to S$ provides the solution
to the master equation.
A straightforward but not necessarily simple
procedure is available for
obtaining $S$ given the classical action $S_0$
and its gauge invariances
for a finite-reducible system.
When quantum effects are incorporated,
$W$ must satisfy
the more singular quantum master equation.
However, there is currently no known method
that guarantees the construction of $W$.
The problem is that
the field-antifield formalism does not
automatically provide
the functional integration measure.
These issues are linked with
those associated with unitarity,
renormalization, quantum gauge invariance,
and anomalies.
Because these aspects of gauge theories
are inherently difficult,
it is not surprising
that the field-antifield formalism
does not provide a simple solution.

Another, less serious weakness,
is that the antibracket formalism
involves quite a bit of mathematical machinery.
Sometimes, a gauge theory is expressed
in a form
which is more complicated than necessary.
This can make computations
somewhat more difficult.

The organization of this article is as follows.
Sect.\ \ref{s:ssgt}
discusses gauge structure.
Some notation is presented during the process of
introducing gauge transformations.
The distinction between irreducible
and reducible gauge theories is made.
The latter involve a redundant set of gauge invariances
so that there are relations among the gauge generators.
As a result,
there exists gauge invariances
for gauge invariances,
and ghosts for ghosts.
A theory is $L$th-stage reducible
if there are gauge invariances
for the gauge invariances for the gauge invariances, etc.,
$L$-fold times.
The general form of the gauge structure
for a first-stage reducible case
is determined.
In Sect.\ \ref{s:egt},
specific gauge theories are presented to illustrate
the concepts
of Sect.\ \ref{s:ssgt}.
The spinless relativistic particle,
non-abelian Yang-Mills theories,
topological Yang-Mills theory,
the antisymmetric tensor field,
free abelian $p$-form theories,
open bosonic string field theory,
the massless relativistic spinning particle,
and the first-quantized bosonic string
are treated.
The spinless relativistic particle
of Sect.\ \ref{ss:srp}
is also used to exemplify notation.
The massless relativistic spinning particle
provides an example of a simple supergravity theory,
namely a theory
with supersymmetric gauge invariances.
This system is used to illustrate
the construction of
supersymmetric and supergravity theories.
A review of the construction
of general-coordinate-invariant theories
is given in the subsection
on the first-quantized open bosonic string.
These mini-reviews should be useful
to the reader who is new to these subjects.

The key concepts of
the field-antifield formalism are elucidated
in Sect.\ \ref{s:faf}.
Antifields are introduced
and the antibracket is defined.
The latter is used to define canonical transformations.
They can be quite helpful in simplifying computations.
Next, the classical master equation
$(S,S)=0$ is presented.
When appropriate boundary conditions are imposed,
it reproduces, in a compact way,
the gauge structure
of Sect.\ \ref{s:ssgt}.
A suitable action $S$ satisfying the master equation
is called a proper solution.
Given the gauge-structure tensors
of a first-stage reducible theory,
Sect.\ \ref{ss:psga}
presents the generic proper solution.
The last part
of Sect.\ \ref{s:faf}
defines and discusses the classical BRST symmetry.
Examples of proper solutions
are provided
in Sect.\ \ref{s:eps}
for the gauge field theories
presented
in Sect.\ \ref{s:egt}.

Sect.\ \ref{s:gff} begins
the passage from the classical to the quantum aspects
of the field-antifield formalism.
The gauge-fixing procedure is discussed.
The gauge-fixing fermion $\Psi$ is a key concept.
It is used as a means of eliminating antifields
in terms of functions of fields.
The result
is an action that is suitable
for use in the path integral.
Only in this context
and in performing standard perturbative computations
are antifields eliminated.
It is shown that results are independent
of the choice of $\Psi$,
if the quantum action $W$ satisfies
the quantum master equation.
To implement gauge-fixing,
more fields and their antifields
must be introduced.
How this works
for irreducible and first-stage reducible theories
is treated first.
Then, for reference purposes,
the general $L$th-stage reducible case is considered.
Delta-function type gauge-fixing is treated
in Sect.\ \ref{ss:dfgfp}.
Again, irreducible and first-stage reducible cases
are presented first.
Again, for reference purposes,
the general $L$th-stage reducible case is treated.
Gauge-fixing by a gaussian averaging process
is discussed
in Sect.\ \ref{ss:ogfp}.
After gauge-fixing,
a classical gauge-fixed BRST symmetry can be defined.
See Sect.\ \ref{ss:gfcbrsts}.
The freedom to perform canonical transformations
permits one to work in any appropriate field basis.
This freedom can be quite useful.
Concepts tend to have different
interpretations in different bases.
One basis, associated with $\Psi$
and called the gauge-fixed basis,
is the last topic
of Sect.\ \ref{s:gff}.
Examples of gauge-fixing procedures are provided
in Sect.\ \ref{s:gfe}.
With the exception of the free $p$-form theory,
the theories are the ones considered
in Sects.\ \ref{s:egt} and \ref{s:eps}.

Quantum effects and possible gauge anomalies
are analyzed in Sect.\ \ref{s:qea}.
The key concepts are quantum-BRST transformations
and the quantum master equation.
Techniques for assisting in finding solutions
to the quantum master equation
are provided
in Sects.\ \ref{ss:sqme},
\ref{ss:qmevg} and \ref{ss:ctqme}.
The generating functional $\Gamma$
for one-particle-irreducible diagrams
is generalized to the field-antifield case
in Sect.\ \ref{ss:eazje}.
This allows one to treat the quantum system
in a manner similar to the classical system.
The Zinn-Justin equation is shown
to be equivalent to the quantum master equation.
When unavoidable violations of the latter occur,
the gauge theory is anomalous.
See Sect.\ \ref{ss:qmevg}.
Explicit formulas
at the one-loop level are given.
In Sect.\ \ref{s:sac},
sample anomaly calculations are presented.
It is shown that the spinless relativistic particle
does not have an anomaly.
In Sect.\ \ref{ss:acsm},
the field-antifield treatment
of the two-dimensional chiral Schwinger model
is presented.
Violations of the quantum master equation
are obtained.
This is expected since the theory is anomalous.
A similar computation is performed
for the open bosonic string.
For $D \ne 26$,
the theory is anomalous,
as expected.
Some of the details of the calculations
are relegated to Appendix C.

Section \ref{s:bdot}
briefly presents several additional topics.
The application of the field-antifield formalism
to global symmetries is presented.
A review is given of the geometric interpretation
of E. Witten \ct{witten90a}.
The next topic is the role of locality.
This somewhat technical issue is important
for renormalizability and for cohomological aspects.
A summary of cohomological methods is given.
Next, the relation
between the hamiltonian and antibracket approaches
is discussed.
The question of unitarity is the subject
of Sect.\ \ref{ss:u}.
One place
where the field-antifield formalism
has played an essential role
is in the $D=26$ closed bosonic string field theory.
This example is rather complicated
and not suitable for pedagogical purposes.
Nevertheless,
general aspects of the antibracket formalism
for the closed string field theory
are discussed.
Finally,
an overview is given of how
to handle anomalous systems
using an extended set of fields and antifields.

Appendix A reviews the mathematical aspects
of left and right derivatives,
integration by parts,
and chain rules for differentiation.
Appendix B discusses in more detail
the regularity condition,
which is a technical requirement
of the antibracket formalism.

At every stage of development of the formalism,
there exists some type of BRST operator.
In the space of fields and antifields
before quantization,
a classical nilpotent BRST transformation $\delta_B$
is defined by using the action $S$
and the antibracket: $\delta_B F = ( F , S )$.
{}From $\delta_B$,
a gauge-fixed version $\delta_{B_\Psi}$
is obtained by imposing the conditions
on antifields
provided by the gauge-fixing fermion $\Psi$.
At the quantum level,
a quantum version $\delta_{\hat B}$
of $\delta_B$ emerges.
In the context of the effective action formulation,
a transformation $\delta_{B_{cq}}$,
acting on classical fields, can be defined
by using $\Gamma$ in lieu of $S$.
Several subsections are devoted
to the BRST operator,
its properties and its utility.

The existence of a BRST symmetry
is crucial to the development.
Observables
are those functionals which are BRST invariant
and cannot be expressed as the BRST variation
of something else.
In other words,
observables correspond to the elements
of the BRST cohomology.
The nilpotency of $\delta_B$ and $\delta_{\hat B}$
are respectively
equivalent to the classical and quantum master equations.
The traditional treatment
of gauge theories using BRST invariance
is reviewed in \ct{baulieu85a}.
For this reason,
we do not discuss
BRST quantization in detail.

The antibracket formalism is rather versatile
in that one can use any set of fields (and antifields)
related to the original fields (and antifields)
by a canonical transformation.
However, under such a change,
the meaning of certain concepts change.
For example,
the gauge structure,
as determined by the master equation,
has a different interpretation
in the gauge-fixed basis than in the original basis.
Most of this review
uses the second viewpoint.
The treatment in the gauge-fixed basis
is handled in Sects.\ \ref{ss:gfb} and \ref{ss:eazje}.

The material in each section
strives to fulfill one of three purposes.
A key purpose
is to present computations
that lead to understanding and insight.
Sections \ref{s:ssgt}, \ref{s:faf},
\ref{ss:g}, \ref{ss:gfcbrsts}, \ref{ss:gfb},
\ref{s:qea} and \ref{s:bdot}
are mainly of this character.
The second purpose
is pedagogical.
This Introduction falls into this category
in that it gives
a quick overview of the formalism
and the important concepts.
Sections \ref{s:egt},
\ref{s:eps}, \ref{s:gfe},
and \ref{s:sac}
analyze specially chosen gauge theories
which
allow the reader
to understand the field-antifield formalism
in a concrete manner.
Finally,
some material is included for technical completeness.
Sections \ref{ss:gfaf}--\ref{ss:ogfp}
present methods
for gauge-fixing
the generic gauge theory.
Parts of sections \ref{ss:irgt},
\ref{ss:gs} and \ref{ss:aoll}
are also for reference purposes.
Probably the reader should not initially
try to read these sections in detail.
Many sections serve a dual role.

A few new results on the antibracket formalism
are presented in this review.
They are included
because they provide insight for the reader.
We have tried
to have a minimum overlap
with other reviews.
In particular,
cohomological aspects
are covered in
\ct{brs74a,dixon76a,bc83a,baulieu85a,dtv85a,band86a,%
henneaux90a,ht92a,tpbook}
and more-detailed aspects of anomalies
are treated
in \ct{tpbook}.
Pedagogical treatments are given in
references \ct{ht92a,tpbook}.
In certain places,
material from
reference \ct{paris92a}
has been used.

This review
focuses on the key points and concepts
of antibracket formalism.
There is some emphasis
on applications to string theory.
Our format is to first present the material abstractly
and then to supply examples.
The reader who is new to this subject
and mainly interested in learning
may wish to reverse this order.
Exercises can be generated
by verifying the abstract results
in each of the sample gauge theories
of Sects.\ \ref{s:egt},
\ref{s:eps}, \ref{s:gfe},
and \ref{s:sac}.
Other systems,
which have been treated
by field-antifield quantization
and may be of use to the reader,
are
the free spin $\frac52$ field
\ct{bv83b},
the spinning string
\ct{gpr90a},
the $10$-dimensional
Brink-Schwarz superparticle and superstring
\ct{gglrsnv89a,gh89a,kallosh89a,lrsnv89a,%
rsnv89a,bk90a,siegel90a,bkp92a,sezgin93a}%
{\footnote{
See
\ct{sezgin93a}
for additional references.
}},
chiral gravity
\ct{jst93a},
$W_3$ gravity
\ct{hull91a,bss92a,bg93a,hull93a,dayi94a,vp94a},
general topological field theories
\ct{bms88a,lp88a,brt89a,bbrt91a,getzler92a,%
ikemori93a,lps93a,horava94a,jv94a},
the supersymmetric Wess-Zumino model
\ct{bbow90a,hlw90a}
and
chiral gauge theories in four-dimensions
\ct{tnp90a}.
The antibracket formalism has found
various interpretations
in mathematics
\ct{gerstenhaber62a,gerstenhaber64a,getzler92a,ps92a,%
lz93a,nersesian93a,schwarz93a,schwarz93b,stasheff93a}.
Some other recent relevant work
can be found in
\ct{verlinde92a,bt93a,bt93b,bcov93a,hata93a}.
The referencing in this review
is thorough but not complete.
A restriction has been made
to only cite works directly relevant
to the issues addressed in each section.
Multiple references are done first chronologically
and then alphabetically.
The titles of references are provided
to give the reader a better indication
of the content of each work.

We work in Minkowski space throughout this article.
Functional integrals are defined
by analytic continuation using Wick rotation.
This is illustrated in the computations
of Appendix C.
We use $\eta_{\mu \nu}$ to denote the flat-space metric
with the signature convention
$ ( -1, 1, 1, \dots , 1 )$.
Flat-space indices are raised and lowered with this metric.
The epsilon tensor
$\varepsilon_{\mu_0 \mu_1 \mu_2 \dots \mu_{d-1} }$
is determined by the requirement that it be
antisymmetric in all indices
and that $\varepsilon_{0 1 2 \dots {d-1} } = 1$,
where $d-1$ and $d$ are respectively
the dimension of space and space-time.
We often use square brackets to indicate a functional
of fields and antifields to avoid confusion
with the antibracket, i.e.,
$S [ \Phi , \Phi^* ]$ in lieu of $S ( \Phi , \Phi^* )$.

\vfill\eject

\section{Structure of the Set of Gauge Transformations}
\label{s:ssgt}

\hspace{\parindent}
The most familiar example of a gauge structure
is the one associated
with a non-abelian Yang-Mills theory
\ct{ym54a},
namely a Lie group.
The commutator of two Lie-algebra generators
produces a Lie-algebra generator.
When a basis is used,
this commutator algebra is determined
by the structure constants of the Lie group.
For example, for the Lie algebra $su(2)$
there are three generators
and the structure constant is the anti-symmetric tensor
on three indices $\varepsilon^{\alpha \beta \gamma}$.
A commutator algebra,
as determined by a set of abstract structure constants,
does not necessarily lead to a Lie algebra.
The Jacobi identity,
which expresses the associativity of the algebra,
must be satisfied
\ct{vannieuwenhuizen81a}.

Sometimes, in more complicated field theories,
the transformation rules involve
field-dependent structure constants.
Such cases are sometimes
referred to as ``soft algebras''
\ct{batalin81a,sohnius83a}.
In such a situation,
the determination of the gauge algebra is more
complicated than in the Yang-Mills case.
The Jacobi identity must be appropriately generalized
\ct{batalin81a,bv81a,dewitt84a}.
Furthermore, new structure tensors
beyond commutator structure constants may appear
and new identities need to be satisfied.

In other types of theories,
the generators of gauge transformations
are not independent.
This occurs when there is ``a gauge invariance''
for gauge transformations.
One says the system is {\it reducible}.
A simple example is a theory constructed using
a three-form $F$ which is expressed
in terms of a two-form $B$
by applying the exterior derivative $F = dB$.
The gauge invariances are
given by the transformation rule $\delta B = dA$
for any one-form $A$.
The theory is invariant under such transformations
because the lagrangian is a functional of $F$ and
$F$ is invariant: $\delta F = d \delta B = ddA = 0$.
However, the gauge invariances are not all independent
since modifying $A$ by $\delta A = d \lambda$ for some
zero-form $\lambda$ leads to no change
in the transformation for $B$.
When $A = d \lambda$, $\delta B = d A = d d \lambda = 0$.
The structure of a gauge theory is more complicated
than the Yang-Mills case
when there are gauge invariances for gauge transformations.

Another complication occurs
when the commutator of two gauge transformations
produces a term that vanishes on-shell,
i.e., when the equations of motion are used.
When equations of motion appear in the gauge algebra,
how should one proceed?

In this section we discuss the above-mentioned complications
for a generic gauge theory.
The questions are
(i) what are the relevant
gauge-structure tensors and
(ii) what equations
do they need to satisfy.
The answers to these questions
lead us to the gauge structure of a theory.

This section constitutes
a somewhat technical but necessary prelude.
A reader might want to consult the examples
in Sect.\ \ref{s:egt}.
The more interesting development
of the field-antifield formalism
begins in Sect.\ \ref{s:faf}.

\subsection{Gauge Transformations}
\label{ss:gt}

\hspace{\parindent}
This subsection introduces the notions
of a gauge theory and a gauge transformation.
It also defines notation.
The antibracket approach employs
an elaborate mathematical formalism.
Hence, one should try to become quickly
familiar with notation and conventions.

Consider a system whose dynamics is governed by
a classical action $S_0 [ \phi ]$,
which depends on $n$ different fields $\phi^i(x)$, $i=1,\cdots,n$.
The index $i$ can label space-time indices $\mu$, $\nu$
of tensor fields,
the spinor indices of fermion fields,
and/or an index distinguishing different types of generic fields.
At the classical level, the fields are functions of space-time.
In the quantum system, they are promoted to operators.
In this section, we treat the classical case only.

Let $\epsilon (\phi^i) = \epsilon_i$ denote
the statistical parity of $\phi^i$.
Each $\phi^i$ is either a commuting field ($\epsilon_i = 0$) or
an anticommuting field ($\epsilon_i = 1$).
One has
$\phi^i (x) \phi^j (y) =
 (-1)^{\epsilon_i \epsilon_j } \phi^j (y) \phi^i (x) $.

Let us assume that the action is invariant
under a set of $m_0$ ($m_0\leq n$)
non-trivial gauge transformations,
which, when written in infinitesimal form, read
\be
\delta\phi^i (x) =
\left( R^i_{\alpha} (\phi)\varepsilon^{\alpha} \right) (x) \quad ,
\quad {\rm where}
\ \alpha=1 \ {\rm or} \ 2  \ \ldots \ {\rm or} \ m_0 \quad .
\label{trans gauge}
\ee
Here, $\varepsilon^{\alpha} (x)$
are infinitesimal gauge parameters,
that is, arbitrary functions of the space-time variable $x$,
and $R^i_{\alpha}$ are
the generators of gauge transformations.
These generators are operators
that act on the gauge parameters.
In kernel form,
$\left( R^i_{\alpha} (\phi)\varepsilon^{\alpha} \right) (x)$
can be represented as
$\int {\dif y R_{\alpha }^i\left( {x,y} \right)
\varepsilon ^{\alpha }\left( y \right)}$.

It is convenient to adopt
the following compact notation
\ct{dewitt64a,dewitt67a}.
Unless otherwise stated, the appearance of a discrete index
also indicates the presence of a space-time variable.
We then use a generalized summation convention in which
a repeated discrete index implies not only a sum over that index
but also an integration over
the corresponding space-time variable.
As a simple example,
consider the multiplication of two matrices $g$ and $h$,
written with explicit matrix indices.
In compact notation,
\be
 {f^A}_B = {g^A}_C {h^C}_B
\label{gsc1}
\ee
becomes not only a matrix product in index space
but also in function space.
Eq.\bref{gsc1} represents
\be
 {f^A}_B \left( { x , y } \right) = \sum_C \int \dif z \,
 {g^A}_C \left( { x , z } \right)
 {h^C}_B \left( { z , y } \right)
\label{gsc2}
\ee
in conventional notation.
In other words, the index $A$
in Eq.\bref{gsc1}
stands for $A$ and $x$
in Eq.\bref{gsc2}.
Likewise, $B$ and $C$
in Eq.\bref{gsc1}
represent
$\{ B, y \}$ and $\{ C, z \}$.
The generalized summation convention for $C$
in compact notation
yields a sum over the discrete index $C$
and an integration over $z$
in conventional notation
in Eq.\bref{gsc2}.
The indices $A$, $B$ and $C$
in compact notation
implicitly represent space-time variables $x$, $y$, $z$, etc.,
and explicitly can be
field indices $i, j, k ,$ etc.,
gauge index $\alpha, \beta , \gamma ,$ etc.,
or any other discrete index
in the formalism.

With this convention, the transformation laws
\be
   \delta\phi^i (x) =
   \sum\limits_{\alpha }
   \int {\dif y R_{\alpha }^i\left( { x,y } \right)
   \varepsilon ^{\alpha }\left( y \right)}
\label{transformation rule long}
\ee
can be written succinctly as
\be
   \delta\phi^i=R^i_{\alpha} \varepsilon^{\alpha} \quad .
\label{transformation rule}
\ee
The index $\alpha$ in Eq.\bref{transformation rule} corresponds
to the indices $y$ and $\alpha$ in Eq.\bref{transformation rule long}.
The index $i$ in Eq.\bref{transformation rule} corresponds
to the indices $x$ and $i$ in Eq.\bref{transformation rule long}.
The compact notation is illustrated
in the example of
Sect.\ \ref{ss:srp}.
Although this notation might seem confusing at first,
it is used extensively in the antibracket formalism.
In the next few paragraphs,
we present equations in both notations.

Each gauge parameter
$\varepsilon^{\alpha}$ is either commuting,
$\epsilon (\varepsilon^{\alpha}) \equiv \epsilon_{\alpha}= 0$,
or is anti-commuting, $\epsilon_{\alpha}= 1$.
The former case corresponds to an ordinary symmetry
while the latter is a supersymmetry.
The statistical parity of $R^i_{\alpha}$,
$\epsilon (R^i_{\alpha})$, is determined from Eq.\bref{trans gauge}:
$\epsilon (R^i_{\alpha}) =
\left( \epsilon_i + \epsilon_{\alpha} \right)
\  \  ({\rm mod } \ 2)$.

Let $S_{0,i} \left( { \phi , x} \right) $
denote the variation of the action
with respect to $\phi^i (x)$:
\be
   S_{0,i}\left( { \phi , x} \right) \equiv
  \rder{S_0 [ \phi ] }{\phi^i (x) } \quad ,
\label{def s0i}
\ee
where the subscript $r$
indicates that the derivative is
to be taken from the right (see Appendix A).
Henceforth, when a subscript index $i$, $j$, etc.,
appears after a comma
it denotes the right derivative
with respect to the corresponding field $\phi^i$, $\phi^j$, etc..
In compact notation, we write Eq.\bref{def s0i}
as $S_{0,i} = \rder{S_0}{\phi^i} $ where the index $i$ here
stands for both $x$ and $i$ in Eq.\bref{def s0i}.

The statement that the action is invariant
under the gauge transformation in Eq.\bref{trans gauge} means that
the Noether identities
\be
  \int {\dif x} \sum\limits_{i=1}^n S_{0,i}
  \left( { x} \right)
  R^i_{\alpha}\left( { x,y } \right) = 0
\label{ident noether long form}
\ee
hold, or equivalently, in compact notation
\be
   S_{0,i} R^i_{\alpha} =0 \quad .
\label{ident noether}
\ee
Eq.\bref{ident noether} is derived by varying $S_0$
with respect to right variations
of the $\phi^i$ given by Eq.\bref{trans gauge}.
When using right derivatives,
the variation $\delta S_0$ of $S_0$,
or of any other object,
is given by
$
 \delta S_0 = S_{0,i} \delta \phi^i
$.
If one were to use left derivatives,
the variation of $S_0$ would read
$
 \delta S_0 =
 \delta \phi^i { {\partial_l S_{0} }  \over {\partial \phi^i} }
$.
Eq.\bref{ident noether long form}
is sometimes zero because the integrand
is a total derivative.
We assume that surface terms can be dropped in such integrals --
this is indeed the case when
Eq.\bref{ident noether long form} is applied
to gauge parameters that fall off sufficiently fast
at spatial and temporal infinity.
The Noether identities in Eq.\bref{ident noether}
are the key equations of this subsection
and can even be thought of
as the definition of when a theory is invariant
under a gauge transformation
of the form in Eq.\bref{trans gauge}.

To commence perturbation theory, one searches for solutions
to the classical equations of motion,
$S_{0,i}\left( { \phi , x} \right)=0$,
and then expands about these solutions.
We assume there exists at least one such
stationary point $\phi_0 = \{ \phi^j_0 \}$ so that
\be
   \restric{S_{0,i}}{\phi_0} = 0 \quad .
\label{saddle point}
\ee
Equation \bref{saddle point}
defines a surface $\Sigma$
in function space,
which is infinite dimensional
when gauge symmetries are present.

As a consequence of the Noether identities,
the equations of motion are not independent.
Furthermore, new saddle point solutions
can be obtained by performing
gauge transformations on any particular solution.
These new solutions should not be regarded as representing
new physics however --
fields related by local gauge transformations
are considered equivalent.

The Noether identities also imply that propagators do not exist.
By differentiating the identities from the left
with respect $\phi^j$,
one obtains
$$
   \frac{\partial_l}{\partial\phi^i}
    \left (S_{0,j} R^j_{\alpha}\right ) =
   \left(\frac{\partial_l \partial_r S_0}
   {\partial\phi^i \partial\phi^j}\right) R^j_{\alpha} +
   S_{0,j} \frac{\partial_l
     R^j_{\alpha}}{\partial\phi^i}
  (-1)^{\epsilon_i \epsilon_j } = 0
\ ,
$$
\be
   \Rightarrow\restric{
   \left(\frac{\partial_l \partial_r S_0}
   {\partial\phi^i \partial\phi^j}\right) R^j_{\alpha}}
   {\phi_0} = 0
\quad ,
\label{hessian degeneracy}
\ee
i.e., the hessian
$\left(\frac{\partial_l \partial_r S_0}
   {\partial\phi^i \partial\phi^j}\right)
$
of $S_0$ is degenerate
at any point on the stationary surface $\Sigma$.
The $R^i_\alpha$ are on-shell null vectors of this hessian.
Since propagators involve the inverse of this hessian,
propagators do not exist for certain combinations of fields.
This means that the standard loop expansion cannot be
straightforwardly applied.
A method is required to overcome this problem.

Technically speaking,
to study the structure of the set of gauge transformations
it is necessary to assume certain {\it regularity conditions}
on the space for which the equations of motion
$S_{0,i} = 0 $ hold.
The interested reader can find these conditions in Appendix B.
A key consequence of the regularity conditions is that
if a function $F(\phi)$ of the fields $\phi$
vanishes on-shell,
that is, when the equations of motion are implemented,
then $F$ must be a linear combination of the equations of motion,
i.e.,
\be
\restric{F( \phi )}{ \Sigma } = 0 \Rightarrow
F(\phi ) = S_{0,i} \lambda^i (\phi )
\quad ,
\label{consequence of rc}
\ee
where
$\restric{}{\Sigma}$
indicates the restriction to the surface
where the equations of motion hold
\ct{wh79a,bv85a,fhst89a,fisch90a,fh90a}.
Eq.\bref{consequence of rc} can be thought of
as a completeness relation
for the equations of motion.
We shall make use of Eq.\bref{consequence of rc} frequently.

Throughout Sect.\ \ref{s:ssgt},
we assume that the gauge generators are
fixed once and for all.
One could take linear combinations of the generators
to form a new set.
This would change the gauge-structure tensors
presented below.
This non-uniqueness is not essential
and is discussed
in Sect.\ \ref{ss:eu}.

To see explicit examples
of the abstract formalism that follows,
one may want to glance
from time to time
at the examples
of Sect.\ \ref{s:egt}.

\subsection{Irreducible and Reducible Gauge Theories}
\label{ss:irgt}

\hspace{\parindent}
It is important to know any dependences
among the gauge generators.
Only with this knowledge is possible
to determine the independent degrees of freedom.
The purpose of this subsection
is to analyze this issue in more detail
for the generic case.

The simplest gauge theories,
for which all gauge transformations
are independent, are called {\it irreducible}.
When dependences exist,
the theory is {\it reducible}.
In reducible gauge theories,
there is a ``kind of gauge invariance for gauge transformations''
or what one might call ``level-one'' gauge invariances.
If the level-one gauge transformations
are independent,
then the theory is called {\it first-stage reducible}.
This may not happen.
Then, there are ``level-two'' gauge invariances, i.e.,
gauge invariances for the level-one gauge invariances
and so on.
This leads to the concept
of an {\it $L$-th stage reducible theory}.
In what follows we let $m_s$ denote
the number of gauge generators at the $s$-th stage
regardless of whether they are independent.

Let us define more precisely the above concepts.
Assume that all gauge invariances of a theory are known
and that the regularity condition described in Appendix B
is satisfied.
Then, the most general solution to the
Noether identities \bref{ident noether}
is a gauge transformation, up to terms proportional
to the equations of motion:
\be
  S_{0,i} \lambda^i = 0 \Leftrightarrow
  \lambda^i = R^i_{0\alpha_0} \lambda^{\prime \alpha_0} +
  S_{0,j} T^{ji}
\quad ,
\label{completesa}
\ee
where $T^{ij}$ must satisfy the graded symmetry property
\be
  T^{ij} = -(-1)^{\eps_i \eps_j } T^{ji}
\quad .
\label{symmetry of Tij}
\ee
The $R^i_{0\alpha_0}$ are the gauge generators
in Eq.\bref{trans gauge}.
For notational convenience,
we have appended a subscript $0$ on the gauge generator
and the gauge index $\alpha$.
This subscript indicates
the level of the gauge transformation.
The second term $S_{0,j}T^{ji}$ in Eq.\bref{completesa}
is known as a trivial gauge transformation.
Such transformations are discussed in the next subsection.
It is easily checked that the action is invariant under
such transformations due to the trivial commuting or
anticommuting properties of the $S_{0,j}$.
The first term $R^i_{0\alpha_0} \lambda^{\prime \alpha_0}$
in Eq.\bref{completesa} is similar
to a non-trivial gauge transformation
of the form of Eq.\bref{trans gauge} with
$\varepsilon^{\alpha_0} = \lambda^{\prime \alpha_0}$.
The key assumption in Eq.\bref{completesa}
is that the set of functionals $R^i_{0\alpha_0}$
exhausts on-shell the relations among the equations of motion,
namely the Noether identities.
In other words, the gauge generators
are on-shell a complete set.
This is essentially equivalent
to the regularity condition.

If the functionals $R^i_{0\alpha_0}$
are independent on-shell
then the theory is {\it irreducible}.
In such a case,
\be
\restric{{\rm rank}\; R^i_{0\alpha_0}}{\Sigma} = m_0
\quad ,
\label{indp generadors}
\ee
where $m_0$ is the number of gauge transformations.
The rank of the hessian
\be
   \mbox{\rm rank}\restric{
   \left(\frac{\partial_l \partial_r S_0}
   {\partial\phi^i \partial\phi^j}\right)} {\Sigma}=
   n-\mbox{\rm rank}\restric{R^i_\alpha}{\Sigma}
\label{rank hessian}
\ee
is
$ n - m_0 $.
Define the
net number of degrees of freedom
$n_{\rm dof}$ to be the number of fields
that enter dynamically in $S_0$,
regardless of whether they propagate.%
{\footnote{
In electromagnetism,
$n_{\rm dof} =3$,
but there are only two propagating
degrees of freedom
corresponding to the two physical polarizations.}}
Then for an irreducible theory
$n_{\rm dof}$ is $n - m_0$
since there are $m_0$ gauge degrees of freedom.
Note that $n_{\rm dof}$ matches the rank of the hessian
in Eq.\bref{rank hessian}.

If, however, there are dependences among the gauge
generators, and the rank of the generators
is less than their number,
$\restric{{\rm rank}\; R^i_{0\alpha_0}}{\Sigma}<m_0$,
then the theory is {\it reducible}.
If $m_0-m_1$ of the generators are independent on-shell,
then there are $m_1$ relations among them
and there exist
$m_1$ functionals $R^{\alpha_0}_{1\alpha_1}$ such that
\bea
    R^i_{0\alpha_0} R^{\alpha_0}_{1\alpha_1}&=&
    S_{0,j} V^{ji}_{1\alpha_1} \ ,
    \quad\quad\quad \alpha_1=1,\ldots,m_1 \quad ,
\nonumber\\
    \epsilon(R^{\alpha_0}_{1\alpha_1})&=&
    \epsilon_{\alpha_0} +
     \epsilon_{\alpha_1}
  \ \ ({\rm mod \ 2}) \quad ,
\label{depen generadors}
\eea
for some $V^{ji}_{1\alpha_1}$,
satisfying
$V^{ij}_{1\alpha_1} = -(-1)^{\eps_i \eps_j } V^{ji}_{1\alpha_1}$.
Here, $\epsilon_{\alpha_1}$
is the statistical parity
of the level-one gauge parameter.
The $R^{\alpha_0}_{1\alpha_1}$
are the on-shell null vectors for
$R^i_{0\alpha_0}$ since
$\restric{R^i_{0\alpha_0} R^{\alpha_0}_{1\alpha_1}}{\Sigma}=0$.
The presence of $V^{ji}_{1\alpha_1}$ in Eq.\bref{depen generadors}
is a way of extending this statement off-shell.
Here and elsewhere,
when a combination of field equations
appears on the right-hand side of an equation,
it indicates the off-shell extension of an on-shell statement;
such an extension can be performed
using the regularity postulate of Appendix B.
Note that,
if
$
  \varepsilon^\alpha = R^{\alpha}_{1\alpha_1} \varepsilon^\alpha_1
$
for any $\varepsilon^\alpha_1$,
then $\delta \phi^i$
in Eq.\bref{transformation rule}
is zero on-shell,
so that no gauge transformation is produced.
In Eq.\bref{depen generadors}
it is assumed that the reducibility of the $R^i_{0\alpha_0}$
is completely contained in $R^{\alpha_0}_{1\alpha_1}$, i.e.,
$R^{\alpha_0}_{1\alpha_1}$ also constitute a complete set
\be
   R^i_{0\alpha_0}\lambda^{\alpha_0}=
   S_{0,j} M_0^{ji} \Rightarrow
   \lambda^{\alpha_0}=
   R^{\alpha_0}_{1\alpha_1}\lambda^{\prime \alpha_1}
   +S_{0,j} T_0^{j\alpha_0}
\quad ,
\label{consequence of regularity}
\ee
for some $\lambda^{\prime \alpha_1}$ and some $T_0^{j\alpha_0}$.

If the functionals $R^{\alpha_0}_{1\alpha_1}$
are independent on-shell
$$
   \restric{{\rm rank}\; R^{\alpha_0}_{1\alpha_1}}{\Sigma}=m_1 \quad ,
$$
then the theory is called {\it first-stage reducible}.
One also has $\restric{{\rm rank}\; R^i_{0\alpha_0}}{\Sigma}=m_0 - m_1$
and the net number of degrees of freedom
in the theory is $n-m_0+m_1$.
Since true and gauge degrees of freedom have been determined,
$$
\restric{{\rm rank}\;
 {{\partial_l \partial_r S_0 }
  \over{\partial \phi^i \partial \phi^j }} }
{\Sigma} =
n - m_0 + m_1 \quad .
$$

If the functionals $R^{\alpha_0}_{1\alpha_1}$
are not all independent on-shell,
relations exist among them
and the theory is second-or-higher-stage reducible.
Then, the on-shell null vectors of $R^{\alpha_0}_{1\alpha_1}$
and higher $R$-type tensors
must be found.

One continues the above construction until it terminates.
A theory is
$L$-{\it th stage reducible}
\ct{bv83b}
if there exist functionals
\be
   R^{\alpha_{s-1}}_{s\alpha_s},\quad\quad
   \alpha_s=1,\ldots,m_s \ , \quad \quad s=0,\ldots,L \quad  ,
\label{generadors r}
\ee
such that $R^i_{0\alpha_0}$ satisfies Eq.\bref{ident noether},
i.e., $S_{0 , i} R^i_{0\alpha_0} = 0$,
and such that, at each stage, the $R^{\alpha_{s-1}}_{s\alpha_s}$
constitute a complete set, i.e.,
$$
   R^{\alpha_{s-1}}_{s\alpha_s} \lambda^{\alpha_s} =
    S_{0,j} M_s^{j\alpha_{s-1}}\Rightarrow
   \lambda^{\alpha_s} = R^{\alpha_s}_{s+1,\alpha_{s+1}}
    \lambda^{\prime \alpha_{s+1}} + S_{0,j} T_s^{j\alpha_s} \quad ,
$$
$$
    R^{\alpha_{s-2}}_{s-1,\alpha_{s-1}}R^{\alpha_{s-1}}_{s\alpha_s}
    =S_{0,i} V^{i\alpha_{s-2}}_{s\alpha_s} \ ,
  \quad\quad\quad s=1,\ldots,L \quad ,
$$
\be
\restric{{ \rm rank}\; R^{\alpha_{s-1}}_{s\alpha_s}}{\Sigma}=
\sum_{t=s}^L (-1)^{t-s} m_t \ ,
\quad\quad s=0,\ldots,L \quad ,
\label{reduc l 1}
\ee
where we have defined
$
R^{\alpha_{-1}}_{0\alpha_0}\equiv R^i_{0\alpha_0}$
and
$\alpha_{-1}\equiv i$.
The $R^{\alpha_{s-1}}_{s\alpha_s}$
are the on-shell null vectors for
$R^{\alpha_{s-2}}_{s-1 \alpha_{s-1}}$.
The statistical parity of $R^{\alpha_{s-1}}_{s\alpha_s}$ is
$\left(
 \epsilon_{\alpha_{s-1}} + \epsilon_{\alpha_{s}}
\right) $ (mod 2),
where $\epsilon_{\alpha_s}$ is the statistical parity
of the $s$-level gauge transformation
associated with the index $\alpha_s$.
Finally,
\be
 n_{\rm dof} =
 \restric{{\rm rank}\;
 {{\partial_l \partial_r S_0 } \over
  {\partial \phi^i \partial \phi^j }} }
{\Sigma} =
n - \sum_{s=0}^L(-1)^s  m_s
\label{dof}
\ee
is the net number of degrees of freedom.

\subsection{Trivial Gauge Transformations}
\label{ss:tgt}

\hspace{\parindent}
As mentioned in the last subsection,
trivial gauge transformations exist.
Since they are proportional to the equations of motion
they do not lead to conservation laws.
This subsection discusses their role
in the gauge algebra.

Given that the finite invertible gauge transformations
satisfy the group axioms,
their infinitesimal counterparts
necessarily form an algebra.
Besides the usual gauge
transformations \bref{trans gauge},
there are the trivial transformations, defined as
\be
   \delta_\mu\phi^i = S_{0,j}\mu^{ji}
\ ,\quad \quad
   \mu^{ji} = - (-1)^{\epsilon_i \epsilon_j}\mu^{ij} \quad ,
\label{trivials}
\ee
where $\mu^{ji}$ are arbitrary functions.
It is easily demonstrated that,
as a consequence
of the symmetry properties of $\mu^{ji}$,
the transformations in Eq.\bref{trivials}
leave the action invariant.
In studying the structure of the gauge transformations,
it is necessary to take into consideration
the presence of such transformations.

To determine their effect on the gauge algebra,
consider the commutator of a trivial transformation
with any other transformation.
Calling the latter
$\delta_r \phi^i = r^i$, one has
$$
   [\delta_\mu,\delta_r]\phi^i= r^i_{,k} S_{0,j} \mu^{jk}-
   S_{0,j}\mu^{ji}_{\;,k} r^k - S_{0,jk} r^k \mu^{ji}
\quad .
$$
Given that $\delta_r$ is a symmetry transformation of $S_0$,
it follows by differentiation by $\phi^j$ that
$$
  S_{0,k} r^k=0\Rightarrow
  S_{0,jk} r^k + S_{0,k} r^k_{,j} = 0
\quad ,
$$
so that the commutator becomes
$$
   [\delta_\mu, \delta_r] \phi^i =
   S_{0,j} \left(
    r^j_{,k}\mu^{ki}
    -(-1)^{\epsilon_i\epsilon_j} r^i_{,k}\mu^{kj}
     -\mu^{ji}_{\;,k} r^k
    \right) =
   S_{0,j} \tilde\mu^{ji}
\quad ,
$$
from which one concludes
that the commutator of a trivial transformation
with any other transformation is a trivial transformation.
Hence, the trivial transformations are a normal subgroup $H$
of the full group of gauge transformations, $\bar G$.

The trivial gauge transformations are of no
physical significance:
They neither lead to conserved currents
nor do they prevent the development
of a perturbative expansion about a stationary point.
They are simply a consequence of
having more than one degree of freedom.
On these grounds, it would seem sensible to
dispense with them and
restrict oneself to the quotient
$G=\bar G/H$.
However,
this is only possible in certain cases.
In general,
the commutator of two non-trivial gauge transformations
produces trivial gauge transformations.
Furthermore, for reasons of convenience,
particularly when it is desirable to have
manifest covariance or preserve locality,
one sometimes wants
to include trivial transformations.
Hence, the full group $\bar G$ is used
for studying the gauge structure of the theory.

\subsection{The Gauge Structure}
\label{ss:gs}

\hspace{\parindent}
In this section we  restrict ourselves to the simpler cases
of irreducible and first-stage-reducible gauge theories.
To avoid cumbersome notation,
we use $R_{\alpha}^i$ for $R_{0 \alpha_0}^i$,
$Z_a^{\alpha}$ for $R_{1 \alpha_1}^{\alpha_0}$,
and in Eq.\bref{depen generadors} we use
$V^{ji}_a$ for $V^{ji}_{1 \alpha_1 }$,
so that the indices $\alpha_0$ and $\alpha_1$
respectively correspond to $\alpha$ and $a$.

The general strategy in obtaining the gauge structure
is as follows \ct{wh79a}.
The first gauge-structure tensors are
the gauge generators themselves,
and the first gauge-structure equations
are the Noether identities
\bref{ident noether}.
One computes commutators, commutators of commutators, etc.,
of gauge transformations.
Graded symmetrization produces identity equations
for the structure tensors
that must be satisfied.
Generic solutions are obtained
by exploiting the consequences of the regularity conditions,
namely, completeness.
In using completeness,
additional gauge-structure tensors appear.
They enter in higher-order
symmetrized commutator identity equations.
The process is continued until it terminates.

Although this section provides some insight,
it is somewhat technical
so that the reader may wish to skip it
at first.
If one is only interested in the irreducible case,
one should read to Eq.\bref{jac2}.
For reasons of space,
many details of the algebra are omitted.
As an exercise,
the reader can provide the missing steps.

Consider the commutator of two gauge
transformations of the type in Eq.\bref{trans gauge}.
On one hand, a direct computation leads to
$$
   [\delta_1,\delta_2]\phi^i=
 \left( R^i_{\alpha,j} R^j_\beta -
  (-1)^{\epsilon_\alpha \epsilon_\beta}
 R^i_{\beta,j} R^j_\alpha\right)
   \varepsilon_1^\beta\varepsilon_2^\alpha
\quad .
$$
On the other hand,
since this commutator is also a gauge symmetry of the action
it satisfies the Noether identity so that,
factoring out the gauge parameters
$\varepsilon_1^\beta$ and $\varepsilon_2^\alpha$,
one may write
$$
 S_{0,i}\left(R^i_{\alpha,j}R^j_\beta
 - (-1)^{\epsilon_\alpha \epsilon_\beta}
  R^i_{\beta,j}R^j_\alpha\right) = 0
\quad .
$$
Taking into account Eq.\bref{completesa}
the above equation implies
the following important relation among the generators
\be
 R^i_{\alpha,j}R^j_\beta -
  (-1)^{\epsilon_\alpha \epsilon_\beta}
 R^i_{\beta,j}R^j_\alpha
   = R^i_\gamma T^\gamma_{\alpha\beta} -
   S_{0,j} E^{ji}_{\alpha\beta}
\quad ,
\label{algebra oberta}
\ee
for some gauge-structure tensors
$T^\gamma_{\alpha\beta}$ and
$E^{ji}_{\alpha\beta}$.
This equation defines
$T^\gamma_{\alpha\beta}$ and $E^{ji}_{\alpha\beta}$.
Restoring the dependence on the gauge parameters
$\varepsilon_1^\beta$ and $\varepsilon_2^\alpha$,
the last two equations imply
\be
   [\delta_1,\delta_2]\phi^i \equiv
    R^i_\gamma T^\gamma_{\alpha\beta}
   \varepsilon_1^\beta\varepsilon_2^\alpha -
   S_{0,j} E^{ji}_{\alpha\beta}
   \varepsilon_1^\beta\varepsilon_2^\alpha
\quad ,
\label{com of two gens}
\ee
where $T^\gamma_{\alpha\beta}$ are known
as the ``structure constants''
of the gauge algebra.
The words {\it structure constants} are in quotes
because in general the $T^\gamma_{\alpha\beta}$
depend on the fields of the theory and are not ``constant''.
The possible presence
of the $E^{ji}_{\alpha\beta}$ term is due
to the fact that the commutator of two gauge
transformations may give rise to trivial gauge transformations
\cite{wh79a,bv81a,bv84a}.

The gauge algebra
generated by the $R^i_\alpha$ is said to be {\it open}
if $E^{ij}_{\alpha\beta}\neq 0$,
whereas the algebra is said to be {\it closed}
if $E^{ij}_{\alpha\beta}=0$.
Moreover, Eq.\bref{algebra oberta} defines a {\it Lie algebra}
if the algebra is closed, $E^{ij}_{\alpha\beta}=0$, and
the $T^\gamma_{\alpha\beta}$ do not depend on the fields $\phi^i$.

The gauge-structure tensors have the following
symmetry properties under the interchange of indices
$$
E^{ij}_{\alpha\beta} =
   -(-1)^{ \epsilon_i \epsilon_j }
   E^{ji}_{\alpha\beta} =
   - (-1)^{\epsilon_\alpha \epsilon_\beta}
   E^{ij}_{\beta\alpha}
\quad ,
\nonumber
$$
\be
T^\gamma_{\alpha\beta}= - (-1)^{\epsilon_\alpha \epsilon_\beta}
T^\gamma_{\beta\alpha}
\quad .
\ee
In other words, $E^{ij}_{\alpha\beta}$
is graded-antisymmetric both in lower indices and in upper indices and
$T^\gamma_{\alpha\beta}$ is graded-antisymmetric in lower indices.
The statistical parity of structure tensors
is determined by the sum of the parities of the tensor indices,
so that
$\epsilon \left( R_\alpha^i \right) =
 \left( \epsilon_\alpha + \epsilon_i \right) $
 (mod 2),
$\epsilon { \left( T^\gamma_{\alpha\beta} \right) } =
  \left(
  \epsilon_\alpha + \epsilon_\beta + \epsilon_\gamma
  \right)
$
(mod 2),
and
$
 \epsilon { \left( E^{ij}_{\alpha\beta} \right) } =
 \left(
  \epsilon_i + \epsilon_j + \epsilon_\alpha + \epsilon_\beta
 \right)
$
(mod 2).

The next step determines
the restrictions imposed by the Jacobi identity.
In general, it leads to
new gauge-structure tensors and equations
\ct{kallosh78a,vannieuwenhuizen81a,dewitt84a,bv85a}.
The identity
$$
   \sum_{\rm cyclic \  over \ 1,\ 2,\ 3}
[\delta_1,[\delta_2,\delta_3]]=0 \quad ,
$$
produces the following relations among
the tensors $R$, $T$ and $E$
\be
   \sum_{\rm cyclic \ over \ 1,\ 2,\ 3 }
   \left(R^i_\delta A^\delta_{\alpha\beta\gamma}-
   S_{0,j} B^{ji}_{\alpha\beta\gamma}\right)
   \veps^\gamma_1\veps^\beta_2\veps^\alpha_3=0 \quad ,
\label{cons jacobi}
\ee
where we have defined
$$
   3 A^\delta_{\alpha\beta\gamma} \equiv
   \left( T^\delta_{\alpha\beta,k}R^k_\gamma -
   T^\delta_{\alpha\eta} T^\eta_{\beta\gamma} \right) +
\nonumber
$$
\be
  (-1)^{\epsilon_\alpha ( \epsilon_\beta + \epsilon_\gamma )}
   \left( T^\delta_{\beta\gamma,k} R^k_\alpha -
   T^\delta_{\beta\eta} T^\eta_{\gamma\alpha}\right) +
   (-1)^{\epsilon_\gamma (\epsilon_\alpha + \epsilon_\beta )}
   \left( T^\delta_{\gamma\alpha,k}R^k_\beta -
   T^\delta_{\gamma\eta} T^\eta_{\alpha\beta} \right)
\quad ,
\label {Adef}
\ee
and
$$
   3B^{ji}_{\alpha\beta\gamma} \equiv \left(
   E^{ji}_{\alpha\beta,k} R^k_\gamma -
   E^{ji}_{\alpha\delta} T^\delta_{\beta\gamma} -
   (-1)^{\epsilon_i \epsilon_\alpha }
   R^j_{\alpha,k} E^{ki}_{\beta\gamma} +
   (-1)^{\epsilon_j (\epsilon_i + \epsilon_\alpha )}
   R^i_{\alpha,k} E^{kj}_{\beta\gamma}
\right)
\nonumber
$$
\be
   + (-1)^{\epsilon_\alpha (\epsilon_\beta + \epsilon_\gamma )}
\left( {\rm RHS \ of \ above \ line \ with \
{{\scriptstyle \alpha \rightarrow \beta} \atop
{ {\scriptstyle \beta \rightarrow \gamma \atop \gamma \rightarrow \alpha }}
}
} \right)
\label{Bdef}
\ee
$$
   + (-1)^{\epsilon_\gamma (\epsilon_\alpha + \epsilon_\beta )}
\left( {\rm \rm RHS \ of \ first \ line \ with \
{{\scriptstyle \alpha \rightarrow \gamma} \atop
{ {\scriptstyle \beta \rightarrow \alpha \atop \gamma \rightarrow \beta }}
}
} \right)
\quad .
\nonumber
$$
A useful but lengthy exercise is to derive
Eq.\bref{cons jacobi}.

If the theory is irreducible,
the on-shell independence of the generators \bref{indp generadors}
and their completeness \bref{completesa}
leads to the following solution of Eq.\bref{cons jacobi}
\be
   A^\delta_{\alpha\beta\gamma} =
   S_{0,j}D^{j\delta}_{\alpha\beta\gamma}
\quad ,
\label{jacobi1}
\ee
where $D^{j\delta}_{\alpha\beta\gamma}$ are new structure functions.

On the other hand, using this solution in the original equation
\bref{cons jacobi}, one obtains the following condition
on the $D^{j\delta}_{\alpha\beta\gamma}$
\be
   \sum_{\rm cyclic \ over \ \epsilon_1,\ \epsilon_2,\ \epsilon_3 }
   S_{0,j}\left(B^{ji}_{\alpha\beta\gamma} -
   (-1)^{\epsilon_j ( \epsilon_i + \epsilon_\delta ) }
   R^i_\delta D^{j\delta}_{\alpha\beta\gamma}\right)
   \epsilon_1^\gamma \epsilon_2^\beta \epsilon_3^\alpha = 0
\quad .
\label{cond sup}
\ee
Again, the completeness of the generators implies that
the general solution of the preceding equation is of the form
\be
   B^{ji}_{\alpha\beta\gamma} +
   (-1)^{\epsilon_i \epsilon_\delta }
   R^j_\delta D^{i\delta}_{\alpha\beta\gamma} -
   (-1)^{\epsilon_j (\epsilon_i + \epsilon_\delta ) }
   R^i_\delta D^{j\delta}_{\alpha\beta\gamma} =
   - S_{0,k} M^{kji}_{\alpha\beta\gamma}
\quad ,
\label {jacobi2}
\ee
where $M^{kji}_{\alpha\beta\gamma}$
is graded antisymmetric in $i$, $j$, and $k$.
In this way, the Jacobi identity leads to
the existence of two new gauge-structure tensors
$D^{j\delta}_{\alpha\beta\gamma}$ and
$M^{kji}_{\alpha\beta\gamma}$
which, for a generic theory, are
different from zero and must satisfy
Eqs.\bref{jacobi1} and \bref{jacobi2}.

Continuing in the same way,
that is to say, commuting more and more gauge transformations,
new structure tensors with increasing numbers of indices are obtained.
These tensors are the so-called structure functions of the gauge algebra
and they determine the nature of the set of gauge transformations
of the theory.
The reader may not be aware of the higher-order tensors
because in the simplest gauge theories,
such as Yang-Mills,
they vanish.

The tensors
$A^\delta_{\alpha\beta\gamma}$,
$B^{ji}_{\alpha\beta\gamma}$,
$D^{j\delta}_{\alpha\beta\gamma}$ and
$M^{kji}_{\alpha\beta\gamma}$
are all graded-antisymmetric in
$\alpha$, $\beta$ and $\gamma$.
In addition
$B^{ji}_{\alpha\beta\gamma}$
is graded-antisymmetric in
$i$ and $j$
while
$M^{kji}_{\alpha\beta\gamma}$
is graded antisymmetric in
$i$, $j$ and $k$.
Here we summarize these properties
$$
   A^\delta_{\alpha\beta\gamma} =
   -(-1)^{\epsilon_\alpha \epsilon_\beta}
   A^\delta_{\beta\alpha\gamma} =
   -(-1)^{\epsilon_\beta \epsilon_\gamma}
   A^\delta_{\alpha\gamma\beta}
\quad ,
$$
$$
   B^{ji}_{\alpha\beta\gamma} =
   -(-1)^{\epsilon_\alpha \epsilon_\beta}
   B^{ji}_{\beta\alpha\gamma} =
   -(-1)^{\epsilon_\beta \epsilon_\gamma}
   B^{ji}_{\alpha\gamma\beta} =
    -(-1)^{\epsilon_i \epsilon_j }
   B^{ij}_{\alpha\beta\gamma}
\quad ,
$$
\be
   D^{j\delta}_{\alpha\beta\gamma} =
   -(-1)^{\epsilon_\alpha \epsilon_\beta}
   D^{j\delta}_{\beta\alpha\gamma} =
   -(-1)^{\epsilon_\beta \epsilon_\gamma}
   D^{j\delta}_{\alpha\gamma\beta}
\quad ,
\ee
$$
   M^{kji}_{\alpha\beta\gamma} =
   -(-1)^{\epsilon_\alpha \epsilon_\beta}
   M^{kji}_{\beta\alpha\gamma} =
   -(-1)^{\epsilon_\beta \epsilon_\gamma}
   M^{kji}_{\alpha\gamma\beta} =
    -(-1)^{\epsilon_j \epsilon_k }
   M^{jki}_{\alpha\beta\gamma}   =
    -(-1)^{\epsilon_i \epsilon_j }
   M^{kij}_{\alpha\beta\gamma}
\  .
$$
The statistical parity of any of the above tensors is given
by the sum of the statistical parities of the indices of that tensor.

A useful device in the study of gauge-structure relations
is to introduce ghost fields $\gh C^\alpha$
with opposite statistics
to those of the gauge parameters $\veps^\alpha$,
\be
   \eps(\gh C^\alpha)=\eps_\alpha+1
\quad ,
\ee
and to replace gauge parameters by ghosts,
as is done in the BRST formalism
\ct{baulieu85a}.
The ghost fields obey
the same boundary conditions as gauge parameters.
The ghosts can be used as a compact way of writing
the gauge-structure equations.
However, in order to do this,
the symmetry properties of
$T^\gamma_{\alpha \beta}$,
$E^{ij}_{\alpha \beta}$,
$D^{j\delta}_{\alpha\beta\gamma}$,
$M^{kji}_{\alpha\beta\gamma}$, etc.,
need to be correctly incorporated.
Note that these tensors are graded anti-symmetric
in lower-index gauge indices
$\alpha$, $\beta$, etc.,
whereas the ghosts satisfy
$
   \gh C^\alpha \gh C^\beta =
   (-1)^{(\eps_\alpha+1)(\eps_\beta+1)}
   \gh C^\beta \gh C^\alpha
$.
If one is given a graded anti-symmetric tensor
$T_{\alpha_1 \alpha_2 \alpha_3 \alpha_4 \dots} \ $,
then a way to make it into a graded symmetric tensor
with symmetry factors
$\eps_{\alpha_1} +1$,
$\eps_{\alpha_2} +1$,
etc., associated with indices $\alpha_1$, $\alpha_2$,
etc., is to multiply by a factor of
$(-1)^{\eps_{\alpha_i}}$
for every other index $\alpha_i$ in
$T_{\alpha_1 \alpha_2 \alpha_3 \alpha_4 \dots} \ $.
In other words, one replaces
$T_{\alpha_1 \alpha_2 \alpha_3 \alpha_4 \dots}$
by
$(-1)^{\eps_{\alpha_2} + \eps_{\alpha_4} + \dots}
T_{\alpha_1 \alpha_2 \alpha_3 \alpha_4 \dots} \ $.
Using this device,
one arrives at a compact way of writing
the Noether identity \bref{ident noether},
the gauge commutator relation \bref{algebra oberta},
as well as Eqs.\bref{jacobi1} and \bref{jacobi2} which arise
from the Jacobi identity:
\be
   S_{0,i} R^i_{\alpha} \gh C^\alpha =0
\quad ,
\label {noether}
\ee
\be
  \left( {
   2 R^i_{\alpha,j}R^j_\beta
   - R^i_\gamma T^\gamma_{\alpha\beta} + S_{0,j} E^{ji}_{\alpha\beta}
  } \right) (-1)^{\epsilon_\alpha} \gh C^\beta \gh C^\alpha = 0
\label {commutator algebra}
\quad ,
\ee
\be
   \left(A^\delta_{\alpha\beta\gamma} -
    S_{0,j}D^{j\delta}_{\alpha\beta\gamma}\right)
   (-1)^{\epsilon_\beta}
   \gh C^\gamma \gh C^\beta \gh C^\alpha = 0
\quad ,
\label {jac1}
\ee
\be
   { \left(
   B^{ji}_{\alpha\beta\gamma} +
   (-1)^{\epsilon_i \epsilon_\delta }
   R^j_\delta D^{i\delta}_{\alpha\beta\gamma} -
   (-1)^{\epsilon_j (\epsilon_i + \epsilon_\delta ) }
   R^i_\delta D^{j\delta}_{\alpha\beta\gamma} +
   S_{0,k} M^{kji}_{\alpha\beta\gamma}
   \right) }
   (-1)^{\epsilon_\beta}
   \gh C^\gamma \gh C^\beta \gh C^\alpha = 0
\quad ,
\label {jac2}
\ee
where $A^\delta_{\alpha\beta\gamma}$ and
$B^{ji}_{\alpha\beta\gamma}$
are defined in Eqs.\bref{Adef} and \bref{Bdef}.
The graded-anticommuting nature of the ghosts automatically
produces the appropriate graded-cyclic sums.

Equations \bref{noether} through \bref{jac2} are
key equations for an irreducible algebra.

Now let us consider a first-stage reducible gauge theory.
In this case,
the existence of non-trivial relations among the generators
in Eq.\bref{depen generadors} leads to
the appearance of new tensor quantities.

For first-stage reducible theories
there are on-shell null vectors
for the generators $R^i_\alpha$.
Let $Z^\alpha_a$ denote these  null vectors.
In Eq.\bref{depen generadors},
the $Z^\alpha_a$ are called $R_{1 a}^{\alpha}$
when $\alpha=\alpha_0$ and $a=\alpha_1$.
The null vectors are independent on-shell.
Their presence modifies the solutions of the
Jacobi identities in Eqs.\bref{jac1} and \bref{jac2} as well as
higher-commutator structure equations.
In addition there are new structure equations.
One of these is Eq.\bref{depen generadors} itself:
\be
 R_\alpha^i Z_b^\alpha = S_{0,j} V_b^{ji}
\quad .
\label{off shell null vector}
\ee
Another is derived as follows.
Take relation \bref{algebra oberta}
and multiply it by $Z^\beta_a$ to obtain
$$
\left(
   {R_{\alpha ,j}^i R_\beta^j-
   \left( {-1} \right)^{\epsilon_\alpha \epsilon _\beta }
   R_{\beta ,j}^i R_\alpha^j -
   R_\gamma^i T_{\alpha \beta }^\gamma +
   S_{0,j}E_{\alpha \beta }^{ji}}
\right) Z_a^\beta =0
\quad .
$$
Use Eq.\bref{off shell null vector}
to express $R_\beta^j  Z_a^\beta$
as a term proportional to equations of motion.
Also do the same with
$R_{\beta ,j}^i R_\alpha^j Z_a^\beta$
and make use of the Noether identity
in Eq.\bref{ident noether}.
After a little algebra, one finds that the previous equation
can be written in the form
$$
   R_\gamma^i
\left(
   {\left( {-1} \right)^{\epsilon_a \epsilon_\beta }
   Z_{a,j}^\gamma R_\beta ^j -
   T_{\beta \delta }^\gamma Z_a^\delta }
\right) =
    S_{0,j} M_{\beta a}^{ji}
\quad ,
$$
for some quantity $M^{ji}_{\beta a}$.
Terms proportional to the
equations of motion
have been collected into $M^{ji}_{\beta a}$.
Using the completeness of the null vectors
$Z^\alpha_a$,
the general solution to this equation is
\be
  \left( {-1} \right)^{\epsilon _a\epsilon _\beta }
   Z_{a,j}^\gamma R_\beta^j -
   T_{\beta \delta }^\gamma Z_a^\delta =
   - Z_d^\gamma A_{a\beta }^d - S_{0,j} G_{a\beta }^{j\gamma }
\quad .
\label{deriv zeta}
\ee
Eq.\bref{deriv zeta} is a new gauge-structure equation
for the first-stage reducible case.
Two new structure tensors
$A^d_{a\beta}$ and $G^{j\gamma}_{a\beta}$ arise.

The null vectors also lead to modifications of the solution
of the Jacobi identity.
Eq.\bref{cons jacobi} still holds but its solution is different.
Instead of Eq.\bref{jac1},
one obtains
\be
\left(
     {A_{\alpha \beta \gamma }^\delta +
     Z_c^\delta F_{\alpha \beta \gamma }^c -
     S_{0,j}D_{\alpha \beta \gamma }^{j\delta }}
\right)
    \left( {-1} \right)^{\epsilon _\beta }
    \gh C^\gamma \gh C^\beta \gh C^\alpha =0
\quad ,
\label{first stage jac1}
\ee
where we have made use of the completeness
of the null vectors $Z^\alpha_a$.
In this equation $A_{\alpha \beta \gamma }^\delta$
stands for the combination of terms in Eq.\bref{Adef}.

Multiplying Eq.\bref{first stage jac1} by
$R_\delta^i$ and using the Jacobi identity result
of Eq.\bref{cons jacobi}
lead to a modification
of Eq.\bref{cond sup} involving
$B^{ij}_{\alpha\beta\gamma}$.
The new result reads
$$
   S_{0,j}
\left( {B_{\alpha \beta \gamma }^{ji} -
   \left( {-1} \right)^{\epsilon _j
    \left( {\epsilon _i+\epsilon _\delta } \right)}
   R_\delta^i D_{\alpha \beta \gamma }^{j\delta } +
   V_c^{ji}F_{\alpha \beta \gamma }^c} \right)
   \left( {-1} \right)^{\epsilon _\beta }
   \gh C^\gamma \gh C^\beta \gh C^\alpha =0
\quad ,
$$
when written using ghosts.
The general solution is
$$
   { \left(
   B^{ji}_{\alpha\beta\gamma} +
   (-1)^{\epsilon_i \epsilon_\delta }
   R^j_\delta D^{i\delta}_{\alpha\beta\gamma} -
   (-1)^{\epsilon_j (\epsilon_i + \epsilon_\delta ) }
   R^i_\delta D^{j\delta}_{\alpha\beta\gamma} + \right. }
$$
\be
   { \left.
   V_c^{ji} F_{\alpha \beta \gamma }^c +
   S_{0,k} M^{kji}_{\alpha\beta\gamma}
   \right) }
   (-1)^{\epsilon_\beta}
   \gh C^\gamma \gh C^\beta \gh C^\alpha = 0
\quad ,
\label {first stage jac2}
\ee
where $B^{ji}_{\alpha\beta\gamma}$
is given in Eq.\bref{Bdef}.

By taking more and more
commutators of gauge transformations,
more structure functions and equations appear,
some of which involve graded symmetrizations
in the first-stage gauge indices $a$, $b$, etc..
As in the irreducible case,
it is useful to introduce ghosts
$\eta^a$ to automatically incorporate
graded symmetrization.
Equations \bref{off shell null vector}
and \bref{deriv zeta} can then be written as
\be
\left(
   {R_\beta ^iZ_a^\beta -S_{0,j}V_a^{ji}}
\right)
   \eta^a=0
\quad ,
\label{first stage eq1}
\ee
and
\be
\left(
   {\left( {-1} \right)^{\epsilon _a\epsilon _\beta }
   Z_{a,j}^\gamma R_\beta^j -
   T_{\beta \delta }^\gamma Z_a^\delta +
   Z_d^\gamma A_{a\beta }^d +
   S_{0,j}G_{a\beta }^{j\gamma }}
\right)
   \eta^a C^\beta =0
\quad .
\label{first stage eq2}
\ee
To summarize,
key equations for first-stage reducible theories
are Eqs.\bref{noether}, \bref{commutator algebra} and
\bref{first stage jac1} -- \bref{first stage eq2}.
Besides the null vectors
$Z_a^\beta$,
the new structure tensors are
$V^{ji}_a$, $A^d_{a\beta}$,
$G^{j\gamma}_{a\beta}$, $F^a_{\alpha\beta\gamma}$
as well as higher-level tensors.

Needless to say,
for a higher-order reducible theory
the number of quantities and equations increases considerably.
The complexity of the formalism makes the study
of the gauge structure at higher levels quite complicated.
A more sensible approach is to have a generating functional
whose expansion
in terms of auxiliary fields
produces the generic gauge-structure tensors.
In addition,
it is desirable to have a simple single equation
which, when expanded in terms of auxiliary fields,
generates the entire set of gauge-structure equations.
The field-antifield method
\ct{bv81a,bv83a,bv83b}
provides such a formalism.
The generating functional for structure tensors is
a generalized action subject to certain boundary conditions
and the classical master equation
contains all the gauge-structure equations.
Before presenting the abstract machinery,
it is of pedagogical value to consider some examples
of the formalism of this section.

\vfill\eject

\section{Examples of Gauge Theories}
\label{s:egt}

\hspace{\parindent}%
This section presents eight gauge theories,
which will be used in Sect.\ \ref{s:eps}
to illustrate the antibracket formalism.
The theories are
(1) the spinless relativistic particle,
(2) Yang-Mills theories,
(3) four-dimensional topological Yang-Mills theories,
(4) the four-dimensional antisymmetric tensor field,
(5) abelian $p$-form theories,
(6) the open bosonic string field theory,
(7) the massless relativistic spinning particle, and
(8) the first-quantized bosonic string.
Models (1), (2), (7) and (8) are closed and irreducible.
Models (3) and (4) are first-stage reducible.
Model (5) is $p$-stage reducible
and model (6) is an infinitely reducible open system.
For each theory,
the classical action $S_0$
and its gauge symmetries are first presented.
Then,
the non-zero gauge-structure tensors are obtained.
The determination of the structure tensors
is the first computational step
in the antibracket formalism.
The results in this section are used
in Sect.\ \ref{s:eps}
to obtain proper solutions $S$.

In the first subsection
on the spinless relativistic particle,
we illustrate the compact notation
of Sect.\ \ref{ss:gt}.
Models (1) and (8) are respectively
one and two-dimensional gravity theories.
Model (8), the first-quantized bosonic string,
is used to explain the construction
of general-coordinate-invariant theories, i.e.,
gravities.
In the subsection\ \ref{ss:mrsp}
on the massless relativistic spinning particle,
we provide a mini-review
of supersymmetry and supergravity.
A brief introduction to string field theory
is given
in Sect.\ \ref{ss:obsft}.

As exercises for the reader,
we suggest the following three computations.
(i) Verify that the gauge transformations
leave $S_0$ invariant.
(ii) Given  $S_0$ and its gauge symmetries,
obtain the results
presented for the gauge-structure tensors.
(iii) Verify
Eqs.\bref{noether}--\bref{jac2}
for the irreducible theories,
and verify
Eqs.\bref{noether}, \bref{commutator algebra} and
\bref{first stage jac1} -- \bref{first stage eq2}
for the first-stage reducible theories.

\subsection{The Spinless Relativistic Particle}
\label{ss:srp}

\hspace{\parindent}%
One of the simplest examples of a model
with a gauge invariance
is the free relativistic particle.
It actually corresponds
to a $0+1$ dimensional gravity theory
with scalar fields.
The supersymmetric generalization
of the spinless relativistic particle
is presented
in Sect.\ \ref{ss:mrsp}.

Let us use this system
to illustrate the formalism
in Sect.\ \ref{s:ssgt}.
The degrees of freedom are
a particle coordinate $x^\mu$
and an einbein $e$
both of which are functions
of a single  proper time variable $\tau$.
The action is given by
\be
  S_0 [x^\mu,e] = \int\dif\tau \ \frac12
 \left( \frac{ \dot x_\mu \dot x^\mu } e - m^2 e \right)
\ ,  \quad\quad
   \phi^i=(x^\mu,e)
\label{srp action}
\quad ,
\ee
where a dot over a variable
indicates a derivative
with respect to proper time.
The variations of the action
with respect to the fields,
Eq.\bref{def s0i},
are
\be
 S_{0,\mu } =
    - {d \over {d\tau }}
  \left( {{{\dot x_\mu } \over e}} \right)
\ , \quad \quad
  S_{0,e}={1 \over 2}
    \left( { - {{\dot x^2} \over {e^2}} - m^2} \right)
\quad ,
\label{srp eom}
\ee
where $S_{0,e}$ is the variation
of the action with respect to field $e$;
in other words, we also use $e$
as a field index for the einbein $e$.
If the equation of motion
$S_{0,e} = 0 $ is used
to solve for $e$,
and this solution is substituted
into the action in Eq.\bref{srp action},
one finds that the action
becomes the familiar one:
${S}_0 = - \int\dif\tau\; m \sqrt{-\dot x^2}$.
Classically, this action and the one
in Eq.\bref{srp action}
are equivalent.

The infinitesimal gauge transformations
for this system can be written as
\be
   \delta x^\mu= {{\dot x^\mu } \over e}
    \varepsilon \ ,
\nonumber\\
   \quad \quad
  \delta e = \dot \varepsilon \quad .
\nonumber
\label{srp trans law}
\ee
It is straightforward to verify that
Eq.\bref{srp trans law} is a symmetry
of Eq.\bref{srp action}.
The Noether identity
in Eq.\bref{ident noether}
reads
\be
  \int {\dif\tau \left\{ {
     - \left( { {{\dot x_\mu } \over e}} \right)
   {d \over {d\tau }}
    \left( {{{\dot x^\mu } \over e}} \right) -
   { {1} \over {2} }
  \left( { {{\dot x^2} \over {e^2}} + m^2} \right)
      {d \over {d\tau }}}
  \right\}\delta \left( {\tau -\tau '} \right)} = 0
\quad  ,
\label{srp ident noether}
\ee
which is verified using integration by parts.

The transformations laws in Eq.\bref{srp trans law}
in the form of Eq.\bref{transformation rule} are
\be
  R^\mu \varepsilon = {{\dot x^\mu } \over e}
    \varepsilon
\ , \quad \quad
  R^e\varepsilon =
   {d \over {d\tau }}\varepsilon  \quad .
\label{srp rdef}
\ee
Eq.\bref{srp rdef} says that
$R^\mu$ is the operator
that is multiplication
by ${{\dot x^\mu } \over e}$
and $R^e$ is ${d \over {d\tau }}$.
In kernel form using compact index notation,
they are
$$
  R_{\sigma}^{\mu \tau} =
  {{\dot x^\mu (\tau) } \over e}
  \delta \left( {\tau - \sigma} \right)
\ , \quad \quad
  R_{\sigma}^{e \tau} =
   {d \over {d\tau }}\delta
   \left( {\tau - \sigma} \right)
\quad .
$$
Recall that in using compact notation
the index $\alpha $ of $R_{\alpha}^i$
in Eq.\bref{transformation rule}
represents not only a discrete index
labelling the different gauge transformations
but also a space-time index.
Since there is only one type of gauge transformation
the discrete index takes on only one value,
which we drop for convenience.
Hence the index $\alpha $ of $R_{\alpha}^i$
is replaced
by the space-time variable $\sigma$.
In this subsection, we use the Greek letters
$\rho$, $\sigma$, $\tau$ and $\upsilon$
to denote proper time variables.
Likewise the index $i$ on $R_{\alpha}^i$
in Eq.\bref{transformation rule}
represents not only a field index $\mu$ or $e$
but also a proper time variable $\tau$.

The algebra of the gauge transformations is simply
$$
\left[ \delta_1 , \delta_2 \right] =
\left[\delta( \varepsilon_1 ) \ ,
 \delta( \varepsilon_2 )\right] = 0
\quad ,
$$
where $\delta( \varepsilon_1 )$
indicates a gauge transformation
with parameter $\varepsilon_1$.
These abelian gauge transformations
\bref{srp trans law}
are related to
the standard reparametrization transformations
\be
   \delta_R x^\mu= \dot x^\mu \varepsilon  \ ,
\quad \quad
    \delta_R e = {d \over {d\tau }}
   \left( e \varepsilon \right)
\label{srp rtrans}
\ee
through the following redefinition
of the gauge parameter
$$
   \varepsilon \longrightarrow \varepsilon e
\quad .
$$
The algebra of reparametrization transformations
reads
\be
\left[\delta_R (\varepsilon_1),
\delta_R (\varepsilon_2)\right] =
\delta_R (\varepsilon_{12}) \quad ,
\label{srp commutatora}
\ee
with the parameter $\varepsilon_{12}$ given by
\be
  \varepsilon_{12}=
\dot \varepsilon_1 \varepsilon_2 -
\varepsilon_1 \dot \varepsilon_2 \quad .
\label{srp commutatorb}
\ee
Eqs.\bref{srp commutatora} and
\bref{srp commutatorb}
correspond to the usual diffeomorphism algebra.
Since the commutator of two gauge transformations
is a gauge transformation,
the algebra is closed and
$E_{\alpha \beta}^{ji}$
in Eq.\bref{com of two gens}
is zero.
This example illustrates
the effect of field-dependent
redefinitions of the gauge parameters
or, equivalently, of the gauge generators:
An abelian algebra
can be transformed into a non-abelian one.
The converse of this also holds.
One can transform any given non-abelian algebra
into an abelian algebra using
field-dependent redefinitions,
a result known
as the abelianization theorem
\ct{bv84a}.
The fact that this process can spoil
the locality of the transformations
is one of the reasons
for using the non-abelian version.

It may appear unusual
that a single family of gauge transformations
produces non-abelian commutation relations.
This is due to the local non-commutativity
of reparametrization transformations
that arises from the time derivatives
in Eq.\bref{srp rtrans}.
Indeed, when $\varepsilon_1$ and $\varepsilon_2$
have non-overlapping support, i.e.,
$\varepsilon_1 (\tau) =0$ where
$\varepsilon_2 (\tau) \not= 0$ and vice-versa,
$\varepsilon_{12} =0$.

It is instructive to see
how a non-zero structure constant
$T_{\alpha \beta}^{\gamma}$
for the diffeomorphism algebra
arises using compact notation.
In what follows,
gauge indices, $\alpha$, $\beta$,
etc.\ are replaced
by proper time variable
$\rho$, $\sigma$, $\tau$, $\upsilon$.
{}From Eq.\bref{srp rtrans} one sees that
the transformation operators $R$
for reparametrizations are
$$
\displaylines{
R_{\sigma}^{\mu \tau } =
\dot x^\mu \left( \tau  \right)
\delta \left( {\tau -\sigma} \right) \quad , \cr
  R_{\sigma}^{e\tau }=e\left( \tau  \right)
{d \over {d\tau }}\delta \left( {\tau -\sigma} \right) +
\dot e\left( \tau  \right)\delta \left( {\tau -\sigma} \right) =
{d \over {d\tau }}\left[ e\left( \tau  \right)
\delta \left( {\tau -\sigma} \right)\right]
 \quad .  \cr
}
$$
For the $x^\mu$ degrees of freedom,
a straightforward computation yields
$$
\displaylines{
R_{\sigma,\nu \upsilon}^{\mu \tau }=
\delta_\nu^\mu \delta \left( {\tau - \sigma} \right)
{d \over {d\tau }}\delta
\left( {\tau - \upsilon} \right)
\quad , \cr
R_{\sigma,\nu \upsilon}^{\mu \tau }
R_{\rho}^{\nu \upsilon} =
\sum\limits_\nu^{} {}\int_{}^{} \ {\dif \upsilon}
\delta_\nu^\mu \delta
\left( {\tau -\sigma} \right)
{d \over {d\tau }}\delta
\left( {\tau -\upsilon} \right)
\dot x^\nu \left( {\upsilon} \right)
\delta \left( {\upsilon-\rho} \right)
\cr
= \ddot x^\mu \left( \tau  \right)
\delta \left( {\tau -\sigma} \right)
\delta \left( {\tau -\rho} \right)+
\dot x^\mu \left( \tau  \right)\delta
\left( {\tau -\sigma} \right)
{d \over {d\tau }}\delta
\left( {\tau -\rho} \right)
\quad . \cr
}
$$
Antisymmetrizing in $\rho$ and $\sigma$
and comparing with
Eq.\bref{algebra oberta} one finds
\be
  T_{\sigma\rho}^\tau =
\delta \left( {\tau -\sigma} \right)
{d \over {d\tau }}\delta
\left( {\tau -\rho} \right) -
\delta \left( {\tau - \rho} \right)
{d \over {d\tau }}\delta
\left( {\tau -\sigma} \right)
\quad ,
\label{srp teq}
\ee
which is in agreement
with Eqs.\bref{srp commutatora}
and \bref{srp commutatorb}.
For $e$,
straightforward computation produces
$$
\displaylines{
R_{\sigma , e \upsilon}^{e\tau } =
\delta \left( {\tau -\upsilon} \right)
{d \over {d\tau }}\delta
\left( {\tau -\sigma} \right) +
\delta \left( {\tau -\sigma} \right)
{d \over {d\tau }}
\delta \left( {\tau -\upsilon} \right) =
{d \over {d\tau }}\left[ \delta \left( {\tau -\upsilon} \right)
\delta \left( {\tau -\sigma} \right) \right]
 \quad , \cr
R_{\sigma,e\upsilon}^{e\tau }R_{\rho}^{e\upsilon} =
e\left( \tau  \right)\left[ {{d \over {d\tau }}
\delta \left( {\tau -\sigma} \right)} \right]
\left[ {{d \over {d\tau }}
\delta \left( {\tau -\rho} \right) } \right] +
\dot e\left( \tau  \right)\delta
\left( {\tau -\rho} \right)
{d \over {d\tau }}
\delta \left( {\tau -\sigma} \right) +
\cr
2\dot e\left( \tau  \right)
\delta \left( {\tau -\sigma} \right){d \over {d\tau }}
\delta \left( {\tau -\rho} \right) +
e\left( \tau  \right)\delta
\left( {\tau -\sigma} \right)
{{d^2} \over {d\tau^2}}\delta
\left( {\tau -\rho} \right) +
\ddot e\left( \tau  \right)
\delta \left( {\tau -\sigma} \right)
\delta \left( {\tau -\rho} \right)
\ \  . \cr }
$$
Antisymmetrizing in $\rho$ and $\sigma$
and using Eq.\bref{com of two gens}, one finds
$T$ is again given by Eq.\bref{srp teq}.

Although compact notation is useful to
represent the formalism of gauge theories
in full generality,
it is cumbersome for specific theories,
especially for those
in which more natural notation
has already been established.
In the examples that follow,
we do not explicitly display equations in compact form
but use more conventional notation.

\subsection{Yang-Mills Theories}
\label{ss:ymt}

\hspace{\parindent}
Yang-Mills theories
\cite{ym54a}
are perhaps
the most familiar gauge theories.
For each Lie algebra ${\cal G}$
there is different theory.
The fundamental fields are
gauge potentials $A_\mu^a$
where there is an index $a$ for each generator
$T_a$ of ${\cal G}$.
In a matrix representation,
the generators are antihermitian matrices
which are conventionally normalized
so that
$Tr \left( T_a T_b \right) =
  - {1 \over 2} \delta_{ab}$.
The generators satisfy
\be
 \left[ T_a, \ T_b \right] = {f_{ab}}^{c} T_c
\quad ,
\ee
where ${f_{ab}}^{c}$ are
the structure constants of ${\cal G}$.
They are real and antisymmetric in lower indices
${f_{ab}}^{c} = - {f_{ba}}^{c}$
and they must satisfy the Jacobi identity
\be
 {f_{ab}}^{e} {f_{ec}}^{d} +
 {f_{ca}}^{e} {f_{eb}}^{d} +
 {f_{bc}}^{e} {f_{ea}}^{d} = 0
\quad .
\label{lie algebra jacobi identity}
\ee

The Yang-Mills action is
\be
    {S}_0 [A_\mu^a] =
    - { {1 } \over {4} } \int {\rm d}^d x
    \ F_{\mu\nu}^a (x) F_a^{\mu\nu} (x) =
     { {1 } \over {2} } \int {\rm d}^d x
     \ Tr \left[F_{\mu\nu} (x) F^{\mu\nu} (x)\right]
\quad ,
\label{action Y-M}
\ee
where $d$ is the dimension of space-time,
$ F_{\mu \nu}(x) \equiv F^a_{\mu \nu}(x) T_a$,
and
where the field strengths $F^a_{\mu \nu}$ are
\be
    F^a_{\mu \nu}(x) \equiv
    \partial_\mu A^a_\nu(x)-\partial_\nu A^a_\mu(x)
    - {f_{bc}}^{a} A^b_\mu(x) A^c_\nu(x)
\quad .
\label{field strengths Y-M}
\ee

The equations of motion,
gauge transformations and gauge algebra are
\be
      \left( D^\mu F_{\mu \nu} \right)_a \equiv
       { {D^{\mu }}_a}^b F_{b \mu \nu}  = 0
   \quad ,
\label{eom Y-M}
\ee
\be
     \delta A^a_\mu = \left(D_\mu \Lambda \right)^a
      \equiv D^{a}_{\mu b} \Lambda^b
   \quad ,
\label{gauge transf Y-M}
\ee
\be
    [\delta ( \Lambda_1 ),
  \delta ( \Lambda_2 ) ] A^c_\mu  =
    \delta ( \Lambda_{12} ) A^c_\mu =
   D^{c}_{\mu d}
 \left( {f_{ab}}^{d} \Lambda_1^b \Lambda^a_2 \right)
      \quad ,
\label{gauge algebra Y-M}
\ee
so that
$\Lambda_{12}^c =
{f_{ab}}^{c} \Lambda_1^b \Lambda_2^a $.
The covariant derivatives
$D^{a}_{\mu b}$ and ${D_{\mu a} }^b$
in the adjoint representation are
$$
    D^{a}_{\mu b} = {\delta^{a}}_{b} \partial_\mu -
   {f_{cb}}^{a} A^c_\mu
\quad ,
$$
\be
  {D_{\mu a} }^b =
  {\delta_{a}}^{b} \partial_\mu +
   {f_{ca}}^{b} A^c_\mu
\quad ,
\label{cov der Y-M}
\ee
where $D^{a}_{\mu b}$ is applied to fields $\phi^b$
with an upper index $b$
and
${D_{\mu a} }^b$ is applied to fields $\phi_b$
with a lower index $b$.
One has
\be
  \int \dif^d x \phi_a D^{a}_{\mu b} \phi^b =
  - \int \dif^d x      \left( {
   { {D_{\mu b}} }^a \phi_a } \right) \phi^b
\quad .
\label{cov diff by parts}
\ee
The operator $R^i_\alpha$
in Eq.\bref{transformation rule}
corresponds to ${ D^{ \mu a} }_{b}$.
The covariant derivative satisfies
$$
  { \left[ D_\mu , \ D_\nu \right]^a}_{b} =
  -{f_{cb}}^a F^c_{\mu \nu}
\quad ,
$$
\be
  { \left[ D_\mu , \ D_\nu \right]_a}^{b} =
   {f_{ca}}^b F^c_{\mu \nu}
\label{cov der commutator}
\quad .
\ee

In using compact notation,
the spatial dependence
as well as index dependence of tensors
needs to be specified.
For local theories,
the spatial dependence is proportional to delta
functions or a finite number of derivatives
acting on delta functions.
When the spatial-temporal part of a tensor structure
is a delta function,
it is proportional to the identity operator in $x$-space
when regarded as an operator.
In such cases, it is convenient to drop explicitly
such identity operators.

Eq.\bref{gauge algebra Y-M} is
in the form of Eq.\bref{com of two gens}
with $E_{\alpha \beta}^{ji}=0$ and
$T_{a b}^{c} = {f_{a b}}^{c}$
where two identity operators
or delta functions are implicit.
One concludes that this example constitutes
a closed, irreducible gauge algebra.

It is useful to verify the key equations
for an irreducible closed algebra
given in Eqs.\bref{noether}-\bref{jac2}.
The generator of gauge transformations is
the covariant derivative in Eq.\bref{cov der Y-M}.
Using Eqs.\bref{eom Y-M} and \bref{gauge transf Y-M},
the Noether identity in Eq.\bref{noether} reads
$$
 \int {\rm d}^d x \
 \left( D^\mu F_{\mu \nu} \right)_b
     \left( D^\nu \gh C \right)^b = 0
\quad .
$$
To verify this equation, integrate by parts,
use the antisymmetry of
$F_{\mu \nu}$ in $\mu$ and $\nu$,
use Eq.\bref{cov der commutator},
and then
make use of the antisymmetry
of ${f_{cd}}^{a}$ in $c$ and $d$:
$$ \int {\rm d}^d x \
\left( D^\mu F_{\mu \nu} \right)_b
     \left( D^\nu \gh C \right)^b =
 - \int {\rm d}^d x \
   \left(  D^\nu D^\mu F_{\mu \nu} \right)_b
     \gh C^b =
$$
$$
 - { 1 \over 2 } \int {\rm d}^d x \
   \left( \left[  D^\nu ,\ D^\mu \right]
      F_{\mu \nu} \right)_b \gh C^b =
  - { 1 \over 2 } \int {\rm d}^d x \
   {f_{bd}}^a F_a^{\nu \mu} F_{\mu \nu}^d
     \gh C^b = 0
\quad .
$$
A straightforward computation of the
$2 R^i_{\alpha,j} R^j_\beta \gh C^\beta \gh C^\alpha $
term
in the commutator algebra equation
of Eq.\bref{commutator algebra}
produces
$ -2 {f_{ba}}^c
  \left(
   \partial^\mu \gh C^b - A^{d \mu} {f_{de}}^b \gh C^e
   \right) \gh C^a
$
which, after a little algebra
that makes use
of Eq.\bref{lie algebra jacobi identity},
leads to
$
 {D^{\mu c}}_d
  \left( {f_{ab}}^d \gh C^b \gh C^a \right)
$.
Using this for
$2 R^i_{\alpha,j}R^j_\beta \gh C^\beta \gh C^\alpha $
in  Eq.\bref{commutator algebra},
one concludes, as expected,
that $T_{a b}^{c} = {f_{a b}}^{c}$
and $E_{\alpha \beta}^{ji}=0$.
When $T_{ab}^{c}={f_{ab}}^{c}$ is used,
the gauge-structure Jacobi equation
in Eq.\bref{jac1}
leads to
$3 A^d_{abc}= - ($
LHS of Eq.\bref{lie algebra jacobi identity}$)$
so that $A^d_{abc}=0$.
Finally, the other consequence
of the Jacobi identity,
namely Eq.\bref{jac2}, produces the tautology $0=0$.
All terms are zero because the tensors
$B^{ji}_{\alpha\beta\gamma}$,
$D^{i\delta}_{\alpha\beta\gamma}$
and $M^{kji}_{\alpha\beta\gamma}$
are all zero.
Higher-level equations
(which were not displayed in Sect.\ \ref{ss:gs})
are automatically satisfied
because higher-level tensors are identically zero.

\subsection{Topological Yang-Mills Theory}
\label{ss:tymt}

\hspace{\parindent}
In four-dimensions, the action
for topological Yang-Mills theory
\ct{witten88a,bs88a}
is proportional to the Pontrjagin index
\be
 S_0 = \frac14 \int\dif^4x
 \  F_{\mu\nu}^a \dual F_a^{\mu \nu}
\quad ,
\label{action top Y-M}
\ee
where the dual field strength
$\dual F_{\mu\nu}^a$ is given by
$
\dual F_{\mu\nu}^a =
\frac12\eps_{\mu\nu\rho\sigma}F^{a\rho\sigma}
$,
and
where $\eps_{0123} =1$.
The interest in this system
is its connection
\ct{witten88a}
to Donaldson theory
\ct{donalson83a}.

This action is invariant under the gauge transformations
in Eq.\bref{gauge transf Y-M} because the lagrangian
is constructed as a group invariant of the field strength.
In addition, since the theory is topological,
the transformation
\be
\delta A_\mu^a=\veps_\mu^a \quad ,
\label{top transf Y-M}
\ee
leaves the action invariant,
as a short calculation verifies.
The two gauge transformations form a closed algebra since
$$
 [\delta (\Lambda_1 ,\veps_1 ) ,
 \delta (\Lambda_2 ,\veps_2 ) ] A_\mu =
\delta (\Lambda_{12} , \veps_{12} ) A_\mu \quad ,
$$
\be
\Lambda_{12}^c \equiv {f_{ab}}^c \Lambda_1^b \Lambda_2^a \ ,
\quad \quad
\veps_{\mu 12}^c \equiv {f_{ab}}^c
 \left(
   \veps_{\mu 1}^b \Lambda_2^a + \Lambda_1^b \veps_{\mu 2}^a
 \right)
\quad .
\label{gauge algebra top Y-M}
\ee

However, the gauge generators are obviously
off-shell linearly dependent
since the $\veps_\mu$ transformations include
ordinary gauge transformations
when $\veps_\mu=D_\mu \Lambda$.
Since the gauge transformations are not all independent,
one has a reducible gauge theory
and the coefficients $Z^\alpha_a$
in Eq.\bref{off shell null vector}
and $A^b_{a\alpha}$
introduced in Eq.\bref{deriv zeta}
are non-zero.
Of course, the theory can be made irreducible
by eliminating ordinary gauge transformations
from the set of all transformations.
However, for other theories
it is not so easy to reduce
the full set to an irreducible subset
without spoiling locality or relativistic covariance
and often it is convenient to formulate
the theory as a reducible system.

Let us use topological Yang-Mills theory
to illustrate the gauge-structure formalism
of Sect.\ \ref{ss:gs}.
The field index $i$ corresponds to both a gauge index
and a vector Lorentz index
since the field is $A^{a \mu}$:
$i \leftrightarrow a \mu $.
There are two types of gauge transformations so that
the gauge index $\alpha$
of Sect.\ \ref{ss:gs} corresponds to the group index $b$
in the case of an ordinary gauge transformation
or to the pair $c \mu$
in the case of a topological gauge transformation:
$\alpha \leftrightarrow \left( b , c \mu \right)$.
The generator of ordinary gauge transformations
is the covariant derivative in Eq.\bref{cov der Y-M}
and the generator of topological transformations
is a delta function:
\be
 R^{a \mu}_b = {D^{ \mu a}}_b \ ,
 \quad \quad
 R^{a \mu}_{b \nu} = \delta^a_b \delta^\mu_\nu
\quad .
\label{Y-M gauge generators}
\ee
The null vectors, denoted as $Z^\alpha_b$
in Sect.\ \ref{ss:gs},
are
\be
Z^a_b = \delta^a_b
\ , \quad \quad
Z^{a \mu}_b = -{D^{\mu a}}_b
\quad .
\label{Y-M null vectors}
\ee
The number of null vectors is
equal to the number of gauge generators.
It is easily verified that
Eq.\bref{first stage eq1} holds
since
$
 R^{b \mu}_c Z^c_a \eta^a + R^{b \mu}_{c \nu}
  Z^{c \nu}_a \eta^a =
 {D^{\mu b}}_c \delta^c_a \eta^a +
 \delta^b_c \delta^\mu_\nu
 \left( - {D^{\nu c}}_a \eta^a \right) = 0
$.
This computation
implies that $V_a^{ji} = 0$
in Eq.\bref{first stage eq1}.

The non-zero structure constants are
\be
 T_{ab}^c = {f_{ab}}^c \ ,
 \quad \quad
 T_{a \mu \  b}^{c \nu} = T_{a \ b \mu}^{c \nu} =
 \delta^\nu_\mu {f_{ab}}^c
\quad ,
\label{top Y-M T tensor}
\ee
whereas $T_{a \mu \  b}^{c } = T_{a \  b \mu}^{c }
 = T_{a b}^{c  \nu } =
T_{a \mu \  b \nu}^{c \lambda}
 = T_{a \mu \  b \nu}^{c } = 0$.
As in the example
of Sect.\ \ref{ss:ymt},
$A_{\alpha \beta \gamma}^{\delta}$ is
zero since
$
 3 A_{abc}^d = -
 \left( {
 {f_{ae}}^{d} {f_{bc}}^{e}  +
 {f_{be}}^{d} {f_{ca}}^{e}  +
 {f_{ce}}^{d} {f_{ab}}^{e}
  } \right) = 0
$.
One also finds
$
 A_{a \mu \ b \ c}^{d \nu} = A_{a \ b \mu \ c}^{d \nu} =
 A_{a \ b \ c \mu}^{d \nu} = \delta_\mu^\nu A_{abc}^d = 0
$.

Eq.\bref{first stage jac1} holds because
$
 A_{\alpha \beta \gamma }^\delta =
 F_{\alpha \beta \gamma }^c =
 D_{\alpha \beta \gamma }^{j\delta }
 = 0
$.
Eq.\bref{first stage jac2} holds because
$B^{ji}_{\alpha\beta\gamma}$ and
$ M^{kji}_{\alpha\beta\gamma}$
are also zero.
Finally,
Eq.\bref{first stage eq2}
is valid as long as $ G_{a\beta }^{j\gamma } = 0 $ and
\be
 A_{ab}^c = - {f_{ab}}^c \ ,
 \quad \quad A_{a \ b \mu}^c = 0
 \quad .
\label{top Y-M A tensor}
\ee
In verifying Eq.\bref{first stage eq2}
it is useful to note that
$Z^{c \nu}_{a,b \mu} = - \delta^\nu_\mu {f_{ab}}^c$,
and $Z^c_{a,j} = 0$
both when $j = b$ and when $j = b \mu$.
For Eq.\bref{first stage eq2},
there are four cases to verify:
(i)   $\gamma =c    $ and $\beta = b $,
(ii)  $\gamma =c    $ and $\beta = b \mu$,
(iii) $\gamma =c \nu$ and $\beta = b$ and
(iv)  $\gamma =c \nu$ and $\beta = b \mu$.
Case (i) gives
$-{f_{bd}}^c \delta^d_a - \delta^c_d A_{ab \mu}^d =0$.
Case (ii) is automatically zero because each term
in Eq.\bref{first stage eq2} is zero.
Case (iii), after a little algebra,
results in the expression
$
{f_{ba}}^c \partial^\nu \left( \eta^a \gh C^b \right) +
{f_{ab}}^c \partial^\nu \left( \eta^a \gh C^b \right) +
 \left(
  {f_{ad}}^c {f_{be}}^d + {f_{bd}}^c {f_{ea}}^d +
  {f_{ed}}^c {f_{ab}}^d
 \right) \eta^a \gh C^b A^{e \nu}
$
which is zero because of the antisymmetry
of ${f_{ab}}^c$ in $ab$
and the Jacobi identity
for the Lie algebra structure constants
in Eq.\bref{lie algebra jacobi identity}.
For case (iv), one finds
$
  \delta^\nu_\mu {f_{ab}}^c -
  \delta^\nu_\mu {f_{ba}}^c = 0
$.
In short,
the important non-zero gauge structure tensors
are given in
Eqs.\bref{Y-M gauge generators} -- \bref{top Y-M A tensor}.
Eqs.\bref{noether}, \bref{commutator algebra} and
\bref{first stage jac1} -- \bref{first stage eq2}
are all satisfied.

In a topological theory,
the number of local degrees of freedom is zero.
One finds that $n_{\rm dof}$ is $4N$ (for $A^{a\mu}$)
minus $4N$ (for $\veps_\mu^a$)
minus $N$ (for $\Lambda^a$)
plus $N$ (for the null vectors
in Eq.\bref{Y-M null vectors}).
Hence, the net number of
local degrees of freedom is zero.

\subsection{The  Antisymmetric Tensor Field Theory}
\label{ss:atft}

\hspace{\parindent}
Another example of first-stage reducible theory is
the  antisymmetric tensor gauge theory.
Consider a tensor field $B^a_{\mu\nu}$ in four dimensions
satisfying
$B^a_{\nu\mu} = - B^a_{\mu\nu}$
whose dynamics is described by the action
\ct{ft81a,thierrymieg90a}
\be
 {S}_0 \left[ A_\mu^a , \ B^a_{\mu\nu} \right] =
  \int {\rm d}^4 x
  \left(
  \frac12 A^a_\mu A_a^{\mu} -
    \frac12 B^a_{\mu\nu}F_a^{\mu\nu}
  \right)
\quad ,
\label{AT action}
\ee
where $A^a_\mu$ is an auxiliary vector field.
The field strength $F^a_{\mu\nu}$
is given in terms of $A^a_\mu$ as
in the Yang-Mills case
via Eq.\bref{field strengths Y-M}.
The action is invariant
under the gauge transformations
\be
 \delta B^a_{\kappa\lambda} =
 \epsilon_{\kappa\lambda\mu\nu}
{D^{\mu a}}_b \Lambda^{b \nu} \ ,
 \quad \quad
 \delta A^{a\mu}=0
\quad .
\label{AT gauge transformations}
\ee
The covariant derivative ${D^{\mu a}}_b$ is given
in Eq.\bref{cov der Y-M}.
The equations of motion derived
from Eq.\bref{AT action} are
\be
 { {\partial_r S_0 } \over { \partial B^a_{\mu\nu} } } =
    -  F_a^{\mu\nu} = 0 \ ,
\quad \quad
  \rder{S_0}{A_a^\mu} = A^a_\mu + {D^{\nu a}}_b B^b_{\nu\mu} = 0
\quad .
\label{AT eom}
\ee
In spite of the presence of
Lie-algebra structure constants ${f_{ab}}^c$,
the model has an abelian gauge algebra,
i.e., $T^\gamma_{\alpha \beta} =0 $,
due to the fact that the vector field,
which appears in the covariant derivative,
does not transform.

The gauge transformations
\bref{AT gauge transformations} have
an on-shell null vector.
Indeed, taking
\be
\Lambda^{b \nu} = {D^{\nu b}}_c \xi^c
\quad ,
\label{AT null vector}
\ee
one finds
using Eq.\bref{cov der commutator} that
$$
 \delta B^a_{\kappa\lambda} =
 \epsilon_{\kappa\lambda\mu\nu}
  {\left(
  D^\mu D^\nu
   \right)^a}_b \xi^{b } =
 - \frac12 \epsilon_{\kappa\lambda\mu\nu}
  {f_{cb}}^a F^{c \mu \nu } \xi^{b }
\quad ,
$$
which vanishes on-shell,
since $ F^{c \mu \nu } = 0$
when the equations of motion \bref{AT eom} are used.
Since the null vectors are independent,
this theory is on-shell first-stage reducible.

It is instructive to determine
the gauge-structure tensors
and verify
Eqs.\bref{noether},
\bref{commutator algebra} and
\bref{first stage jac1} -- \bref{first stage eq2}.
The field index $i$
in Sect.\ \ref{s:ssgt}
corresponds to $a \mu \nu$
in the case of the antisymmetric field
$B^{a \mu \nu}$
and corresponds to $a \mu$
in the case of $A^{a \mu}$.
The gauge index $\alpha$
of Sect.\ \ref{s:ssgt}
corresponds to $b \nu$
which are the indices of $\Lambda^{b \nu}$.
The null index $a$
of Sect.\ \ref{s:ssgt}
is a Lie algebra generator index.
The generator of gauge transformations
$R^i_\beta$ is
\be
 R^{a \mu \nu}_{b \lambda} =
 {\epsilon^{\mu\nu\rho\sigma}} \eta_{\sigma\lambda}
  {D_{\rho b}^a} \ ,
 \quad \quad
 R^{a \mu }_{b \lambda} = 0
\quad ,
\label{AT gauge generators}
\ee
where the second equation holds
because $A^{a \mu }$ does not transform
and $\eta_{\sigma\lambda}$
is the flat space-time metric.
The null vectors $ Z^{\alpha}_a$ are
the covariant derivative operators
\be
 Z^{b \lambda}_a = {D^{\lambda b}}_a
\quad .
\label{AT null vectors}
\ee
Using Eq.\bref{first stage eq1},
one finds
$
 R^{c \mu \nu}_{b \lambda} Z^{b \lambda}_a =
  \frac12 \epsilon^{\mu\nu\rho\sigma}
   {f_{ab}}^c F^b_{\rho\sigma}
$
so that
$$
 V^{b \rho\sigma \  c \mu \nu }_a =
  - \epsilon^{\mu\nu\rho\sigma}
  \delta^{bd} {f_{ad}}^c
\quad ,
$$
\be
 V^{ji }_a = 0  \ ,
  \quad {\rm if} \ j=b \nu \ {\rm or} \  i=b \nu
\quad ,
\label{AT Vjia}
\ee
that is, $ V^{ji }_a = 0$ if $i$ or $j$ corresponds
to the field index of $A^{b \nu}$.
The derived quantities
$A_{\alpha \beta \gamma }^\delta$
and $B^{ij}_{\alpha\beta\gamma}$ are zero.
Other gauge-structure tensors also vanish:
\be
 T^{\gamma}_{\alpha \beta} = E^{ji}_{\alpha\beta} =
 D^{j\delta}_{\alpha\beta\gamma} =
 F_{\alpha \beta \gamma }^c =
 M^{kji}_{\alpha\beta\gamma} = A^d_{a\beta} = 0
\quad .
\label{AT zero tensors}
\ee

The gauge-structure equations are all satisfied.
Eq.\bref{noether} holds because
the action is invariant
under gauge transformations,
as is easily checked.
Eq.\bref{first stage eq1} is satisfied.
In fact, it was used above in Eq.\bref{AT Vjia}
to obtain $V^{ji }_a$.
Eqs.\bref{first stage jac1}
and \bref{first stage jac2}
are satisfied because
all the tensors entering these
equations are zero.
Each term in Eq.\bref{commutator algebra} is zero:
The first term
$R^i_{\alpha , j} R^j_\beta$
is zero because
$R^i_{\alpha , j} = 0 $ when $j = a \mu \nu$, i.e.,
when $j$ is a field index of $B^{a \mu \nu}$,
and $R^j_\beta = 0$ when $j = a \mu $, i.e.,
when $j$ is a field index of $A^{a \mu }$.
The other terms vanish because
the structure tensors vanish.
Likewise each term
in Eq.\bref{first stage eq2} is zero:
The first term
$Z^\gamma_{\alpha , j} R^j_\beta$
is zero
because $Z^\gamma_{\alpha , j} $
like $R^i_{\alpha , j}$ is zero
when $j = a \mu \nu$.

Let $N$ be the number of generators of ${\cal G}$, i.e.,
the dimension of the Lie algebra ${\cal G}$.
The number of degrees of freedom $n_{\rm dof}$
is $4 N$ for $A^{a \mu}$ plus $6 N $ for
$B^{a}_{ \mu \nu}$ minus
the number of gauge transformations
$4 N$ plus the number of null vectors $N$,
so that Eq.\bref{dof} reads
$
  n_{\rm dof} = 7 N
$.

\subsection{Abelian $p$-Form Theories}
\label{ss:apft}

\hspace{\parindent}%
It is not hard to find
an example of an $L$-th stage
reducible theory.
Let $A$ be a $p$-form and
define $F$ to be its field strength:
$F=dA$ where $d$ is
the exterior derivative.
For $p+1$ less than
the dimension $d$ of spacetime,
an action for this theory is
\be
  S_0 = - \frac12 \int \  F \wedge {}^\ast F
 \quad ,
\label{p-form action}
\ee
where $\ast$ is the dual star operation
that takes
a $q$-form into a $d-q$ form and
$\wedge$ is the wedge product.
On basis $q$-forms,
it is defined by
$$
  {}^\ast \left( {dx^{\mu_1} \wedge dx^{\mu_2} \wedge
    \ldots \wedge dx^{\mu_q}} \right) =
$$
$$
  {1 \over {\left( {n-p} \right)! } }
   \varepsilon^{\mu_1 \mu_2\ldots \mu_d}
    \eta _{\mu_{q+1} \nu_{q+1}}\eta_{\mu_{q+2} \nu_{q+2}}
     \ldots \eta_{\mu_d \nu_d} dx^{\nu_{q+1}}
    \wedge dx^{\nu_{q+2}} \wedge \ldots \wedge dx^{\nu_d}
\quad ,
$$
where $\eta_{\mu \nu}$ is the flat space-time metric
and $\varepsilon^{\mu_1 \mu_2\ldots \mu_d}$
is the antisymmetric tensor symbol.
The case $p=1$ corresponds
to abelian Yang-Mills theory.
Using $d d = 0$,
one sees that
the action is invariant
under the gauge transformation
\be
 \delta A = d \lambda_{p-1}
\quad ,
\label{p-form gauge transformations}
\ee
where $\lambda_{p-1}$ is a $p-1$ form.
This gauge transformation has
its own gauge invariance.
In fact,
there is a tower of gauge invariances for gauge
invariances given by
\bea
\delta \lambda_{p-1}& =&  d \lambda_{p-2} \quad ,
\nonumber\\
  &\vdots&
\nonumber\\
 \delta \lambda_{1}& =&  d \lambda_{0} \quad ,
\nonumber
\eea
where $\lambda_{q}$ is a $q$-form.
Hence, the theory is $p-1$ stage reducible.
The number of degrees of freedom is
\be
 n_{\rm dof} = {d \choose p} -
   {d \choose {p-1} } + {d \choose {p-2} } -
   \dots - (-1)^{p} {d \choose {1} } +(-1)^p {d \choose {0} } =
  {{d-1} \choose {p} }
\quad .
\label{p-form ndof}
\ee
Note that ${{d} \choose {q} }$ is the dimension
of the space of $q$-forms
in $d$ dimensional space-time.

The gauge generators
at the $s$-th stage,
$R^{\alpha_s -1}_{s \alpha_s}$, correspond
to the exterior derivative $d$
acting on the space of ${(p-1-s)}$ forms, $d^{(p-1-s)}$:
\bea
 R_0& \leftrightarrow&   d^{(p-1)} \quad ,
\nonumber\\
 R_1 &\leftrightarrow&   d^{(p-2)} \quad ,
\nonumber\\
 &\vdots&
\nonumber\\
 R_{p-1} &\leftrightarrow&   d^{(0)} \quad .
\nonumber
\eea

Because $d d = 0$,
Eq.\bref{reduc l 1} holds off-shell and
all $V^{i\alpha_{s-2}}_{s\alpha_s}$ are zero.
With the exception of the gauge generators,
all gauge-structure tensors are zero.

If $n (p+1) = d =$ dimension of space time,
a topological term
can be added to the action
\be
 \Delta S_0 = \int
 \underbrace{ F \wedge F \wedge
   \dots \wedge F}_{ n {\rm {\ terms }}}
\quad .
\label{p-form top term}
\ee
For $n=2$ one can consider the quantization of
$\Delta S_0$ alone if $d = 2 \left( p + 1 \right)$.
This would be another example of a topological theory.

Abelian $p$-form theories provide a good background
for covariant open string field theory,
a non-abelian generalization of $p$-form theory
which is infinite-stage reducible.

\subsection{Open Bosonic String Field Theory}
\label{ss:obsft}

\hspace{\parindent}
The covariant $d=26$ open string field theory was obtained
by E. Witten \ct{witten86a}.
It resembles a Chern-Simons theory.
The fundamental object is a string field $A$.
Although one can proceed
without a detailed understanding of $A$,
$A$ can be expanded as a series
in first-quantized string states
whose coefficients are ordinary particle fields.
Each member of this infinite tower of states corresponds
to a particular vibrational mode of the string.
In this manner,
string theory is able to incorporate collectively
many particles.
For example,
the open bosonic string possesses a tachyonic scalar,
a massless vector field,
and numerous massive states of all possible spins.
For reviews on open bosonic string field theory,
see refs.\ct{samuel87a,kaku88a,thorn89a}.
Also useful is the discussion
in Sect.\ \ref{ss:fqbs2},
where the first quantization
of the bosonic string is treated.

Covariant open string field theory
can be formulated axiomatically
\ct{witten86a,witten86b}.
Fields are classified according to
their string ghost number.
If the string ghost number of $B$ is $g(B)=p$
then we say that
$B$ is a string $p$-form in a generalized sense.
The ingredients of abstract string theory are
a derivation $Q$, a star operation $\ast$
which combines pairs of fields
to produce a new field,
and an integration operation $\int$
which yields a complex number $\int B$
for each integration over a string field $B$.
These objects satisfy five axioms:

\medskip

\begin{quote}

(1) The nilpotency of $Q$: $QQ=0$.

\medskip
(2)  Absence of surface terms in integration:
$\int Q A = 0$.
This axiom is equivalent to
an integration-by-parts rule.

\medskip
(3)  Graded distributive property of
$Q$ across $\ast$: \hspace{\fill}
\linebreak
$
 Q \left( A \ast B \right) =
 QA \ast B + (-1)^{g(A)} A \ast QB
$.

\medskip
(4)  Associativity of the star product:
$
 \left( A \ast B \right) \ast C =
 A \ast \left(B  \ast C \right)
$.

\medskip
(5)  Graded commutativity
of the star product under the integral:
\hspace{\fill}
\linebreak
 $\int A \ast B = (-1)^{g(A)g(B)} \int B \ast A$.

\end{quote}

\medskip

The ghost number of a star product of fields
is the sum of their ghost numbers:
$g(A \ast B) = g(A) + g(B) $.
The derivation $Q$ increases the ghost number by $1$:
$g(QA) = g(A) +1$.
In some circles, including
refs.\ct{witten86a,witten86b},
the ghost number is shifted by $-3/2$
so that $A$ has ghost number $-1/2$
instead of $1$.

The axioms are satisfied
for non-abelian Chern-Simons theory
in three dimensions.
The field is a non-abelian vector potential
which is converted into a Lie-algebra-valued $1$-form
by multiplying by $dx^\mu$:
${A^a}_b \equiv A^a_{\mu b} dx^\mu \equiv
A^{s}_{\mu} {\left( {T_s} \right)^a}_b dx^\mu$,
where the $T_s$ are
a set of matrix generators for the Lie group.
The derivation $Q$ is the exterior derivative $d$.
The star product is the wedge product
and a matrix multiplication:
${\left( A*B \right)^a}_b =
{A^a}_c \wedge {B^c}_b $.
Integration is an integral
over a three-dimensional manifold $M$
without boundary and a trace over the Lie-algebra indices:
$\int A \equiv \int_M {A^a}_a $.
Axiom (1) is satisfied because $dd=0$.
Axiom (2) holds because $M$ has no boundary.
Axiom (3) is satisfied because
the exterior derivative $d$
is graded distributive across the wedge product.
Axiom (4) holds because both the wedge product
and matrix multiplication are associative.
Finally, axiom (5) is satisfied
because of the graded antisymmetry of the wedge product
and the cyclic property
of a trace: $Tr(MN) = Tr(NM)$
for any two matrices $M$ and $N$.
In this example and in open string field theory,
$\int B = 0$ unless $B$ is a string 3-form.

The action for the string 1-form $A$ is given by
\be
   S_0 \left[ A \right] =
  \frac12 \int A \ast Q A +
  \frac13 \int A \ast A \ast A
\quad .
\label{obs action}
\ee
Using the above five axioms
it is straightforward to show
that the action is invariant
under the gauge transformation
\be
 \delta A=Q\Lambda_0 + A \ast \Lambda_0 -
  \Lambda_0 \ast A
\quad ,
\label{obs gauge transformation}
\ee
where $\Lambda_0$ is any string $0$-form.
The equation of motion for $A$ reads
\be
  F\equiv  { QA + A \ast A }  = 0
\quad .
\label{obs eom}
\ee

For the open bosonic string,
$Q$ is the  BRST charge $Q_{BRST}$
of the first-quantized theory
promoted to an operator by second quantization
(see Eq.\bref{fqbs gf BRST} with $\rho = 1$).
The $\ast$ product is intuitively described as follows.
Let $\sigma$ be the parameter
that determines a point on the string,
so that $\sigma = 0$ corresponds to one endpoint
and $\sigma = \pi$ corresponds to the other endpoint.
Divide the string in two at $\sigma = \pi /2 $
and call the two halves the left and right halves.
Let $C = A \ast B$.
Then $\ast$ ``glues''
the left half of $A$ to the right half of $B$
by delta functions and what remains is $C$
so that the left half of $C$ is the right half of $A$
and the right half of $C$ is the left half of $B$.
See Fig.\ 1.
The star operator can be thought of
as matrix multiplication
if the range $0 \le \sigma \le \pi /2 $
of points of the string
is associated with one matrix index
and the range
$\pi /2 \le \sigma \le  \pi $ is associated
with the other matrix index.
The integral operator is a delta function
equating the left and right halves.
See Fig.\ 2.
In the matrix analogy, it is the trace.
Precise definitions of $Q$, $\ast$ and $\int$
in terms of the vibrational modes of the string, i.e.,
particle excitations, can be found
in refs.\ct{cst86a,samuel86a,gj87a,gj87b}.

\begin{picture}(400,150)
\put(132,108){$\sigma_A = { {\pi} }$}
\put(180,108){$\sigma_B = { 0 }$}
\put(170,105){\line(0,-1){40}}
\put(175,105){\line(0,-1){40}}
\put(148,65){A}
\put(190,65){B}
\put(170,65){\line(-1,-1){30}}
\put(175,65){\line(1,-1){30}}
\put(170,40){C}
\put(102,35){$\sigma_A = 0$}
\put(133,25){$\sigma_C = \pi$}
\put(210,35){$\sigma_B = \pi $}
\put(180,25){$\sigma_C = 0$}
\put(95,5){\rm Figure\ 1.\ The\ String\ Star\ Product}
\end{picture}

\font\bigsymbol=cmsy10 scaled\magstep4

\begin{picture}(400,110)
\put(155,85){$\sigma = { {\pi} \over {2}}$}
\put(170,30){\oval(5,100)[t]}
\put(136,27){$\sigma = 0$}
\put(175,27){$\sigma = \pi$}
\put(105,55){ {\bigsymbol \char'163} \ \ {$\longleftrightarrow$} }
\put(100,10){\rm Figure\ 2.\ The\ String\ Integral}
\end{picture}

Unlike the three-dimensional Chern-Simons theory,
the open bosonic string theory possesses $q$-forms
for $q < 0$.
As a consequence there are on-shell gauge invariances
for gauge invariances at all levels:
$$
 \displaylines{
  \delta \Lambda_0=Q\Lambda_{-1} +
     A \ast \Lambda_{-1}+\Lambda_{-1} \ast A \quad , \cr
  \delta \Lambda_{-1} = Q\Lambda_{-2} +
     A \ast \Lambda_{-2}-\Lambda_{-2} \ast A \quad , \cr
  \vdots \cr
 }
$$
\be
  \delta \Lambda_{-n}=Q\Lambda_{-n-1} +
     A \ast \Lambda_{-n-1} -
     \left( {-1} \right)^{n+1}\Lambda_{-n-1} \ast A
    \quad ,
\label{obs gauge invariance tower}
\ee
$$
  \vdots
$$
where $\Lambda_{q}$ is a string $q$-form.
To verify these gauge invariances, note that if
$
 \Lambda_{-n+1} = Q\Lambda_{-n} +
 A \ast \Lambda_{-n} -
  \left( {-1} \right)^n \Lambda_{-n} \ast A
$
and if one changes $\Lambda_{-n}$ by
$
 \delta \Lambda_{-n} =
  Q\Lambda_{-n-1} + A \ast \Lambda_{-n-1} -
  \left( {-1} \right)^{n+1}\Lambda_{-n-1} \ast A
$
then $\Lambda_{-n+1}$ changes by
$
 \delta \Lambda_{-n+1} =
 \left( {QA+A \ast A} \right) \ast \Lambda_{-n-1} -
  \Lambda_{-n-1} \ast \left( {QA + A \ast A} \right)
$,
as a little algebra reveals.
As a consequence, this change
$\delta \Lambda_{-n+1}$ in $\Lambda_{-n+1}$ is
zero when the equations of motion in Eq.\bref{obs eom}
are invoked.
One of the confusing aspects
of quantizing the open bosonic
string is the following.
In the free theory,
the second term
in Eq.\bref{obs action}
is dropped.
Then the theory is invariant under
the tower of gauge transformations
$\delta \Lambda_{-n}=Q\Lambda_{-n-1}$
for $n \ge -1$ where
$\Lambda_{1} \equiv A$.
The gauge invariances of gauge invariances
hold off-shell because of the nilpotency of $Q$.
So the quantization of the free-theory proceeds
without any non-zero on-shell structure tensors.
In contrast,
for the interacting theory
there exist non-zero on-shell tensors
in the gauge structure.

Indeed,
let us determine the gauge structure of the theory.
The algebra is closed because
the commutator of two gauge transformations
$\Lambda_0^{(1)}$ and $\Lambda_0^{(2)}$
is equivalent to
a single gauge transformation given by
$
 \Lambda_0^{(12)} =
    \Lambda_0^{(1)} \ast \Lambda_0^{(2)}
  - \Lambda_0^{(2)} \ast \Lambda_0^{(1)}
$.
Hence, $E^{ji}_{\alpha \beta} = 0$
in Eq.\bref{com of two gens}
and $T^\gamma_{\alpha \beta}$ is the matrix
element of the $\ast$ operator
among three string states.
This is also known as the three-point vertex function
and has been computed
in refs.\ct{cst86a,samuel86a,gj87a,gj87b}.
The gauge generators $R_s$
in Eq.\bref{reduc l 1} correspond
to the covariant derivative operator $\gh D_{(-s)}$
restricted to the space of $-s$ forms,
\be
 R_s \leftrightarrow \gh D_{(-s)}
\quad ,
\label{obs R generators}
\ee
where $\gh D_{(q)}$
is defined by
\be
 \gh D_{(q)} B_{q} =
 Q B_{q} + A \ast B_{q} - (-1)^q  B_{q} \ast A
\quad ,
\label{obs covariant derivative}
\ee
when acting on any $q$-form $B_{q}$ and
$A$ is the string gauge field.
Eq.\bref{reduc l 1},
when applied to a $-s$ form ghost $\eta^{(s)}$,
reads
\be
 \gh D_{(-s+1)} \gh D_{(-s)} \eta^{(s)} =
 F \ast \eta^{(s)} - \eta^{(s)} \ast F
\quad ,
\ee
where $F$ is given in Eq.\bref{obs eom}.
Hence, the $ V^{i\alpha_{s-2}}_{s\alpha_s}$
in Eq.\bref{reduc l 1} are not zero
and are given by
the commutator of the star operator
restricted to the space of $-s$ forms:
\be
  S_{0,i} V^{i}_{s} \eta^{(s)}
  \leftrightarrow   F \ast \eta^{(s)} -
    \eta^{(s)} \ast F
\quad .
\label{obs V's}
\ee

\subsection{The Massless Relativistic Spinning Particle}
\label{ss:mrsp}

\hspace{\parindent}
In this subsection, we consider an example
with an anti-commuting gauge parameter, i.e,
a system with local supersymmetry.
Local supersymmetric theories are also known
as supergravity theories.
In such theories,
the lagrangians
and transformation laws
are usually complicated.
Since our purpose is not to review supersymmetry,
we treat one of the simplest examples:
the massless relativistic spinning particle
\ct{bcl76a,bm77a,bszvh77a}.
It is a $(0+1)$-dimensional supergravity theory.

We begin with the massless flat-space version
of the spinless relativistic particle,
namely, the theory discussed
in Sect.\ \ref{ss:srp}
and governed by the action
in Eq.\bref{srp action} with $e=1$ and $m=0$.
To obtain local supersymmetry,
one can proceed in two ways.
One can first implement general coordinate invariance
and then supersymmetry,
or
one can first implement supersymmetry
and then general coordinate invariance:

\bigskip

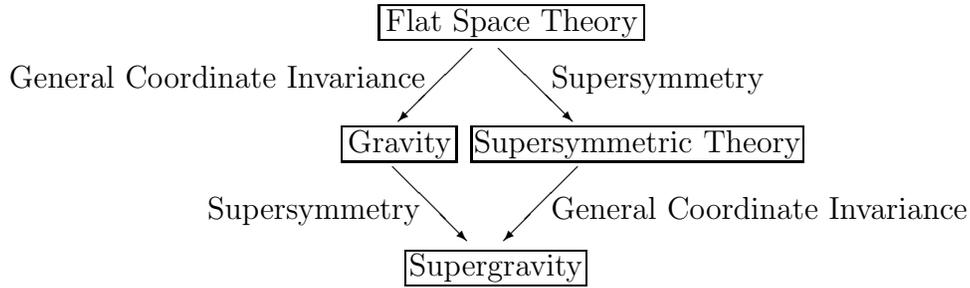
\begin{picture}(300,100)

\put(160,93) {\framebox(100,13) {Flat Space Theory}}
\put(195,90){\vector(-1,-1){28}}
\put(205,90){\vector(1,-1){28}}
\put(20,75) {General Coordinate Invariance}
\put(225,75) {Supersymmetry}
\put(146,47) {\framebox(43,13) {Gravity}}
\put(195,47) {\framebox(125,13){Supersymmetric Theory}}
\put(165,45){\vector(1,-1){28}}
\put(235,45){\vector(-1,-1){28}}
\put(225,25) {General Coordinate Invariance}
\put(95,25) {Supersymmetry}
\put(170,0) {\framebox(68,13) {Supergravity}}

\end{picture}

\medskip
\noindent
The result of such a program is a supergravity.

Consider the first step of the left path
of the above diagram.
To implement general coordinate invariance,
the Poincar\'e group must be realized locally.
In $0+1$-dimensions, the Poincar\'e group consists
of translations: $\tau \rightarrow \tau + \varepsilon$.
The local version of this transformation is
$\tau \rightarrow \tau + \varepsilon (\tau)$, that is,
$\varepsilon$ becomes a function of $\tau$.
The fields for the spinless relativistic particle
are the $x^\mu (\tau)$
of Sect.\ \ref{ss:srp}.
Under local translations they change as
$
 x^\mu \left( {\tau +
   \varepsilon \left( \tau  \right)} \right) -
 x^\mu \left( \tau  \right) =
 \varepsilon \left( \tau  \right)
  \dot x^\mu \left( \tau  \right) + \ldots
$.
Hence, the transformation law for
$x^\mu \left( \tau  \right)$ is
the one given
in Eq.\bref{srp rtrans}.

As in any gauge theory,
it is necessary to introduce a gauge field and
promote ordinary derivatives to covariant derivatives
\ct{utiyama54a,ym54a,kibble61a}:
$
 \partial_\tau \rightarrow {\cal D}_\tau
$
so that covariant derivatives transform
in the same way as fields, namely
\be
  \delta \left( {{\cal D}_\tau x^\mu
   \left( \tau  \right)} \right) =
  \varepsilon \left( \tau  \right)
  \partial_\tau \left( {{\cal D}_\tau x^\mu
    \left( \tau  \right)} \right)
\quad .
\label{trans law for cov der}
\ee
The gauge field is contained in the einbein $e$
of Sect.\ \ref{ss:srp}.
It transforms as
in Eq.\bref{srp rtrans}.
The covariant time derivative ${\cal D}_\tau $ is
\be
  {\cal D}_\tau  x^\mu \left( \tau  \right)
  \equiv {1 \over e}
   \partial_\tau x^\mu \left( \tau  \right)
\quad .
\label{cov time der}
\ee
Using Eq.\bref{srp rtrans}
it is a simple exercise to verify that
${\cal D}_\tau  x^\mu \left( \tau  \right)$ transforms
as in Eq.\bref{trans law for cov der}.

Any function $F$ of $x^\mu$ and ${\cal D}_\tau x^\mu$
transforms in the same way as $x^\mu$, i.e.,
\be
  \delta F =
 \varepsilon \left( \tau  \right)\partial_\tau  F
\quad ,
\ee
as is easily checked.
Given any such $F$, the action
\be
  \int {\dif \tau } e F
 \left( {x^\mu, {\cal D}_\tau x^\mu } \right)
\ee
is invariant since
$$
  \int {\dif \tau } \delta \left( {e F} \right) =
  \int {\dif \tau } \left( \partial_\tau
   \left( {\varepsilon e} \right) F +
  \varepsilon e\partial_\tau  F \right) =
  \int {\dif \tau } \partial_\tau
  \left( {\varepsilon e F} \right)
   \longrightarrow 0
\quad ,
$$
if $\varepsilon (\tau) $
goes to zero at $\tau = \pm \infty$.
The action
in Eq.\bref{srp action} corresponds to
\be
  F =  {1 \over 2} \left(
     {\cal D}_\tau x^\mu \left( \tau  \right)
     {\cal D}_\tau x_\mu \left( \tau  \right)  -
     m^2 \right)
\quad .
\label{F for srp}
\ee
One could also add to Eq.\bref{F for srp}
an arbitrary potential term
$V\left( {x^\mu \left( \tau  \right)} \right)$ and
the theory would still be
locally coordinate invariant;
however, the global symmetries
$x^\mu \rightarrow x^\mu + a^\mu$,
where $a^\mu$ are constants,
would be broken.
Requiring such a symmetry forces
$V\left( {x^\mu \left( \tau  \right)} \right)$
to be constant.

The above describes the general idea in constructing
theories with general coordinate invariance.
In higher-dimensions, $e$ has more components
and a spin connection must be introduced
for fields with spin.
In short,
the procedure labelled ``general coordinate invariance''
in the above diagram corresponds
to introducing gravitational gauge fields and
promoting ordinary derivatives to covariant derivatives.
For more details,
see the example
in Sect.\ \ref{ss:fqbs}.

Let us consider the process of supersymmetrizing
the massless version
of the spinless relativistic particle,
the example presented
in Sect.\ \ref{ss:srp}.
The action is given
in Eq.\bref{srp action}
with $m=0$ and $e=1$.
It is necessary to have $m=0$ because
there does not exist a supersymmetric generalization
of the $m^2$ term.
The fields of the spinless relativistic particle
are $x^\mu$ which are
functions of the proper time variable $\tau$.
To supersymmetrize the theory,
one goes from ordinary space to superspace
by enlarging the coordinate system from $\tau$
to $\tau$ and $\theta$:
$
  \tau \to \left( {\tau ,\theta } \right)
$,
where the anticommuting coordinate $\theta$ is
the supersymmetric partner of $\tau$.
Fields are then promoted to superfields
by making them functions of $\theta$:
\be
  x^\mu \left( \tau  \right)
  \to X^\mu \left( {\tau ,\theta } \right)
  \equiv
  x^\mu \left( \tau  \right) +
  \theta \psi^\mu \left( \tau  \right)
\quad ,
\label{theta exp}
\ee
where the right-hand side
of Eq.\bref{theta exp} is a Taylor series expansion
in $\theta$.
It terminates at $\theta$ because $\theta \theta =0$.
The $\theta$-component of the superfield
$X^\mu$ is $\psi^\mu$.
It is the fermionic partner of $x^\mu$.

Although our purpose is not to review supersymmetry,
we present enough background to make
this section self-contained.
For extensive reviews on supersymmetry
see refs.\ct{ggrs83a,wb83a}.
Superfields form an algebra.
This means that they can be added and multiplied.
Given two superfields
$
  A_1=a_1+\theta b_1
$
and
$
  A_2=a_2+\theta b_2
$
their sum is
$
  A_1+A_2=\left( {a_1+a_2} \right) +
  \theta \left( {b_1+b_2} \right)
$
and their product is
$
  A_1A_2=a_1a_2+a_1\theta b_2+\theta b_1a_2 =
  a_1a_2+\theta \left( {b_1a_2 +
  \left( {-1} \right)^{\epsilon
  \left( {a_1} \right)}a_1b_2} \right)
$.
Superfields can be manipulated
in a manner similar to complex numbers.
One only has to be careful about minus signs
arising from quantities that anticommute.
Supersymmetry calculus is also straightforward.
In addition to taking derivatives
with respect to $\tau$ one
can take derivatives with respect to $\theta$.
In a natural way, one defines
$
  \partial_\theta \equiv
 {\partial  \over {\partial \theta }}
$
so that
$
  \partial_\theta A = b
$
when
$
  A = a + \theta b
$.
Note that $\partial_\theta$ is a left derivative.
Integration is defined as the operation
that selects the $\theta$ component of a superfield
and serves the same function as $\partial_\theta$:
$
  \int {\dif \theta }A = \partial_\theta A = b
$.
For more on Grassmann integration
and supermanifolds,
see refs.\ct{berezin66a,bl75a,leites80a,dewitt84b,berezin87a}.

The generator
of supersymmetry transformations is denoted by $Q$
(not to be confused with the BRST operator
of the previous subsection).
It is also convenient
to use instead of $\partial_\theta$
a generalized $\theta$-derivative $D_\theta$.
More explicitly $D_\theta$ and $Q$ are
$$
  D_\theta \equiv \partial_\theta -
    i\theta \partial_\tau
\quad ,
$$
\be
  Q\equiv \partial_\theta +
   i\theta \partial_\tau
\quad .
\label{def of D and Q}
\ee
The $1$-dimensional super-algebra is
$$
  \lbrace Q, \ Q \rbrace = 2QQ =
   2i\partial_\tau  \equiv 2H
\quad ,
$$
\be
  \left[ {H,Q} \right] = 0
\quad .
\label{flat space susy algebra}
\ee
These equations are easily verified
using Eq.\bref{def of D and Q}.
In theories
for which a supersymmetry charge $Q$ exists,
the hamiltonian necessarily has a non-negative spectrum
as a consequence of $H = Q Q \ge 0$.
In addition, $ \left[ {H,Q} \right] = 0 $ implies that
there is a fermionic state for every bosonic state
and vice-versa, except possibly for a zero energy state.
The (anti)commutators involving $D_\theta$ are
$  \left\{ {Q,D_\theta } \right\}=0 $,
$  \left[ {H,D_\theta } \right]=0 $,
and
$  \left\{ { D_\theta , D_\theta } \right\} = - 2 H $.

The supersymmetry transformations are defined by
$\delta A = i\xi Q A$ where $\xi$
is an anticommuting parameter.
For $ X^\mu$, a simple calculation gives
$  \delta X^\mu =i\xi QX^\mu =i\xi \psi^\mu +
  \theta \xi \partial_\tau x^\mu  $.
Given that $\delta X^\mu$ is defined by
$
 \delta X^\mu \equiv \delta x^\mu +
 \theta \delta \psi^\mu
$,
one obtains for the transformation rules
for the components of $ X^\mu$
$$
  \delta x^\mu =i\xi \psi^\mu
\quad ,
$$
\be
  \delta \psi^\mu =\xi \partial_\tau x^\mu
\label{flat space ss trans}
\quad .
\ee
The reason for using $D_\theta$
in lieu of $\partial_\theta$ is that
$D_\theta A$ transforms as a superfield
if $A$ transforms as such,
since
$D_\theta$ commutes with $Q$:
$
  \delta \left( {D_\theta A } \right) =
  D_\theta \left( {i \xi Q A } \right) =
  i\xi Q\left( {D_\theta A } \right)
$.
The same statement holds for $\partial_\tau  A$ since
$
\left[ {\partial_\tau ,Q} \right] =
  -i \left[ {H,Q} \right] = 0
$
so that
one does not need to generalize $\partial_\tau $.

Supersymmetric-invariant actions
are constructed from the
$ \theta$ component of a superfield:
Note that
$
  \delta \left( {\left. A
    \right|_{\theta \ {\rm component}}} \right) =
  \xi \partial_\tau
  \left( {\left. A \right|_{\theta =0}} \right)
$
so that if one integrates this component over $\tau$
to obtain an action ${S}_0$,
this action is invariant if fields
fall off sufficiently fast at $\tau \to \pm \infty$.
Given that $\partial_\tau  X^\mu $, $D_\theta X^\mu$
and $X^\mu$ are all superfields
and that superfields form an algebra,
the $\theta$ component of any function of these
fields transforms as a total $\tau$ derivative:
$
  \delta \left( {\left.
  {F\left(
  {\partial_\tau  X^\mu ,D_\theta X^\mu ,X^\mu }
     \right)}
             \right|_{\theta \ {\rm component}}}
         \right) =
   \xi \partial_\tau
  \left( {\left. F \right|_{\theta =0}} \right)
$.
Since $\int {\rm d} \theta$
picks out the $\theta$ component
of a superfield, an invariant action is
\be
  { S}_0 = \int {\dif \tau \int {\dif \theta }}
  F \left(
   {\partial_\tau X^\mu ,D_\theta X^\mu ,X^\mu }
   \right)
\quad .
\label{gen flat space ss action}
\ee
The supersymmetric generalization
of the action for the massless relativistic particle
consists in taking $F$ in
\bref{gen flat space ss action} as
$
  F= {i \over 2}
  D_\theta X^\mu \partial_\tau X_\mu
$
since, when written in component fields,
it contains the correct kinetic energy term for $x^\mu$:
\be
 { S}_0 =
   {i \over 2}
 \int {\dif \tau \int {\dif \theta }}
  D_\theta X^\mu \partial_\tau X_\mu =
 \frac12 \int {\rm d} \tau
 \left( \partial_\tau x^\mu \partial_\tau x_\mu -
 i \psi^\mu \partial_\tau \psi_\mu
 \right)
\label{flat space ss action}
\quad .
\ee
It is easily checked that
Eq.\bref{flat space ss action}
is invariant under the transformations
in Eq.\bref{flat space ss trans}.
One could add interactions by using a more general
$
 F \left(
 {\partial_\tau X^\mu ,D_\theta X^\mu ,X^\mu }
 \right)
$.

The final goal of this subsection
is to implement both local translational
and local supersymmetry and
construct a supergravity theory.
Following the procedures above,
we promote the einbein $e (\tau)$
to a superfield $E$:
$
  e\left( \tau  \right) \to
  E\left( {\tau ,\theta } \right)
$
where
\be
 E\left( {\tau ,\theta } \right)
 \equiv
  e\left( \tau  \right) +
 \theta \chi \left( \tau  \right)
\quad .
\label{def of E}
\ee
Here, $\chi$ is the superpartner of $e$.
The infinitesimal gauge parameters associated
with general coordinate invariance and supersymmetry
are grouped into one superfield $\eta^t$ via
\be
  \eta^t \left( {\tau ,\theta } \right) \equiv
  \varepsilon \left( \tau  \right) +
  \theta \xi \left( \tau  \right)
\quad .
\ee
Using $\eta^t$ and
the $\theta$-partner to $\eta^t$,
given by
\be
  \eta^\theta \left( {\tau ,\theta } \right) \equiv
  iE\partial_\theta \eta^t=ie\xi \left( \tau  \right) +
  i\theta \chi \xi \left( \tau  \right)
\quad ,
\ee
one writes the transformation rule
for a superfield such as
$A\left( {\tau ,\theta } \right)$ as
\be
  \delta A =
  \eta^t \left( {\tau ,\theta }
   \right)\partial_\tau A +
  \eta^\theta \left( {\tau ,\theta }
       \right)\partial_\theta A
\quad .
\label{local susy trans law}
\ee
Notice that
Eq.\bref{local susy trans law} reduces
to the correct transformation law
$
 \delta A =
  \varepsilon \left( \tau  \right) \partial_\tau A
$
when $\xi=0$.
Furthermore,
when
$\chi \rightarrow 0$,
$\varepsilon \left( \tau  \right) \rightarrow 0$,
$e \rightarrow 1$,
and $\xi$ is not longer a function of $\tau$,
the right-hand side of
Eq.\bref{local susy trans law} becomes
$
  \left( i\xi \partial_\theta +
  \theta \xi \partial_\tau \right) A =
  i\xi Q A
$,
which is the flat space supersymmetry transformation.
Hence Eq.\bref{local susy trans law}
reduces to expected results
in these two limits.
The transformation law for $E$ must be generalized
from
$\delta e =
\partial_{\tau} \left( \varepsilon e \right)$.
As will become clear below
when we construct invariant actions,
one needs
\be
  \delta E =
  \partial_\tau \left( {\eta^tE} \right) -
  \partial_\theta \left( {\eta^\theta E} \right)
\quad .
\label{trans law for E}
\ee

With this transformation law for $E$,
it is straightforward to find
covariant derivative operators
${\cal D}_\tau$ and ${\cal D}_\theta$ that
maintain the supersymmetry transformation law
in Eq.\bref{local susy trans law}.
One requires
$
  \delta \left( {{\cal D}_\tau A } \right) =
  \eta^t\partial_\tau
  \left( {{\cal D}_\tau A } \right) +
  \eta^\theta \partial_\theta
    \left( {{\cal D}_\tau A } \right)
$
and
$
  \delta \left( {{\cal D}_\theta A } \right) =
  \eta^t\partial_\tau
  \left( {{\cal D}_\theta A } \right) +
  \eta^\theta \partial_\theta
  \left( {{\cal D}_\theta A } \right)
$.
A solution is
$$
  {\cal D}_\tau A \equiv
  {1 \over E}\left( {\partial_\tau A -
  i\partial_\theta E\partial_\theta A } \right)
\quad ,
$$
\be
  {\cal D}_\theta A \equiv \partial_\theta A -
  i\theta {\cal D}_\tau A
\quad .
\label{def of susy cov ders}
\ee
When acting on the superfield $X^\mu$,
one finds
$$
  {\cal D}_\tau X^\mu  =
  {1 \over e}
    \left( {\partial_\tau x^\mu -i\chi \psi^\mu }
   \right) +
  {\theta  \over e}\left( {\partial_\tau \psi^\mu -
  {\chi  \over e}\partial_\tau x^\mu } \right)
\quad ,
$$
\be
  {\cal D}_\theta X^\mu  =
  \psi^\mu - i{\theta  \over e}
  \left( {\partial_\tau x^\mu -
  i\chi \psi^\mu } \right)
\quad .
\label{susy component cov ders}
\ee
Again, superfields form an algebra
so that any function $F$
of ${\cal D}_\tau X^\mu$,
${\cal D}_\theta X^\mu$
and $X^\mu$
transforms as a superfield in curved space:
$$
  \delta F\left(
  {{\cal D}_\tau X^\mu ,{\cal D}_\theta X^\mu ,X^\mu }
         \right) =
  \eta^t \partial_\tau F + \eta^\theta \partial_\theta F
\quad .
$$

It is now straightforward to show that
\be
 { S}_0 = \int {\dif \tau }\int {\dif \theta } E  F
\quad ,
\label{gen susy action}
\ee
is an invariant action since
$$
  \delta \int {\dif \tau }\int {\dif \theta } E  F =
$$
$$
  \int {\dif \tau }\int {\dif \theta }
 \left( {\left( {\partial_\tau
  \left( {\eta^t E} \right) -
  \partial_\theta
  \left( {\eta^\theta E} \right)} \right) F +
  E \left( {\eta^t\partial_\tau F +
  \eta^\theta \partial_\theta F} \right)}
 \right)
$$
$$
  = \int {\dif \tau }\int {\dif \theta }
  \left( {\partial_\tau \left( {\eta^t EF} \right) -
  \partial_\theta \left( {\eta^\theta EF} \right)}
  \right)
$$
is a total derivative.
The transformation rule for $E$
in Eq.\bref{trans law for E},
including the minus sign on the right-hand side,
was deliberately chosen so that $EF$ would lead
to an invariant super-lagrangian.
Taking
$
 F = { i \over 2}
 {\cal D}_\theta X^\mu {\cal D}_\tau X_\mu
$
produces the extension
of Eq.\bref{flat space ss action}
with general coordinate invariance:
$$
  { S}_0 = {i \over 2}
  \int {\dif \tau }\int {\dif \theta }
  E{\cal D}_\theta X^\mu {\cal D}_\tau X_\mu =
$$
\be
  {1 \over 2}\int {\dif \tau }
  \left( {{1 \over e}
  \partial_\tau x^\mu \partial_\tau x_\mu -
  i\psi^\mu \partial_\tau \psi_\mu -
  {{2i\chi } \over e}
  \psi^\mu \partial_\tau x_\mu } \right)
\quad .
\label{mrsp action}
\ee

By construction, one is guaranteed
that Eq.\bref{mrsp action} is invariant
under the transformations
in Eqs.\bref{local susy trans law} (when $A = X^\mu$)
and \bref{trans law for E}.
In components, these equations read
$$
  \delta x^\mu =
  \varepsilon \partial_\tau x^\mu +ie\xi \psi^\mu
\quad ,
$$
$$
  \delta \psi^\mu =
  \varepsilon \partial_\tau \psi^\mu +
  \xi \left(
  {\partial_\tau x^\mu -i\chi \psi^\mu } \right)
\quad ,
$$
\be
  \delta e =
  \partial_\tau \left( {\varepsilon e} \right) +
    2ie\xi \chi
\label{component susy trans}
\quad ,
\ee
$$
  \delta \chi =
  \partial_\tau \left( {\varepsilon \chi +
    \xi e} \right)
\quad .
$$
A good exercise is to check explicitly that
Eq.\bref{mrsp action} is invariant under
Eq.\bref{component susy trans}.

Since the transformations rules
in Eqs.\bref{local susy trans law}
and \bref{trans law for E} are defined
independent of an action,
the algebra should close off-shell.
The only thing to check is that
the commutator of two transformations produces
an effect on $E$ similar to the one on $X^\mu$.
A little algebra reveals that
\be
 \left[ \delta_1 , \  \delta_2 \right] \phi^i =
   \delta_{12} \phi^i
\quad ,
\label{sg algebra}
\ee
where $\phi^i = x^\mu$, $\psi^\mu$, $e$ or $\chi$,
and $\delta_{12}$
corresponds to a transformation with
$$
 \varepsilon_{12} =
     \varepsilon_2 \dot \varepsilon_1
   - \varepsilon_1 \dot \varepsilon_2
  + 2 i e \xi_2 \xi_1
\quad ,
$$
\be
\xi_{12} =
  \varepsilon_2 \dot \xi_1 -
  \varepsilon_1 \dot \xi_2 +
 \dot \varepsilon_1  \xi_2 -
 \dot \varepsilon_2  \xi_1 +
  2 i \chi \xi_2 \xi_1
\quad .
\label{sg commutator parameter}
\ee
The algebra closes off-shell but has field-dependent
structure constants due to
the presence of $e$ and $\xi$
in Eq.\bref{sg commutator parameter}.
In the non-supersymmetric limit
for which $\xi_1 = \xi_2 = \chi =0$,
the spinless-relativistic-particle algebra in
Eq.\bref{srp commutatora}
is reproduced.
In the flat-space limit for which $\chi=0$, $e=1$,
and gauge parameters
do not depend on $\tau$,
Eq.\bref{sg commutator parameter} reduces
to the flat space supersymmetry algebra
in Eq.\bref{flat space susy algebra}.

In four dimensions, the construction
of supergravity theories
is considerably more complicated.
Several gravitational superfields enter.
The interested reader can find introductions
to four-dimensional supergravity
in refs.\ct{vannieuwenhuizen81a,ggrs83a,wb83a}.
The formalism presented here is the one-dimensional analog
of the covariant-$\Theta$ formalism of Chapters XX and XXI
of ref.\ct{wb83a}.

\subsection{The First-Quantized Bosonic String}
\label{ss:fqbs}

\hspace{\parindent}
The open bosonic string is
the two-dimensional generalization
of the relativistic particle considered
in Sect.\ \ref{ss:srp}.
It is a general-coordinate-invariant theory.
As such, it can be regarded as a certain type
of two-dimensional gravity theory.

Let us review the construction
of general-coordinate-invariant theories
involving some scalar fields $A^i$.
In what follows, all the $A^i$ behave similarly
and we drop the superscript index $i$.
We use the vielbein formulation.
The gauge fields which implement local coordinate
invariances are vielbeins
${e_{a}}^m$,
where $a = 0, \dots , d-1$
and $m =  0, \dots , d-1$.
These indices take on $d$ values corresponding
to time and space coordinates.
The index $m$ is the Einstein index associated
with local coordinate transformations.
The index $a$ is the tangent-space Lorentz index
used for implementing local Lorentz transformations.
The inverse metric $g^{m n}$
is the inverse of the metric $g_{m n}$:
$
  g^{m p} g_{p n} \equiv \delta^m_n
$.
It is related to the vielbein via
$g^{m n}\equiv \eta^{ab} {e_a}^m {e_b}^n$,
where $\eta^{ab}$ is the flat-space metric.
For $d=2$,
$
  \eta^{a b} = \eta_{a b} =
    \left( {
    \matrix { -1 & 0 \cr
               0 & 1 \cr }
     } \right)
$.
Lorentz indices $a$, $b$, $c$, etc.\ are raised
and lowered using $\eta^{ab}$ and $\eta_{ab}$.
Einstein indices are raised and lowered using
$g^{m n}$ and $g_{m n}$.
Hence,
the inverse vielbein ${e^a}_m$ is
$
 {e^a}_m \equiv g_{m n} \eta^{ab}{e_b}^n
$
and satisfies
${e^a}_n {e_a}^m = \delta^m_n$ and
${e^a}_m {e_b}^m = \delta^a_b$,
as one can easily check.
The metric can be expressed
in terms of ${e^a}_m$ via
$g_{m n} = \eta_{ab}{e^a}_m {e^b}_n$.

Under an infinitesimal local translation
in the $m$-th direction
by $\varepsilon^m$,
a scalar field $A$
changes by
$
  \delta A =
  A \left(x^m + \varepsilon^m \right) -
  A \left(x^m \right)
$ or
\be
\delta A = \varepsilon^m \partial_m A
\quad .
\label{gc var of scalar}
\ee
One would like to find covariant derivatives
${\cal D}_a A$ that transform in the same way:
$
\delta ( {\cal D}_a A) =
\varepsilon^m \partial_m
\left( {{\cal D}_a A} \right)
$.
The vielbein ${e_a}^m$ allows one to do this.
Letting
\be
  {\cal D}_a A = {e_a}^m\partial_m A
\quad ,
\label{gc cov der}
\ee
and requiring the correct transformation law
for ${\cal D}_a A$,
one finds that ${e_a}^m$ must transform as
\be
  \delta {e_a}^m =
 \varepsilon^n \partial_n {e_a}^m -
 {e_a}^n \partial_n \varepsilon^m
\quad .
\label{gc var of vielbein}
\ee
Any function $F$ of $A$ and ${\cal D}_a A$
will transform like $A$:
$
  \delta F \left( {A,{\cal D}_a A} \right)
  = \varepsilon^m \partial_m F
$.
Suppose one can find a density $e$
whose transformation law is
\be
  \delta e =
 \partial_m \left( {\varepsilon^m e} \right)
\quad .
\label{gc var of density}
\ee
Then $eF$ transforms as a total derivative
$
  \delta eF = \partial_m
 \left( {\varepsilon^m eF} \right)
$
and is a candidate for an action density.
When the action density
is integrated over space-time,
it is invariant
under local coordinate transformations
as long as gauge parameters vanish at infinity.

The density $e$ can be constructed
from the determinant
of the vielbein $\det \left( {{e_.}^.} \right)$.
Note that $\det \left( {{e_.}^.} \right)$
transforms as
$$
  \delta \det \left( {{e_.}^.} \right) =
  \delta {e_a}^m \times
 {\mbox {cofactor matrix of }}  {e_a}^m =
 \delta {e_a}^m {e^a}_m \det \left( {{e_.}^.} \right)
\quad .
$$
The derivative of $\det \left( {{e_.}^.} \right)$ is
$$
  \partial_n \det \left( {{e_.}^.} \right)
  = \partial_n {e_a}^m \times
 {\mbox {cofactor matrix of }}  {e_a}^m
   = \partial_n {e_a}^m
  {e^a}_m \det \left( {{e_.}^.} \right)
\quad .
$$
Using these results, a short calculation reveals that
\be
  e \equiv {1 \over {\det \left( {{e_.}^.} \right)}} =
  \det \left( {{e^.}_.} \right)
\label{def of density}
\ee
transforms as
in Eq.\bref{gc var of density}.
The relativistic particle,
presented as the example
in Sect.\ \ref{ss:srp},
is a $(0+1)$-dimensional theory
where the field $e$ in that section
is the same as the density $e$ defined here:
$
  e = \det \left( {{e^.}_.} \right) = {e^0}_0 = 1 / {{e_0}^0}
$.

In contrast to the metric $g_{m n}$,
which has $d(d+1)/2$ degrees of freedom,
the vielbein ${e_a}^m$ has $d^2$ degrees of freedom.
To compensate for this difference,
one requires that the theory be invariant
under local Lorentz transformations
that act on Lorentz indices $a$, $b$, etc..
The transformation laws are
\be
  \delta {e_a}^m =
 {\left( {\varepsilon M} \right)_a}^b {e_b}^m
\ , \quad \quad
  \delta A = 0
\label{local Lorentz trans}
\quad ,
\ee
where ${\left( {\varepsilon M} \right)_a}^b$
stands for
$
  {1 \over 2}
 {( \varepsilon^{cd} M_{cd} )_a}^b
$.
Here, the $d(d-1)/2$ parameters $\varepsilon^{cd}$ satisfy
$
  \varepsilon^{cd} = - \varepsilon^{dc}
$
and characterize the size
of the infinitesimal-local-Lorentz transformations.
The $d(d-1)/2$ matrix generators $ M_{ab}$ satisfy
the Lorentz algebra
$
  \left[ {M_{ab}, M_{cd}} \right] =
  -\eta_{ac} M_{bd} + \eta_{ad} M_{bc} +
  \eta_{bc} M_{ad} - \eta_{bd} M_{ac}
$
and are antisymmetric in the lower indices $c$ and $d$:
$
  M_{cd} = - M_{dc}
$.
When we write
${ (  M_{cd} )_a}^b$,
$a$ and $b$ are matrix indices whereas
$c$ and $d$ label the different Lorentz transformations:
$ M_{cd}$ produces an infinitesimal transformation
in the $c$--$d$ plane.
One explicit realization of $ M_{cd}$ is
$
  { ( M_{cd} )_a}^b =
  \eta_{ac} \delta_d^b - \eta_{ad} \delta_c^b
$.
The extra $d(d-1)/2$ gauge invariances guarantees
that the physical numbers of degrees of freedom
in ${e_a}^m$
and $g_{m n}$ are the same.

Given a number of fields
$V_a^i$,
with Lorentz indices,
local invariants are constructed
by using the flat space metric $\eta^{ab}$ via
$
  \eta^{ab} V_a^i V_b^j
$
or the flat space antisymmetric
$\varepsilon^{a_0 a_1 \ldots a_{d-1}}$ symbol via
$
  \varepsilon^{a_0 a_1 \ldots a_{d-1}}
  V_{a_0}^{i_0} V_{a_1}^{i_1} \ldots
    V_{a_{d-1}}^{i_{d-1}}
$.
A locally general coordinate and
Lorentz invariant action
for a scalar field $A$ is
\be
  { S}_0 = - {1 \over 2}\int {\dif t \ \dif^{d-1}x} \ e
  \eta^{ab} {\cal D}_a A {\cal D}_bA
\quad .
\label{gen rel scalar action}
\ee

For the rest of this subsection, we set $d=2$.
When the dimension of spacetime is $1+1$,
the action
Eq.\bref{gen rel scalar action} is
also classically invariant
under local conformal transformations
that
scale the vielbein by a spacetime dependent function
$\varepsilon$:
\be
  \delta {e_a}^m = \varepsilon {e_a}^m
  \ , \quad \quad
  \delta A = 0
\quad .
\label{conformal trans}
\ee
We thus have three types of local transformations:
translations, Lorentz rotations and scaling.
Let us respectively associate
$\varepsilon^.$, $\varepsilon^{..}$ and
$\varepsilon$ with each of these.
In other words,
we use the number of indices on the infinitesimal
parameter to distinguish the transformations.
The algebra of gauge transformations
is as follows.
The commutator of two local translations
is a local translation so that they form a subalgebra.
Likewise the commutator
of two local Lorentz transformations
is a local Lorentz transformation.
The commutator of a translation and
a Lorentz rotation is
a Lorentz rotation.
Local translations and Lorentz rotations
also form a subalgebra.
Local scaling transformations commute.
They also commute with local Lorentz transformations.
Finally, the commutator of a scaling transformation
and a translation produces a scaling transformation.
These last few statements correspond to
$$
  \left[ {\delta \left( {\varepsilon_1^.} \right) ,
   \delta \left( {\varepsilon_2^.} \right)}
  \right] =
   \delta \left( {\varepsilon_{12}^.} \right)
 \ , \quad \quad {\rm where} \quad
  \varepsilon_{12}^m = \varepsilon_2^n
   \partial_n \varepsilon_1^m -
   \varepsilon_1^n \partial_n \varepsilon_2^m
\quad ,
$$
$$
  \left[ {\delta \left( {\varepsilon_1^{..}} \right),
  \delta \left( {\varepsilon_2^{..}} \right)}
   \right] =
   \delta \left( {\varepsilon_{12}^{..}} \right)
\ , \quad \quad {\rm where} \quad
  \varepsilon_{12}^{ab} =
   \left( {\varepsilon_1^{ac}\varepsilon_2^{bd} -
   \varepsilon_2^{ac}\varepsilon_1^{bd}} \right)\eta_{cd}
\quad ,
$$
$$
  \left[ {\delta \left( {\varepsilon_1^{..}} \right),
  \delta \left( {\varepsilon_2^.} \right)} \right] =
  \delta \left( {\varepsilon_{12}^{..}} \right)
\ , \quad \quad {\rm where} \quad
  \varepsilon_{12}^{ab} =
  \varepsilon_2^n \partial_n \varepsilon_1^{ab}
\quad ,
$$
\be
  \left[ {\delta \left( {\varepsilon_1} \right),
  \delta \left( {\varepsilon_2} \right)} \right]=0
\quad ,
\label{bosonic string algebra}
\ee
$$
  \left[ {\delta \left( {\varepsilon_1} \right),
   \delta
  \left( {\varepsilon_2^{..}} \right)} \right] = 0
\quad ,
$$
$$
  \left[ {\delta \left( {\varepsilon_1} \right),
  \delta \left( {\varepsilon_2^.} \right)} \right] =
  \delta \left( {\varepsilon_{12}} \right)
\ , \quad \quad {\rm where} \quad
  \varepsilon_{12} = \varepsilon_2^n
     \partial_n \varepsilon_1
\quad .
$$
Equation \bref{bosonic string algebra}
is the gauge algebra
for the bosonic string.
It is a closed irreducible algebra
with field-independent
commutator structure constants
and hence a Lie algebra.

The scalar fields for the $D=26$ bosonic string
are target space coordinate $X^\mu$
where $\mu$ ranges from $0$ to $25$.
The action is Eq.\bref{gen rel scalar action}
for each of these fields
\be
  { S}_0 =
  - {1 \over 2}
   \int {\dif \tau }\,\int_{0}^{\pi} {\dif \sigma } e
   \eta^{ab} \eta_{\mu \nu }
   {\cal D}_a X^\mu {\cal D}_b X^\nu
\quad ,
\label{bosonic string action}
\ee
where
$\eta_{\mu \nu}$ is the flat-space $25+1$ metric
with signature
$\left( { -1,\ 1,\ 1,\ \dots \ ,\ 1 } \right)$,
${\cal D}_a$ is given
in Eq.\bref{gc cov der},
and
$   e =
   { {1} / { \left( {e_\tau}^\tau {e_\sigma}^\sigma -
              {e_\tau}^\sigma {{e_\sigma}^\tau } \right) } }
$.
The world-sheet coordinates are denoted
by $\tau$ and $\sigma$
instead of $t$ and $x$.
Hence the first-quantized bosonic string
is a scale-invariant two-dimensional gravity theory
on a finite spatial region
since $\sigma$ ranges from $0$ to $\pi$.
For the closed bosonic string,
the fields $X^\mu$ are periodic in $\sigma$,
whereas for the open bosonic string,
one requires
$\partial_\sigma X^\mu$ to vanish at
$\sigma = 0$ and $\sigma = \pi$.
Eq.\bref{bosonic string action} is invariant
under the general coordinate transformations
in Eqs.\bref{gc var of scalar} and
\bref{gc var of vielbein},
under the Lorentz
gauge transformations
in Eq.\bref{local Lorentz trans},
and under the local scale transformations
in Eq.\bref{conformal trans}
where $A$ stands for $X^\mu$;
however,
the local translation parameters $\varepsilon^{m}$
must obey the same boundary conditions as $X^\mu$.
It is a useful exercise
to check explicitly the gauge invariances
of Eq.\bref{bosonic string action}.

There are four gauge invariances, two translations,
one Lorentz boost and one scaling transformation.
It is necessary for the target space to be
$26$ dimensions in order to avoid
an anomaly in one of these invariances.
This is discussed
in Sect.\ \ref{ss:aobs}.
Hence in $D=26$, the components of ${e_a}^m$
can be fixed to constant values.
Often
$
  {e_a}^m=
 \left( {\matrix{ 1 & 0 \cr
                  0 & 1 \cr}}
  \right)
$ is used.
Then Eq.\bref{bosonic string action}
becomes a free field theory.
However, in gauge fixing,
one should introduce ghosts.
These ghosts play a somewhat minor role
in the first-quantized theory.
However, in the second-quantized formulation,
namely the string field theory
of Sect.\ \ref{ss:obsft},
the ghosts as well as $X^\mu ( \sigma )$
become coordinates of the string field $A$
in Eq.\bref{obs action}.

Eq.\bref{bosonic string action} is classically
equivalent to the usual Nambu-Goto action
\ct{nambu70a,goto71a}.
Using the equations of motion of ${e_a}^m$
to eliminate the vielbein,
one finds,
after some algebra, that
$$
  {1 \over 2}e\eta^{ab}\eta_{\mu \nu }
   {\cal D}_aX^\mu {\cal D}_bX^\nu \to
   \sqrt {\left(
   {\partial_\tau X^\mu
       \partial_\sigma X_\mu } \right)^2
   - (\partial_\tau X^\mu \partial_\tau X_\mu)
    (\partial_\sigma X^\nu \partial_\sigma X_\nu) }
\quad ,
$$
which is the familiar Nambu-Goto action density.

\vfill\eject

\section{The Field-Antifield Formalism}
\label{s:faf}

\hspace{\parindent}
Consider the classical system defined
in Sect.\ \ref{s:ssgt},
described by the action $S_0 [ \phi^i ]$
and having gauge invariances.
The ultimate goal is
to quantize this theory in a covariant way.
The field-antifield formalism
was developed to achieve this aim.
In this section, we present the field-antifield formalism
at the classical level.
In brief, it involves five steps:

\medskip

\begin{quote}

(i) The original configuration space,
consisting of the $\phi^i$,
is enlarged to include additional fields such as
ghost fields, ghosts for ghosts, etc..
One also introduces the antifields of these fields.

\medskip
(ii) On the space of fields and antifields,
one defines an odd symplectic structure $( \ , \ )$
called the antibracket.

\medskip
(iii) The classical action $S_0$ is extended to
include terms involving ghosts and antifields and is
denoted by $S$.

\medskip
(iv) The {\it classical master equation}
is defined to be $(S,\ S)=0$.

\medskip
(v) Finally, one finds solutions
to the classical master equation
subject to certain boundary conditions.

\end{quote}

\medskip

The key result of this section
is that the solution in steps (iv) and (v)
leads to a set of equations
containing all relations defining the gauge algebra
and its solution.
The action $S$ is the generating functional
for the structure functions.
Hence, the field-antifield formalism is a compact
and efficient way of obtaining
the gauge structure derived in Sect.\ \ref{s:ssgt}.

\subsection{Fields and Antifields}
\label{ss:fa}

\hspace{\parindent}
Suppose a theory is irreducible
with $m_0$ gauge invariances.
Then, at the quantum level, $m_0$ ghost fields
are needed.
As in Sect.\ \ref{ss:gs},
it is useful to introduce
these ghost fields at the classical level.
Hence, the field set $\Phi^A$ is
$\Phi^A=\left\{\phi^i, \ {\cal C}^{\alpha_0}_0 \right\}$
where $\alpha_0=1,\ldots,m_0$.
If the theory is first-stage reducible,
there are gauge invariances for gauge invariances
and hence there are ghosts for ghosts.
If there are $m_1$ first-level gauge invariances
then, to the above set,
one adds the ghost-for-ghost fields
$ {\cal C}^{\alpha_1}_1 $ where
$\alpha_1=1,\ldots,m_1$.
In general for an $L$-th stage reducible theory,
the set of fields $\Phi^A$, $A=1, \ldots , N$, is
\be
   \Phi^A=\left\{ \; \phi^i, \  {\cal C}^{\alpha_s}_s; \ \
   s=0,\ldots,L; \ \  \alpha_s=1,\ldots,m_s\right\}
\quad .
\label{field set}
\ee
An additive conserved charge,
called ghost number,
is assigned to each of these fields.
The classical fields $\phi^i$ are assigned ghost number zero,
whereas ordinary ghosts have ghost number one.
Ghosts for ghosts,
i.e., level-one ghosts,
have ghost number two, etc..
Similarly, ghosts have opposite statistics
of the corresponding gauge parameter,
but ghosts for ghosts have the same statistics
as the gauge parameter, and so on,
with the statistics alternating for higher-level ghosts.
More precisely,
\be
   {\rm gh} \left[ {\cal C}^{\alpha_s}_s \right] = s+1 \ ,\quad\quad
   \eps({\cal C}^{\alpha_s}_s) =
    \eps_{\alpha_s}+s+1  \ ( {\rm mod \ } 2 )
\quad .
\label{ghost numbers and statistics}
\ee
Next, one introduces an antifield $\Phi^*_A$,
$A=1, \ldots , N$,
for each field $\Phi^A$.
The ghost number and statistics of $\Phi^*_A$ are
\be
   {\rm gh} \left[ \Phi^*_A \right] =
 - {\rm gh} \left[ \Phi^A   \right] - 1 \ , \quad \quad
   \eps \left( \Phi^*_A \right) =
    \eps \left( \Phi^A   \right) +1
   \ ( {\rm mod \ } 2 )
\quad ,
\label{antifield ghost numbers and statistics}
\ee
so that the statistics of $\Phi^*_A$
is opposite to that of $\Phi^A$.

At this stage,
the antifields do not have
any direct physical meaning.
They are only used as a mathematical tool
to develop the formalism.
However, they can be interpreted
as source coefficients
for BRST transformations.
This is made clear
from the discussion of the effective action.
See Sect.\ \ref{ss:eazje}.
For computing correlation functions, $S$-matrix elements
and certain quantities,
antifields are eliminated by a gauge-fixing procedure.

The set of fields and antifields just introduced
is called the {\it classical basis}.%
{\footnote{
This definition differs somewhat
from the one
of refs.\ct{tp93a,vp94a}.
}}
When gauge-fixing is considered
in Sect.\ \ref{s:gff},
another basis,
known as the gauge-fixed basis,
can be introduced.
Then, the implications
of the classical master equation change.
This is discussed
in Sect.\ \ref{ss:gfb}.

\subsection{The Antibracket}
\label{ss:a}

\hspace{\parindent}
In the space of fields and antifields, the antibracket
ref.\ct{zinnjustin75a,bv81a}
is defined by
\be
 (X,Y) \equiv \frac{\partial_rX}{\partial\Phi^A}\frac{\partial_lY}
  {\partial\Phi^*_A}-\frac{\partial_rX}{\partial\Phi^*_A}
  \frac{\partial_lY}{\partial\Phi^A}
\quad .
\label{antibracket def}
\ee
Many properties of $( X , Y )$ are
similar to a graded version of the Poisson bracket,
with the grading of $X$ and $Y$
being $\epsilon_X+1$ and $\epsilon_Y+1$
instead of $\epsilon_X$ and $\epsilon_Y$.
The antibracket satisfies
$$
 (Y,X) = -(-1)^{(\epsilon_X+1)(\epsilon_Y+1)}(X,Y)
\quad ,
$$
$$
  \left( \left( X, Y \right) , Z \right) +
  (-1)^{( \epsilon_X + 1 )( \epsilon_Y + \epsilon_Z )}
  \left( \left( Y, Z \right) , X \right) +
  (-1)^{( \epsilon_Z + 1 )( \epsilon_X + \epsilon_Y )}
  \left( \left( Z, X \right) , Y \right) = 0
\ ,
$$
$$
 {\rm gh} [(X,Y)] = {\rm gh}[X] + {\rm gh}[Y] + 1
\quad ,
$$
\be
 \epsilon[(X,Y)] =
  \epsilon_X + \epsilon_Y + 1
  \ ( {\rm mod \  } 2 )
\quad .
\label{antibracket properties}
\ee
The first equation says that $( \ , \ )$ is graded antisymmetric.
The second equation shows that $( \ , \ )$ satisfies
a graded Jacobi identity.
The antibracket ``carries'' ghost number one and
has odd statistics.
{}From these properties
and the definition of right and left derivatives,
one concludes that
$$
 (B,B) = 2\frac{\partial_rB}{\partial\Phi^A}\frac{\partial_lB}
            {\partial\Phi^*_A} \ , \quad\quad
            \mbox{\rm for $B$ bosonic}
\quad ,
$$
$$
 (F,F) = 0 \ , \quad\quad
            \mbox{\rm for $F$ fermionic}
\quad ,
$$
\be
    ((X,X),X) =  0 \ , \quad\quad
         \mbox{\rm for any $X$}
\quad .
\label{() properties}
\ee
The above expression for $(B,B)$ is derived in Appendix A.
The antibracket $(X,Y)$ is also a graded derivation
with ordinary statistics for $X$ and $Y$:
$$
  ( X , YZ ) = ( X , Y ) Z + (-1)^{\eps_Y \eps_Z } ( X , Z ) Y
\quad ,
$$
\be
  ( XY , Z ) = X ( Y , Z ) + (-1)^{\eps_X \eps_Y } Y ( X , Z )
\quad .
\label{bracket derivation}
\ee

The antibracket defines an odd symplectic structure
because it can be written as
\be
 (X,Y) = \frac{\partial_r X}{\partial z^a } \zeta^{ab}
         \frac{\partial_l Y}{\partial z^b }
\ , \quad \quad {\rm where } \quad
  \zeta^{ab} \equiv \left(
  \begin{array}{cc}
   0 & \delta^A_B\\
   -\delta^A_B & 0
  \end{array}\right)
\quad ,
\label{sym form of antibracket}
\ee
when one groups the fields and antifields collectively
into $z^a$: $z^a = \{\Phi^A , \Phi^*_A \}$,
$a = 1, \dots , 2N$.
The expression for the antibracket
in Eq.\bref{sym form of antibracket}
is sometimes useful in abstract proofs.
The antibracket formalism
can be developed
in an arbitrary coordinate system,
in which case $\zeta^{ab}$
is replaced
by an odd closed field-dependent two-form.
For more details,
see Sect.\ \ref{ss:afgc}.
However, locally there always exists a basis
for which $\zeta^{ab}$ is of the form
of Eq.\bref{sym form of antibracket}
\ct{schwarz93a}.

The antibracket in the space of fields and antifields
plays a role analogous to the Poisson bracket.
Whereas the Poisson bracket is used at the classical level
in a hamiltonian formulation,
the antibracket is used at the classical or quantum level
in a lagrangian formalism.
One can use the antibracket in a manner similar to the
Poisson bracket.
The antifield $\Phi^*_A$ can be thought of
as the conjugate variable to $\Phi^A$ since
\be
 \left( \Phi^A , \Phi^*_B \right) = \delta^A_B
\quad .
\label{canonical relation}
\ee
Infinitesimal canonical transformations
\ct{bv81a}
in which $\Phi^A \rightarrow \bar \Phi^A$ and
$\Phi^*_A \rightarrow \bar \Phi^*_A$,
where $\bar \Phi^A$ and $\bar \Phi^*_A$ are functions
of the $\Phi$ and $\Phi^*$,
are defined by
\be
  \bar \Phi^A = \Phi^A +
    \varepsilon \left( \Phi^A , F \right)
    + O(\varepsilon^2)
\ , \quad \quad
  \bar \Phi^*_A = \Phi^*_A +
    \varepsilon \left( \Phi^*_A , F \right)
    + O(\varepsilon^2)
\quad ,
\label{canonical transformation}
\ee
where $F$ is an arbitrary function of the fields and antifields
with ${\rm gh} [F] = - 1 $ and $ \eps (F) = 1 $.
A short calculation
reveals that the canonical structure
in Eq.\bref{canonical relation} is maintained:
\be
 \left( \bar \Phi^A , \bar \Phi^*_B  \right) = \delta^A_B
    + O(\varepsilon^2) \ ,
\quad \quad
 \left( \bar \Phi^A , \bar \Phi^B  \right) = 0
    + O(\varepsilon^2) \ ,
\quad \quad
 \left( \bar \Phi^*_A , \bar \Phi^*_B  \right) = 0
    + O(\varepsilon^2)
\quad .
\label{bracket structure}
\ee
Canonical transformations preserve the antibracket, i.e.,
$$
 \frac{\partial_r X}{\partial \bar \Phi^A}\frac{\partial_lY}
  {\partial \bar \Phi^*_A}-\frac{\partial_rX}{\partial \bar \Phi^*_A}
  \frac{\partial_lY}{\partial \bar \Phi^A}
  = \frac{\partial_rX}{\partial\Phi^A}\frac{\partial_lY}
  {\partial\Phi^*_A}-\frac{\partial_rX}{\partial\Phi^*_A}
  \frac{\partial_lY}{\partial\Phi^A}     + O(\varepsilon^2)
\quad .
$$
Under Eq.\bref{canonical transformation},
an arbitrary scalar function $G$ of fields and antifields
transforms as
$G \to G + \delta G$ where
\be
  \delta G = \varepsilon ( G , F ) + O(\varepsilon^2)
\quad .
\label{canonical trans on function}
\ee
An alternative approach
\ct{vt82a,bv84a}
uses a function
which is a combination of original and transformed fields,
$
  F_2 [ {\Phi ,\tilde \Phi^*} ]
$,
to produce a canonical transformation via
\be
  \tilde \Phi^A =
  {{\partial_l F_2\left[ {\Phi ,\tilde \Phi^*} \right] }
   \over {\partial \tilde \Phi_A^*}}
\quad ,\quad \quad
  \Phi_A^* ={{\partial_r F_2
   \left[ {\Phi ,\tilde \Phi^*} \right] } \over {\partial \Phi^A}}
\quad ,
\label{f2 can trans}
\ee
where
$\epsilon \left( F_2 \right) = 1 $ and
$ {\rm gh} \left[ F_2 \right] = - 1$.
Of course, one must require this change
of variables to be non-singular.
The advantage of using $F_2$
is that the antibracket structure is exactly preserved
\be
  \left( {\tilde \Phi^A,\tilde \Phi_B^*} \right) =
  \delta_B^A \quad ,\quad \quad
   \left( {\tilde \Phi^A,\tilde \Phi^B} \right) =
  \left( {\tilde \Phi_A^*,\tilde \Phi_B^*} \right) = 0
\quad .
\label{F2 preserve bracket}
\ee
so that $F_2$ generates a finite transformation.
A proof of
Eq.\bref{F2 preserve bracket}
is given
in ref.\ct{tnp90a}.
In using Eq.\bref{f2 can trans},
one must solve for
$\Phi^A$ and $\Phi^*_A$
in terms of
$\tilde \Phi^A$ and $\tilde \Phi^*_A$, i.\ e.,
$\Phi^A = \bar \Phi^A [ \tilde \Phi , \tilde \Phi^* ]$
and
$\Phi^*_A = \bar \Phi^*_A [ \tilde \Phi , \tilde \Phi^* ]$.
Then the canonical transformation
corresponds to the shifts
$\Phi^A \to \bar \Phi^A [ \Phi , \Phi^* ]$
and
$\Phi^*_A \to \bar \Phi^*_A [ \Phi , \Phi^* ]$.
Infinitesimal transformations are recovered using
\be
  F_2 = \tilde \Phi_A^* \Phi^A +
  \varepsilon f_2\left[ {\Phi ,\tilde \Phi^*} \right]
\quad .
\label{infinitesimal f2}
\ee
When $f_2 = - F$,
Eq.\bref{canonical transformation}
is reproduced to order $\varepsilon$.

The role of the antibracket at quantum level
is discussed
in Sect.\ \ref{s:qea}.

\subsection{Classical Master Equation and Boundary Conditions}
\label{ss:cmebc}

\hspace{\parindent}
Let $S[ \Phi, \Phi^* ]$ be an arbitrary functional of
fields and antifields with
the dimensions of action and
with ghost number zero and even statistics:
$\eps (S) = 0$ and ${\rm gh} [S] = 0$.
The equation
\be
   (S,S)  = 2{{\partial_r S}
  \over {\partial \Phi^A}}{{\partial_l S}
   \over {\partial \Phi_A^*}} =0
\label{master equation}
\ee
is called the {\it classical master equation}.
This simple looking equation is the main topic
of this subsection.

Not every solution to Eq.\bref{master equation} is of interest.
It is necessary to satisfy certain boundary conditions.
A relevant solution plays a double role.
On one hand, a solution $S$ is
the generating functional for the structure
functions of the gauge algebra.
All relations among structure functions
are contained in Eq.\bref{master equation},
thereby reproducing the equations
in Sect.\ \ref{ss:gs}
and generalizing them
to the generic $L$-th stage reducible theory.
On the other hand, $S$ is the starting action
to quantize covariantly the theory.
After a gauge-fixing procedure is implemented,
one can commence perturbation theory.
The latter aspects are treated
in Sect.\ \ref{s:gff}.

Regard $S$ as the action
for fields and antifields.
The variations of $S$ with respect
$\Phi^A$ and $\Phi_A^*$
are the equations of motion:
\be
  \rder{S}{z^a} = 0
\quad ,
\label{eqs of motion for S}
\ee
where
the collective variables $z^a$
in Eq.\bref{sym form of antibracket}
are used.
We assume there exists a least one stationary point
for which Eq.\bref{eqs of motion for S} holds.
We let $\Sigma$ denote this subspace of stationary points
in the full space of fields and antifields:
\be
  \Sigma  \longleftrightarrow
  \left\{ \rder{S}{z^a} = 0 \right\}
\quad .
\label{def of Sigma}
\ee
It turns out that,
given a classical solution $\phi_0$ of $S_0$
as in Eq.\bref{saddle point},
one possible stationary point is
\be
   \phi^j = \phi^j_0
\ , \quad \quad
   {\cal C}^{\alpha_s}_s=0 ;
\ \
   s=0, \ldots,L;
\ \  \alpha_s=1,\ldots,m_s
\ , \quad\quad
   \Phi^*_A=0
\quad .
\label{field-antifield saddle point}
\ee

An action $S$,
satisfying the master equation \bref{master equation},
possesses its own set of gauge invariances.
Indeed, by differentiating Eq.\bref{master equation}
with respect to $z^b$,
one finds after a little algebra that
\be
  \rder{S}{z^a} {\cal R}^a_b=0 \ ,\quad \quad a=1,\ldots,2N
\quad ,
\label{gauge invariances for S}
\ee
where
\be
  {\cal R}^a_b \equiv
  \zeta^{ac}\frac{\partial_l\partial_r S}
  {\partial z^c\partial z^b}
\quad .
\label{def of cal R}
\ee
Although there appears to be $2N$ gauge invariances,
not all of them
are independent on-shell as we now demonstrate.
Differentiating Eq.\bref{gauge invariances for S}
with respect to $z^d$,
multiplying by $\zeta^{cd}$,
using the definition of ${\cal R}^a_b$
in Eq.\bref{def of cal R}
and imposing the stationary condition
in Eq.\bref{eqs of motion for S},
one finds that
$$
 \restric{ {\cal R}^c_a {\cal R}^a_b }
  { \Sigma } = 0
\quad ,
$$
where $\Sigma$ denotes the space satisfying
the on-shell condition
in Eq.\bref{def of Sigma}.
One concludes that ${\cal R}^a_b$ is on-shell nilpotent.
It is an elementary result of matrix theory
that a nilpotent $2N \times 2N $ matrix has rank
less than or equal to $N$.
Hence, at the stationary point
there exist at least $N$ relations
among the gauge generators ${\cal R}^a_b$
and therefore the number $2 N - r$
of independent gauge transformations
on-shell is greater than or equal to $N$,
where $r$ is the rank of the hessian of $S$
at the stationary point:
\be
   r \equiv {\rm rank \ }
    \restric{
       \frac{\partial_l \partial_r S}
        {\partial z^a \partial z^b} }
  { \Sigma }
\quad .
\label{rank of hessian of S}
\ee
Necessarily,
$$
 r \leq N
\quad .
$$
A solution to the master equation is called {\it proper}
if
\be
 r = N
\quad ,
\label{def of proper solution}
\ee
where $r$ is given in Eq.\bref{rank of hessian of S}.
When $r < N$, there are solutions
to $\rder{S}{z^a} \delta z^a = 0 $
other than the ones given
by Eqs.\bref{gauge invariances for S}
and \bref{def of cal R}.

Usually, only proper solutions are of interest.
The reason is simple.
The action $S$ contains the physical fields $\phi^i$
and the ghost fields necessary for quantization.
However, the antifields $\Phi^*_A$,
$A = 1, \dots , N$, are unphysical.
If $r=N$, then
the number of independent gauge invariances
of the type in Eq.\bref{gauge invariances for S}
is the number of antifields.
As a consequence,
one can remove
the $N$ non-physical variables $\Phi^*_A$,
while maintaining the $N$ fields $\Phi^A$.
At a later stage,
the antifields can be eliminated
through a gauge-fixing procedure.
Throughout the rest of this article,
we restrict the discussion to the proper case.

We have not yet specified the relation between
$S_0$ and $S$.
To make contact with the original theory,
one requires the proper solution to contain
the original action $S_0 [ \phi ]$.
This requirement ensures the correct classical limit.
It corresponds to the following boundary condition on $S$
\be
  \restric{S \left[ \Phi,\Phi^* \right] }{\Phi^*=0} =
   S_0 \left[ \phi \right]
\quad .
\label{classical bc}
\ee
An additional boundary requirement is
\be
   \restric{\frac{\partial_l \partial_r S}
    {\partial {\cal C}^*_{s-1,\alpha_{s-1}}
    \partial {\cal C}^{\alpha_s}_s}}{\Phi^*=0} =
    R^{\alpha_{s-1}}_{s\alpha_s}(\phi)
\ ,\quad \quad s=0,\ldots,L
\quad ,
\label{hessian bc}
\ee
where ${\cal C}^*_{s-1,\alpha_{s-1}}$ is the antifield of
${\cal C}^{\alpha_{s-1}}_{s-1}$:
\be
    {\cal C}^*_{s,\alpha_{s}} \equiv
    \left( {\cal C}^{\alpha_{s}}_{s} \right)^*
\quad .
\label{def of C*}
\ee
In Eq.\bref{hessian bc}, for notational convenience we define
\be
  {\cal C}^{\alpha_{-1}}_{-1} \equiv \phi^i
 \ , \quad \quad
  {\cal C}^*_{-1,\alpha_{-1}} \equiv \phi^*_i
\ , \quad
 {\rm with \ } \alpha_{-1} = i
\quad .
\ee
Actually, Eq.\bref{hessian bc} does not need to be imposed
as a separate boundary condition.
Although it is not obvious,
the requirement of being proper
\bref{def of proper solution} and
the classical boundary condition in Eq.\bref{classical bc}
necessarily imply that a solution $S$ must
satisfy Eq.\bref{hessian bc}
\ct{bv85a}.

The proper solution $S$ can be expanded in a power series
in antifields.
Given the ghost number restriction on $S$
and the boundary conditions in Eqs.\bref{classical bc}
and \bref{hessian bc},
the expansion necessarily begins as
\be
     S \left[ \Phi,\Phi^* \right] = S_0 \left[ \phi \right]
     +\sum_{s=0}^L {\cal C}^*_{s-1,\alpha_{s-1}}
       R^{\alpha_{s-1}}_{s\alpha_s}
      {\cal C}^{\alpha_s}_s +O(C^{*2})
\quad .
\label{beg fsr proper solution}
\ee
In the next subsection
we obtain the leading terms in the solution
for a first-stage reducible theory.
We also
demonstrate how the gauge algebra is encoded
in the classical master equation,
when the classical basis is used.

\subsection{The Proper Solution and the Gauge Algebra}
\label{ss:psga}

\hspace{\parindent}
This subsection establishes
the connection between a proper solution
to the classical master equation
and the equations that the gauge-structure tensors
must satisfy.
For a first-stage reducible theory
the minimum set of fields is $\phi^i$,
${\cal C}^{\alpha_0}_0$ (which we call ${\cal C}^\alpha$)
and ${\cal C}^{\alpha_1}_1$ (which we call $\eta^a$).
One expands the action $S [ \Phi,\Phi^* ] $ as a Taylor
series in ghosts and antifields,
while requiring the total ghost number to be zero.
The result is
$$
  S\left[ {\Phi ,\Phi^*} \right] =
  S_0\left[ \phi  \right] +
  \phi_i^*R_\alpha^i{\cal C}^\alpha +
  {\cal C}_\alpha^*
  \left( Z_a^\alpha \eta^a +
  \frac12 T_{\beta\gamma }^\alpha
   \left( {-1} \right)^{\eps_\beta }{\cal C}^\gamma
    {\cal C}^\beta\right)
$$
$$
    +\eta_a^*
    \left( {A_{b\alpha }^a\eta^b{\cal C}^\alpha +
    \frac12
     F_{\alpha \beta \gamma }^a
    \left( {-1} \right)^{\eps_\beta }
    {\cal C}^\gamma {\cal C}^\beta {\cal C}^\alpha } \right)
$$
$$
  +\phi_i^*\phi_j^*
  \left( {-1} \right)^{\eps_i}
   \left( {\frac12  V_a^{ji}\eta^a + \frac14
    E_{\alpha \beta }^{ji}
    \left( {-1} \right)^{\eps_\alpha }
    {\cal C}^\beta {\cal C}^\alpha } \right)
$$
\be
  +{\cal C}_\delta^*\phi_i^*
  \left( {-1} \right)^{\eps_\delta }
   \left( {G_{a\beta }^{i\delta }\eta^a{\cal C}^\beta -
   \frac12  D_{\alpha \beta \gamma }^{i\delta }
    \left( {-1} \right)^{\eps_\beta }
    {\cal C}^\gamma {\cal C}^\beta {\cal C}^\alpha } \right)
\label{fsr proper solution}
\ee
$$
  +\frac{1}{12} \phi_i^*\phi_j^*\phi_k^*
   \left( {-1} \right)^{\eps_j}
    M_{\alpha \beta \gamma }^{kji}
    \left( {-1} \right)^{\eps_\beta }
   {\cal C}^\gamma {\cal C}^\beta {\cal C}^\alpha +\ldots
\quad ,
$$
where,
with the exception of $R_\alpha^i$ and $Z_a^\alpha$,
which are fixed by the boundary condition
in Eq.\bref{beg fsr proper solution},
one should, at this stage, think of the tensors,
$T_{\alpha \beta }^\gamma$,
$E_{\alpha \beta }^{ji}$,
etc.\ in Eq.\bref{fsr proper solution}
as having no relation to the tensors
found in the gauge algebra structure
of Sect.\ \ref{ss:gs}.
The minus signs and fractions,
${1 \over 2}$, ${1 \over 4}$, and ${1 \over {12} }$,
in Eq.\bref{fsr proper solution}
are put in for convenience.

The goal is to find a solution
to Eq.\bref{master equation}.
It is necessary to compute derivatives of $S$
with respect to fields and antifields.
One finds
$$
  {{\partial_r S} \over {\partial \phi^i}}=
  S_{0,i}+\phi_j^*\left( {R_\alpha^j{\cal C}^\alpha } \right)_{,i} +
  {\cal C}_\alpha^*\left( {Z_a^\alpha \eta^a +
   \frac12  T_{\beta \gamma }^\alpha
   \left( {-1} \right)^{\eps_\beta }
    {\cal C}^\gamma {\cal C}^\beta } \right)_{,i}+
$$
$$
  \frac14 \phi_j^*\phi_k^*
   \left( {-1} \right)^{\eps_j}
    \left( {E_{\alpha \beta }^{kj}
   \left( {-1} \right)^{\eps_\alpha }
   {\cal C}^\beta {\cal C}^\alpha } \right)_{,i}+\ldots
\quad ,
$$
$$
  {{\partial_l S} \over {\partial \phi_i^*}} =
   R_\alpha^i{\cal C}^\alpha +\phi_j^*
   \left( {-1} \right)^{\eps_i}
    \left( {V_a^{ji}\eta^a + \frac12
    E_{\alpha \beta }^{ji}
   \left( {-1} \right)^{\eps_\alpha }
    {\cal C}^\beta {\cal C}^\alpha } \right)
$$
$$
  +\left( {-1} \right)^{\eps_\delta \eps_i}
   {\cal C}_\delta^* \left( {G_{a\beta }^{i\delta }
   \eta^a{\cal C}^\beta -
   \frac12   D_{\alpha \beta \gamma }^{i\delta }
   \left( {-1} \right)^{\eps_\beta }
    {\cal C}^\gamma {\cal C}^\beta {\cal C}^\alpha } \right) +
$$
$$
  + \frac14 \left( {-1} \right)^{\eps_j}
    \phi_j^*\phi_k^*M_{\alpha \beta \gamma }^{kji}
   \left( {-1} \right)^{\eps_\beta }
   {\cal C}^\gamma {\cal C}^\beta {\cal C}^\alpha +\ldots
\quad ,
$$
$$
  {{\partial_r S} \over {\partial {\cal C}^\alpha }}=
  \phi_i^*R_\alpha^i+{\cal C}_\gamma^*
   T_{\alpha \beta }^\gamma
    \left( {-1} \right)^{\eps_\alpha }
    {\cal C}^\beta +
  \left( {-1} \right)^{\eps_i}
   \frac12  \phi_i^*\phi_j^*
    E_{\alpha \beta }^{ji}
    \left( {-1} \right)^{\eps_\alpha }
   {\cal C}^\beta +\ldots
\quad ,
$$
$$
  { {\partial_l S} \over { \partial {\cal C}_\alpha^* } }=
    Z_a^\alpha \eta^a +
   \frac12 T_{\beta \gamma }^\alpha
    \left( {-1} \right)^{\eps_\beta }
     {\cal C}^\gamma {\cal C}^\beta
   - \frac12 \left( {-1} \right)^{\eps_\alpha }
   \phi_i^* D_{\beta \gamma \delta }^{i\alpha }
    \left( {-1} \right)^{\eps_\gamma }
    {\cal C}^\delta {\cal C}^\gamma {\cal C}^{\beta }  + \ldots
\quad ,
$$
$$
   {{\partial_r S} \over {\partial \eta^a}}=
    {\cal C}_\alpha^*Z_a^\alpha + \phi_i^*\phi_j^*
    \left( {-1} \right)^{\eps_i}
    \frac12  V_a^{ji}+\ldots
\quad ,
$$
$$
    { {\partial_l S} \over {\partial \eta_a^*} } =
     A_{b \alpha }^a \eta^b {\cal C}^\alpha +
    \frac12   F_{\alpha \beta \gamma }^a
     \left( {-1} \right)^{\eps_\beta }
    {\cal C}^\gamma {\cal C}^\beta {\cal C}^\alpha + \ldots
\quad .
$$
Although ghosts do not depend on $\phi^i$,
we write $( \ \ \ )_{,i}$ in the first equation
to avoid some minus signs.

Given the above equations, it is straightforward
to form the sum over products of derivatives
in Eq.\bref{master equation}.
We leave this step to the reader
and simple quote the result
\ct{bv81a,bv83b}:
The classical master equation \bref{master equation}
is satisfied if the tensors,
$T_{\alpha \beta }^\gamma$,
$E_{\alpha \beta }^{ji}$, etc.\ in Eq.\bref{fsr proper solution}
are the ones
in Sect.\ \ref{ss:gs}.
In other words,
Eq.\bref{fsr proper solution}
with the tensors identified
as the ones
in Sect.\ \ref{ss:gs}
is a proper solution to the master equation.
One finds that $(S,S)=0$ implies
the gauge structure in equations
Eqs.\bref{noether}, \bref{commutator algebra} and
\bref{first stage jac1} -- \bref{first stage eq2}.
The reason why one equation $(S, S)=0$ is able to reproduce
many equations is that the coefficients
of each ghost and antifield term must separately be zero.
Up to overall factors,
Eq.\bref{noether} is the term independent of antifields,
Eq.\bref{commutator algebra}
is the coefficient of $\phi_i^* $
which is bilinear in the ${\cal C}^\alpha$,
Eq.\bref{first stage jac1}
is the coefficient linear in ${\cal C}_\delta^* $
and trilinear in the ${\cal C}^\alpha$,
Eq.\bref{first stage jac2}
is the coefficient of $\phi_i^* \phi_j^*  $
which is trilinear in the ${\cal C}^\alpha$,
Eq.\bref{first stage eq1}
is the coefficient linear in $\phi_i^* $
and in $\eta^a$,
and Eq.\bref{first stage eq2}
is the coefficient of ${\cal C}_\gamma^* $
which is linear in both $\eta^a$ and ${\cal C}^\beta$.
The coefficients of higher-order terms produce
the higher-order gauge structure equations.
For an irreducible system,
the proper solution is the above
with $\eta^a$ and $\eta^*_a$ set to zero.

Summarizing,
the antibracket formalism
using fields and antifields
allows a simple determination of the relevant
gauge structure tensors.
The proper solution to the classical master equation
is a compact way of
expressing the relations among the structure tensors.

\subsection{Existence and Uniqueness}
\label{ss:eu}

\hspace{\parindent}
At this point,
there are two obvious and interesting questions.
Does there always exist a proper solution
to the classical master equation
and is the proper solution unique?
Given reasonable conditions,
which include the regularity postulates of Appendix B,
there always exists a proper solution.
This result was obtained
in refs.\ct{vt82a,bv85a}
for the case of an irreducible theory.
For a general $L$-th stage reducible theory,
the theorem was proven in ref.\ct{fh90a}.

To gain insight into the question of uniqueness,
assume that a proper solution $S$ has been found
so that $(S,S)=0$.
Suppose that one performs
a canonical transformation
via Eq.\bref{f2 can trans}.
Then the transformed $S$
also obeys the classical master equation
because such canonical transformations
preserve the antibracket.
For an infinitesimal canonical transformation,
the transformed proper solution $S^\prime$
is,
according to Eq.\bref{canonical trans on function},
given by
$S^\prime = S + \varepsilon (S, F) + O(\varepsilon^2) $.
This is the only ambiguity in the proper solution.
Given the minimal set of fields in Eq.\bref{field set},
the proper solution
to the classical master equation is unique
up to canonical transformations.
This result was obtained
in refs.\ct{vt82a,bv85a,fh90a}.
Since our aim is pedagogical,
we do not present the proof here.
The interested reader can consult the above references.
Below we illustrate the non-uniqueness question
using the example of the spinless relativistic particle.
Canonical transformations
which lead to field redefinitions of ghosts
correspond to the freedom
of redefining the gauge generators.
This leads to changes in the structure tensors
of Sect.\ \ref{s:ssgt}
and corresponds to the non-uniqueness
mentioned at the end
of Sect.\ \ref{ss:gt}.

When gauge-fixing and
path integral quantization is considered,
it is necessary to enlarge the minimal set of fields
in Eq.\bref{field set} to a non-minimal set.
It turns out that
trivial variable pairs can be added to the theory
while maintaining the classical master equation
and its properness.
Let $ \Lambda$ and $\Pi$ be two new fields
and let $ \Lambda^* $ and $\Pi^*$
be the corresponding antifields.
Choose the ghost numbers and statistics so that
$$
  {\rm gh} \left[ \Pi \right] =
  {\rm gh} \left[ \Lambda \right] + 1
\quad ,
$$
\be
 \eps \left( \Pi \right) =
   \eps \left( \Lambda \right) + 1
  \ ( {\rm mod \ } 2 )
\quad .
\label{Pi gh numbers and eps}
\ee
Then
$     {\rm gh} \left[ \Lambda^* \right] =
  -   {\rm gh} \left[ \Lambda \right] -1
  = - {\rm gh} \left[ \Pi \right]
$
and
$
 \epsilon \left( \Lambda^* \right) =
  \epsilon \left( \Lambda \right) + 1
 \ ( {\rm mod \ } 2 )
= \epsilon \left( \Pi \right)
$,
so that
\be
  S_{\rm trivial} = \int \dif x \Lambda^* \Pi
\label{trivial action term}
\ee
can be added to $S$.
Adding such a bilinear to $S$ does not ruin
the classical master equation $( S, S) =0$.
When a non-minimal set of fields is employed,
the proper solution to the master equation is unique
up to canonical transformations and the addition
of trivial pairs.

\subsection{The Classical BRST Symmetry}
\label{ss:brsts}

\hspace{\parindent}
An important concept in gauge theories
is BRST symmetry.
The BRST symmetry is
what remains of gauge invariance
after gauge-fixing has been implemented.
In this sense,
it can be regarded as a substitute
for gauge invariance.
There are three important features
governing the BRST transformation:
nilpotency, graded derivation,
and invariance of the action $S$.

Even before gauge-fixing,
the field-antifield formalism
has BRST symmetry.
Via the antibracket,
the generator $\delta_B$ of this symmetry is
the proper solution $S$ itself.
Define the classical BRST transformation
of a functional $X$ of fields and antifields by
\be
  \delta_B X \equiv \left( {X,S} \right)
\quad .
\label{def of BRST}
\ee
The transformation rule for fields
and antifields is therefore
\be
  \delta_B \Phi^A={{\partial_l S} \over {\partial \Phi_A^*}}
\ , \quad \quad
  \delta_B \Phi_A^* =
   -{{\partial_l S} \over {\partial \Phi^A}} =
   \left( {-1} \right)^{\epsilon_A+1}
   {{\partial_r S} \over {\partial \Phi^A}}
\quad .
\label{BRST trans of fields and antifields}
\ee
Note that the field-antifield action $S$ is
classically BRST symmetric
\be
  \delta_B S=0
\quad ,
\label{BRST inv of action}
\ee
as a consequence of
$
  \left( {S,S} \right) = 0
$.

The BRST operator $\delta_B$
is a nilpotent graded derivation%
\footnote{
An alternative definition for $\delta_B$ given by
$$
  \delta_B X \equiv (-1)^{\epsilon_X}\left( {X,S} \right)
\quad ,
$$
is used by some authors.
In this case,
the $\delta_B$ continues to be a
nilpotent graded derivation,
but acts from the left to right, i.e.,
\bref{der prop of BRST} is replaced by
$$
   \delta_B \left( {XY} \right) =
  (\delta_B X)  Y+
   (-1)^{\epsilon_X}X\delta_B Y
\quad .
$$}:
Given two functionals $X$ and $Y$,
\be
  \delta_B \left( {XY} \right) =
   X\delta_B Y +
   \left( {-1} \right)^{\epsilon_Y}
   \left( { \delta_B X } \right)  Y
\quad ,
\label{der prop of BRST}
\ee
and
\be
\delta_B^2 X=0
\quad .
\label{nilpotency of BRST}
\ee
The nilpotency follows from two properties
of the antibracket: the graded Jacobi identity
and graded antisymmetry
(see Eq.\bref{antibracket properties}).
These properties imply
$  \left( {\left( {X,S} \right),S} \right) =
  -\left( {\left( {S,S} \right),X} \right) +
   \left( {-1} \right)^{\epsilon_X+1}
   \left( {\left( {S,X} \right),S} \right) =
   -\left( {\left( {S,S} \right),X} \right) -
  \left( {\left( {X,S} \right),S} \right)
$
which leads to
$
  \left( {\left( {X,S} \right),S} \right) =
   -{1 \over 2}\left( {\left( {S,S} \right),X} \right) = 0
$.
Therefore,
$
  \delta_B^2 X =
  \left( {\left( {X,S} \right),S} \right) = 0
$.

A functional ${\cal O}$ is
a {\it classical observable}
if $\delta_B {\cal O} = 0$ and ${\cal O} \ne \delta_B Y$
for some $Y$.
Two observables are considered equivalent
if they differ by a BRST variation
\ct{fisch90a,fh90a}.
A linear combination of observables is an observable.
Because of the graded derivation property of $\delta_B$
in Eq.\bref{der prop of BRST},
the product of two classical observables
is BRST invariant.
Thus, observables form an algebra.

For a closed irreducible theory,
the BRST transformation rules
for $\Phi^A$ and $\Phi_A^*$,
which depend on the original action
$S_0$ and the structure tensors
$R$ and $T$,
can be obtained
using Eqs.\bref{fsr proper solution} and
\bref{BRST trans of fields and antifields}:
$$
  \delta_B\phi^i=R_\alpha^i{\cal C}^\alpha + \ldots
\quad ,
$$
$$
  \delta_B{\cal C}^\alpha =
  {1 \over 2}T_{\beta \gamma }^\alpha
   \left( {-1} \right)^{\epsilon_\beta }
   {\cal C}^\gamma {\cal C}^\beta +
   \ldots
\quad ,
$$
$$
  \delta_B\phi_i^* =
  - \left( {-1} \right)^{\epsilon_i}S_{0,i} -
   \left( {-1} \right)^{\epsilon_\alpha \epsilon_i}
   \phi_j^*R_{\alpha ,i}^j{\cal C}^\alpha
  - {1 \over 2}\left( {-1} \right)^{\left( {\epsilon_\alpha +
    \epsilon_\beta +1} \right) \epsilon_i +
   \epsilon_\alpha }
   {\cal C}_\gamma^*T_{\alpha \beta ,i}^\gamma
     {\cal C}^\beta {\cal C}^\alpha +\ldots
\ ,
$$
\be
  \delta_B{\cal C}_\alpha^* =
   \left( {-1} \right)^{\epsilon_\alpha }
    \phi_i^*R_\alpha^i +
    {\cal C}_\gamma^*T_{\alpha \beta }^\gamma {\cal C}^\beta + \ldots
\quad ,
\label{BRST trans for fsr}
\ee
where we have displayed the terms
involving $S_0$, $R$ and $T$.
When the gauge-structure equations hold off-shell
at all levels,
then the tensors
$D_{\alpha \beta \gamma }^{i\delta }$,
$ M_{\alpha \beta \gamma }^{kji}$,
etc.\ in Eq.\bref{fsr proper solution}
are zero, and
the proper action is linear in antifields.
In this case,
there are no additional terms
in Eq.\bref{BRST trans for fsr}
for irreducible systems.
We illustrate BRST symmetry
in a few of the examples in the next section.

Because Sect.\ \ref{s:faf} has introduced
many ideas,
it is worth enumerating the most important ones.
After reading this section, one should know
the following tools and concepts:
antifields and the antibracket,
canonical transformations,
the classical master equation,
properness and proper solution,
classical BRST symmetry, and
classical observables.

\vfill\eject

\section{Examples of Proper Solutions}
\label{s:eps}

\hspace{\parindent}
In this section,
we present proper solutions to the master equation
for the examples considered
in Sect.\ \ref{s:egt}.
Given the gauge-structure tensors,
Eq.\bref{fsr proper solution}
immediately provides the proper solution $S$
for an irreducible or first-stage-reducible theory.
Since all but two of the example field theories
fall into one of these two cases,
the construction of $S$ is straightforward.
For the abelian $p$-form theory,
Eq.\bref{beg fsr proper solution} is needed.
The infinitely reducible open bosonic string
requires some guesswork to obtain $S$.
Given $S$,
the BRST transformations $\delta_B$ can be determined
from Eq.\bref{def of BRST}.
For reasons of space,
we display these transformations
only for the spinless relativistic particle,
Yang-Mills theory
and the open bosonic string field theory.
In Sect.\ \ref{sss:srp},
it is shown how two proper solutions
to the spinless relativistic particle
are related by a canonical transformation.

For exercises,
we suggest the following
(i) verify
$ \left( { S, S } \right) = 0 $,
(ii) determine the action of $\delta_B$
on the fields and antifields in each example,
(iii) use the results of (ii) to verify that $S$
is invariant under BRST transformations.
Since (iii) and (i) are related,
the calculations are similar
and the reader may want to do only one or the other.

\subsection{The Spinless Relativistic Particle}
\label{sss:srp}

\hspace{\parindent}
The spinless relativistic particle, considered
in Sect.\ \ref{ss:srp},
is an example of a closed irreducible theory.
The minimal set of fields and antifields are
\be
\Phi^A=\left\{ {x^\mu ,e,{\cal C}} \right\}  \ ,
\quad \quad
\Phi_A^*=\left\{ {x_\mu^*,e^*,{\cal C}^*} \right\}
\quad ,
\ee
with statistical parities
\be
  \epsilon(x^\mu) = \epsilon(e) = \epsilon(\gh C^*)=0
\ ,
\quad \quad
  \epsilon(x^*_\mu) = \epsilon(e^*) = \epsilon(\gh C) =1
\quad ,
\label{rsp field parities}
\ee
and ghost numbers
\be
  {\rm gh} [ x^\mu ] = {\rm gh} [e] = 0 \ ,
\quad
  {\rm gh} [\gh  C] = 1 \ ,
\quad
  {\rm gh} [x^*_\mu] = {\rm gh} [e^*] = -1 \ ,
\quad
  {\rm gh} [\gh C^*] = -2
\  .
\label{rsp ghost numbers}
\ee

Two versions of the gauge transformations were presented.
In the first, given
in Eq.\bref{srp trans law},
the only non-zero structure tensors
are the gauge generators $R^i_\alpha$
given in Eq.\bref{srp rdef}.
Using this and the general solution
in Eq.\bref{fsr proper solution},
one finds
\be
  S\left[ {\Phi ,\Phi^*} \right] =
  S_0\left[ {x^\mu ,e} \right] +
   \int {\dif \tau }
    \left( {x_\mu^*\dot x^\mu {{{\cal C}} \over e} +
     e^*\dot {\cal C}} \right)
\quad ,
\label{rsp ps1}
\ee
where $S_0 \left[ {x^\mu ,e} \right] $
is given in Eq.\bref{srp action}.
In the second version, given
in Eq.\bref{srp rtrans},
the gauge generators $R^i_\alpha$ and
the commutator structure constant $T_{\beta \gamma}^{\alpha}$
are non-zero.
Using Eq.\bref{fsr proper solution}, one obtains
\be
  S\left[ {\Phi ,\Phi^*} \right] =
  S_0\left[ {x^\mu ,e} \right] +
  \int {\dif \tau }\left( {x_\mu^*\dot x^\mu {\cal C} +
   e^*e\dot {\cal C} + e^*\dot e{\cal C} +
  {\cal C}^*\dot {\cal C}{\cal C}} \right)
\quad .
\label{rsp ps2}
\ee

By the uniqueness theorem of the previous section,
the solutions in Eqs.\bref{rsp ps1} and \bref{rsp ps2}
must be related by a canonical transformation.
In fact, Eq.\bref{rsp ps1} is mapped to Eq.\bref{rsp ps2}
if $x^\mu$, $x_\mu^*$ and $e$ are unchanged but
$$
  {\cal C} \to e{\cal C} \ ,
\quad \quad
  {\cal C}^*\to {1 \over e} {\cal C}^* \ ,
\quad \quad
  e^*\to e^*-{1 \over e} {\cal C}^*{\cal C}
\quad .
$$
One can check that this is a canonical transformation
by verifying that the antibracket structure
in Eq.\bref{bracket structure} is preserved.
The infinitesimal version of the transformation is
generated by the fermion
$$
  F= \ln \left( e \right) {\cal C}^* {\cal C}
\quad ,
$$
when $F$ is used in Eq.\bref{canonical transformation}.
Infinitesimally,
$$
  \delta {\cal C} =
  \varepsilon \ln \left( e \right){\cal C} \ ,
\quad
  \delta {\cal C}^* =
  -\varepsilon \ln \left( e \right){\cal C}^* \ ,
\quad
  \delta e^* = -\varepsilon {1 \over e}{\cal C}^*{\cal C}
\quad .
$$
The finite transformation is obtained by iterating
the infinitesimal transformation $N$ times,
requiring $N \varepsilon =1$
and then letting $N \to \infty$.
Alternatively, the full transformation
can be generated using
\be
  F_2 = \int \dif \tau
 \left( {
   \tilde x_\mu^* x^\mu +
   \tilde e^* e + { {\tilde {\cal C}^* {\cal C}} \over e}
 } \right)
\label{srp f2}
\ee
and
Eq.\bref{f2 can trans}.

It is straightforward to obtain
the BRST transformation rules
using Eq.\bref{BRST trans of fields and antifields}.
{}From Eq.\bref{rsp ps1},
one finds
$$
  \delta_B e = \dot {\cal C}
  \ , \quad \quad
  \delta_B x^\mu = {{\dot x^\mu {\cal C}} \over e}
  \ , \quad \quad
  \delta_B {\cal C} = 0
\quad ,
$$
$$
  \delta_B e^* =
    {1 \over 2}
  \left( {{{\dot x^2} \over {e^2}} + m^2} \right) +
  {{x_\mu^*\dot x^\mu {\cal C}} \over {e^2}}
\quad ,
$$
\be
  \delta_B x_\mu^* =
  {d \over {d\tau }}
  \left( {{{ \dot x_\mu + x_\mu^*
  {\cal C}} \over e}} \right)
\quad ,
\label{BRST trans rsp ps1}
\ee
$$
  \delta_B {\cal C}^*={{x_\mu^*\dot x^\mu } \over e}-\dot e^*
\quad .
$$
It is a useful exercise to verify the nilpotency of
$\delta_B $
when acting on any field or antifield.
One must be careful of signs.
In this regard
Eq.\bref{der prop of BRST} is useful.

\subsection{Yang-Mills Theories}
\label{sss:ymt}

\hspace{\parindent}
Yang-Mills theories have
a closed irreducible algebra.
The number of gauge parameters is the rank of the group.
There is a ghost field ${\cal C}^a$ for each gauge parameter
$\Lambda^a$ in Eq.\bref{gauge transf Y-M}.
Hence the fields and antifields are
\be
  \Phi^A = \left\{ {A^{a\mu },{\cal C}^a} \right\} \ ,
\quad \quad
  \Phi_A^* = \left\{ {A_{a\mu }^*,{\cal C}_a^*} \right\}
\quad .
\ee
{}From Eq.\bref{fsr proper solution},
the proper solution is
\be
   S=\int {\dif^d x} \left\{
   { -{1 \over 4} F_{\mu \nu }^a F_{a}^{\mu \nu } +
    A_{a\mu }^*{D^{\mu a}}_b {\cal C}^b +
   {1 \over 2}{\cal C}_c^*{f_{ab}}^c {\cal C}^b {\cal C}^a} \right\}
\quad .
\label{YM's proper solution}
\ee
The first term is the classical action \bref{action Y-M} and
the second and third terms correspond respectively
to the $R^i_\alpha $
and $T^\gamma_{\alpha \beta}$ terms
in Eq.\bref{fsr proper solution}.

As a check, let us verify the classical master equation
$(S,S)=0$.
Using Eq.\bref{master equation}, one finds
$$
  (S,S)=\int {\dif^d x} \left( {D^\nu F_{\nu \mu }} \right)_a
     {D^{\mu a}}_b {\cal C}^b
$$
$$
  -\int {\dif^d x} A_{d\mu }^*{f_{ac}}^d {\cal C}^c
    {D^{\mu a}}_b {\cal C}^b +
    {1 \over 2}\int {\dif^d x} A_{d\mu }^*
     {D^{\mu d}}_c {f_{ab}}^c {\cal C}^b {\cal C}^a
$$
$$
  + \frac{1}{4} \int {\dif^d x}
    \, {\cal C}_e^* {f_{dc}}^e {\cal C}^c
   {f_{ab}}^d {\cal C}^b {\cal C}^a
\quad .
$$
The first term vanishes when one integrates by parts,
uses the antisymmetry of $F_{\mu \nu }$ in $\mu$ and $\nu $,
employs Eq.\bref{cov der commutator} and
makes use of the antisymmetry of ${f_{ab}}^c$ in $a$ and $b$.
The second and third terms cancel when one makes use
of the anticommuting nature
of ${\cal C}^a$ and ${\cal C}^b$,
of the antisymmetry of ${f_{ab}}^c$ in $a$ and $b$,
and of the Jacobi identity
for the Lie group structure constants
in Eq.\bref{lie algebra jacobi identity}.
The last term vanishes for similar reasons.
Hence, $(S,S)=0$.

The BRST transformation rules
are obtained from
Eqs.\bref{BRST trans of fields and antifields}
and \bref{YM's proper solution}:
$$
\delta_B A^{a\mu  }= {D^{\mu a}}_b {\cal C}^b
\quad ,
$$
$$
  \delta_B {\cal C}^a =
 {1 \over 2}{f_{bc}}^a{\cal C}^c {\cal C}^b
\quad ,
$$
\be
  \delta_B A_{a\mu }^* =
   - \left( {D^\nu F_{\nu \mu }} \right)_a +
  {f_{ab}}^c A_{c\mu }^* {\cal C}^b
\quad ,
\label{BRST trans ymt}
\ee
$$
  \delta_B{\cal C}_a^* =
  - \left( {D^\mu A_\mu^*} \right)_a +
   {\cal C}_c^* {f_{ab}}^c{\cal C}^b
\quad .
$$

\subsection{Topological Yang-Mills Theory}
\label{sss:tymt}

\hspace{\parindent}
This theory is first-stage reducible.
The action given in Eq.\bref{action top Y-M}
is invariant under two types of gauge transformations.
The gauge parameters
associated with these transformations are
$\Lambda^a$ of Eq.\bref{gauge transf Y-M} and
$\varepsilon^{a \mu}$ of Eq.\bref{top transf Y-M}.
Correspondingly,
one needs to introduce ghosts
${\cal C}^a \leftrightarrow \Lambda^a $ and
${\cal C}^{a\mu } \leftrightarrow \varepsilon^{a \mu }$.
For the gauge invariances of the gauge invariances,
the ghost-for-ghost field $\eta^a$ must be introduced.
Hence the fields and antifield are
\be
  \Phi^A =
    \left\{ {A^{a\mu },{\cal C}^a,{\cal C}^{a\mu }, \eta^a} \right\} \ ,
\quad \quad
  \Phi_A^* =
   \left\{ {A_{a\mu }^*,{\cal C}_a^*,{\cal C}_{a\mu }^*, \eta^*_a} \right\}
\quad .
\ee
The non-zero gauge tensors are
the gauge generators $R^i_\alpha $
given in Eq.\bref{Y-M gauge generators},
the commutator structure constants $T^\gamma_{\alpha \beta}$
given in Eq.\bref{top Y-M T tensor},
the null vectors $Z^\alpha_a$
given in Eq.\bref{Y-M null vectors},
and $A^a_{b \alpha}$ given in Eq.\bref{top Y-M A tensor}.
Other tensors are zero.
Inserting the non-zero tensors in Eq.\bref{fsr proper solution},
one finds
\ct{gr90a}
$$
  S=\int {\dif^4x}\left\{ {{1 \over 4}
   F_{\mu \nu }^a \ {}^* F_a^{\mu \nu } +
   A_{a\mu }^* \left( {{D^{\mu a}}_b {\cal C}^b +
   {\cal C}^{a\mu }} \right)}
   +{1 \over 2}{\cal C}_c^*
       {f_{ab}}^c {\cal C}^b {\cal C}^a \right.
$$
\be
  \left. {+
    {\cal C}_{c\mu }^* {f_{ab}}^{c}
      {\cal C}^{b\mu }{\cal C}^a +
    {\cal C}_b^*  \eta^b  -
   {\cal C}_{a\mu }^* {{D^{\mu a}}_b} \eta^b +
   \eta_c^*{f_{ba}}^c\eta^a{\cal C}^b} \right\}
\quad .
\label{tym proper solution}
\ee

\subsection{The Antisymmetric Tensor Field Theory}
\label{sss:atft}

\hspace{\parindent}
This system has been treated
by the antifield formalism
in refs.\ct{agm88a,batlle88a,bbs88a,bg88a,diaz88a,%
fisch90a,bcg92a}.
The theory is on-shell first-stage reducible.
The fields and antifields are
\be
  \Phi^A = \left\{ {
    A^{ a \mu} , B^{a\mu \nu }, {\cal C}^{a\mu }, \eta^a
                    } \right\} \ ,
\quad \quad
  \Phi_A^* = \left\{ {
   A_{a\mu}^*, B_{a\mu \nu }^*, {\cal C}_{a\mu }^*,\eta_a^*
                     } \right\}
\quad .
\label{astft fields}
\ee
The non-zero structure tensors $R^i_\alpha$,
$Z^\alpha_a$ and $V^{ij}_a$ are given
in Eqs.\bref{AT gauge generators} -- \bref{AT Vjia}.
The proper solution is
\be
   S = S_0 + \int \dif^4 x \left( {
   \frac{1}{2} B_{a}^{* \kappa \lambda}
    \varepsilon_{\kappa \lambda \mu \nu }
    {D^{\mu a}}_b {\cal C}^{b\nu } +
     {\cal C}_{b\mu }^* {D^{\mu b}}_a \eta^a -
    {1 \over 8} B_{c\rho \sigma }^* B_{\mu \nu}^{*b}
   \varepsilon^{\mu \nu \rho \sigma } {f_{ab}}^{c} \eta^a
      } \right)
\label{astft proper solution}
\ ,
\ee
where $S_0$ is given in Eq.\bref{AT action}.
An effect of the on-shell reducibility
is the appearance of
terms quadratic
in the antifields.

\subsection{Abelian $p$-Form Theories}
\label{sss:apft}

\hspace{\parindent}
Recall that these are examples of $p-1$ stage
off-shell reducible theories.
Consequently there are $p$ different types of ghosts,
${\cal C}_0,{\cal C}_1,\ldots ,{\cal C}_{p-1}$,
where ${\cal C}_s$ is a $p-1-s$ form:
\be
  \Phi^A = \left\{ {A,{\cal C}_0,
   {\cal C}_1,\ldots ,{\cal C}_{p-1}} \right\} \ ,
\quad \quad
   \Phi_A^* = \left\{ {A^*,{\cal C}_0^*,
   {\cal C}_1^*,\ldots ,{\cal C}_{p-1}^*} \right\}
\quad .
\label{apft fields}
\ee
The proper solution is
\be
  S=\int \left\{ { - \frac{1}{2} F\wedge {}^* F
   + {}^*(A^*)\wedge d{\cal C}_0 +
  \sum\limits_{i=1}^{p-1}
   {}^*\left( {{\cal C}_{i-1}^*} \right)\wedge d{\cal C}_i}
  \right\}
\quad ,
\label{apft proper solution}
\ee
where $\wedge$ is the wedge product
and a ``$\, {}^* \, $'' in front
of a field or antifield
indicates the dual star operation.
In Eq.\bref{apft proper solution},
some antifields have been redefined by a minus sign
factor compared to the definitions
in Sect.\ \ref{s:ssgt}.
As a consequence of integration by parts
and $dd=0$,
as well as the definitions of the Hodge star operation
and the wedge product,
one can check that $(S,S)=0$.
Indeed,
$$
 {{\partial_r S} \over {\partial A_{\mu_1\mu_2\ldots \mu_p}}}
 {{\partial_l S} \over {\partial A^{*\mu_1\mu_2\ldots \mu_p}}}
  \propto \int {} \left( { {}^* d {}^*F} \right)
  \wedge {}^*d{\cal C}_0
  \propto \int {} d
  \left( { {}^*F\wedge d{\cal C}_0 } \right) = 0
\ ,
$$
$$
  {{\partial_r S} \over {\partial \left( {{\cal C}_i}
    \right)_{\mu_1\mu_2\ldots \mu_{p-i-1}}}}
    {{\partial_l S} \over {\partial \left( {{\cal C}_i}
    \right)^{*\mu_1\mu_2\ldots \mu_{p-i-1}}}}
   \propto
$$
$$
  \int {}
  \left( { {}^*d {}^* \left( {{\cal C}_i^*} \right)} \right)
  \wedge {}^*d{\cal C}_{i+1}
  \propto \int {} d \left( { {}^* \left( {{\cal C}_i^*} \right)
  \wedge d {\cal C}_{i+1} } \right) = 0 \ ,
  \quad \quad i=0,1,\ldots , p-1
\ ,
$$
where
$
  {\cal C}_{-1}^* \equiv A^*
$.

\subsection{Open String Field Theory}
\label{sss:osft}

\hspace{\parindent}
This is an infinite-stage reducible theory.
Hence, there are ghosts, ghosts for ghosts,
ghosts for ghosts for ghosts, etc..
The fields are the string field $A$ and the
infinite tower ${{\cal C}_s}, \  s=0, \dots , \infty$:
\be
  \Phi^A = \left\{ {A,{\cal C}_0,{\cal C}_1,
    {\cal C}_2,\ldots } \right\} \ ,
\quad \quad
  \Phi^*_A = \left\{ {A^*,{\cal C}_0^*,{\cal C}_1^*,
    {\cal C}_2^*,\ldots } \right\}
\quad ,
\label{obsft fields}
\ee
where $A$ is a string $1$-form,
${\cal C}_0$ is a string $0$-form,
${\cal C}_1$ is a string $-1$-form, etc..
Likewise $A^*$ is a string $1$-form,
${\cal C}_0^*$ is a string $0$-form,
${\cal C}_1^*$ is a string $-1$-form, etc..
In Chern-Simons string field theory,
odd forms have odd grading and even forms
have even grading.
In addition, according to
Eq.\bref{ghost numbers and statistics} of
the field-antifield formalism,
$\varepsilon \left( {\cal C}_s \right)$ is
zero for $s$ odd and one for $s$ even.
With respect to the calculus of
string differential forms,
the total statistics is important.
Consequently, for any fields and/or antifields
$\varphi^i$ and $\varphi^j$,
axioms (3) and (5)
of Sect.\ \ref{ss:obsft}, namely
the graded distributive property of $Q$
across the star product
and the graded commutativity of the star product
under the integral,
become
$$
  Q\left( {\varphi^i*\varphi^j} \right) =
  Q\varphi^i*\varphi^j+
  \left( {-1} \right)^{s_i}\varphi^i*Q\varphi^j
\quad ,
$$
\be
  \int {\varphi^i*\varphi^j}=
   \left( {-1} \right)^{ {s_i} {s_j} }
   \int {\varphi^j*\varphi^i}
\quad ,
\label{integral exchange}
\ee
where $s_i$ is the total statistics of $\varphi^i$:
\be
  s_i \equiv s \left( \varphi^i \right) =
    g\left( {\varphi^i} \right) +
   \epsilon \left( {\varphi^i} \right)
   \ ( {\rm mod \ } 2 )
\quad .
\ee
Here, $g\left( {\varphi^i} \right)$ is the ghost number
of $\varphi^i$
which is the same as the order of the string form,
and $\epsilon \left( {\varphi^i} \right)$
is given by Eq.\bref{ghost numbers and statistics}.

The quantization of the bosonic string was carried out
by C. Thorn \ct{thorn87a}
and by M. Bochicchio \ct{bochicchio87a}.
Since this subject has been reviewed in
ref.\ct{thorn89a},
we keep the discussion brief.
To gain some insight in finding the proper solution,
let us compute the non-zero terms in Eq.\bref{fsr proper solution}.
The classical action $S_0$ is given in Eq.\bref{obs action}.
The $R^i_\alpha$, $Z^\alpha_a$, $V^{ij}_a$
and $T^\gamma_{\alpha \beta}$ terms
in Eq.\bref{fsr proper solution} are
respectively
$$
  \int {}^*A^* *\left( {Q{\cal C}_0 +
   A*{\cal C}_0 + {\cal C}_0*A} \right)
\quad ,
$$
$$
  \int {}^*{\cal C}_0^* * \left( {Q{\cal C}_1 + A*{\cal C}_1
   + {\cal C}_1*A} \right)
\quad ,
$$
\be
  \int {}^*A^* * {}^*A^* * {\cal C}_1
\quad ,
\label{some terms}
\ee
$$
  \int {}^*{\cal C}_0^* * {\cal C}_0 * {\cal C}_0
\quad ,
$$
where ${}^*$ before a field
is the string analog of the dual star operation
determined by the bilinear form $\int A * B$.
It takes $p$-forms into $3-p$ forms.
Note that the structure of the above terms is similar to
the classical action $S_0$ evaluated using various fields
and antifields.
The sign of the term ${\cal C}_0*A$ in the first equation
is opposite to that of
Eq.\bref{obs gauge invariance tower}
because ${\cal C}_0$ has odd statistics.
The field-antifield statistics, ghost number and total
statistics for the fields are
$$
  \epsilon \left( {{\cal C}_i} \right) =
  i+1  \ ( {\rm mod \ } 2 )  \ ,
\quad
  g\left( {{\cal C}_i} \right)=-i  \ ,
\quad
  {\rm gh} \left[ {{\cal C}_i} \right] = i+1  \ ,
\quad
  s\left( {{\cal C}_i} \right)=1 \ ,
$$
\be
  \epsilon \left( {^*{\cal C}_i^*} \right) = i
  \ ( {\rm mod \ } 2 )  \ ,
\quad
  g\left( {^*{\cal C}_i^*} \right) = i+3   \ ,
\quad
 {\rm gh} \left[ {^*{\cal C}_i^*} \right] = -i - 2  \ ,
\quad
  s\left( {^*{\cal C}_i^*} \right)=1 \ ,
\ee
for $i \ge -1$,
where we have defined
\be
  {\cal C}_{-1}\equiv A \ , \quad \quad
  {\cal C}_{-1}^*\equiv A^*
\quad .
\ee
Note that the total statistics of ${\cal C}_i$ and
the ${}^*{\cal C}_i^*$ is odd,
so that it makes sense to define a field
which is the formal sum
$$
 \Psi \equiv \ldots +
   \stackrel{{\textstyle{s+3}}}{{}^*{\cal C}_s^*}
   + \ldots +
   \stackrel{{\textstyle{3}}}{{}^*{\cal C}_0^*}
 + \stackrel{{\textstyle{2}}}{{}^*A^*}
 + \stackrel{{\textstyle{1}}}{A}
 + \stackrel{{\textstyle{0}}}{{\cal C}_0}
 + \ldots +
   \stackrel{{\textstyle{-s}}}{{\cal C}_s}
 + \ldots
\quad ,
$$
\be
 \Psi = \sum\limits_{p=-\infty }^\infty  {\psi_p} \ ,
\ \  {\rm where \ }
  \psi_{-p} \equiv {\cal C}_p \ , \ {\rm for \ } \ p\ge -1 \ ,
\quad
  \psi_p \equiv {}^*{\cal C}_{p-3}^* \ , {\rm for \ } \ p\ge 2
\ .
\label{psi sum}
\ee
In the first equation,
the order of the string form is denoted above the field.
The total statistics of the field-antifield $\Psi$ is odd:
$s\left( \Psi  \right)=1$.
The terms in Eq.\bref{some terms}
and the classical action
are both contained in the following
ansatz for the proper solution
$$
  S={1 \over 2}\int \Psi *Q\Psi +{1 \over 3}\int \Psi *\Psi *\Psi =
$$
\be
  {1 \over 2}\sum\limits_{p=-\infty }^\infty
   {} \int {\psi_{2-p}}*Q\psi_p +
    {1 \over 3}
   \sum\limits_{
     { {p=-\infty } \atop
     {q=-\infty } }
               }^\infty
  {}\int {\psi_p}*\psi_q*\psi_{3-p-q}
\quad .
\label{obsft proper solution}
\ee

By obtaining the $L$-th stage reducible proper solution,
computing the non-zero structure tensors,
one could derive Eq.\bref{obsft proper solution}.
Instead let us check the ansatz.
One simply needs to verify the classical master equation
\be
 0 \mathrel{\mathop=^{\rm ?}} (S,S) \propto
  \int \left( Q\Psi + \Psi * \Psi \right) *
       \left( Q\Psi + \Psi * \Psi \right)
\quad .
\label{obsft master equation}
\ee
Several terms are produced.
The first is
$$
 \int {}Q\Psi *Q\Psi =\int {}Q\left( {\Psi *Q\Psi } \right) = 0
\quad ,
$$
where axioms (1), (2) and (3) of
Sect.\ \ref{ss:obsft} have been used.
Note that
$$
  0=\int {} Q\left( {\Psi *\Psi *\Psi } \right)=
$$
$$
  \int {} \left( {Q\Psi *\Psi *\Psi } \right) -
  \int {} \left( {\Psi *Q\Psi *\Psi } \right) +
  \int {} \left( {\Psi *\Psi *Q\Psi } \right)=
$$
$$
  3 \int {}\left( {Q\Psi *\Psi *\Psi } \right) =
  3 \int {}\left( {\Psi *\Psi *Q\Psi } \right)
\quad ,
$$
where axioms (2)--(5) are used.
This implies that
$
  \int {}\left( {Q\Psi *\Psi *\Psi } \right) =
  \int {}\left( {\Psi *\Psi *Q\Psi } \right) = 0
$,
leading to the vanishing of
two of the terms in Eq.\bref{obsft master equation}.
The last term in Eq.\bref{obsft master equation}
is zero when axioms (4) and (5) are combined to give
$$
\int {}\left( {\Psi *\Psi } \right)*\left( {\Psi *\Psi } \right) =
\int {}\left( {\Psi *\Psi *\Psi *\Psi } \right) =
 -\int {}\left( {\Psi *\Psi *\Psi *\Psi } \right)
\quad ,
$$
where
Eq.\bref{integral exchange}
is used in the last step.
Since the master equation is satisfied
and the suitable boundary conditions
are correctly implemented,
Eq.\bref{obsft proper solution}
is a proper solution.
The classical action in Eq.\bref{obs action}
and the proper solution in Eq.\bref{obsft proper solution},
although structurally identical,
differ in that the field entering the action is different.
In $S_0$ it is the string one-form $A$,
while in $S$
it is the tower $\Psi$,
given in Eq.\bref{psi sum},
which includes ghosts and antighosts
as well as $A$.

The BRST transformation rules
for the fields and antifields are
\be
\delta_B \Psi =
  Q\Psi +\Psi *\Psi
\quad .
\label{BRST trans obsft}
\ee
The transformation rule
for a particular component $\psi_p$ of $\Psi$
in Eq.\bref{psi sum}
is obtained by selecting the $p$-form term
of Eq.\bref{BRST trans obsft}.

Let us verify the nilpotency of $\delta_B$:
$$
  \delta_B^2 \Psi =
  Q\delta_B\Psi + \Psi *\delta_B\Psi -\delta_B\Psi *\Psi =
$$
$$
  Q^2\Psi + Q\left( {\Psi *\Psi } \right) +
  \Psi *Q\Psi +\Psi *\left( {\Psi *\Psi } \right) -
  Q\Psi *\Psi -\left( {\Psi *\Psi } \right)*\Psi = 0
\quad .
$$
The last equality holds due to the axioms of
open string field theory:
The first term is zero because of axiom (1),
the nilpotency of $Q$;
the second, third and fifth terms cancel
because of axiom (3),
the graded distributive property of $Q$;
and the fourth and sixth terms cancel by axiom (4),
the associativity of the star product.

\subsection{The Massless Relativistic Spinning Particle}
\label{sss:mrsp}

\hspace{\parindent}
This system has a closed irreducible algebra but
possesses local supersymmetry.
It has been treated by the antifield formalism
in ref.\ct{gprr90a}.
To the original fields,
$x^\mu $, $\psi^\mu$, $e$, and $\chi$, one adds
the ghosts ${\cal C}$ and $\Gamma$ respectively
for general coordinate transformations and
for local supersymmetry transformations.
The fields and antifields are
\be
  \Phi^A = \left\{ {x^\mu , \psi^\mu, e, \chi , {\cal C}, \Gamma } \right\} \ ,
\quad \quad
   \Phi_A^* = \left\{ { x_\mu^*, \psi_\mu^*, e^*,
    \chi^*, {\cal C}^* , \Gamma^*} \right\}
\quad ,
\label{mrsp fields}
\ee
where $x^\mu $, $e$, $\Gamma $, $\psi_\mu^*$, $\chi^*$ and ${\cal C}^*$
are commuting while
$\psi^\mu $, $\chi $, ${\cal C}$, $\Gamma^*$, $x_\mu^*$ and $e^*$
are anticommuting.

The only non-zero terms in Eq.\bref{fsr proper solution}
are $S_0$, $\phi_i^*R_\alpha^i{\cal C}^\alpha$ and
${1 \over 2}{\cal C}_\gamma^*T_{\alpha \beta }^\gamma
 \left( {-1} \right)^{\epsilon_\alpha }{\cal C}^\beta {\cal C}^\alpha$.
These quantities are determined
in Eqs.\bref{mrsp action}--\bref{sg commutator parameter}.
The result is
$$
  S = S_0 + \int {\dif \tau }
  \left\{ {x_\mu^*\left( {-ie\psi^\mu \Gamma
  + \dot x^\mu {\cal C} } \right) +
  \psi_\mu^*\left( {\left( {\dot x^\mu -
  i \chi \psi^\mu } \right)\Gamma +
  \dot \psi^\mu {\cal C}} \right) + } \right.
$$
$$
  e^*\left( {-2ie\chi \Gamma +
   \dot e{\cal C} + e \dot {\cal C}} \right) +
   \chi^*\left( {e\dot \Gamma + \dot e\Gamma +
   \chi \dot {\cal C} + \dot \chi {\cal C}} \right)+
$$
\be
  \left. {
  {\cal C}^* \left( { \dot {\cal C} {\cal C} +
  i e \Gamma \Gamma
                    } \right) +
  \Gamma^*\left( {{\cal C}\dot \Gamma - \dot {\cal C} \Gamma +
   i\chi \Gamma \Gamma } \right)
         } \right\}
\quad .
\label{mrsp proper solution}
\ee
where $S_0$ is given in Eq.\bref{mrsp action}

\subsection{The First-Quantized Bosonic String}
\label{sss:fqbs}

\hspace{\parindent}
This system has a closed irreducible algebra.
There are three types of gauge invariances:
local translations, local Lorentz boosts
and scaling transformations
to which we respectively assign the ghosts
${\cal C}^m$, ${\cal C}^{ab}$ and ${\cal C}$
where $a$, $b$ and $m$ take on the values
$\sigma$ and $\tau$
and ${\cal C}^{ba} = - {\cal C}^{ab}$.
Combining these ghosts with the original fields
$X^\mu$ and ${e_a}^m$,
one arrives at the following set
of fields and antifields:
\be
  \Phi^A=\left\{ {X^\mu ,{e_a}^m,{\cal C}^m,{\cal C}^{ab},{\cal C}} \right\}
\quad , \quad \quad
  \Phi_A^*=\left\{ {X_\mu^*,{e^{*a}}_m,
  {\cal C}_m^*,{\cal C}_{ab}^*,{\cal C}^*} \right\}
\quad ,
\ee
where $\mu$ ranges from $0$ to $D-1=25$.
The only non-zero terms in Eq.\bref{fsr proper solution}
are $S_0$, $\phi_i^*R_\alpha^i{\cal C}^\alpha$ and
${1 \over 2}{\cal C}_\gamma^*T_{\alpha \beta }^\gamma
 \left( {-1} \right)^{\epsilon_\alpha }
{\cal C}^\beta {\cal C}^\alpha$.
These tensors are computed
in Sect.\ \ref{ss:fqbs}.
One thus obtains
$$
  S = S_0 +
   \int {\dif \tau }\int_0^\pi  {\dif \sigma }
   \left\{ {X_\mu^*{\cal C}^m\partial_m X^\mu +
    }
    \right.
$$
$$
  {e^{*a}}_m\left( {{\cal C}^n \partial_n {e_a}^m -
    {e_a}^n \partial_n {\cal C}^m} +
    {e_a}^m {\cal C} + {e_b}^m
   \eta_{ac} {\cal C}^{cb}\right)
$$
\be
  \left. {
  - {\cal C}_m^* {\cal C}^n \partial_n {\cal C}^m +
 {\cal C}_{ab}^* \left({\cal C}^{ac} {\cal C}^{bd}
   \eta_{cd} -
  {\cal C}^n \partial_n {\cal C}^{ab}\right) -
    {\cal C}^*{\cal C}^n \partial_n{\cal C}} \right\}
\quad ,
\label{fqbs S}
\ee
where $S_0$ is given in Eq.\bref{bosonic string action}.

\vfill\eject


\section{The Gauge-Fixing Fermion}
\label{s:gff}

\hspace{\parindent}
This section discusses the process
of gauge-fixing in the field-antifield formalism
\ct{bv81a,bv83b},
thereby paving the way
to the quantization of gauge theories
via the path integral approach.
The important concept is a gauge-fixing fermion $\Psi$.
It is a Grassmann-odd functional with ghost number $-1$.
In Sect.\ \ref{ss:g},
it is shown how antifields
can be eliminated to obtain an action
suitable for the computation of correlation functions
in standard perturbation theory.
Results are independent of $\Psi$
if the action satisfies
the quantum master equation
in Eq.\bref{quantum master equation}
\ct{bv77a,bv81a}.
To construct an appropriate $\Psi$,
additional fields and their antifields
are needed.
See Sect.\ \ref{ss:gfaf}.
Details of the gauge-fixing procedure
for a general theory
are presented
in Sects.\ \ref{ss:dfgfp} and \ref{ss:ogfp}.
Since Sects.\ \ref{ss:gfaf}--\ref{ss:ogfp}
are somewhat technical,
the reader may wish to read only
the irreducible case,
which is discussed at the beginning of each subsection.
In particular instances,
gauge-fixing can be done without $\Psi$
by performing a cleverly chosen canonical transformation
\ct{fh89a,siegel89a,vanproeyen91a,bkp92a,tp93a,vp94a}.

After gauge-fixing is completed,
the theory is still invariant
under gauge-fixed BRST transformations $\delta_{B_\Psi}$
\ct{bv81a}.
The nilpotency of $\delta_{B_\Psi}$ is not guaranteed off-shell
\ct{ts86a,batlle88a,bg88a,tnp90a}.
However, when the equations of motion for the gauge-fixed
action are used, $\delta^2_{B_\Psi}=0$.
An attractive way of viewing the effects of gauge-fixing
associated with $\Psi$ and the content of $\delta_{B_\Psi}$,
is to perform a canonical transformation
to the so-called gauge-fixed basis.
In this basis,
one retains antifields as sources
of gauge-fixed BRST transformations,
and
the classical master equation
reproduces the algebraic structure
associated with $\delta_{B_\Psi}$.

\subsection{Generalities}
\label{ss:g}

\hspace{\parindent}
Although ghost fields
have been incorporated into the theory,
the field-antifield action
$S$
still possesses gauge invariances
(See Eq.\bref{gauge invariances for S}).
Hence it is not yet suitable for quantization
via the path integral approach.
A gauge-fixing procedure is needed.
The theory also contains many antifields
that usually one wants to eliminate
before computing amplitudes and $S$-matrix elements.
One cannot simply set the antifields to zero
because the action would reduce to the original
classical action
$S_0$,
which is not appropriate
for commencing perturbation theory
due to gauge invariances.
Following ref.\ct{bv81a},
antifields are eliminated
by using a gauge-fixing fermion
$\Psi \left( \Phi \right) $ via
\be
   {\Phi_A^* = {{\partial \Psi } \over {\partial \Phi^A}}}
\quad .
\label{fermion gauge fixing}
\ee
Note that $\Psi$ is a functional of fields only.
It does not matter
whether right or left derivatives are used
in Eq.\bref{fermion gauge fixing}
because
$
  {{\partial_l \Psi } \over {\partial \Phi^A}}
  ={{\partial_r \Psi } \over {\partial \Phi^A}}
$
since
$ \epsilon \left( \Psi  \right) = 1 $
(see Eq.\bref{l and r der relation}
of Appendix A).
We denote the ``surface'' in functional space
determined by
Eq.\bref{fermion gauge fixing} by $\Sigma_\Psi$,%
{\footnote{
Do not confuse $\Sigma_\Psi$ with $\Sigma$
of Eq.\bref{def of Sigma}.
The latter ($\Sigma$ without a subscript $\Psi$)
corresponds to the on-shell condition.
}}
\be
 \Sigma_\Psi \longleftrightarrow
   \left\{ {
    \Phi_A^*
  = {{\partial_r \Psi } \over {\partial \Phi^A}}
           }  \right\}
\label{def of Sigma sub psi}
\quad ,
\ee
so that
\be
  \left. X \right|_{\Sigma_\Psi } \equiv
    X\left[ {\Phi ,{{\partial \Psi }
   \over {\partial \Phi }}} \right]
\quad .
\label{surface Sigma sub Psi}
\ee
The corresponding gauge-fixed action is
$\left. S \right|_{\Sigma_\Psi } \equiv S_{\Psi}$.
Matching the statistics and ghost numbers
of
$\Phi_A^*$
and
${ {\partial \Psi } \over {\partial \Phi^A} }$
leads to
\be
  \epsilon \left( \Psi  \right)=1
\quad , \quad \quad
  {\rm gh} \left[ \Psi  \right]=-1
\quad .
\label{fgns}
\ee

To quantize the theory,
let us use the path integral approach
with the constraint in
Eq.\bref{fermion gauge fixing}
implemented by a delta function:
\be
  I_\Psi \left( X \right) = \int {\left[ {\dif \Phi } \right]
  \left[ {\dif \Phi^*} \right]}
  \ \delta
  \left( {\Phi_A^*-
   {{\partial \Psi } \over {\partial \Phi^A}}} \right)
   \exp \left( {{i \over {\hbar }}W
   \left[ {\Phi ,\Phi^*} \right]} \right)
  X\left[ {\Phi ,\Phi^*} \right]
\quad ,
\label{path integral}
\ee
where $X$
is a correlation function of interest.
Here $W$ is the quantum generalization of
the field-antifield action $S$.
By the correspondence principle,
the two should be equal in the
$\hbar \rightarrow 0$
limit.
One must choose $ \Psi$
so that the theory is non-degenerate,
that is,
when the action is expanded
about a solution of the equations of motion,
propagators exist.
Such a $\Psi$ is called {\it admissible}.
Conditions on $\Psi$ assuring admissibility
are given
in Sect.\ \ref{ss:dfgfp}
for certain gauge-fixing schemes.
When $S_0$ is local,
it is desirable
that $\Psi$ be a local functional of the fields
so as to preserve
the locality of the gauge-fixed action.
Selecting a ``good'' $\Psi$
is a matter of skill.
A judicious choice can greatly simplify
a particular computation, and
the choice can depend on the type of computation,
e.g.\ a correlation function,
a proof of unitarity,
a proof of renormalizability, etc..

The freedom in choosing
$\Psi$ corresponds to
the choice of the gauge-fixing procedure.
One would like results to be independent
of gauge fixing.
Let us determine when this is the case.
Define the integrand in Eq.\bref{path integral}
to be ${\cal I}$, i.e.,
\be
  {\cal I}\left[ {\Phi ,\Phi^*} \right] \equiv
   \exp \left( {{i \over {\hbar}}W
   \left[ {\Phi ,\Phi^*} \right] } \right) X
   \left[ {\Phi ,\Phi^*} \right]
\quad .
\label{integrand}
\ee
Under an infinitesimal change
$\delta \Psi$
of $\Psi$,
$I_\Psi \left( X \right)$
changes by
$$
  I_{\Psi +\delta \Psi }
  \left( X \right)-I_\Psi \left( X \right)=
  \int {\left[ {\dif \Phi } \right]}
   \left( {
   {\cal I} \left[ {\Phi ,{{\partial \Psi }
          \over {\partial \Phi }} +
   {{\partial \delta \Psi } \over {\partial \Phi }}} \right]  -
   {\cal I}\left[ {\Phi ,
   {{\partial \Psi } \over {\partial \Phi }}} \right]
    } \right)
$$
$$
  =\int {\left[ {\dif \Phi } \right]}
   {{\partial_r {\cal I}} \over {\partial \Phi_A^*}}
   {{\partial_l \delta \Psi } \over {\partial \Phi^A}} +
   O\left( {\left( {\delta \Psi } \right)^2} \right)
$$
$$
  =\int {\left[ {\dif \Phi } \right]}
  \Delta {\cal I} {\delta \Psi } +
  O\left( {\left( {\delta \Psi } \right)^2} \right)
\quad ,
$$
where integration by parts
(see Eq.\bref{int by parts} in Appendix A)
has been used,
and where the operator $\Delta$ is defined to be
\ct{bv81a,bv83b}
\be
  \Delta \equiv \left( {-1} \right)^{\epsilon_A+1}
  {{\partial_r } \over {\partial \Phi^A}}
  {{\partial_r } \over {\partial \Phi_A^*}}
\quad .
\label{def Delta}
\ee
It is a kind of ``nilpotent divergence operator''
in the space of fields and antifields.
It formally satisfies
$$
    \Delta \Delta = 0
 \quad , \quad \quad
$$
$$
  \Delta \left( X Y \right) = X \Delta Y +
  (-1)^{\epsilon_Y } \left( \Delta X \right) Y +
  (-1)^{\epsilon_Y } \left( X, Y \right)
\quad ,
$$
\be
  \Delta \left( \ (X, Y) \ \right) =
    \left( X, \Delta Y \right) -
  (-1)^{\epsilon_Y } \left( \Delta X,  Y \right)
\quad .
\label{Delta properties}
\ee
Note that
\be
 {\rm gh} \left[ \Delta \right] =
  \epsilon \left( \Delta \right) = 1
  \quad .
\label{Delta stats}
\ee
According to the above calculation,
the integral
$I_\Psi \left( X \right)$
is infinitesimally independent of $\Psi$ if
\be
 \Delta {\cal I} = 0
\label{good integrand}
\quad .
\ee
Eq.\bref{good integrand} is a requirement
of a ``good'' integrand.

The path integral itself
should be gauge independent.
This corresponds to setting $X = 1 $
in $I_\Psi \left( X \right)$,
from which one finds
\be
  \Delta \exp \left( {{i \over {\hbar}}W } \right) =
  \exp \left( {i \over {\hbar} } W \right)
  \left( {   {i \over {\hbar} } \Delta W
    - {1 \over {2 \hbar^2 } } \left( {W,W} \right)
          } \right)
   = 0
\quad ,
\label{del on exp W}
\ee
so that one needs
\be
   {1 \over 2}\left( {W,W} \right) =
   i\hbar \Delta W
\quad .
\label{quantum master equation}
\ee
Eq.\bref{quantum master equation}
is known as the {\it quantum master equation}
\ct{bv81a,bv83b}.
When $W$ satisfies
Eq.\bref{quantum master equation},
$X$ in
Eq.\bref{integrand}
must satisfy
\be
\left( {X,W} \right) = i\hbar \Delta X
\label{qme for correlation}
\ee
to produce gauge-invariant correlation functions.
Eqs.\bref{quantum master equation}
and \bref{qme for correlation}
summarize the conditions
so that a computation
does not depend on the choice of $\Psi$
\ct{fisch90a,henneaux90a}.

If $W$ is expanded in powers of $\hbar$
(the loop expansion) via
\be
  W = S + \sum\limits_{p=1}^\infty  {\hbar^p} M_p
\quad ,
\label{hbar expansion of W}
\ee
the quantum master equation
(Eq.\bref{quantum master equation})
implies that the first term $S$
and quantum correction terms $M_p$
must satisfy
$$
  \left( {S,S} \right) = 0
\ , \quad \quad
  \left( {M_1,S} \right) =
 i \Delta S
\quad ,
$$
\be
  \left( {M_n,S} \right)=
 i \Delta M_{n-1}-{1 \over 2}\sum\limits_{p=1}^{n-1} {}
 \left( {M_p,M_{n-p}} \right)
\ , \ {\rm  for \ } n \ge 2
\quad .
\label{conditions of loop terms}
\ee
By the correspondence principle, $S$
in Eq.\bref{hbar expansion of W}
is identified as the classical field-antifield action
and is consistent with
Eqs.\bref{quantum master equation} and
\bref{conditions of loop terms}
because it satisfies the classical master equation
$\left( S, S \right) = 0$.

Likewise, when $X$ in
Eq.\bref{qme for correlation}
is expanded in an $\hbar$ series
\be
X=\sum\limits_{p=0}^\infty \hbar^p X_p
\quad ,
\label{X hbar series}
\ee
one finds that
Eq.\bref{qme for correlation} implies
$$
  \left( {X_0,S} \right)=0
\quad ,
$$
\be
  \left( {X_n,S} \right) =
   i \Delta X_{n-1} - \sum\limits_{p=1}^n
    \left( {X_{n-p},M_p} \right)
\ , \ {\rm for \ } n \ge 1
\quad .
\label{X hbar BRST invariance}
\ee
The first equation
says that the ``classical'' part
of the quantum operator $X$
is classically BRST invariant.

\subsection{Gauge-Fixing Auxiliary Fields}
\label{ss:gfaf}

\hspace{\parindent}
To eliminate antighost fields
by Eq.\bref{fermion gauge fixing},
$\Psi$ must be a functional of fields
and have ghost number $-1$.
However,
the fields of the minimal sector
introduced
in Sect.\ \ref{s:faf}
all have non-negative ghost numbers.
Hence, it is impossible to construct an acceptable $\Psi$
unless one introduces additional fields.
At this point, one takes advantage
of adding trivial pairs
(c.f. Sect.\ \ref{ss:eu})
to the theory.
This subsection enumerates
the auxiliary trivial pairs
needed to obtain an admissible $\Psi$
\ct{bv83b,fs90b}.

For an irreducible theory,
one introduces one trivial pair
\be
 \bar {\cal C}_{0\alpha_0}
  \ , \quad  \quad
  \bar \pi_{0\alpha_0}
\quad ,
\label{aux fields for irr theory}
\ee
with statistics
and ghost numbers equal to
$$
 \epsilon \left( \bar {\cal C}_{0\alpha_0} \right) =
  \epsilon_{\alpha_0} + 1
  \ , \quad  \quad
 {\rm gh} \left[ \bar {\cal C}_{0\alpha_0} \right] = -1
\quad ,
$$
\be
 \epsilon \left( \bar \pi_{0\alpha_0} \right) =
  \epsilon_{\alpha_0}
  \ , \quad  \quad
 {\rm gh} \left[ \bar \pi_{0\alpha_0} \right] = 0
\quad .
\ee
The fields $\bar {\cal C}_{0\alpha_0}$ are
the Faddeev-Popov antighosts of
${\cal C}_{0}^{\alpha_0}$.

For a first-stage reducible theory,
one needs, in addition to the above pair,
two more trivial pairs
$$
 \bar {\cal C}_{1\alpha_1}
 \ , \quad \quad
 \bar \pi_{1\alpha_1}
 \quad ,
$$
\be
{\cal C}_1^{1\alpha_1}
 \ , \quad \quad
  \pi_1^{1\alpha_1}
\quad .
\label{aux fields for first stage red theory}
\ee
Their statistics
and ghost numbers are
$$
  \epsilon \left( \bar {\cal C}_{1\alpha_1} \right) =
  \epsilon \left( {\cal C}_1^{1\alpha_1} \right) =
  \epsilon_{\alpha_1}
  \quad ,
$$
$$
 {\rm gh} \left[ \bar {\cal C}_{1\alpha_1} \right] = -2
  \ , \quad \quad
 {\rm gh} \left[ {\cal C}_1^{1\alpha_1} \right] = 0
  \quad ,
$$
\be
  \epsilon \left( \bar \pi_{1\alpha_1}  \right) =
  \epsilon \left( \pi_1^{1\alpha_1} \right) =
  \epsilon_{\alpha_1} + 1
  \quad  ,
\ee
$$
 {\rm gh} \left[ \bar \pi_{1\alpha_1} \right] = -1
  \ , \quad \quad
 {\rm gh} \left[ \pi_1^{1\alpha_1} \right] = 1
  \quad .
$$

In general,
for an $L$th-stage reducible theory,
for each integer $s$ ranging from $0$ to $L$
one introduces $s+1$ trivial pairs
\ct{bv83b}:
$$
 \bar {\cal C}_{s\alpha_s}^k \ , \ \bar \pi_{s\alpha_s}^k
 \ ,\quad \quad k=0, \ 2, \ 4, \  \ldots ,
 \ 2 \left[ {{s \over 2}} \right]
\quad ,
$$
\be
  {\cal C}_s^{k\alpha_s} \  , \ \pi_s^{k\alpha_s}
  \ ,\quad \quad k=1, \ 3, \ 5, \  \ldots ,
  \ 2\left[ {{{s-1} \over 2}} \right]+1
 \quad ,
\label{auxiliary fields}
\ee
where $\alpha_s = 1, \dots , m_s$.
The total number of auxiliary trivial pairs is
$\sum\limits_{s=0}^L (s+1) =
\left( L+1 \right) \left( L+2 \right) / 2 $.
The ghosts
$\bar {\cal C}_{s\alpha_s}^k$,
${\cal C}_s^{k\alpha_s}$,
in Eq.\bref{auxiliary fields}
for $k \ge 1 $ are known as {\it extraghosts},
while the fields
$\bar \pi_{s\alpha_s}^k$, $\pi_s^{k\alpha_s}$,
are usually called Lagrange multipliers
for reasons which become clear below.
The original set of fields
$\left\{ \phi^i, \ C_s^{\alpha_s} \right\}$
of Sect.\ \ref{s:faf}
is called the
{\it minimal set}.
The minimal sector fields together with the auxiliary fields
in Eq.\bref{auxiliary fields}
are called the
{\it non-minimal set}.
In Eq.\bref{auxiliary fields},
$\left[ \  \right] $
stands for the greatest integer so that
$$
 \left[ {{m \over 2}} \right] =
 \left\{ \matrix{{m \over 2} \ ,
  \quad {\rm if} \ m \ {\rm is \  even} \ , \hfill\cr
  {{m-1} \over 2} \  , \quad
   {\rm if} \ m \ {\rm is \ odd} \ . \hfill\cr} \right.
$$
In general,
the subscript $s$ on
${\cal C}_s^{k\alpha_s}$ and
$\bar {\cal C}_{s\alpha_s}^k$
indicate the level of the field,
whereas the superscript $k$ distinguishes
different fields at the same level.
It is convenient to associate
$ \alpha_{-1}$ with the index $i$
on $\phi^i$
and define
$$
 {\cal C}_s^{-1\alpha_s} \equiv {\cal C}_s^{\alpha_s}
\ , \quad
  {\cal C}_{-1}^{-1\alpha_{-1}}\equiv \phi^i
\ , \quad
  \alpha_{-1} \equiv i
\quad ,
$$
\be
  { \bar {\cal C}}_{s\alpha_s}^0 \equiv \bar {\cal C}_{s\alpha_s}
   \ ,\quad \quad \bar \pi_{s\alpha_s}^0
  \equiv \bar \pi_{s\alpha_s}
\quad .
\label{field equivalences}
\ee
The following equations
also apply to the minimal sector fields
when $k=-1$ and $0$.

The statistics of
$\bar {\cal C}_{s\alpha_s}^k$ and
${\cal C}_s^{k\alpha_s}$
are the same as those of
${\cal C}_s^{ \alpha_s}$, namely,
\be
 \epsilon \left( \bar {\cal C}_{s\alpha_s}^k \right) =
  \epsilon \left( {\cal C}_s^{k\alpha_s} \right) =
   \epsilon_{\alpha_s} + s + 1  \ ( {\rm mod} \ 2 )
\quad ,
\label{aux eps}
\ee
and their ghost numbers are
$$
{\rm gh} \left[ \bar {\cal C}_{s \alpha_s}^k \right] = k - s - 1
  \ , \quad \quad
  L \ge s \ge 0 \ ,
 \quad  s \ge k \ge 0 \ , \quad {\rm for} \ k \ {\rm even}
 \quad ,
$$
\be
 {\rm gh} \left[ {\cal C}_s^{k \alpha_s} \right] = s - k
 \ , \quad \quad
 L \ge s \ge -1 \ ,
 \quad  s \ge k \ge -1 \ ,
 \quad {\rm for} \ k \ {\rm odd}
 \quad .
\label{aux gh numbers and eps}
\ee
The statistics and ghost numbers of
$ \bar \pi_{s\alpha_s}^k $
and
$ \pi_s^{k\alpha_s} $
are determined by the requirement
that they form a trivial pair respectively
with
$ \bar {\cal C}_{s\alpha_s}^k $
and
$ {\cal C}_s^{k \alpha_s} $.
Using Eq.\bref{Pi gh numbers and eps},
$$
 \epsilon \left( \bar \pi_{s\alpha_s}^k \right) =
  \epsilon \left( \pi_s^{k\alpha_s} \right) =
   \epsilon_{\alpha_s} + s   \ ({\rm mod} \ 2 )
	\quad ,
$$
$$
{\rm gh} \left[  \bar \pi_{s\alpha_s}^k \right] = k - s
 \ , \quad \quad
 L \ge s \ge 0 \ ,
 \quad  s \ge k \ge 0 \ ,
 \quad {\rm for} \ k \ {\rm even}
 \quad ,
$$
\be
 {\rm gh} \left[ \pi_s^{k\alpha_s} \right] = s - k + 1
 \ , \quad \quad
 L \ge s \ge 1 \ ,
 \quad  s \ge k \ge 1 \ ,
 \quad {\rm for} \ k \ {\rm odd}
 \quad .
\label{aux Pi gh numbers and eps}
\ee
The statistics and ghost numbers of the antifields
are determined from
Eq.\bref{antifield ghost numbers and statistics}.

The minimal sector fields along with
the auxiliary
$ \bar {\cal C}_{s\alpha_s}^k $
and
$ {\cal C}_s^{k \alpha_s} $
can be arranged
into the following useful triangular tableau
\ct{bv83b}:

\begin{picture}(400,240)
\put(170,220){$\phi^i$}
\put(150,190){\vector(3,4){15}}
\put(171,218){\line(-3,-4){15}}
\put(200,190){\line(-3,4){20}}
\put(140,180){$\bar {\cal C}_{0\alpha_0}$}
\put(200,180){${\cal C}^{\alpha_0}_0$}
\put(123,154){\vector(3,4){15}}
\put(141,178){\vector(-3,-4){15}}
\put(183,154){\vector(3,4){15}}
\put(201,178){\line(-3,-4){15}}
\put(230,150){\line(-3,4){20}}
\put(110,140){${\cal C}^{1\alpha_1}_1$}
\put(170,140){$\bar {\cal C}_{1\alpha_1}$}
\put(230,140){${\cal C}^{\alpha_1}_1$}
\put(93,114){\vector(3,4){15}}
\put(111,138){\vector(-3,-4){15}}
\put(153,114){\vector(3,4){15}}
\put(171,138){\vector(-3,-4){15}}
\put(213,114){\vector(3,4){15}}
\put(231,138){\line(-3,-4){15}}
\put(260,110){\line(-3,4){20}}
\put(80,100){$\bar {\cal C}^2_{2\alpha_2}$}
\put(140,100){${\cal C}^{1\alpha_2}_2$}
\put(200,100){$\bar {\cal C}_{2\alpha_2}$}
\put(260,100){${\cal C}^{\alpha_2}_2$}
\put(63,74){\vector(3,4){15}}
\put(81,98){\vector(-3,-4){15}}
\put(123,74){\vector(3,4){15}}
\put(141,98){\vector(-3,-4){15}}
\put(183,74){\vector(3,4){15}}
\put(201,98){\vector(-3,-4){15}}
\put(243,74){\vector(3,4){15}}
\put(261,98){\line(-3,-4){15}}
\put(290,70){\line(-3,4){20}}
\put(50,60){${\cal C}^{3\alpha_3}_3$}
\put(110,60){$\bar {\cal C}^2_{3\alpha_3}$}
\put(170,60){${\cal C}^{1\alpha_3}_3$}
\put(230,60){$\bar {\cal C}_{3\alpha_3}$}
\put(290,60){${\cal C}^{\alpha_3}_3$}
\put(170,35){.}
\put(170,30){.}
\put(170,25){.}
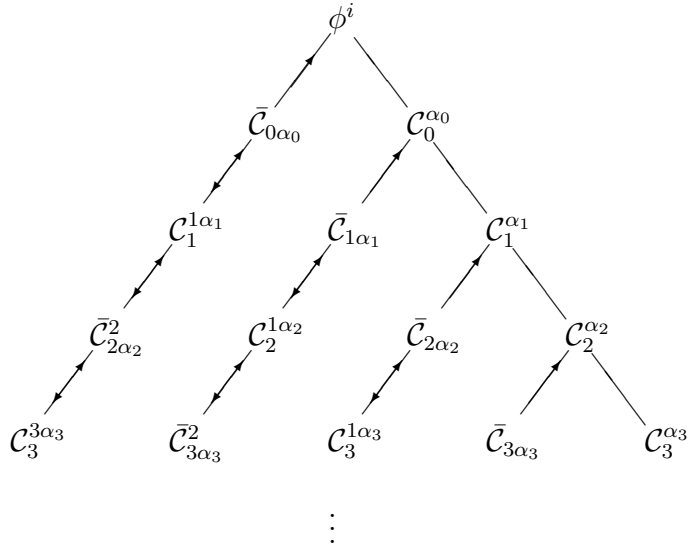
\put(70,5){\rm Diagram\ 2.\ The\ Triangular\ Field\ Tableau}
\end{picture}

The index $s$ of ${\cal C}^{k \alpha_s}_{s}$ or
$\bar {\cal C}^{k}_{s \alpha_s}$
labels different horizontal rows,
whereas the index $k$ labels different rows
slanting to the right and downward.

The new proper solution $S_{\rm nm}$
of the classical master equation,
involving the non-minimal set of fields,
is given by
\be
  S_{\rm nm} = S + S_{\rm aux}
\quad ,
\label{nm proper sol}
\ee
where $S$ is the proper solution
for the minimal set of fields presented
in Sect.\ \ref{s:faf}
and $S_{\rm aux}$ is the trivial-pair proper solution
for the auxiliary fields given by
$$
 S_{\rm aux} = \bar \pi_{0\alpha_0}\bar {\cal C}_0^{*\alpha_0}
 + \ \ \ \ \ \ \
$$
$$
 {\cal C}_{1\alpha_1}^{1*} \pi_1^{1\alpha_1} +
 \bar \pi_{1\alpha_1}\bar {\cal C}_1^{*\alpha_1} +
$$
$$
 \bar \pi_{2\alpha_2}^2\bar {\cal C}_2^{2*\alpha_2} +
 {\cal C}_{2\alpha_2}^{1*} \pi_2^{1\alpha_2} +
 \bar \pi_{2\alpha_2}\bar {\cal C}_2^{*\alpha_2} +
$$
$$
 {\cal C}_{3\alpha_3}^{3*} \pi_3^{3\alpha_3} +
 \bar \pi_{3\alpha_3}^2 \bar {\cal C}_3^{2*\alpha_3} +
 {\cal C}_{3\alpha_3}^{1*} \pi_3^{1\alpha_3} +
 \bar \pi_{3\alpha_3}\bar {\cal C}_3^{*\alpha_3} +
$$
$$
\dots
$$
or
\be
  S_{{\rm aux}} =
  \sum \limits^L_{\scriptstyle {k = 0} \hfill\atop
  \scriptstyle {k\ {\rm even}}\hfill} {}\
   \sum\limits_{s = k}^L {}
   \bar \pi_{s\alpha_s}^k \bar {\cal C}_s^{k*\alpha_s} +
   \sum\limits^L_{\scriptstyle {k\ = 1}\hfill\atop
   \scriptstyle {k\ {\rm odd}}\hfill} {}
  \ \sum\limits_{s = k}^L {}
  {\cal C}_{s\alpha_s}^{k*} \pi_s^{k\alpha_s}
\quad .
\label{aux S}
\ee

The auxiliary antighost fields at level $s$
are eliminated from the theory using
Eq.\bref{fermion gauge fixing}
which leads to
$$
 \bar {\cal C}_s^{k*\alpha_s} =
  {{\partial \Psi } \over
   {\partial \bar {\cal C}_{s\alpha_s}^k}}
 \ ,\quad \quad k=0, \ 2, \ 4, \  \ldots ,
 \ 2 \left[ {{s \over 2}} \right]
  \quad ,
$$
\be
  {\cal C}_{s\alpha_s}^{k*} =
  {{\partial \Psi } \over
   {\partial {\cal C}_s^{k\alpha_s}}}
  \ ,\quad \quad k= -1, \ 1, \ 3, \ 5, \  \ldots ,
  \ 2\left[ {{{s-1} \over 2}} \right]+1
\quad ,
\label{fixing aux antighosts}
\ee
at each level $s$.

\subsection{Delta-Function Gauge-Fixing Procedure}
\label{ss:dfgfp}

\hspace{\parindent}
Let us consider a gauge-fixing scheme
for which $\Psi$ is not a function of
$ \bar \pi_{s\alpha_s}^k  $ or
$ \pi_s^{k\alpha_s}$
(for $ k \ge 0 $).
This is called a
$\delta$-function gauge-fixing procedure
because when
Eq.\bref{fixing aux antighosts}
is substituted
in $S_{\rm aux}$ in
Eq.\bref{aux S},
the conditions
\be
  \bar {\cal C}_s^{k*\alpha_s} =
  {{\partial \Psi } \over
   {\partial \bar {\cal C}_{s\alpha_s}^k}} = 0
 \ , \quad \quad
       {\cal C}_{s\alpha_s}^{k*} =
  {{\partial \Psi } \over
   {\partial {\cal C}_s^{k\alpha_s}}} = 0
 \quad ,
\label{gf delta functions}
\ee
are implemented
because the $\pi_s^{k\alpha_s}$
and $\bar \pi_{s\alpha_s}^k$
in Eq.\bref{aux S}
act as Lagrange multipliers for these equations.
Indeed,
in a functional integral approach,
integration over
$\bar \pi_{s\alpha_s}^k$ and
$\pi_s^{k\alpha_s}$
leads to the insertion of
$
  \delta \left( {{\partial \Psi } \over
   {\partial \bar {\cal C}_{s\alpha_s}^{k} }^{\ } } \right)
$
and
$
 \delta \left(   {{\partial \Psi } \over
   {\partial {\cal C}_s^{k\alpha_s}}^{\ } }  \right)
$
in the integrand.
Since $\Psi$ does not depend on
$ \bar \pi_{s\alpha_s}^k  $ or
$ \pi_s^{k\alpha_s}$,
$
\bar \pi_{s}^{k*\alpha_s} =
 {{\partial \Psi } \over
   {\partial \bar \pi_{s\alpha_s}^k }}  = 0
$ and
$
 \pi_{s\alpha_s}^{k*} =
  {{\partial \Psi } \over
   {\partial  \pi_s^{k\alpha_s} }}   = 0
$.

Not every $\Psi$ is acceptable.
For example,
if $\Psi$ is identically zero,
all antifields are set to zero
and the gauge-fixed proper solution reduces
to the original action $S_0$.
This $S_0$ could not be used as the starting point
for perturbation theory
because propagators would not exist.
In the rest of this section,
we determine the conditions on $\Psi$
so that an admissible gauge-fixing procedure
arises.
The discussion is somewhat technical
so that a reader may wish to skip
to Sect.\ \ref{ss:gfcbrsts}.
For the irreducible case,
one should read
to Eq.\bref{irr conditions on Psi}.

It is useful to define $n_s$
as the net number of non-gauge degrees of freedom
for the ghost $C^{\alpha_s}_s$ at level $s$:
\be
n_s\equiv \sum\limits_{t=s}^L {\left( {-1} \right)^{t-s}m_t}
\quad .
\label{def of ns}
\ee
The number of gauge degrees of freedom
of ${\cal C}^{\alpha_s}_s$
is $m_s - n_s = n_{s+1}$.
If we define $m_{-1} = n$
and use the notation in
Eq.\bref{field equivalences},
these statements also apply to $\phi^i$
when $s=-1$:
$n_{-1} = m_{-1} - n_0 = n - n_0 = n_{\rm dof}$
is the number of non-gauge degrees of freedom
in $\phi^i$.

In this paragraph,
the general strategy is discussed.
The basic idea is that the conditions in
Eq.\bref{gf delta functions}
fix the gauge degrees of freedom of level $s-1$ fields
and non-propagating independent degrees of freedom
of level $s+1$ fields.
In the Triangular Field Tableau of Diagram 2,
arrows indicate how
setting the antifield of one field to zero
via Eq.\bref{gf delta functions}
eliminates degrees of freedom in another field.
An upward-slanting arrow
going from $\Phi^B$ to $\Phi^A$ such as

\medskip

\begin{picture}(400,70)
\put(170,60){$\Phi^A$}
\put(153,34){\vector(3,4){15}}
\put(171,58){\line(-3,-4){15}}
\put(140,20){$\Phi^B$}
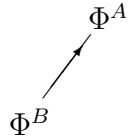
\put(50,5){Diagram 3.\ Elimination of Gauge Degrees of Freedom}
\end{picture}

\medskip

\noindent
indicates that the gauge degrees of freedom in $\Phi^A$
are eliminated by the equation
${{\partial \Psi} \over {\partial \Phi^*_B}}=0$.
To ensure their elimination,
two points on the gauge slice determined by $\Psi$
should not be related by an infinitesimal gauge transformation.
This leads to conditions on $\Psi$
that are presented below.
A downward-slanting arrow going from $\Phi^A$ to $\Phi^B$ such as

\medskip

\begin{picture}(400,70)
\put(170,60){$\Phi^A$}
\put(153,34){\line(3,4){15}}
\put(171,58){\vector(-3,-4){15}}
\put(140,20){$\Phi^B$}
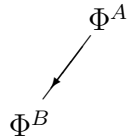
\put(35,5){Diagram 4.\ Elimination of Non-Gauge Degrees of Freedom}
\end{picture}

\medskip

\noindent
indicates that non-propagating
independent degrees of freedom
in $\Phi^B$
are eliminated by the equation
${{\partial \Psi} \over {\partial \Phi^*_A}}=0$.
When two arrows appear on a line,
there is the fixing
of both independent and gauge degrees of freedom.
In this delta-function gauge-fixing procedure,
only $\phi^i$,
${\cal C}^{\alpha_s}_s$ and $\bar {\cal C}_{s \alpha_s}$
contain propagating fields.
In the Triangular Field Tableau of Diagram 2,
there are no downward-sloping lines terminating
on these fields to indicate that non-gauge degrees
of freedom are eliminated.
In contrast,
there are upward-sloping lines meaning that
the gauge degrees of freedom in
$\phi^i$,
${\cal C}^{\alpha_s}_s$ and $\bar {\cal C}_{s \alpha_s}$
are fixed.
The fields
${\cal C}_s^{k\alpha_s}$,
for $k \ge 1$ and odd $k$, and
$\bar {\cal C}_{s\alpha_s}^k$,
for $k \ge 2$ and even $k$,
are non-propagating, i.e.,
there is no quadratic form in the action for these fields.
When gauge-fixing is fully implemented,
they are completely fixed.
Like
${\cal C}^{\alpha_s}_s$ and $\bar {\cal C}_{s \alpha_s}$,
it will be useful to think of
${\cal C}_s^{k\alpha_s}$ and
$\bar {\cal C}_{s\alpha_s}^k$
as having $n_s$
independent (or non-gauge) degrees of freedom
and $n_{s+1}$ gauge degrees of freedom.
Since there are both
upward-sloping and downward-sloping lines
terminating on these fields,
both gauge and independent degrees of freedom are fixed.
Hence, all $m_s = n_{s+1} + n_s $
fields are removed from the theory.
The difference between
${\cal C}^{\alpha_s}_s$ and $\bar {\cal C}_{s \alpha_s}$
and the other ghosts and antighosts
is that the independent degrees of freedom in these fields
are propagating.
The fields $\bar {\cal C}_{s \alpha_s}$ play a dual role:
They fix the gauge degrees of freedom in
${\cal C}^{\alpha_{s-1}}_{s-1}$
and serve as the antighosts of
${\cal C}^{\alpha_s}_s$,
meaning that
they enter quadratically in the action with these fields.

Let $\Phi_0$ be a solution to the equations of motion
for the action $S_{\Psi}$
fixed by the gauge conditions in
Eq.\bref{fermion gauge fixing}.
Since one is interested in performing
a perturbative expansion
about this solution, write
\be
  \Phi^A = \Phi_0^A + \delta \Phi^A
\quad ,
\label{pert exp}
\ee
where $\delta \Phi^A$ is the quantum fluctuation of $\Phi^A$.
According to the Triangular Field Tableau of Diagram 2,
the gauge degrees of freedom of $\phi^i$
should be fixed by
the equation $0 = \bar {\cal C}_0^{*\alpha_0}$.
Expanding about the perturbative solution,
one obtains
\be
 0 = \bar {\cal C}_0^{*\alpha_0} =
 {{\partial \Psi } \over {\partial \bar {\cal C}_{0\alpha_0}}} =
  \left( {{{\partial \Psi } \over
  {\partial \bar {\cal C}_{0\alpha_0}}}} \right)_0
  + \left( {
  {{ \partial_l \partial_r \Psi } \over
   { \partial \bar {\cal C}_{0\alpha_0} \partial \phi^i}}
           } \right)_0 \delta \phi^i + \ldots
\quad ,
\label{eq fixing gauge modes of phi i}
\ee
where we use the abbreviation
\be
  \left( {} \right)_0 \equiv
  \left. {\left( {} \right)}
   \right|_{ \left\{ \Phi^A=\Phi_0^A \right\} }
\quad .
\label{0 notation}
\ee
The matrix
$
 \left( {
  {{ \partial_l \partial_r \Psi } \over
   { \partial \bar {\cal C}_{0\alpha_0} \partial \phi^i}}
       } \right)_0
$
that multiplies
$\delta \phi^i$
must eliminate
$n_0$ gauge degrees of freedom.
This requires that $\Psi$ be such that
\be
  {\rm rank \ }
 \left( {
  {{ \partial_l \partial_r \Psi } \over
   { \partial \bar {\cal C}_{0\alpha_0} \partial \phi^i}}
         } \right)_0
    = n_0
\quad .
\label{psi level 0 cond}
\ee
The gauge modes of $\phi^i$ are proportional to
$R_{\alpha_0}^i$.
Hence one also needs
\be
 {\rm rank \ }
  \left( {
  {{ \partial_l \partial_r \Psi } \over
   { \partial \bar {\cal C}_{0\alpha_0} \partial \phi^i}}
    R_{\beta_0}^i    } \right)_0 = n_0
\quad ,
\label{psi level 0 condz}
\ee
to ensure that the correct degrees of freedom
in $\phi^i$ are fixed.
Using Eq.\bref{beg fsr proper solution}
and
$\phi_i^* = { {\partial \Psi} \over {\partial \phi^i} }$,
one sees that
the operator in Eq.\bref{psi level 0 condz}
is the quadratic form for
$\bar {\cal C}_{0\alpha_0}$ and ${\cal C}^{\beta_0}_0$
in the action $S_\Psi$:
\be
 S_{\Psi} = \dots +
  \bar {\cal C}_{0\alpha_0}
  \left( {
  {{ \partial_l \partial_r \Psi } \over
   { \partial \bar {\cal C}_{0\alpha_0} \partial \phi^i}}
    R_{\beta_0}^i    } \right)_0
 {\cal C}^{\beta_0}_0 + \dots
\quad .
\label{qf for level 0 ghosts}
\ee
Hence Eq.\bref{psi level 0 condz}
is consistent with the fact that
$\bar {\cal C}_{0\alpha_0}$ and ${\cal C}^{\beta_0}_0$
should have $n_0$ propagating degrees of freedom.

If the system is irreducible,
$n_0 =m_0$ and
no further constraints
on the gauge-fixing fermion $\Psi$
are necessary.
Eqs.\bref{psi level 0 cond}
and \bref{psi level 0 condz}
are the only requirements on $\Psi$.
There are $n-m_0$ propagating modes for $\phi^i$
and $m_0$ propagating modes for
$\bar {\cal C}_{0\alpha_0}$ and ${\cal C}^{\alpha_0}_0$.
Of the original $n$ degrees of freedom of $\phi^i$,
$m_0$ have been gauge fixed.
The simplest gauge-fixing fermion $\Psi_\delta$ is
\ct{bv83b}
\be
  \Psi_\delta = \bar {\cal C}_{0\alpha_0}
  \chi^{\alpha_0}\left( \phi \right)
\quad ,
\label{irr Psi}
\ee
where $\chi^{\alpha_0}$ is
an arbitrary functional of $\phi$.
The subscript $\delta$ on $\Psi$
indicates that this is a delta-function
gauge-fixing scheme.
The condition
$
 0=\bar {\cal C}_0^{*\alpha_0}
$
in Eq.\bref{eq fixing gauge modes of phi i}
leads to
\be
\chi^{\alpha_0}\left( \phi  \right)=0
\quad ,
\label{gauge condition on phi}
\ee
so that $\chi^{\alpha_0}$
is the gauge-fixing condition on $\phi$.
Eqs.\bref{psi level 0 cond}
and \bref{psi level 0 condz}
respectively become
$$
  {\rm  rank \ } \left( {\chi_{,i}^{\alpha_0}
    \left( {\phi_0} \right)} \right) = n_0
\quad ,
$$
\be
  {\rm  rank \ } \left( {\chi_{,i}^{\alpha_0}
   R_{\beta_0}^i\left( {\phi_0} \right)} \right)
   = n_0
\quad ,
\label{irr conditions on Psi}
\ee
where $n_0 =m_0$ for the irreducible case.

Suppose the theory is at least first-stage reducible.
Then
$\bar {\cal C}_{0\alpha_0}$ and ${\cal C}^{\alpha_0}_0$
have $m_0$ degrees of freedom but
$n_1$ of these are gauge modes and thus
$m_0 - n_1 = n_0$ are independent.
According to the
Triangular Field Tableau of Diagram 2,
the equation
$\bar {\cal C}^{* \alpha_1 }_{1 } = 0$
is used to fix the $n_1$ gauge degrees of freedom
in ${\cal C}^{\alpha_0}_0$,
and
$ {\cal C}_{1 \alpha_1 }^{1 * } = 0$
is used to fix the $n_1$ gauge degrees of freedom
in $\bar {\cal C}_{0\alpha_0}$.
Expanding the former condition
about the perturbative solution,
one obtains
\be
  0=\bar {\cal C}_1^{*\alpha_1} =
   {{\partial \Psi } \over
       {\partial \bar {\cal C}_{1\alpha_1}}} =
    \left( {{{\partial \Psi } \over
     { \partial \bar {\cal C}_{1\alpha_1}}}} \right)_0 +
     \left( {
      { { \partial_l \partial_r \Psi } \over
       { \partial \bar {\cal C}_{1\alpha_1}
         \partial {\cal C}_0^{\alpha_0} } }
             } \right)_0
    \delta {\cal C}_0^{\alpha_0}+\ldots
\quad .
\label{eq fixing gauge modes of C 0 0}
\ee
Actually,
the first term on the right-hand side
of Eq.\bref{eq fixing gauge modes of C 0 0}
vanishes,
but we display it to emphasize
the idea that we are expanding
about the perturbative solution.
Since one wants to fix $n_1$ gauge modes of
${\cal C}^{\alpha_0}_0$,
a necessary condition on $\Psi$ is that
\be
  {\rm rank \ }
 \left( {
      { { \partial_l \partial_r \Psi } \over
       { \partial \bar {\cal C}_{1\alpha_1}
         \partial {\cal C}_0^{\alpha_0} } }
         } \right)_0 = n_1
\quad .
\label{psi level 1 conda}
\ee
The modes to gauge-fix are those
that are non-propagating
in Eq.\bref{qf for level 0 ghosts}.
These modes are proportional to
\be
 {Z_{1\beta_1}^{-1 \alpha_0}} \equiv
\left. {R_{1\beta_1}^{\alpha_0}} \right|_0
\quad ,
\label{def of Z level 1}
\ee
of Eq.\bref{depen generadors}.
Since there are $n_1$ such modes,
\be
  {\rm rank \ }
  \left(  {Z_{1\beta_1}^{-1 \alpha_0}}  \right) = n_1
\quad .
\label{rank of Z level 1}
\ee
Because these modes need to be eliminated
by the
$0=\bar {\cal C}_1^{*\alpha_1}$ condition
in Eq.\bref{eq fixing gauge modes of C 0 0},
they should not be annihilated by the quadratic form
$
 \left( {
      { { \partial_l \partial_r \Psi } \over
       { \partial \bar {\cal C}_{1\alpha_1}
         \partial {\cal C}_0^{\alpha_0} } }
         } \right)_0
$
in Eq.\bref{psi level 1 conda}.
This leads to the condition on $\Psi$ that
\be
  {\rm rank \ }
  \left( {
      { { \partial_l \partial_r \Psi } \over
       { \partial \bar {\cal C}_{1\alpha_1}
         \partial {\cal C}_0^{\alpha_0} } }
   Z_{1\beta_1}^{-1 \alpha_0}
        } \right)_0 = n_1
\quad .
\label{psi level 1 condaz}
\ee
Using
$
  { {\cal C}_{0 \alpha_0}^{*} } =
  { {\partial_r \Psi} \over { \partial {\cal C}_0^{\alpha_0} } }
$
in Eq.\bref{beg fsr proper solution}
and Eq.\bref{def of Z level 1},
one notices that the operator in
Eq.\bref{psi level 1 condaz}
is the quadratic form for
$\bar {\cal C}_{1 \alpha_1}$ and
${\cal C}_1^{\beta_1}$.
Since these fields have $n_1$
propagating degrees of freedom,
the rank of the quadratic form should be $n_1$,
which is consistent with
Eq.\bref{psi level 1 condaz}.

To fix the $n_1$ gauge modes of
$\bar {\cal C}_{0\alpha_0}$,
use
\be
  0 = {\cal C}_{1\alpha_1}^{1*} =
  {{\partial \Psi } \over {\partial {\cal C}_1^{1\alpha_1}}} =
   \left( {{{\partial \Psi } \over
     {\partial {\cal C}_1^{1\alpha_1}}}} \right)_0 +
   \delta \bar {\cal C}_{0\alpha_0}
  \left( {
   { { \partial_l \partial_r \Psi } \over
     { \partial \bar {\cal C}_{0\alpha_0}
       \partial {\cal C}_1^{1\alpha_1}  }  }
          } \right)_0
    + \ldots
\quad .
\label{eq fixing gauge modes of C bar 0 0}
\ee
A necessary condition that these
gauge degrees of freedom can be fixed is
\be
  {\rm rank \ }
  \left( {
   { { \partial_l \partial_r \Psi } \over
     { \partial \bar {\cal C}_{0\alpha_0}
       \partial {\cal C}_1^{1\alpha_1}  }  }
          } \right)_0
   = n_1
\quad .
\label{psi level 1 condb}
\ee
Next, note that, since
$
 {\rm rank \ }
   \left( {
  { { \partial_l \partial_r \Psi } \over
    { \partial \bar {\cal C}_{0\alpha_0} \partial \phi^i } }
          } \right)_0
  = n_0
$
(see Eq.\bref{psi level 0 cond})
and the index $\alpha_0$ takes on $m_0$ values,
there must be $m_0 - n_0 = n_1$ left zero modes
${\bar Z_{1\alpha_0}^{0 \alpha_1}}$
for the operator
$
   \left( {
  { { \partial_l \partial_r \Psi } \over
    { \partial \bar {\cal C}_{0\alpha_0} \partial \phi^i } }
          } \right)_0
$:
\be
  \bar Z_{1\alpha_0}^{0 \alpha_1}
   \left( {
  { { \partial_l \partial_r \Psi } \over
    { \partial \bar {\cal C}_{0\alpha_0} \partial \phi^i } }
          } \right)_0
   = 0
\quad ,
\label{def of Z bar level 1}
\ee
where we use a redundant superscript $\alpha_1$ on
$\bar Z_{1\alpha_0}^{0 \alpha_1}$
but compensate for this by requiring
\be
  {\rm rank \ }
  \left( {\bar Z_{1\alpha_0}^{0 \alpha_1}} \right) = n_1
\quad .
\label{rank of Z bar level 1}
\ee
The $\bar Z_{1\alpha_0}^{0 \alpha_1}$
are the non-propagating modes of
$\bar {\cal C}_{0\alpha_0}$
in the quadratic form
in Eq.\bref{qf for level 0 ghosts}.
They
must be fixed from the
$0 = {\cal C}_{1\alpha_1}^{1*}$ condition
in Eq.\bref{eq fixing gauge modes of C bar 0 0},
so that
\be
  {\rm rank \ }
 \left( {
   \bar Z_{1\beta_0}^{0 \alpha_1}
   { { \partial_l \partial_r \Psi } \over
     { \partial \bar {\cal C}_{0\alpha_0}
       \partial {\cal C}_1^{1\alpha_1}  }  }
       } \right)_0
   = n_1
\quad .
\label{psi level 1 condbz}
\ee

Finally one needs to fix
the  $n_1$ independent or ``non-gauge'' modes
of $ {\cal C}_1^{1\alpha_1}$.
{}From Diagram 2,
one sees that
this is done using
\be
 0 = \bar {\cal C}_0^{*\alpha_0} =
   {{\partial \Psi } \over
     {\partial \bar {\cal C}_{0\alpha_0}}} =
   \left( {{{\partial \Psi } \over
    {\partial \bar {\cal C}_{0\alpha_0}}}} \right)_0 +
 \left( {
   { { \partial_l \partial_r \Psi } \over
     { \partial \bar {\cal C}_{0\alpha_0}
       \partial {\cal C}_1^{1\alpha_1}  }  }
       } \right)_0
  \delta {\cal C}_1^{1\alpha_1} + \ldots
\quad ,
\label{eq fixing gauge modes of C 1 1}
\ee
for which one needs
$$
 {\rm rank \ }
 \left( {
   { { \partial_l \partial_r \Psi } \over
     { \partial \bar {\cal C}_{0\alpha_0}
       \partial {\cal C}_1^{1\alpha_1}  }  }
       } \right)_0
  = n_1
\quad ,
$$
which is the same condition as
Eq.\bref{psi level 1 condb}.

If the system is first-stage reducible
then $n_1 = m_1$ and there are no gauge degrees of freedom
in $ {\cal C}_1^{1\alpha_1}$
so that the above condition eliminates
all the modes of ${\cal C}_1^{1\alpha_1}$.
For $\phi^i$,
one has $n - m_0 + m_1$ propagating modes
and $m_0 - m_1$ gauge modes fixed by gauge conditions.
For
$\bar {\cal C}_{0\alpha_0}$ and
${\cal C}^{\alpha_0}_0$,
there are $m_0 - m_1$ propagating modes
and the $m_1$ gauge degrees of freedom are fixed.
Lastly, all
$m_1$ degrees of freedom of
$\bar {\cal C}_{1\alpha_1}$
and of
${\cal C}^{\alpha_1}_1$
are propagating.
Summarizing,
for a first-stage reducible theory
the conditions on $\Psi$ are
given in Eqs.\bref{psi level 0 cond},
\bref{psi level 0 condz},
\bref{psi level 1 conda},
\bref{psi level 1 condaz},
\bref{psi level 1 condb}, and
\bref{psi level 1 condbz}.

An example
of an acceptable $\Psi_\delta$ is
\ct{bv83b}
\be
  \Psi_\delta =\bar {\cal C}_{0\alpha_0}
    \chi^{\alpha_0}\left( \phi  \right) +
      \bar {\cal C}_{1\alpha_1}
   \omega_{1\alpha_0}^{0\alpha_1}\left( \phi  \right)
      {\cal C}_0^{\alpha_0} +
    \bar {\cal C}_{0\alpha_0}
   \bar \omega_{0\alpha_1}^{1\alpha_0}\left( \phi  \right)
    {\cal C}_1^{1\alpha_1}
\quad ,
\label{1st stage Psi}
\ee
where
$
 \chi^{\alpha_0}
$,
$
 \omega_{1\alpha_0}^{0\alpha_1}
$ and
$
 \bar \omega_{0\alpha_1}^{1\alpha_0}
$
are arbitrary functionals of $\phi$
subject to
Eq.\bref{irr conditions on Psi}
and
$$
  {\rm rank \ } \left( {\omega_{1\alpha_0}^{0\alpha_1}
    \left( {\phi_0} \right)} \right)=n_1
\quad ,
$$
$$
  {\rm rank \ } \left( {\omega_{1\alpha_0}^{0\alpha_1}
   \left( {\phi_0} \right)
  Z_{1\beta_1}^{-1 \alpha_0} } \right) = n_1
\quad ,
$$
$$
 {\rm rank \ } \left( {\bar \omega_{0\alpha_1}^{1\alpha_0}
   \left( {\phi_0} \right)} \right)=n_1
\quad ,
$$
\be
  {\rm rank \ }
 \left( {
   \bar Z_{1\alpha_0}^{0 \alpha_1}
   \bar \omega_{0 \alpha_1}^{1 \alpha_0} \left( {\phi_0} \right)
        } \right)=n_1
\quad ,
\label{1st stage conditions on Psi}
\ee
where $n_1 = m_1$ for first-stage reducible theories.
The $Z_{1\beta_1}^{-1 \alpha_0}$ are given
in Eq.\bref{def of Z level 1}
and $\bar Z_{1\alpha_0}^{0 \alpha_1}$
is determined by
$
 \bar Z_{1\alpha_0}^{0 \alpha_1} \chi_{ , i}^{\alpha_0}
  \left( {\phi_0} \right) = 0
$.

For completeness,
we present the conditions on $\Psi$
for the general $L$th-stage reducible theory.
There are fields from levels $s=-1$ to $s=L$.
See Triangular Field Tableau of Diagram 2.

For odd $k$ and $-1 \le k \le s$,
there are conditions fixing
the $n_{s+1}$ gauge degrees
of ${\cal C}_s^{k\alpha_s}$
coming from the diagram

\medskip

\begin{picture}(400,70)
\put(170,60){${\cal C}_s^{k\alpha_s}$}
\put(153,34){\vector(3,4){15}}
\put(171,58){\line(-3,-4){15}}
\put(140,20){$\bar {\cal C}_{s+1\alpha_{s+1}}^{k+1}$}
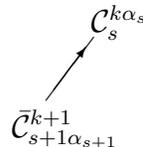
\put(10,5){Diagram 5.\ Elimination of Gauge Degrees of Freedom
in ${\cal C}_s^{k\alpha_s}$}
\end{picture}

\medskip

\noindent
and arising from
\be
  0 = \bar {\cal C}_{s+1}^{k+1*\alpha_{s+1}} =
  {{\partial \Psi } \over
    {\partial \bar {\cal C}_{s+1\alpha_{s+1}}^{k+1}}} =
   \dots +
   \left( {
    { { \partial_l \partial_r \Psi } \over
      { \partial \bar {\cal C}_{s+1\alpha_{s+1}}^{k+1}
        \partial {\cal C}_s^{k\alpha_s} } }
           } \right)_0
   \delta {\cal C}_s^{k\alpha_s} + \ldots
\quad .
\label{eq fixing gauge modes of C k s}
\ee
A necessary condition that the $n_{s+1}$
gauge modes of ${\cal C}_s^{k\alpha_s}$
can be fixed is
\be
  {\rm rank \ }
   \left( {
    { { \partial_l \partial_r \Psi } \over
      { \partial \bar {\cal C}_{s+1\alpha_{s+1}}^{k+1}
        \partial {\cal C}_s^{k\alpha_s} } }
           } \right)_0
   =  n_{s+1}
\quad ,
\label{psi level s conda}
\ee
where $k$ is odd and ranges between
$ s \ge k \ge -1 $,
and $s$ is restricted to $ L-1 \ge s \ge -1$.

When there is a diagram such as

\medskip

\begin{picture}(400,70)
\put(170,60){$\bar {\cal C}_{s-1\alpha_{s-1}}^{k-1}$}
\put(153,34){\line(3,4){15}}
\put(171,58){\vector(-3,-4){15}}
\put(140,20){$ {\cal C}_s^{k\alpha_s}$}
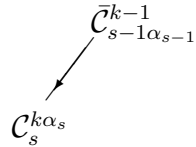
\put(10,5){Diagram 6.\ Elimination of Independent Degrees of Freedom
in ${\cal C}_s^{k\alpha_s}$}
\end{picture}

\medskip

\noindent
the remaining
$n_s$ independent degrees of freedom
in $ {\cal C}_s^{k\alpha_s} $ are fixed using
\be
  0=\bar {\cal C}_{s-1}^{k-1*\alpha_{s-1}} =
    {{\partial \Psi } \over
       {\partial \bar {\cal C}_{s-1\alpha_{s-1}}^{k-1}}} =
 \dots +
   \left( {
    { { \partial_l \partial_r \Psi } \over
      { \partial \bar {\cal C}_{s-1\alpha_{s-1}}^{k-1}
        \partial {\cal C}_s^{k\alpha_s} } }
           } \right)_0
    \delta {\cal C}_s^{k\alpha_s}+\ldots
\quad .
\label{eq fixing ind modes of C k s}
\ee
A necessary condition that these $n_{s}$
non-gauge modes can be fixed is
\be
  {\rm rank \ }
   \left( {
    { { \partial_l \partial_r \Psi } \over
      { \partial \bar {\cal C}_{s-1\alpha_{s-1}}^{k-1}
        \partial {\cal C}_s^{k\alpha_s} } }
           } \right)_0
    = n_s
\quad ,
\label{psi level s condb}
\ee
where $k$ is odd,
$ s \ge k \ge 1 $,
and $L \ge s \ge 1$.

The gauge modes
of
${\cal C}_s^{k\alpha_s}$
are those not fixed by the condition
in Eq.\bref{eq fixing ind modes of C k s}.
They are annihilated by the quadratic form
$
   \left( {
    { { \partial_l \partial_r \Psi } \over
      { \partial \bar {\cal C}_{s-1\alpha_{s-1}}^{k-1}
        \partial {\cal C}_s^{k\alpha_s} } }
           } \right)_0
$
in Eq.\bref{eq fixing ind modes of C k s}.
Since this operator
acts on $m_s$-component vectors
and has rank $n_s$,
it must have exactly
$m_s - n_s = n_{s+1}$ right null vectors.
We denote these by
$Z_{s+1\beta_{s+1}}^{k\alpha_s}$
using the redundant subscript index
$\beta_{s+1}$
but requiring
\be
  {\rm rank \ } \left( {Z_{s+1\beta_{s+1}}^{k\alpha_s}}
        \right) = n_{s+1}
\quad .
\label{rank of Z s k}
\ee
More precisely,
\be
   \left( {
    { { \partial_l \partial_r \Psi } \over
      { \partial \bar {\cal C}_{s-1\alpha_{s-1}}^{k-1}
        \partial {\cal C}_s^{k\alpha_s} } }
           } \right)_0
 Z_{s+1\beta_{s+1}}^{k\alpha_s} = 0
\quad .
\label{def of Z s k}
\ee
A sufficient condition that these gauge modes
$ Z_{s+1\beta_{s+1}}^{k\alpha_s}$
can be removed from
${\cal C}_s^{k\alpha_s}$
via Eq.\bref{eq fixing gauge modes of C k s}
is that
\be
  {\rm rank \ }
  \left( {
    { { \partial_l \partial_r \Psi } \over
      { \partial \bar {\cal C}_{s+1\alpha_{s+1}}^{k+1}
        \partial {\cal C}_s^{k\alpha_s} } }
   Z_{s+1\beta_{s+1}}^{k\alpha_s} } \right)_0 = n_{s+1}
\quad .
\label{psi level s condaz}
\ee
The conditions on $\Psi$
in Eq.\bref{psi level s condaz}
hold for odd $k$,
$ s \ge k \ge -1$ and $L-1 \ge s \ge -1$
if one defines
\be
 Z_{{s+1}\beta_{{s+1}}}^{-1 \alpha_{s}} \equiv
 \left. { R_{{s+1}\beta_{{s+1}}}^{\alpha_{s}} } \right|_0
\quad .
\label{def of Z -1 s}
\ee
When $k=-1$ the operator in
Eq.\bref{psi level s condaz}
is the quadratic form of $S_\Psi$ for
$\bar {\cal C}_{s+1 \alpha_{s+1}}$
and ${\cal C}_{s+1}^{\beta_{s+1}}$,
as can be seen
from Eqs.\bref{beg fsr proper solution}
and \bref{fixing aux antighosts}.
The condition that these fields have $n_{s+1}$
propagating modes
is consistent with
Eq.\bref{psi level s condaz}
for the case $k=-1$.

The discussion for
$\bar {\cal C}_{s\alpha_{s}}^k$ where $k$ is even
works similarly.
The $n_s$ independent degrees of freedom of
$\bar {\cal C}_{s\alpha_{s}}$,
via the diagram

\medskip

\begin{picture}(400,70)
\put(170,60){${\cal C}_{s-1}^{ k-1 \alpha_{s-1}}$}
\put(153,34){\line(3,4){15}}
\put(171,58){\vector(-3,-4){15}}
\put(140,20){$\bar {\cal C}_{s\alpha_{s}}^k$}
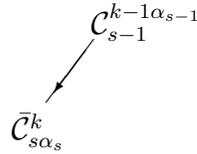
\put(10,5){Diagram 7.\ Elimination of Independent Degrees of Freedom
in $\bar {\cal C}_{s\alpha_{s}}^k$}
\end{picture}

\medskip

\noindent
are fixed by
\be
  0={\cal C}_{s-1\alpha_{s-1}}^{k-1 *} =
  {{\partial \Psi } \over
    {\partial {\cal C}_{s-1}^{k-1 \alpha_{s-1}}}} =
 \dots  +
   \delta \bar {\cal C}_{s\alpha_s}^k
  \left( {
  { {\partial_l \partial_r \Psi } \over
    { \partial \bar {\cal C}_{s\alpha_s}^k
      \partial {\cal C}_{s-1}^{k-1 \alpha_{s-1}} } }
        } \right)_0
   + \ldots
\quad ,
\label{eq fixing ind modes of C bar k s}
\ee
which requires
$$
  {\rm rank \ }
  \left( {
  { {\partial_l \partial_r \Psi } \over
    { \partial \bar {\cal C}_{s\alpha_s}^k
      \partial {\cal C}_{s-1}^{k-1 \alpha_{s-1}} } }
        } \right)_0
   = n_s
\quad .
$$
This is the same condition as
in Eq.\bref{psi level s conda}
when
$s \rightarrow s+1$ and
$k \rightarrow k+1$.

The gauge modes of
$\bar {\cal C}_{s\alpha_{s}}^k$
are the left zero modes
of the operator
$
  \left( {
  { {\partial_l \partial_r \Psi } \over
    { \partial \bar {\cal C}_{s\alpha_s}^k
      \partial {\cal C}_{s-1}^{k-1 \alpha_{s-1}} } }
        } \right)_0
$
in Eq.\bref{eq fixing ind modes of C bar k s}.
Since this operator acts to the left
on $m_s$-component vectors and has rank $n_s$,
there are exactly
$m_s - n_s= n_{s+1}$ such zero modes.
We denote these by
$\bar Z_{s+1\alpha_s}^{k \alpha_{s+1}}$,
using the redundant superscript index
$\alpha_{s+1}$ but requiring
\be
  {\rm rank \ }
   \left( {\bar Z_{s+1\alpha_s}^{k \alpha_{s+1}}} \right) =
    n_{s+1}
\quad .
\label{rank of Z bar s k}
\ee
More precisely,
the $\bar Z_{s+1\alpha_s}^{k \alpha_{s+1}}$
are defined by
\be
  \bar Z_{s+1\alpha_s}^{k \alpha_{s+1}}
  \left( {
  { {\partial_l \partial_r \Psi } \over
    { \partial \bar {\cal C}_{s\alpha_s}^k
      \partial {\cal C}_{s-1}^{k-1 \alpha_{s-1}} } }
        } \right)_0
   = 0
\quad ,
\label{def of Z bar s k}
\ee
where $k$ is even,
$ s-1 \ge k \ge 0$, and  $L-1 \ge s \ge 0 $.
These $n_{s+1}$ gauge degrees of freedom
of $\bar {\cal C}_{s\alpha_{s}}^k$,
via the diagram

\medskip

\begin{picture}(400,70)
\put(170,60){$\bar {\cal C}_{s\alpha_s}^k$}
\put(153,34){\vector(3,4){15}}
\put(171,58){\line(-3,-4){15}}
\put(140,20){${\cal C}_{s+1}^{k+1\alpha_{s+1}}$}
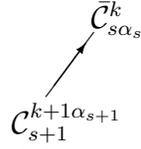
\put(10,5){Diagram 8.\ Elimination of Gauge Degrees of Freedom
in $\bar {\cal C}_{s\alpha_s}^k$}
\end{picture}

\medskip

\noindent
are fixed by
\be
  0 = {\cal C}_{s+1\alpha_{s+1}}^{k+1*} =
 { {\partial \Psi } \over
   {\partial {\cal C}_{s+1}^{k+1\alpha_{s+1}} }
  } =
 \dots +
  \delta \bar {\cal C}_{s\alpha_s}^k
  \left( {
  { {\partial_l \partial_r \Psi } \over
    { \partial \bar {\cal C}_{s\alpha_s}^k
      \partial {\cal C}_{s+1}^{k+1 \alpha_{s+1}} } }
        } \right)_0
   + \ldots
\quad ,
\label{eq fixing gauge modes of C bar k s}
\ee
which requires
$$
  {\rm rank \ }
  \left( {
  { {\partial_l \partial_r \Psi } \over
    { \partial \bar {\cal C}_{s\alpha_s}^k
      \partial {\cal C}_{s+1}^{k+1 \alpha_{s+1}} } }
        } \right)_0
  = n_{s+1}
\quad .
$$
This is the same condition as
in Eq.\bref{psi level s condb}
when
$s \rightarrow s-1$ and $k \rightarrow k-1$.
Since these gauge modes are proportional to
$\bar Z_{s+1\alpha_s}^{k \beta_{s+1}}$,
a sufficient condition
for the elimination of these modes is
\be
  {\rm rank \ }
  \left( {
   \bar Z_{s+1\alpha_s}^{k \beta_{s+1}}
  { {\partial_l \partial_r \Psi } \over
    { \partial \bar {\cal C}_{s\alpha_s}^k
      \partial {\cal C}_{s+1}^{k+1 \alpha_{s+1}} } }
          } \right)_0=n_{s+1}
\quad ,
\label{psi level s condbz}
\ee
where $k$ is even,
$ s-1 \ge k \ge 0$, and $L-1 \ge s \ge 0 $.

As a final remark,
note that all the gauge-fixing conditions
have been used:
The equations $0={\cal C}_{s\alpha_s}^{k*}$
(and $0=\bar {\cal C}_s^{k*\alpha_s}$)
impose $n_s$ constraints
on level $s-1$ gauge fields
and $n_{s+1}$ constraints
on level $s+1$ independent fields.
The total number of constraints imposed
is the number of values of the index
$\alpha_s$.
This number is $m_s$ and equals
$n_s + n_{s+1}$.

For an $L$th-stage reducible theory,
it is necessary for $\Psi$ to satisfy
Eqs.\bref{psi level s conda},
\bref{psi level s condb},
\bref{psi level s condaz}
and \bref{psi level s condbz}.
All degrees of freedom in
${\cal C}_s^{k\alpha_s}$
for $k$ odd and $k \ge 1$ and in
$\bar {\cal C}_{s\alpha_s}^k$
for $k$ even and $k \ge 2$
are fixed.
Of the original $\phi^i$,
$n_0$ of the degrees of freedom are fixed
and the remaining $n-n_0$ modes
are propagating.
For
$\bar {\cal C}_{s\alpha_s}$ and
${\cal C}^{\alpha_s}_s$
with $0 \le s \le L-1$,
$n_s$ fields are propagating
and $n_{s+1}$ fields have been gauge-fixed.
For $s=L$,
all $n_L=m_L$ modes of
$\bar {\cal C}_{L\alpha_L}$ and
${\cal C}^{\alpha_L}_L$
are propagating.

One useful example of a gauge-fixing fermion is
\ct{bv83b}
$$
  \Psi_\delta =\bar {\cal C}_{0\alpha_0}
  \chi^{\alpha_0}\left( \phi \right) +
$$
\be
  \sum\limits_{s=1}^L
   {}\sum\limits^{s}_{\scriptstyle {k = 0}\hfill\atop
     \scriptstyle {k\ {\rm even}} \hfill}
  {\bar {\cal C}_{s\alpha_s}^k
   \omega_{s\alpha_{s-1}}^{k\alpha_s}
   \left( \phi  \right){\cal C}_{s-1}^{k-1\alpha_{s-1}}} +
   \sum\limits_{s=1}^L
    {}\sum\limits^s_{\scriptstyle {k = 1} \hfill\atop
      \scriptstyle {k\ {\rm odd}} \hfill}
  {\bar {\cal C}_{s-1\alpha_{s-1}}^{k-1}
   \bar \omega_{s\alpha_s}^{k\alpha_{s-1}}
   \left( \phi \right){\cal C}_s^{k\alpha_s}}
\quad ,
\label{Lth stage Psi}
\ee
where, for each upward-sloping line
(like in Diagram 5 with $k \rightarrow k-1$
and $s \rightarrow s-1$),
one associates the matrix
$\omega_{s\alpha_{s-1}}^{k\alpha_s}$
and where, for each downward-sloping line
(like in Diagram 6),
one associates the matrix
$\bar \omega_{s\alpha_s}^{k\alpha_{s-1}}$.
In addition
to Eq.\bref{irr conditions on Psi},
one requires
Eqs.\bref{psi level s conda},
\bref{psi level s condaz},
Eq.\bref{psi level s condb},
and \bref{psi level s condbz},
that is,
$$
  {\rm rank \ } \left( {\omega_{s\alpha_{s-1}}^{k\alpha_s}
   \left( {\phi_0} \right)} \right)=n_s
\quad ,
$$
\be
  {\rm rank \ } \left( {\omega_{s\alpha_{s-1}}^{k\alpha_s}
   \left( {\phi_0} \right)
   Z_{s\beta_s}^{k-1\alpha_{s-1}}} \right)=n_s
\quad ,
\label{Lth stage conda}
\ee
for
$k$ even, $s \ge k \ge 0$
and $L \ge s \ge 1$,
as well as
$$
  {\rm rank \ } \left( {\bar \omega_{s\alpha_s}^{k\alpha_{s-1}}
    \left( {\phi_0} \right)} \right)=n_s
\quad ,
$$
\be
  {\rm rank \ } \left( {\bar Z_{s\alpha_{s-1}}^{k-1\beta_s}
   \bar \omega_{s\alpha_s}^{k\alpha_{s-1}}
   \left( {\phi_0} \right)} \right)=n_s
\quad ,
\label{Lth stage condb}
\ee
for
$k$ odd, $ s \ge k \ge 1$
and $L \ge s \ge 1$.

\subsection{Other Gauge-Fixing Procedures}
\label{ss:ogfp}

\hspace{\parindent}
The previous subsection considered the most general
gauge-fixing fermion $\Psi$
which is independent of
$ \bar \pi_{s\alpha_s}^k  $ and
$ \pi_s^{k\alpha_s}$.
By allowing $\Psi$
to be linear in
$ \bar \pi_{s\alpha_s}^k  $ and
$ \pi_s^{k\alpha_s}$,
one is able to have gaussian averaging
over a gauge condition
\ct{bv83b}.
Such possibilities arise when
\be
  \Psi = \Psi_\delta + \Psi_\pi
\quad ,
\label{gaussian Psi}
\ee
where $\Psi_\delta$ is
a $\delta$-function-type gauge-fixing fermion
which is independent of
$ \bar \pi_{s\alpha_s}^k  $ and
$ \pi_s^{k\alpha_s}$,
and where $\Psi_\pi$ is linear in
$ \bar \pi_{s\alpha_s}^k $ and
$ \pi_s^{k\alpha_s}$.

For example,
in an irreducible theory,
one may take
\be
  \Psi_\pi ={1 \over 2}
   \bar {\cal C}_{0\alpha_0}
   \sigma_0^{0\alpha_0\beta_0} \left( \phi \right)
   \bar \pi_{0\beta_0}
\quad ,
\label{irr Psi pi}
\ee
where
$\sigma_0^{0\alpha_0\beta_0}$ is
an arbitrary matrix.
The total gauge-fixing fermion
$\Psi$ is
\be
  \Psi =
  \bar {\cal C}_{0\alpha_0}
  \chi^{\alpha_0} \left( \phi  \right) +
  {1 \over 2}\bar {\cal C}_{0\alpha_0}
   \sigma_0^{0\alpha_0\beta_0} \left( \phi  \right)
    \bar \pi_{0\beta_0}
\quad .
\label{irr gaussian Psi}
\ee
The antifields
$ \bar {\cal C}_0^{*\alpha_0} $
are eliminated
via Eq.\bref{fermion gauge fixing}
to give
\be
  \bar {\cal C}_0^{*\alpha_0} =
  \chi^{\alpha_0} +
   {1 \over 2}\sigma_0^{0\alpha_0\beta_0}
   \bar \pi_{0\beta_0}
\quad ,
\label{gaussian sol for C star 0 0}
\ee
which,
when substituted
into Eq.\bref{aux S},
results in
\be
  \left. {S_{\rm aux}} \right|_{\Sigma_\Psi } =
   \bar \pi_{0\alpha_0}
   \chi^{\alpha_0}\left( \phi  \right)
   + {1 \over 2} \bar \pi_{0\alpha_0}
    \sigma_0^{0\alpha_0\beta_0}
      \bar \pi_{0\beta_0}
\quad .
\label{irr gaussian aux S}
\ee
Without the quadratic functional of
$
\bar \pi_{0\alpha_0}
$,
Eq.\bref{irr gaussian aux S}
would give the gauge-fixing conditions
$\chi^{\alpha_0} \left( \phi  \right) = 0$
on $\phi$.
With the quadratic functional of
$
\bar \pi_{0\alpha_0}
$,
one is able to perform a gaussian averaging
over these gauging-fixing conditions.
If the matrix
$ \sigma_0^{0\alpha_0\beta_0} $
is completely invertible
then the averaging is done over all gauge invariances.
By choosing
$ \sigma_0^{0\alpha_0\beta_0} $
not to have an inverse,
one may obtain delta-function conditions
for some gauge invariances
and a gaussian-averaging of others.
In some quantization schemes,
$
\bar \pi_{0\alpha_0}
$ may become propagating,
in which case
$
\bar \pi_{0\alpha_0}
$ is called a Nielsen-Kallosh ghost
\ct{kallosh78a,nielsen78a,nielsen81a}.

For a first-stage reducible theory,
one can add
\be
  \Psi_\pi ={1 \over 2}\bar {\cal C}_{0\alpha_0}
   \sigma_0^{0\alpha_0\beta_0}
   \bar \pi_{0\beta_0} +
   {1 \over 2}\bar {\cal C}_{1\alpha_1}
   \rho_{1\beta_1}^{0\alpha_1}
   \pi_1^{1\beta_1}+{1 \over 2}
   \bar \pi_{1\alpha_1}
    \rho_{1\beta_1}^{0\alpha_1}
     {\cal C}_1^{1\beta_1}
\label{1st stage Psi pi}
\ee
to the $\delta$-function gauge-fixing fermion
$\Psi_\delta$ in
Eq.\bref{1st stage Psi},
where
$
  \sigma_0^{0\alpha_0\beta_0}
$ and
$
  \rho_{1\beta_1}^{0\alpha_1}
$
are arbitrary matrix functionals of
$ \phi $.
Eliminating antifields by
Eq.\bref{fermion gauge fixing}
and inserting into
Eq.\bref{aux S}
leads to
\be
  \left. {S_{\rm aux}} \right|_{\Sigma_\Psi } =
   \ldots +{1 \over 2}\bar \pi_{0\alpha_0}
   \sigma_0^{0\alpha_0\beta_0}
  \bar \pi_{0\beta_0}+
   \bar \pi_{1\alpha_0}
     \rho_{1\beta_1}^{0\alpha_1}
   \pi_1^{1\beta_1} + \ldots
\quad .
\label{1st stage gaussian aux S}
\ee
For appropriate
$
  \sigma_0^{0\alpha_0\beta_0}
$ and
$
  \rho_{1\beta_1}^{0\alpha_1}
$,
integration over
$\bar \pi_{0\alpha_0}$,
$\pi_1^{1\beta_1}$ and
$\bar \pi_{1\alpha_0}$
leads to a gaussian averaging gauge-fixing procedure.

For an $L$th-stage system,
the most general $\Psi_\pi$,
satisfying
Eq.\bref{fgns}
and being
linear in
$\pi_s^{k\alpha_s}$ and
$\bar\pi_{s\alpha_s}^k$
and linear in the auxiliary ghosts and antighosts,
is
$$
  \Psi_\pi ={1 \over 2}
    \sum\limits_{\scriptstyle {k=0}\hfill\atop
      \scriptstyle {k\ {\rm even} }\hfill}^{L-1} {}
    \sum\limits_{s=k+1}^L {}
  \left( {\bar {\cal C}_{s\alpha_s}^k\rho_{s\beta_s}^{k\alpha_s}
   \left( \phi \right)
   \pi_s^{k+1\beta_s} +
   \bar\pi_{s\alpha_s}^k
   \rho_{s\beta_s}^{k\alpha_s} \left( \phi \right)
   {\cal C}_s^{k+1\beta_s}} \right)
$$
\be
   + {1 \over 2}\sum\limits_{\scriptstyle {k=0}\hfill\atop
    \scriptstyle {k\ {\rm even} }\hfill}^L
    {\bar {\cal C}_{k\alpha_k}^k}
    \sigma_k^{k\alpha_k\beta_k} \left( \phi \right)
   \bar \pi_{k\beta_k}^k
\quad ,
\label{Lth stage Psi pi}
\ee
where
$ \rho_{s\beta_s}^{k\alpha_s} $ and
$ \sigma_k^{k\alpha_k\beta_k} $
are arbitrary matrices.
If one eliminates antifields by
Eq.\bref{fermion gauge fixing}
and inserts the result into
Eq.\bref{aux S},
one obtains the following quadratic terms for $\pi$ fields
$$
  \left. {S_{\rm aux}} \right|_{\Sigma_\Psi } =
   \ldots +
   \sum\limits_{\scriptstyle {k=0}\hfill\atop
     \scriptstyle {k\ {\rm even} }\hfill}^{L-1} {}
   \sum\limits_{s=k+1}^L {}
  \bar \pi_{s\alpha_s}^k
   \rho_{s\beta_s}^{k\alpha_s}
  \pi_s^{k+1\beta_s}
$$
\be
  + {1 \over 2} \sum\limits_{\scriptstyle {k=0}\hfill\atop
     \scriptstyle {k\ {\rm even} }\hfill}^L
   {\bar \pi_{k\alpha_k}^k}
   \sigma_k^{k\alpha_k\beta_k}
   \bar \pi_{k\beta_k}^k+\ldots
\quad ,
\label{Lth stage gaussian aux S}
\ee
thereby allowing a gaussian average procedure
for gauge invariances at all levels.

In some instances,
in lieu of Eq.\bref{fermion gauge fixing},
one might want to eliminate certain fields
rather than antifields.
The simplest way to accomplish this
is to first perform a canonical transformation
that interchanges some fields for antifields
\ct{fh89a,siegel89a,vanproeyen91a,bkp92a,tp93a,vp94a}.
Then,
when Eq.\bref{fermion gauge fixing} is used,
elimination of certain antifields
will correspond to the elimination of
some original fields.
Given the freedom
to first perform canonical transformations,
Eq.\bref{fermion gauge fixing} is quite general.

\subsection{Gauge-Fixed Classical BRST Symmetry}
\label{ss:gfcbrsts}

\hspace{\parindent}
The gauge-fixed theory inherits
a remnant of the original BRST symmetry $\delta_B$
in Eq.\bref{def of BRST}.
Given $\Psi$,
one defines the gauge-fixed BRST operator
$\delta_{B_\Psi }$ by
\ct{bv81a}
\be
   \delta_{B_\Psi} X \equiv
   \left. {\left( {X,S} \right)} \right|_{\Sigma_\Psi }
\quad ,
\label{gauge-fixed BRST}
\ee
where $X$ is any functional of the $\Phi^A$.
For $X = \Phi^A$, one obtains
\be
  \delta_{B_\Psi }\Phi^A =
  \left. {{{\partial_l S} \over
   {\partial \Phi_A^*}}} \right|_{\Sigma_\Psi }
\quad .
\label{gf BRST trans for Phi}
\ee
This is the same result
as first performing
the non-gauge-fixed BRST transformation
in Eq.\bref{def of BRST}
and then imposing the gauge-fixing fermion condition
in Eq.\bref{fermion gauge fixing}.
The BRST transformation is
a global symmetry of the classical gauge-fixed action
since
$$
 \delta_{B_\Psi }S_\Psi =
  \left. {\left( {S_\Psi ,S} \right)} \right|_{\Sigma_\Psi } =
   \left. {{{\partial_r S_\Psi } \over {\partial \Phi^A}}
   {{\partial_l S} \over {\partial \Phi_A^*}}} \right|_{\Sigma_\Psi }=
$$
\be
  \left. {\left( {{{\partial_r S}
     \over {\partial \Phi^A}}
   {{\partial_l S} \over {\partial \Phi_A^*}}} \right)}
    \right|_{\Sigma_\Psi } +
   \left. {{{\partial_r S} \over {\partial \Phi_B^*}}
   {{\partial_r \partial_r \Psi }
     \over {\partial \Phi^A\partial \Phi^B}}
  {{\partial_l S} \over {\partial \Phi_A^*}}}
  \right|_{\Sigma_\Psi }
  = 0
\quad ,
\label{gf BRST inv of S}
\ee
where the first term vanishes
as a consequence of the classical master equation
and the second term vanishes
due to statistical symmetry properties of the factors.

Due to the elimination of antifields,
$\delta_{B_\Psi}$
is not necessarily off-shell nilpotent.
For example,
consider
$$
  \delta_{B_\Psi }^2\Phi^A =
  {{\partial_r \partial_l S} \over
    {\partial \Phi^C \partial \Phi_A^*}}\left.
  {{{\partial_l S} \over {\partial \Phi_C^*}}}
   \right|_{\Sigma_\Psi }+{{\partial_r \partial_l S}
    \over {\partial \Phi_C^* \partial \Phi_A^*}}
  {{\partial_r \partial_l \Psi } \over
   {\partial \Phi^D \partial \Phi^C}}\left.
   {{{\partial_l S} \over {\partial \Phi_D^*}}}
    \right|_{\Sigma_\Psi }
\quad ,
$$
where the second term arises from the chain rule
(see Eq.\bref{chain rules} of Appendix A)
via the implicit dependence on $\Phi^D$ through
$\Phi_C^*$,
which is now a functional of the $\Phi$
via Eq.\bref{fermion gauge fixing}.
By using the identity
$
  0 = {{\partial_l
  \left( {S,S} \right)} \over {\partial \Phi_A^*}}
$,
the first term can be rearranged and
then the two terms combine
if one notes that
\be
  {{\partial_r \left( {S_{\Psi }} \right)}
    \over {\partial \Phi^C}} =
    \left. { { {\partial_r S} \over {\partial \Phi^C} } }
   \right|_{\Sigma_\Psi } +
    \left. { {{\partial_r S} \over
              {\partial \Phi_D^*} }
   { {\partial_r \partial_l \Psi } \over
      {\partial \Phi^C \partial \Phi^D} } }
    \right|_{\Sigma_\Psi }
\quad .
\label{gf eq of motion}
\ee
The final result is
\ct{batlle88a,bg88a,tnp90a}
\be
  \delta_{B_\Psi }^2\Phi^A =
   \left( {-1} \right)^{\epsilon_A}
   {{\partial_r \left( {S_{\Psi }} \right)}
    \over {\partial \Phi^C}}\left.
    {{{\partial_l \partial_l S} \over
  {\partial \Phi_C^* \partial \Phi_A^*}}}
    \right|_{\Sigma_\Psi }
\quad .
\label{nilpot cal of gf BRST}
\ee
The equations of motion
for the gauge-fixed action are
$
   { {\partial_r \left( {S_{\Psi }}
    \right) } \over {\partial \Phi^C} } = 0
$.
Hence, the gauge-fixed BRST symmetry
is on-shell nilpotent.
{}From Eq.\bref{fsr proper solution},
one sees that a necessary condition for off-shell nilpotency
is that the algebra is closed,
$ E_{\alpha \beta }^{ji} = 0$,
and that there is off-shell reducibility,
$V_a^{ji} = 0$.

Another way of obtaining the on-shell nilpotency
is to note that the gauge-fixed BRST transformation
agrees with the ordinary BRST transformation
evaluated at $\Sigma_\Psi$,
up to equations of motion.
For functionals which do not depend on antifields
this follows from Eq.\bref{gauge-fixed BRST}.
For functionals, which depended
on antifields before gauge-fixing
but are now evaluated at $\Sigma_\Psi$,
the result follows because
\be
  \delta_{B_\Psi }\Phi_A^*\left( \Phi  \right) =
  \left. {\left( {\delta_B\Phi_A^*} \right)}
    \right|_{\Sigma_\Psi } +
   {{\partial_l
    \left( {S_{\Psi }} \right)}
    \over {\partial \Phi^A}}
\quad .
\label{gf BRST trans for antifields}
\ee
The last term is a gauge-fixed equation of motion.
In other words,
the processes of
performing the off-shell BRST transformation
in Eq.\bref{def of BRST}
and gauge-fixing via
Eq.\bref{fermion gauge fixing}
commute up to equations of motion.

\subsection{The Gauge-Fixed Basis}
\label{ss:gfb}

\hspace{\parindent}
Another way to gauge-fix the theory
is to first perform a canonical transformation
to the ``gauge-fixed'' basis and then
set antifields to zero
\ct{siegel89a,siegel89b,vanproeyen91a,bkp92a,tp93a}.
The canonical transformation
in Eq.\bref{f2 can trans}
is done with
$ F_2 = \Phi^A \tilde \Phi_A^* + \Psi \left( \Phi \right) $
to give
\be
  \Phi^A = \tilde \Phi^A
\ , \quad \quad
  \Phi_A^* = \tilde \Phi_A^* +
  {{\partial \Psi } \over {\partial \Phi^A}}
\quad ,
\label{gauge-fixed basis}
\ee
so that $\Phi^A \to \Phi^A$ and
$
  \Phi_A^* \to \Phi_A^* +
 {{\partial \Psi } \over {\partial \Phi^A}}
$.
The gauge-fixed basis consists of
the new tilde fields and antifields.

In this new basis, the original gauge invariances
are replaced by the classical gauge-fixed BRST symmetry
$\delta_{B_\Psi}$.
It emerges as the symmetry
of the gauge-fixed action $S_\Psi$.
To understand how this comes about,
expand the original proper solution $S$
in a power series in antifields via
\be
  S \left[ {\Phi ,\Phi^*} \right] =
   S_0 \left[ {\phi^i} \right] +
   \Phi_A^*R^A +
  {1 \over 2}\Phi_A^*\Phi_B^*R^{BA} + \ldots
\quad ,
\label{antifield exp of S}
\ee
and perform the change of variables
in Eq.\bref{gauge-fixed basis} to obtain
$$
  S\left[ {\Phi ,\Phi^*} \right] =
   S_0\left[ {\phi^i} \right] +
   {{\partial \Psi } \over {\partial \Phi^A}}R^A +
   {1 \over 2}{{\partial \Psi } \over {\partial \Phi^A}}
  {{\partial \Psi } \over {\partial \Phi^B}}R^{BA} + \ldots \
$$
$$
  + \tilde \Phi_A^*
    \left( {R^A + {{\partial \Psi } \over
      {\partial \Phi^B}}R^{BA}+\ldots } \right)
$$
$$
  + {1 \over 2}\tilde \Phi_A^*\tilde \Phi_B^*
   \left( {R^{BA}+\ldots } \right) + \dots
$$
\be
  \equiv \tilde S_0\left[ \Phi  \right] +
   \tilde \Phi_A^*\tilde R^A +
  {1 \over 2}\tilde \Phi_A^*\tilde \Phi_B^*\tilde R^{BA} +
  \ldots \equiv \tilde S \left[ { \Phi ,\tilde \Phi^*} \right]
\quad .
\label{antifield exp of S bar}
\ee
The terms in the $\tilde \Phi^*$ expansion
have an interpretation.
The $\tilde \Phi^*$-independent term
\be
 \tilde S_0\left[ \Phi  \right] =
 \left. S \right|_{\Sigma_\Psi } \equiv S_\Psi
\label{S sub Psi}
\ee
is the gauge-fixed action.
The term linear in $\tilde \Phi^*$
$$
  \tilde R^A =
   \left. {{{\partial_l \tilde S}
     \over {\partial \tilde \Phi_A^*}}}
   \right|_{\left\{ {\tilde \Phi_B^*=0} \right\}} =
   \left. {{{\partial_l S} \over {\partial \Phi_A^*}}}
   \right|_{\Sigma_\Psi } \equiv \delta_{B_\Psi }\Phi^A
$$
is the generator of gauge-fixed BRST transformations
since
$
 \delta_{B_\Psi } X =
 {{\partial_r X } \over {\partial \Phi^A}} \tilde R^A
$.
Hence, $\tilde S$
is a theory in which gauge transformations
have been replaced by
the gauge-fixed BRST transformation $\delta_{B_\Psi }$.
As in Sect.\ \ref{s:ssgt},
one can define tensors
associated with the gauge structure.
It is necessary to absorb ghost fields ${\cal C}$
into their definition, i.e.,
$ \tilde R^A \equiv \tilde R^A_B \tilde {\cal C}^B $,
$ \tilde E^{BA}
  \equiv {1 \over 2}\tilde E_{DC}^{BA}
  \tilde {\cal C}^C\tilde {\cal C}^D
$,
etc..
Since $\delta_{B_\Psi}^2$ is zero only on-shell,
the algebra is open
and there exists a non-zero tensor field
$\tilde E^{BA}$.
Comparing
Eqs.\bref{com of two gens} and
\bref{nilpot cal of gf BRST},
one concludes
\be
\tilde E^{BA} =
   \left( {-1} \right)^{\epsilon_A}
   \left. {{{\partial_l \partial_l S} \over
   {\partial \Phi_B^*\partial \Phi_A^*}}} \right|_{\Sigma_\Psi }
\quad .
\label{E tensor for gf BRST}
\ee
For an open algebra,
according to Eq.\bref{fsr proper solution},
$\tilde E^{BA}\left( {-1} \right)^{\epsilon_A}$
is the coefficient of the proper solution
quadratic in antifields.
Hence,
$
 \tilde R^{BA} \equiv
$
$
\left. { { {\partial_l \partial_l \tilde S}
     \over { \partial \tilde \Phi_B^*
      \partial \tilde \Phi_A^*}^{\ } }}
   \right|_{\left\{ {\tilde \Phi_B^*=0} \right\} }
$
$
   =
  \left. {{{\partial_l \partial_l S} \over
    {\partial \Phi_B^*\partial \Phi_A^*}}}
   \right|_{\Sigma_\Psi }
$
should be equal to
$
   \tilde E^{BA}\left( {-1} \right)^{\epsilon_A}
$.
This is indeed true since it is equivalent to
Eq.\bref{E tensor for gf BRST}.
Hence, the coefficient of $\tilde S$
which is quadratic in tilde antifields
is the tensor $\tilde E^{BA}$
associated with the fact
that the gauge-fixed BRST transformations
form an open algebra.

The classical master equation for $\tilde S$
encodes the algebraic structure
of the gauge-fixed BRST transformation
\ct{gp93c}.
A straightforward computation gives
\be
  0={1 \over 2}\left( {\tilde S,\tilde S} \right) =
  {{\partial_r S_\Psi } \over {\partial \Phi^A}}\tilde R^A +
   \tilde \Phi_B^*
  \left( {{{\partial_r\tilde R^B} \over {\partial \Phi^A}}
  \tilde R^A+\tilde R^{BA}{{\partial_r S_\Psi }
   \over {\partial \Phi^A}}} \right)+\ldots
\quad .
\label{classical master eq for S bar}
\ee
Since each coefficient
of the antifield expansion must be zero,
\be
  0 =
 {{\partial_r S_\Psi } \over {\partial \Phi^A}}\tilde R^A
  = \delta_{B_\Psi }S_\Psi
\quad ,
\label{BRST inv of gf action}
\ee
which is the invariance of the gauge-fixed action
under $\delta_{B_\Psi }$.
The vanishing of the second coefficient
in Eq.\bref{classical master eq for S bar}
leads to the equation
\be
  {{\partial_r \tilde R^B} \over {\partial \Phi^A}}\tilde R^A =
  -\tilde R^{BA}{{\partial_r S_\Psi } \over {\partial \Phi^A}}
\quad .
\label{second term}
\ee
Eq.\bref{second term} is related
to the on-shell nilpotency of $\delta_{B_\Psi }$:
$$
  0=\delta_{B_\Psi }\left( {\delta_{B_\Psi }S_\Psi } \right) =
{{\partial_r} \over {\partial \Phi^B}}
  \left( {{{\partial_r S_\Psi } \over
  {\partial \Phi^A}}\tilde R^A} \right)\tilde R^B =
$$
$$
  {{\partial_r S_\Psi } \over {\partial \Phi^A}}
  {{\partial_r \tilde R^A} \over {\partial \Phi^B}}\tilde R^B +
   \left( {-1} \right)^{\left(
       {\epsilon_A+1} \right)\epsilon_B}
    {{\partial_r \partial_r S_\Psi } \over
   {\partial \Phi^B\partial \Phi^A}}\tilde R^A\tilde R^B
\quad .
$$
The second term vanishes by the statistical
properties of the factors.
The first term vanishes
when Eq.\bref{second term} is used
in conjunction with statistical symmetry arguments.
Higher order terms in
$0=\left( {\tilde S,\tilde S} \right)$
represent other consistency requirements
of the gauge-fixed classical BRST transformation.

In addition to the gauge structure,
other concepts generalize to the gauge-fixed basis.
In the effective action
or any other approach for which antifields
remain present,
antifields are interpreted as the sources
for gauge-fixed BRST transformations.
See Sect.\ \ref{ss:eazje}
for more discussion.
In the gauge-fixed basis,
the classical limit
in Eq.\bref{classical bc}
is replaced by
the gauge-fixed condition
\be
  \restric{ \tilde S \left[ { \Phi,\tilde \Phi^* } \right]}
     {\tilde \Phi^*=0} =
     S_\Psi [ \Phi ]
\label{gauge-fixed bc}
\ee
of Eq.\bref{S sub Psi}.
The proper condition
in Eqs.\bref{rank of hessian of S} and
\bref{def of proper solution}
becomes
\ct{vanproeyen91a,bkp92a,tp93a,gp93c}
\be
   {\rm rank \ }
    \restric{ \frac{\partial_l \partial_r
      { S_\Psi \left[ \Phi \right] }  }
     {\partial \Phi^A \partial \Phi^B } }
 { \{ {\rder{S_\Psi}{\Phi^D} = 0 } \} } = N
\quad .
\label{gauge-fixed proper condition}
\ee
Eq.\bref{gauge-fixed proper condition}
ensures that propagators
for the $\Phi^A$ are defined
so that the usual perturbation theory
can be developed.

\vfill\eject

\section{Gauge-Fixing Examples}
\label{s:gfe}

\hspace{\parindent}
In this subsection,
we present some gauge-fixing procedures
for the examples considered
in Sects.\ \ref{s:egt}
and \ref{s:eps}.
The first step is to introduce
the auxiliary fields and antifields
necessary for gauge-fixing,
as specified
in Sect.\ \ref{ss:gfaf}.
The second step is to choose an appropriate $\Psi$.
Here, there is quite a bit of freedom
and we make specific choices.
The third step is to eliminate antifields
using $\Psi$
and substitute the results into $S$
to obtain the gauge-fixed action $S_\Psi$.

The reader interested in doing exercises
can try the following.
(i) Derive the equations of motion of the gauge-fixed action.
(ii) Determine the effect of the gauge-fixed BRST transformation
$\delta_{B_\Psi}$ on the fields.
(iii) Check the gauge-fixed BRST invariance
of the gauge-fixed action.
(iv) Compute $\delta^2_{B_\Psi}$ and
verify that the non-zero terms,
when present, vanish
if the equations of motion are used.
(v) Perform a gauge-fixing procedure
with a different gauge-fixing fermion $\Psi$.

\subsection{The Spinless Relativistic Particle}
\label{ss:srp2}

\hspace{\parindent}
Since this system is irreducible,
one only needs to append
one non-trivial pair $\{ \bar {\cal C}$, $\bar \pi \}$.
The non-minimal proper solution is
\be
 S = \int {\dif \tau }
   \left\{ {{1 \over 2}
     \left( { {{\dot x^2} \over e} - e m^2} \right) +
       x_\mu^*\dot x^\mu {{\cal C} \over e} +
     e^*\dot {\cal C} +
    \bar \pi \bar {\cal C}^* } \right\}
\quad .
\label{nm rsp ps}
\ee
An example of a gauge-fixing fermion is
\be
  \Psi =\int {\dif \tau } \bar {\cal C}\left( {e-1} \right)
\quad ,
\label{rsp psi}
\ee
which,
via Eq.\bref{fermion gauge fixing},
leads to
\be
  \bar {\cal C}^* = e-1
\ ,\quad \quad
  e^* = \bar {\cal C}
\ ,\quad \quad
  x_\mu^*=0
\ ,\quad \quad
  \bar \pi^*=0
\quad .
\label{rsp antifield elimination}
\ee
The gauge-fixed action is
\be
  S_\Psi = \int {\dif \tau }
  \left\{ {{1 \over 2}
  \left( { {{\dot x^2} \over e} - e m^2 } \right) +
  \bar {\cal C}\dot {\cal C} + \bar \pi
   \left( {e-1} \right)} \right\}
\quad .
\label{rsp gf action1}
\ee
Because the original algebra is closed,
the gauge-fixed BRST transformation
$\delta_{B_\Psi}$,
given by
$$
  \delta_{B_\Psi} e = \dot {\cal C}
  \ , \quad \quad
  \delta_{B_\Psi} x^\mu = {{\dot x^\mu {\cal C}} \over e}
  \ , \quad \quad
  \delta_{B_\Psi} {\cal C} = 0
\quad ,
$$
\be
  \delta_{B_\Psi} \bar \pi = 0
  \ , \quad \quad
  \delta_{B_\Psi} \bar {\cal C} = \bar \pi
\quad ,
\label{rsp gf BRST trans}
\ee
satisfies $\delta_{B_\Psi} \delta_{B_\Psi}=0$ off-shell.
After integrating over $\bar \pi$,
the gauge $e=1$ is implemented
and Eq.\bref{rsp gf action1} becomes
\be
  S_\Psi \to \int {\dif \tau }
  \left\{ {{1 \over 2}
   \left( { \dot x^2 - m^2} \right) +
  \bar {\cal C} \dot {\cal C}} \right\}
\quad .
\label{rsp gf action2}
\ee
Since the equations of motion for $\bar \pi$ and $e$
have been used,
Eq.\bref{rsp gf action2}
is no longer invariant under
Eq.\bref{rsp gf BRST trans}.
To derive the modified BRST transformations,
start with Eq.\bref{nm rsp ps},
and perform the shift,
$ \bar {\cal C}^* \to \bar {\cal C}^* + e-1 $,
$ e^* \to e^* + \bar {\cal C} $,
to the gauge-fixed basis.
Then, determine the equation of motion for $\bar \pi$
using
$ { {\partial_r S_\Psi} \over {\partial e} } = 0$
and perform another canonical transformation
to shift $\bar \pi$ by this solution:
$
  \bar \pi \to \bar \pi +
  {1 \over 2}\left( {\dot x^2 / e^2 + m^2} \right)
$.
The gauge-fixed BRST transformations
of the transformed action at the saddle point,
$e=1$, $\bar \pi =0$, given by
\be
  \delta_{\tilde B_\Psi } x^\mu = \dot x^\mu {\cal C}
\ , \quad \quad
  \delta_{\tilde B_\Psi } {\cal C}=0
\ , \quad \quad
  \delta_{\tilde B_\Psi } \bar {\cal C} =
   {1 \over 2} \left( {\dot x^2 + m^2} \right)
\quad ,
\label{rsp gf BRST trans2}
\ee
constitute a symmetry of the action
in Eq.\bref{rsp gf action2},
as can easily be checked.
Because equations of motion
have been used,
$\delta_{\tilde B_\Psi }$
is not longer nilpotent off-shell.
Indeed,
a computation of $\delta_{\tilde B_\Psi }^2$
reveals that it is, in general, nonzero;
it is zero
if the equations of motion,
$\ddot x^\mu =0$ and $\dot {\cal C} = 0$
are used.

\subsection{Yang-Mills Theories}
\label{ss:ymt2}

\hspace{\parindent}
Since this is an irreducible system,
it is necessary to introduce
one trivial pair $\{ \bar {\cal C}_a$, $\bar \pi_a \}$
for each generator index $a$.
Frequently, $\bar \pi_a$ is denoted by $B_a$ and
we adopt this notation.
The non-minimal proper solution is
\be
   S = \int {\dif^d x} \left\{
   { -{1 \over 4} F_{\mu \nu }^a F_{a}^{\mu \nu } +
    A_{a\mu }^*{D^{\mu a}}_b {\cal C}^b +
   {1 \over 2}{\cal C}_c^*{f_{ab}}^c {\cal C}^b {\cal C}^a}
  +  B_a \bar {\cal C}^{*a}
          \right\}
\quad .
\label{ymt nm ps}
\ee

To illustrate
a gaussian-averaging gauge-fixing procedure,
we choose
\be
  \Psi =\int {\dif^d x}
  \bar {\cal C}_a\left( { - {{B^a} \over {2 \xi }} +
    \partial^\mu A_\mu^a} \right)
\quad ,
\label{ymt xi psi}
\ee
where $\xi$ is a parameter.
Elimination of antifields
via Eq.\bref{fermion gauge fixing}
gives
$$
  \bar {\cal C}^{*a}= - {{B^a} \over {2\xi }} +
   \partial^\mu A_\mu^a
\ ,\quad \quad
    B_a^{*} = - {{\bar {\cal C}_a} \over {2\xi }} \ ,
$$
\be
  A_{a\mu }^* = - \partial_\mu \bar {\cal C}_a
\ ,\quad \quad
  {\cal C}^{*}_a = 0
\quad .
\label{ymt antifield elimination}
\ee
The gauge-fixed action $S_\Psi$ is
\be
   S_\Psi = \int {\dif^d x} \left\{
   {
     -{1 \over 4} F_{\mu \nu}^a F_{a}^{\mu \nu} -
   \partial_\mu \bar {\cal C}_a {D^{\mu a}}_b {\cal C}^b -
   \left( { {{B_a} \over {2\xi}} -
   \partial^\mu A_{a\mu} } \right) B^a
      }  \right\}
\quad .
\label{ymt gf action1}
\ee
The gauge-fixed BRST transformations are
$$
\delta_{B_\Psi} A^{a\mu}= {D^{\mu a}}_b {\cal C}^b
\quad ,
$$
$$
  \delta_{B_\Psi} {\cal C}^a =
 {1 \over 2}{f_{bc}}^a{\cal C}^c {\cal C}^b
\quad ,
$$
\be
  \delta_{B_\Psi} \bar {\cal C}_a = B_a
\quad ,
\label{BRST gf trans ymt}
\ee
$$
  \delta_{B_\Psi} B_a = 0
\quad .
$$
The nilpotency of $\delta_{B_\Psi}$
holds off-shell
because the
original gauge
algebra is closed.
The gaussian integration over $B_a$
can be performed
for Eq.\bref{ymt gf action1} to give
\be
   S_\Psi \to \int {\dif^d x}\left\{
    { -{1 \over 4} F_{\mu \nu}^a F_{a}^{\mu \nu} -
   \partial_\mu \bar {\cal C}_a {D^{\mu a}}_b {\cal C}^b
      + { {\xi} \over {2} }
   \partial^\mu A_\mu^a \partial^\nu A_{a \nu}
       } \right\}
\quad ,
\label{ymt gf action2}
\ee
which is known as the Yang-Mills action
fixed in the $R_\xi$ gauge
\ct{al73a}.
The case $\xi=1$ is the Feynman gauge.
When $\xi \to \infty$,
the $\bar \pi_a = B_a$ dependence in $\Psi$
of Eq.\bref{ymt xi psi} disappears
and the Landau gauge
$\partial^\mu A_\mu^a = 0$
is imposed as a delta-function condition.
Because the quadratic forms
in Eq.\bref{ymt gf action2} are non-degenerate,
$\Psi$ is an admissible gauge-fixing fermion.
Propagators exist
and Eq.\bref{ymt gf action2}
can be used as an action
for the Yang-Mills perturbation series
\ct{al73a,iz80a}.

The BRST symmetry
of Eq.\bref{ymt gf action2}
is determined by the procedure
described at the end
of Sect.\ \ref{ss:srp2}.
One performs canonical transformations
to the gauge-fixed basis
and then to the classical solution for $B^a$.
The latter is determined by varying the action
with respect to $B^a$ itself,
and leads to the shift
$
B^a \to B^a + \xi \partial^\mu A_\mu^a
$.
One finds that the BRST symmetry
of Eq.\bref{ymt gf action2}
is given by
\be
  \delta_{\tilde B_\Psi } A^{a\mu } =
      {D^{\mu a}}_b {\cal C}^b
\ , \quad \quad
  \delta_{\tilde B_\Psi }{\cal C}^a =
    {1 \over 2} {f_{bc}}^a {\cal C}^c {\cal C}^b
\ , \quad \quad
  \delta_{\tilde B_\Psi } \bar{\cal C}^a =
     \xi \partial^\mu A_\mu^a
\ .
\label{BRST gf trans ymt2}
\ee
The gauge-fixed BRST generator
$\delta_{\tilde B_\Psi } $
is only nilpotent on-shell.

\subsection{Topological Yang-Mills Theory}
\label{ss:tymt2}

\hspace{\parindent}
By using a redundant set of gauge transformations,
we have rendered this theory first-stage reducible.
The minimal-field proper solution is given
in Eq.\bref{tym proper solution}.
The gauge parameters are
$\Lambda^a$ and
$\varepsilon^{a \mu}$.
The number of level $1$ gauge invariances
is the rank of the group.
Correspondingly,
we need to introduce
the trivial pairs
$\{ \bar {\cal C}_a, \bar \pi_{a } \}$ and
$\{ \bar {\cal C}_{a\mu},\bar \pi_{a\mu} \}$
at level $0$,
and $\{ {\cal C}_1^{1a},\pi_1^{1a} \}$ and
$\{ \bar {\cal C}_{1a},\bar \pi_{1a} \}$
at level $1$.
In keeping with the notation
of Sects.\ \ref{ss:tymt} and \ref{sss:tymt},
we use $\eta$ instead of ${\cal C}$
for level $1$ ghosts:
$\bar {\cal C}_{1a} \to \bar \eta_{a}$ and
${\cal C}_1^{1a} \to \eta^{1a}$.
We also use $B$ instead of $\bar \pi$
at level $0$:
$\bar \pi_{a} \to B_{a}$ and
$\bar \pi_{a\mu} \to B_{a\mu}$,
and
$\Lambda$ instead of $\pi$
at level $1$:
$\bar \pi_{1a} \to \bar \Lambda_{a}$ and
$\pi_1^{1a} \to \Lambda^{1a}$.

To the minimal proper solution
in Eq.\bref{tym proper solution},
we add
\be
  S_{\rm aux} = \int {\dif^4x}
   \left( { B_a\bar {\cal C}^{*a} +
    B_{a\mu }\bar {\cal C}^{*a\mu } +
   \eta_{a}^{1*}\Lambda^{1a} +
   \bar \Lambda_{a}\bar \eta^{*a} } \right)
\quad .
\label{tym aux S}
\ee
For the gauge-fixing fermion,
we select
\be
  \Psi =\int {\dif^4x}
   \left( {
   \bar {\cal C}_a\partial^\mu A_\mu^a +
   \bar {\cal C}_{a\mu }A^{a\mu } +
   \bar \eta_{a} {\cal C}^a +
    \eta^{1a} \bar {\cal C}_a
   } \right)
\quad .
\label{tym psi}
\ee
The elimination of antifields using $\Psi$
gives
$$
  \bar {\cal C}^{*a} = \partial^\mu A_\mu^a + \eta^{1a}
\ , \quad \quad
  \bar {\cal C}^{*a\mu } = A^{a\mu }
\quad ,
$$
$$
  \eta_a^* = 0
\ , \quad \quad
  \eta_{a}^{1*}= \bar {\cal C}_a
\ , \quad \quad
  \bar \eta^{*a}={\cal C}^a
\quad ,
$$
\be
  A_{a\mu }^* = -\partial_\mu \bar {\cal C}_a +
     \bar {\cal C}_{a\mu }
\ , \quad \quad
   {\cal C}_a^* = \bar \eta_{a}
\ , \quad \quad
    \bar {\cal C}^{a*} = \eta^{1a}
\quad .
\label{tym antifield elimination}
\ee
The gauge-fixed action becomes
$$
  S_{\Psi} = \int {\dif^4x} \left\{ {{1 \over 4}
   F_{\mu \nu }^a \ {}^*F^{a\mu \nu } +
   \left( {-\partial_\mu \bar {\cal C}_a + \bar {\cal C}_{a\mu }} \right)
      \left( {{D^{\mu a}}_b {\cal C}^b +
     {\cal C}^{a\mu }} \right)}
   \right.
$$
$$
   + {1 \over 2} \bar \eta_{c}
       {f_{ab}}^c {\cal C}^b {\cal C}^a
  + A_{c\mu } {f_{ab}}^{c}
      {\cal C}^{b\mu }{\cal C}^a  +
    \left( { \bar \eta_{b} -
   A_{a\mu } {{D^{\mu a}}_b}  } \right) \eta^b +
$$
\be
  \left. {
  B_a \left( { \partial^\mu A_\mu^a +
     \eta^{1a} } \right) +
   B_{a\mu} A^{a\mu} +
   \bar {\cal C}_a \Lambda^{1a} +
  \bar \Lambda_{a} {\cal C}_{}^{ a^{{}^{{}{}}} }
  } \right\}
\quad .
\label{tym gf action1}
\ee
The integration over
$B_{a\mu }$
produces
$\delta \left( { A^{a\mu} } \right)$
in the integrand of the functional integral.
This delta function can be used to perform
the $A^{a\mu}$ integral.
Then the integration over
$B_a$, $\Lambda^{1a}$
and $\bar \Lambda_{a}$
produces
$
   \delta \left( { \eta^{1a} } \right)
   \delta \left( { \bar {\cal C}_a } \right)
   \delta \left( {{\cal C}^a} \right)
$
which can be used to do the
$\eta^{1a}$, $\bar {\cal C}_a $
and ${{\cal C}^a}$ integrations.
The gauge-fixed action is reduced to
\be
  S_\Psi \to \int {\dif^4x}
  \left( {\bar {\cal C}_{a\mu } {\cal C}^{a\mu } +
   \bar \eta_{a}\eta^a} \right)
\quad .
\label{tym gf action2}
\ee
The integration over
$\bar {\cal C}_{a\mu }$ and $\bar \eta_{a}$
leads to
$
 \delta \left( {{\cal C}^{a\mu }} \right)
 \delta \left( {\eta^a} \right)
$
which can be used to do
the
$ {\cal C}^{a\mu} $ and $ {\eta^a} $ integrals.
All the functional integrals over fields
have been performed,
leaving no local degrees of freedom in the action.
Not too surprisingly,
the integration over the topological action
leads to a trivial lagrangian.
The functional integral
produces a finite number.
It is important that all the terms
in $\Psi$ in Eq.\bref{tym psi} be present.
If the last term $ \eta^{1a} \bar {\cal C}_a $
were dropped from $\Psi$,
then the integrations over
$\Lambda^{1a}$ and $ \bar {\cal C}_a $
would produce singular contributions.

\subsection{The  Antisymmetric Tensor Field Theory}
\label{ss:atft2}

\hspace{\parindent}
Like the previous example,
this is a first-stage reducible theory,
for which one needs to add one trivial pair
$\{ \bar {\cal C}_{a\mu },\bar \pi_{a\mu } \}$
at level $0$
and two trivial pairs
$\{ {\cal C}_1^{1a},\pi_1^{1a} \}$ and
$\{ \bar {\cal C}_{1a},\bar \pi_{1a} \}$
at level $1$.
As in the previous subsections,
we use $B$ for $\bar \pi$ at level $0$,
and use
$\eta$ and $\Lambda$ respectively for
${\cal C}_1$  and $\pi_1$ at level $1$.
To the minimal-field proper solution,
one adds the auxiliary action
\be
  S_{\rm aux} = \int {\dif^4x}
   \left( {B_{a\mu }\bar {\cal C}^{*a\mu } +
   \bar \Lambda_{a}\bar \eta^{*a} +
   \eta_{a}^{1*}\Lambda^{1a}} \right)
\quad .
\label{atft aux S}
\ee
An appropriate gauge-fixing fermion is
\be
 \Psi = \int {\dif^4x}
     \left( {
  {1 \over 2}\bar {\cal C}_{a\mu }
   \varepsilon^{\mu \nu \rho \sigma }
   \partial_\nu B_{\rho \sigma }^a +
   \bar \eta_{a}\partial^\mu {\cal C}_\mu^a +
   \partial^\mu \bar {\cal C}_{a\mu} \eta^{1a}
       } \right)
\quad .
\label{atft psi}
\ee
The elimination of antifields using $\Psi$
gives
$$
 \bar {\cal C}^{*a\mu } =
   {1 \over 2} \varepsilon^{\mu \nu \rho \sigma }
   \partial_\nu B_{\rho \sigma }^a -
   \partial^\mu \eta^{1a}
\ , \quad \quad
   \eta_a^* = 0
\ , \quad \quad
   \bar \eta^{*a} = \partial^\mu {\cal C}_\mu^a
\ , \quad \quad
  \eta_{a}^{1*} = \partial^\mu \bar {\cal C}_{a\mu}
\ ,
$$
\be
  B_a^{*\rho \sigma } =
  \varepsilon^{ \rho \sigma \mu \nu }
  \partial_\mu \bar {\cal C}_{a\nu }
\ , \quad \quad
   {\cal C}_a^{*\mu } = -\partial^\mu \bar \eta_{a}
\ , \quad \quad
  A_{a\mu }^* = 0
\quad .
\label{atft antifield elimination}
\ee
The gauge-fixed action is
$$
 S_{\Psi} = S_0 +
 \int {\dif^4 x} \left( {
   - {1 \over 2} \partial_\nu \bar {\cal C}_{a\mu }
   \left( { {D^{\nu a}}_b{\cal C}^{\mu b} -
  {D^{\mu a}}_b {\cal C}^{\nu b}} \right)}
             \right.
$$
$$
  -\partial_\mu \bar \eta_{a} {D^{\mu a}}_b\eta^b +
  {1 \over 4}\partial_\mu \bar {\cal C}_{c\nu }
   \partial_\rho \bar {\cal C}_{\sigma }^b
  \varepsilon^{\mu \nu \rho \sigma } {f_{ab}}^c \eta^a +
$$
\be
  \left. {  B_{a\mu }
  \left( {{1 \over 2} \varepsilon^{\mu \nu \rho \sigma}
   \partial_\nu B_{\rho \sigma}^a -
   \partial^\mu \eta^{1a}} \right) +
  \bar \Lambda_{a} \partial^\mu {\cal C}_\mu^a +
  \partial^\mu \bar {\cal C}_{a\mu} \Lambda^{1a}
   } \right)
\quad ,
\label{atft gf action}
\ee
where $S_0$ is given
in Eq.\bref{AT action}.
One interesting feature of this example
is the appearance of a trilinear ghost term.
It originates from the gauge-fixing
of the bilinear term in antifields
in Eq.\bref{astft proper solution}.
The fields $B_{a\mu }$, $\bar \Lambda_{a}$
and $\Lambda^{1a}$
are Lagrange multipliers
for the gauge conditions
$
  {{1 \over 2} \varepsilon^{\mu \nu \rho \sigma}
   \partial_\nu B_{\rho \sigma}^a -
   \partial^\mu \eta^{1a}} = 0
$,
$\partial^\mu {\cal C}_\mu^a = 0$ and
$\partial^\mu \bar {\cal C}_{a\mu} = 0$.
If one wants to perform a gaussian averaging
over these conditions,
one adds
$$
 \Psi_{\pi} =
   -{1 \over 2}\int {\dif^4x}
  \left( {{{\bar {\cal C}_{a\mu }B^{a\mu }} \over \xi } +
 {{\bar \eta_{a}\bar \Lambda^{a}} \over {\bar \xi}} +
  {{\Lambda_{a}^1\eta^{1a}} \over {\xi^1}}} \right)
$$
to $\Psi$ in Eq.\bref{atft psi}.
Here, $\xi$, $\bar \xi$ and $\xi^1$ are constants.
One could also replace ordinary derivatives
$\partial_\mu$
by covariant ones $D_\mu$
in Eq.\bref{atft psi}.
Then $D_\mu$ replaces $\partial_\mu$
in Eqs.\bref{atft antifield elimination}
and \bref{atft gf action}.
In this case,
$ A_{a\mu }^* \ne 0$,
but since $A_{a\mu }^*$ does not enter the proper solution
in Eq.\bref{astft proper solution},
the gauge-fixed action
is as in Eq.\bref{atft gf action}
with $\partial_\mu \to D_\mu$.

\subsection{Open String Field Theory}
\label{ss:osft2}

\hspace{\parindent}
Since the open bosonic string is
an infinite-reducible theory,
one introduces the extraghosts
${\cal C}^k_s$ for $k=1,3,5, \dots$ and $s \ge k$,
and $\bar {\cal C}^k_s$ for $k=0,2,4, \dots$ and $s \ge k$
in the Triangular Field Tableau of Diagram 2
for all levels $s$:

\begin{picture}(400,240)
\put(170,220){$A$}
\put(150,190){\vector(3,4){15}}
\put(171,218){\line(-3,-4){15}}
\put(200,190){\line(-3,4){20}}
\put(140,180){$\bar {\cal C}_{0}$}
\put(200,180){${\cal C}^{}_0$}
\put(123,154){\vector(3,4){15}}
\put(141,178){\vector(-3,-4){15}}
\put(183,154){\vector(3,4){15}}
\put(201,178){\line(-3,-4){15}}
\put(230,150){\line(-3,4){20}}
\put(110,140){${\cal C}^{1}_1$}
\put(170,140){$\bar {\cal C}_{1}$}
\put(230,140){${\cal C}^{}_1$}
\put(93,114){\vector(3,4){15}}
\put(111,138){\vector(-3,-4){15}}
\put(153,114){\vector(3,4){15}}
\put(171,138){\vector(-3,-4){15}}
\put(213,114){\vector(3,4){15}}
\put(231,138){\line(-3,-4){15}}
\put(260,110){\line(-3,4){20}}
\put(80,100){$\bar {\cal C}^2_{2}$}
\put(140,100){${\cal C}^{1}_2$}
\put(200,100){$\bar {\cal C}_{2}$}
\put(260,100){${\cal C}^{}_2$}
\put(63,74){\vector(3,4){15}}
\put(81,98){\vector(-3,-4){15}}
\put(123,74){\vector(3,4){15}}
\put(141,98){\vector(-3,-4){15}}
\put(183,74){\vector(3,4){15}}
\put(201,98){\vector(-3,-4){15}}
\put(243,74){\vector(3,4){15}}
\put(261,98){\line(-3,-4){15}}
\put(290,70){\line(-3,4){20}}
\put(50,60){${\cal C}^{3}_3$}
\put(110,60){$\bar {\cal C}^2_{3}$}
\put(170,60){${\cal C}^{1}_3$}
\put(230,60){$\bar {\cal C}_{3}$}
\put(290,60){${\cal C}^{}_3$}
\put(170,35){.}
\put(170,30){.}
\put(170,25){.}
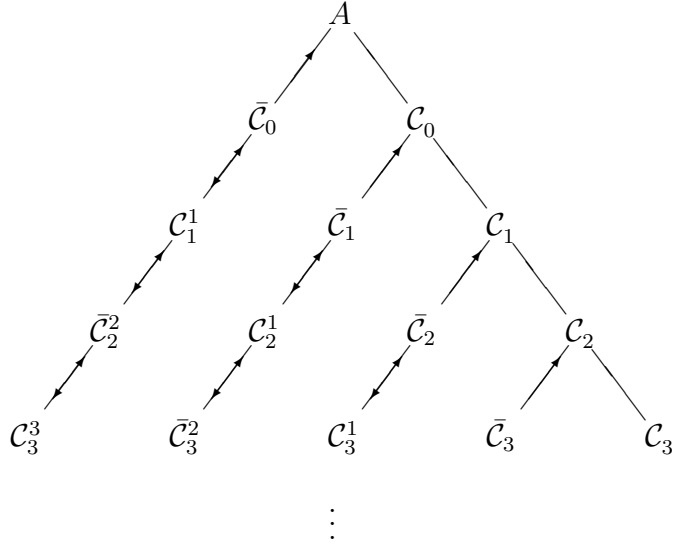
\put(70,5){\rm Diagram 9.\ The Triangular String Field Tableau}
\end{picture}

\noindent
In addition, there are the trivial-pair partners
$\pi^k_s$ for $k=1,3,5, \dots$ and $s \ge k$,
and $\bar \pi^k_s$ for $k=0,2,4, \dots$ and $s \ge k$.
Although there are no $\alpha_i$ type indices
on
${\cal C}^k_s$, $\bar {\cal C}^k_s$,
$\pi^k_s$ and $\bar \pi^k_s$,
they are string fields
and represent an infinite tower
of ordinary particle fields.
The ghost numbers of the fields are
$$
  {\rm gh} \left[ {{\cal C}_s^k} \right] =
  {\rm gh} \left[ {\bar {\cal C}_s^{k*}} \right] = s-k
\ , \quad \quad
  g\left( {{\cal C}_s^k} \right) =
  g\left( {\bar {\cal C}_s^{k*}} \right) = 1+k-s
\quad ,
$$
$$
  {\rm gh} \left[ {\bar {\cal C}_s^k} \right] =
  {\rm gh} \left[ {{\cal C}_s^{k*}} \right] = k-s-1
\ ,\quad \quad
   g\left( {\bar {\cal C}_s^k} \right) =
   g\left( {{\cal C}_s^{k*}} \right) = 1+k-s
\quad ,
$$
$$
  {\rm gh} \left[ {\bar \pi_s^k} \right] = k-s
\ ,\quad \quad
  g\left( {\bar \pi_s^k} \right) = 1+k-s
\quad ,
$$
\be
  {\rm gh} \left[ {\pi_s^k} \right] = \ s-k+1
\ , \quad \quad
  g\left( {\pi_s^k} \right) = 1+k-s
\quad ,
\label{obsft ghost numbers}
\ee
where gh$[ \ ]$ is the field-antifield ghost number
and $g( \ )$ is the string ghost number.
In Eq.\bref{obsft ghost numbers},
we use the abbreviations
in Eq.\bref{field equivalences}
for the cases when $k=-1$ and $k=0$.
All ${\cal C}_s^k$, ${ \bar {\cal C}_s^{k*} }$
and ${\bar \pi_s^k}$
have odd total statistics,
while ${\bar {\cal C}_s^k}$, ${\cal C}_s^{k*}$
and $\pi_s^k$
have even total statistics.

The auxiliary action $S_{\rm aux}$
in Eq.\bref{aux S} is
\be
  S_{\rm aux} = \sum\limits_{\scriptstyle {k=0}\hfill\atop
  \scriptstyle {k\ {\rm even} }\hfill}^\infty
    {}\sum\limits_{s=k}^\infty  {}
   \int \ {}^* \bar \pi_s^k  * \bar {\cal C}_s^{k*} +
  \sum\limits_{\scriptstyle {k=1}\hfill\atop
  \scriptstyle {k\ {\rm odd} }\hfill}^\infty  {}
   \sum\limits_{s=k}^\infty \int \ {}^* {\cal C}_s^{k*}*\pi_s^k
\quad ,
\label{obsft aux S}
\ee
where ${ \ }^*$ before a field is
the string Hodge star operation.
For an arbitrary field or antifield $\varphi$,
$
 {\rm gh} \left[ { \ }^* \varphi \right] =
 {\rm gh} \left[  \varphi \right]
$
but
$
 g\left( { \ }^* \varphi \right) =
 3 - g\left(  \varphi \right)
$.
The action
in Eq.\bref{obsft aux S}
is to be added
to the minimal-field proper solution $S$
in Eq.\bref{obsft proper solution}.

The most convenient gauge
for the open bosonic string field theory
is the Siegel-Feynman gauge
\ct{siegel84a}
which imposes the condition $\bar c_0 A =0$
on the string field $A$.
When implemented,
the quadratic action
of the massless vector $A_\mu$
becomes
$
 { {1} \over {2} }\int \dif^{26} x \
 A_\mu \partial_\nu \partial^\nu A^\mu
$.
Above,
$\bar c_0$ is the zero mode
of the antighost
of first-quantized open string theory
in Sect.\ \ref{ss:fqbs}
and Sect.\ \ref{sss:fqbs}.
It is the conjugate momentum
of the zero mode $c_0$ of the first-quantized ghost.
These modes satisfy
\be
  \left\{ {c_0, \bar c_0} \right\} =
   c_0 \bar c_0 + \bar c_0 c_0  =  1
\ ,\quad \quad
   c_0 c_0 = 0
\ ,\quad \quad \quad
  \bar c_0 \bar c_0=0
\quad ,
\label{c 0 b 0 algebra}
\ee
and have string ghost numbers of
$g\left( { c_0 } \right) = 1$ and
$g\left( { \bar c_0 } \right) = - 1$.
See Sect.\ \ref{ss:fqbs2}
for more discussion.
The gauge fermion $\Psi_{\rm sf}$,
which implements the Siegel-Feynman gauge,
has an expansion that,
for the first few levels,
reads
$$
  \Psi_{\rm sf} = \int \ {}^* \bar {\cal C}_0 * c_0 \bar c_0 A
$$
$$
  \int \ {}^* \bar {\cal C}_1 * c_0 \bar c_0 {\cal C}_0 +
  \int \ {}^* {\cal C}_1^1 * c_0 \bar c_0 \bar {\cal C}_0 +
$$
$$
  \int \ {}^* \bar {\cal C}_2 * c_0 \bar c_0 {\cal C}_1 +
  \int \ {}^* {\cal C}_2^1* c_0 \bar c_0 \bar {\cal C}_1 +
  \int \ {}^* \bar {\cal C}_2^2 * c_0 \bar c_0 {\cal C}_1^1
    + \ldots
\quad .
$$
For all levels,
\be
  \Psi_{\rm sf} = \sum\limits_{\scriptstyle {k=0}\hfill\atop
  \scriptstyle {k\ {\rm even } }\hfill}^\infty
   {}\sum\limits_{s=k}^\infty  \int {}
  \ {}^* \bar {\cal C}_s^k * c_0 \bar c_0 {\cal C}_{s-1}^{k-1} +
  \sum\limits_{\scriptstyle {k=1}\hfill\atop
  \scriptstyle {k\ {\rm odd} }\hfill}^\infty
  \sum\limits_{s=k}^\infty  \int {}
   \ {}^* {\cal C}_s^k * c_0 \bar c_0 \bar {\cal C}_{s-1}^{k-1}
\quad .
\label{sf fermion}
\ee

Since $\Psi_{\rm sf}$ is not a functional of
$\pi^k_s$ or $\bar \pi^k_s$,
a delta-function type gauge-fixing procedure
is implemented.
Elimination of the antifields
via Eq.\bref{fermion gauge fixing}
leads to
$$
  \bar {\cal C}_s^{k* }= c_0 \bar c_0 {\cal C}_{s-1}^{k-1} +
    \bar c_0 c_0{\cal C}_{s+1}^{k+1}
  \ ,\quad k=0,2,4,\ldots
\quad ,
$$
\be
  {\cal C}_s^{k*} = \bar c_0 c_0 \bar {\cal C}_{s-1}^{k-1} +
   c_0 \bar c_0 \bar {\cal C}_{s+1}^{k+1}
 \ ,\quad k=1,3,5,\ldots
\quad .
\label{sf elimination of antifields}
\ee
When the antifields
in Eq.\bref{sf elimination of antifields}
are substituted into $S_{\rm aux}$
of Eq.\bref{obsft aux S}
and the integration
over $\pi^k_s$ or $\bar \pi^k_s$
is performed,
the following conditions are implemented
$$
  \bar {\cal C}_s^{k*}=0\ \Rightarrow
  \ \bar c_0 {\cal C}_{s-1}^{k-1}=0
\ {\rm and}\
  c_0{\cal C}_{s+1}^{k+1}=0
  \ ,\quad k=0,2,4,\ldots
\quad ,
$$
\be
  {\cal C}_s^{k*}=0
\ \Rightarrow \
  c_0 \bar {\cal C}_{s-1}^{k-1}=0
\ {\rm and}\
  \bar c_0 \bar {\cal C}_{s+1}^{k+1} = 0
 \ ,\quad k=1,3,5,\ldots
\quad ,
\label{sf delta functions}
\ee
which follow
from Eqs.\bref{c 0 b 0 algebra}
and \bref{sf elimination of antifields}.
For $k \ge 1$,
$     {\cal C}_{s}^{k}$ and
$\bar {\cal C}_{s}^{k}$
are set equal to zero because
any field $\Phi$,
for which
$
  \bar c_0 \Phi =0
$ and
$
 c_0 \Phi =0
$,
must be identically zero,
as Eq.\bref{c 0 b 0 algebra} implies.
On the other hand,
for ${\cal C}_{s-1}$ and
$\bar {\cal C}_s$,
only half these conditions are imposed:
$$
  \bar c_0 {\cal C}_{s-1}=0
\ ,\quad
  s\ge 0
\quad ,
$$
\be
  c_0 \bar {\cal C}_s=0
\ ,\quad
  s\ge 0
\quad .
\label{sf conditions}
\ee

When these results are used in
$
  {\cal C}_s^{*} \equiv {\cal C}_s^{-1*} =
  c_0 \bar c_0 \bar {\cal C}_{s+1}^0 \equiv
  c_0 \bar c_0 \bar {\cal C}_{s+1}
$, for $s \ge -1$,
one finds
${\cal C}_s^{-1*} = \bar {\cal C}_{s+1}$,
so that the gauge-fixed string field
$\Psi_{{\Sigma}_{{\rm sf}}}$
in Eq.\bref{psi sum} is
\be
 \Psi_{{\Sigma}_{{\rm sf}}} = \ldots +
 {}^*\bar {\cal C}_{s+1} +
  \ldots + {}^*\bar {\cal C}_1 +
  {}^*\bar {\cal C}_0 +
  A + {\cal C}_0 + \ldots + {\cal C}_s + \ldots
\quad .
\label{sf psi sum}
\ee
The condition
$c_0 \bar {\cal C}_s = 0$ implies
$\bar c_0 {}^* \bar {\cal C}_s = 0$.
Hence, the constraints
in Eq.\bref{sf conditions}
can be written succinctly as
\be
 \bar c_0 \Psi_{{\Sigma}_{{\rm sf}}} = 0
\quad .
\label{sf gauge}
\ee
Eq.\bref{sf gauge} implies
that $\Psi_{\Sigma_{\rm sf}}$
has no term proportional to $c_0$.
Since the string integral is zero
unless there is a $c_0$ factor present,
$Q$ must supply it.
Since $Q = c_0 L_0 + \dots$,
\be
  Q \to c_0 L_0
\quad ,
\label{Q reduction}
\ee
where $ L_0$ is the string ``Klein-Gordon'' operator
\be
  L_0 = {1 \over 2}
  \left( {\partial_\mu \partial^\mu - {\cal M}^2} \right)
\quad ,
\label{L 0}
\ee
and ${\cal M}^2$ is the mass-squared operator.
The gauge-fixed action $S_{\Psi_{\rm sf}}$ is
\be
   S = {1 \over 2} \int \Psi_{\Sigma_{\rm sf}} *
      c_0 L_0 \Psi_{\Sigma_{\rm sf}} +
   {1 \over 3}\int \Psi_{\Sigma_{\rm sf}} *
       \Psi_{\Sigma_{\rm sf}} * \Psi_{\Sigma_{\rm sf}}
\quad ,
\label{sf action}
\ee
where $\Psi_{\Sigma_{\rm sf}}$
in Eq.\bref{sf psi sum}
satisfies equation \bref{sf gauge}.
This is the result obtained
in refs.\ct{bochicchio87a,thorn87a}.
Off-shell scattering amplitudes
in the Siegel-Feynman gauge
have been obtained
in refs.\ct{giddings86a,fgst88a,samuel88a,%
bs89a,bs89b,samuel90a}.

\subsection{The Massless Relativistic Spinning Particle}
\label{ss:mrsp2}

\hspace{\parindent}
This irreducible system
has super-reparametrization invariance.
It illustrates the gauge-fixing procedure
for a simple supergravity theory.
The system possesses one ordinary gauge symmetry
and one gauge supersymmetry,
for which
auxiliary trivial pairs
$\{ \bar {\cal C}$, $\bar \pi \}$ and
$\{ \bar \Gamma $, $\bar \Lambda \}$
are needed.
The fields $\bar \pi$, $\bar \Gamma$,
$\bar {\cal C}^*$ and $\bar \Lambda^*$
are commuting,
while
$\bar {\cal C}$, $\bar \Lambda$,
$\bar \pi^*$ and $\bar \Gamma^*$
are anticommuting.
The auxiliary action for the pairs is
\be
  S_{\rm aux} = \int {\dif \tau }
   \left( {\bar \pi \bar {\cal C}^* +
   \bar \Lambda \bar \Gamma^*} \right)
\quad .
\label{mrsp aux S}
\ee

The fermion
\be
  \Psi = \int {\dif \tau }
    \left( {\bar {\cal C}\left( {e-1} \right) +
    \bar \Gamma \chi } \right)
\label{mrsp psi}
\ee
imposes the delta-function conditions
\be
  e=1
\ ,\quad \quad
  \chi = 0
\quad .
\label{mrsp gf conditions}
\ee
A straightforward calculation shows that
the gauge-fixed action becomes
\be
  S_\Psi \to {1 \over 2}\int {\dif \tau }
   \left( {\partial_\tau x^\mu \partial_\tau x_\mu -
   i\psi^\mu \partial_\tau \psi_\mu } \right) +
   \int {\dif \tau }
   \left( {\bar {\cal C}\partial_\tau {\cal C} +
    \bar \Gamma \partial_\tau \Gamma } \right)
\quad .
\label{mrsp gf action}
\ee
For $x^\mu$ and $\psi^\mu$,
this is the flat-space action
in Eq.\bref{flat space ss action}.
Using the superfield formulation
in Eq.\bref{theta exp}
and defining the ghost superfields
$$
  G\equiv {\cal C} + \theta \Gamma
\ ,\quad \quad
   \bar G=\bar \Gamma + \theta \bar {\cal C}
\quad ,
$$
the action
in Eq.\bref{mrsp gf action}
can be written in supersymmetric form as
\be
  i\int {\dif \tau }\int {\dif \theta }
   \left( {{1 \over 2}
  D_\theta X^\mu \partial_\tau X_\mu -
   D_\theta \bar G D_\theta G} \right)
\quad ,
\label{mrsp gf action in sf form}
\ee
where $D_\theta$ is given
in Eq.\bref{def of D and Q}.

\subsection{The First-Quantized Bosonic String}
\label{ss:fqbs2}

\hspace{\parindent}
The BRST quantization
of the first-quantized bosonic string
was carried out
in refs.\ct{fujikawa82a,hwang83a,ko83a}.
The field-antifield treatment is similar.
In the formulation
of Sect.\ \ref{sss:fqbs},
there are three types of gauge transformations
for this irreducible system.
Correspondingly,
one needs to introduce
three level $0$ auxiliary trivial pairs
$\{ \bar {\cal C}$, $\bar \pi \}$,
$\{ \bar {\cal C}_n$, $\bar \pi_n \}$,
and $\{ \bar {\cal C}_{ab}$, $\bar \pi_{ab} \}$,
where $a$, $b$ and $n$ take on the values
$\sigma$ and $\tau$,
and $\bar {\cal C}_{ab}$ and $\bar \pi_{ab}$
are antisymmetric in $a$ and $b$.
The auxiliary action $S_{\rm aux}$ is
\be
  S_{\rm aux} =
  \int {\dif\tau }\int\limits_0^\pi  {\dif\sigma }
   \left\{ {\bar \pi \bar {\cal C}^* +
   \bar \pi_n\bar {\cal C}^{*n} +
   {1 \over 2}\bar \pi_{ab}\bar {\cal C}^{*ab}} \right\}
\quad .
\label{aux S for fqbs}
\ee
This action is to be added to $S$
in Eq.\bref{fqbs S}
to give a total action
$S_{\rm total} = S + S_{\rm aux}$.
For the gauge-fixing fermion,
we use
$$
  \Psi =\int {\dif \tau } \int_0^\pi {\dif \sigma }
  \left\{ {
   \bar {\cal C}_\tau
    \left( {{e_\tau}^\sigma -{e_\sigma}^\tau } \right) +
   \bar {\cal C}_\sigma
    \left( {{e_\tau}^\tau -{e_\sigma}^\sigma } \right) +
 } \right.
$$
\be
 \left. {
   \bar {\cal C}_{\tau \sigma }
    \left( {e_\tau}^\sigma + {e_\sigma}^\tau \right) +
   \bar {\cal C}
    \left( { {e_\tau}^\tau + {e_\sigma}^\sigma -
          2 \rho^{-1/2} } \right)
  } \right\}
\quad .
\label{fqbs psi}
\ee
Such a $\Psi$ leads to the delta-function-type
gauge conditions
\be
  {e_\tau}^\tau ={e_\sigma}^\sigma = \rho^{-1/2}
\ ,\quad \quad
  {e_\tau}^\sigma ={e_\sigma}^\tau = 0
\quad .
\label{fqbs gf conditions}
\ee
This is known as the conformal gauge.
The antifields are
$$
  {e^{*\tau}}_\tau =
    \bar {\cal C} + \bar {\cal C}_\sigma  \ ,
\quad \quad
  {e^{*\sigma}}_\sigma =
    \bar {\cal C} - \bar {\cal C}_\sigma
\quad ,
$$
$$
  {e^{*\tau}}_\sigma =
    \bar {\cal C}_{\tau \sigma } + \bar {\cal C}_\tau \ ,
\quad \quad
  {e^{*\sigma}}_\tau =
    \bar {\cal C}_{\tau \sigma } - \bar {\cal C}_\tau
\quad ,
$$
$$
  {\bar {\cal C}}^{* \tau} =
    {{e_\tau}^\sigma -{e_\sigma}^\tau }  \ ,
\quad \quad
  {\bar {\cal C}}^{* \sigma} =
 {{e_\tau}^\tau -{e_\sigma}^\sigma }
\quad ,
$$
$$
  {\bar {\cal C}}^{* \tau \sigma} =
    {{e_\tau}^\sigma + {e_\sigma}^\tau } \ ,
\quad \quad
  {\bar {\cal C}}^{*} =
    { {e_\tau}^\tau + {e_\sigma}^\sigma -
          2 \rho^{-1/2} }
\quad ,
$$
\be
  X_\mu^*=0
\ ,\quad \quad
  {\cal C}^*={\cal C}_n^*={\cal C}_{ab}^*=0
\quad .
\label{fqbs antifield elimination}
\ee
The gauge-fixed action becomes
$$
  S_\Psi \to
    \int {\dif \tau }\int_0^\pi {\dif \sigma }
  \left\{ {1 \over 2} {\eta_{\mu \nu }
  \left( {\partial_\tau X^\mu \partial_\tau X^\nu -
  \partial_\sigma X^\mu \partial_\sigma X^\nu }
     \right)} \right.
$$
$$
  + \ \bar {\cal C}_\tau \rho^{-1/2}
    \left( { \partial_\sigma {\cal C}^\tau  -
    \partial_\tau {\cal C}^\sigma } \right) +
  \bar {\cal C}_\sigma \rho^{-1/2}
    \left(  \partial_\sigma {\cal C}^\sigma -
  {\partial_\tau {\cal C}^\tau } \right)
$$
$$
   + \ 2 \bar {\cal C}
  \left( { {\cal C}^n \partial_n \rho^{-1/2} } \right)
   - \bar {\cal C}_{\tau \sigma} \rho^{-1/2}
  \left(
     { \partial_\tau {\cal C}^\sigma +
     \partial_\sigma {\cal C}^\tau + 2 {\cal C}^{\tau \sigma}
  } \right)
$$
\be
 \left. {
   + \ \bar {\cal C} \rho^{-1/2}
  \left( {
      2{\cal C} - \partial_\tau {\cal C}^\tau -
                \partial_\sigma {\cal C}^\sigma
  } \right)
 } \right\}
\quad ,
\label{fqbs gf action}
\ee
where the sum over $n$
is over $\tau$ and $\sigma$:
$
 {\cal C}^n \partial_n \rho^{-1/2} =
 {\cal C}^\tau \partial_\tau \rho^{-1/2} +
 {\cal C}^\sigma \partial_\sigma \rho^{-1/2}
$.
Integrating over ${\cal C}$ and
${\cal C}^{\tau \sigma}$ leads to
delta functions for $\bar {\cal C}$ and
$\bar {\cal C}_{\tau \sigma}$,
so that the gauge-fixed action becomes
the first two lines
in Eq.\bref{fqbs gf action}:
$$
  S_\Psi \to
    \int {\dif \tau }\int_0^\pi {\dif \sigma }
  \left\{ {1 \over 2} {\eta_{\mu \nu }
  \left( {\partial_\tau X^\mu \partial_\tau X^\nu -
  \partial_\sigma X^\mu \partial_\sigma X^\nu }
     \right)} \right.
$$
\be
 \left. {
  + \ \bar {\cal C}_\tau \rho^{-1/2}
    \left( { \partial_\sigma {\cal C}^\tau -
    \partial_\tau {\cal C}^\sigma } \right) +
  \bar {\cal C}_\sigma \rho^{-1/2}
    \left( { \partial_\sigma {\cal C}^\sigma -
  \partial_\tau {\cal C}^\tau } \right)
 } \right\}
\quad .
\label{fqbs gf action short}
\ee

At this stage,
one can either define
a new $\bar {\cal C}_n$ field
which absorbs the factor of $\rho^{-1/2}$
or one can set
$\rho \left( \tau , \sigma \right) = 1 $,
as long as $D=26$.%
{\footnote{
When $D \neq 26$, one cannot do this due to
Weyl anomaly.
See Sect.\ \ref{ss:aobs}.
}}
In either case,
since the resulting action is a free theory,
it is straightforward to quantize the system.
In what follows,
we consider the open string case.
The equations of motion are
$$
 \partial_\tau \partial_\tau X^\mu -
 \partial_\sigma \partial_\sigma X^\mu = 0
\quad ,
$$
$$
  \partial_\tau {\cal C}^\tau -
    \partial_\sigma {\cal C}^\sigma = 0
\ ,\quad \quad
  \partial_\tau {\cal C}^\sigma -
      \partial_\sigma {\cal C}^\tau = 0
\quad ,
$$
\be
  \partial_\tau \bar {\cal C}_\tau -
  \partial_\sigma \bar {\cal C}_\sigma =0
\ ,\quad \quad
  \partial_\tau \bar {\cal C}_\sigma -
   \partial_\sigma \bar {\cal C}_\tau = 0
\quad ,
\label{fqbs eom}
\ee
and
\be
  \partial_\sigma X^\mu = {\cal C}^\sigma =
   \bar {\cal C}_\sigma =0
\ ,\quad \quad {\rm at} \ \sigma =0, \pi
\quad ,
\label{fqbs bc}
\ee
which follow
from a careful treatment
of boundary conditions.
The solutions are
$$
  X^\mu \left( {\tau ,\sigma } \right) =
   x^\mu + {{p^\mu \tau } } + i
  \sum\limits_{\scriptstyle {n=-\infty }\hfill\atop
  \scriptstyle {n \ne 0}\hfill}^\infty
   { {\alpha_n^\mu} \over {n} }
  e^{-in\tau } \cos \left( {n\sigma } \right)
\quad ,
$$
$$
  {\cal C}^\tau \left( {\tau ,\sigma } \right) =
   \sum\limits_{\scriptstyle {n=-\infty }}^\infty
  c_n e^{-i n \tau } \cos \left( {n\sigma } \right)
\quad ,
$$
$$
  {\cal C}^\sigma \left( {\tau ,\sigma } \right) =
  -i\sum\limits_{n=-\infty }^\infty
  c_n e^{-i n \tau }\sin \left( {n\sigma } \right)
\quad ,
$$
$$
  \bar {\cal C}_\sigma \left( {\tau ,\sigma } \right) =
   - i \sum\limits_{\scriptstyle {n=-\infty }}^\infty
  \bar c_n e^{-i n \tau }\cos \left( {n\sigma } \right)
\quad ,
$$
\be
  \bar {\cal C}_\tau \left( {\tau ,\sigma } \right) =
   \sum\limits_{n=-\infty }^\infty
   \bar c_n e^{-i n \tau }\sin \left( {n\sigma } \right)
\quad ,
\label{fqobs mode expansions}
\ee
where the modes
$\alpha_n^\mu$, $c_n$ and $\bar c_n$
($n = - \infty \ {\rm to } \ + \infty $, $n \ne 0$)
are harmonic-oscillator-like
raising and lowering operators
satisfying
$$
  \left[  {p^\mu ,x^\nu } \right] = -i\eta^{\mu \nu }
\ ,\quad \quad
   \left[ {\alpha_n^\mu ,\alpha_{-m}^\nu } \right] =
   n\eta^{\mu \nu }\delta_{nm}
\quad ,
$$
\be
  \left\{ {c_n,\bar c_{-m}} \right\} = \delta_{nm}
\ ,\quad \quad
  \left\{ {c_n,c_m} \right\}=0
\ ,\quad \quad
   \left\{ {\bar c_n,\bar c_m} \right\}=0
\quad ,
\label{mode commutation relations}
\ee
where $\eta^{\mu \nu }$ is
the $D= 25 + 1$ flat space metric
with signature $(-,+,+,\dots)$.
The operators $x^\mu$ and $p^\mu$
correspond to the center-of-mass position
and momentum of the string.
The zero-mode
$c_0$ and $\bar c_0$ play an important role
in the Siegel-Feynman gauge
of the open bosonic string
as discussed in Sect.\ \ref{ss:osft2}.
States in the theory
are obtained by applying
creation operators,
corresponding to modes $n$ with $n<0$,
to the vacuum states,
which are eigenfunctions of
$p^\mu$ and the zero-mode ghost system.

In second quantization,
fields become functionals
of the first-quantized variables
$X^\mu \left( { 0 , \sigma } \right)$,
${\cal C}^\tau \left( { 0 , \sigma } \right)$ and
$\bar {\cal C}_\sigma \left( { 0 , \sigma } \right)$.
The string fields
for the open bosonic string field theory
of Sects.\ \ref{ss:obsft}, \ref{sss:osft}
and \ref{ss:osft2}
are such functionals.
Alternatively,
one may expand the string fields
as a linear combination of
first-quantized states
whose coefficients are ordinary particle fields.
For more details,
see the reviews
in refs.\ct{samuel87a,thorn89a}.

The gauge-fixed BRST symmetry of
the action
in Eq.\bref{fqbs gf action short}
is
$$
 \delta_{B_\Psi} X^\mu = {\cal C}^n \partial_n X^\mu \ ,
$$
$$
  \delta_{B_\Psi} {\cal C}^m =
   - {\cal C}^n \partial_n {\cal C}^m \ ,
$$
\be
  \delta_{B_\Psi} \bar {\cal C}_\tau =
    - \rho^{1/2} \partial_\tau X_\mu \partial_\sigma X^\mu +
    \rho^{1/2} \partial_n
    \left( \rho^{-1/2} {\bar{\cal C}_\tau {\cal C}^n} \right) +
    \bar{\cal C}_\sigma \left( {\partial_\sigma {\cal C}^\tau +
   \partial_\tau {\cal C}^\sigma } \right)
\  ,
\label{fqbs gf BRST}
\ee
$$
  \delta_{B_\Psi} \bar {\cal C}_\sigma =
    - \rho^{1/2} {1 \over 2}\left( {
     \partial_\tau X_\mu \partial_\tau X^\mu  +
    \partial_\sigma X^\mu \partial_\sigma X_\mu } \right) +
    \rho^{1/2} \partial_n
     \left( \rho^{-1/2} {\bar{\cal C}_\sigma {\cal C}^n} \right) +
    \bar{\cal C}_\tau \left( {\partial_\sigma {\cal C}^\tau +
    \partial_\tau {\cal C}^\sigma } \right)
\ ,
$$
where $m$ takes on the values $\tau$ and $\sigma$
and the sum over $n$ is over $\tau$ and $\sigma$.
A straightforward computation confirms that
\be
  \delta_{B_\Psi} S_\Psi \equiv
   {{\partial_r S_\Psi } \over {\partial \Phi^A}}
   \delta_{B_\Psi} \Phi^A = 0
\quad ,
\label{fqbs BRST inv of gf action}
\ee
where $S_\Psi$ is given
in Eq.\bref{fqbs gf action short}.
A useful exercise is to perform this computation.
Because some fields have been eliminated
using equations of motion,
$  \delta_{B_\Psi} \delta_{B_\Psi} = 0 $
only on-shell.

\vfill\eject

\section{Quantum Effects and Anomalies}
\label{s:qea}

\hspace{\parindent}
In this section
we discuss some of the quantum aspects
of the field-antifield formalism.
Basically, the classical quantities and concepts
have quantum counterparts:
The proper solution $S$
is replaced by a quantum action $W$.
The quantum master equation is used in lieu of
the classical master equation.
The BRST transformation $\delta_B$
is generalized to a quantum version $\delta_{\hat B}$.
It is nilpotent only if the quantum master equation
is satisfied.
Quantum observables are elements
of the cohomology of $\delta_{\hat B}$.
A violation of the quantum master equation
corresponds to a gauge anomaly
\ct{tnp90a}
and is the subject
of Sects.\ \ref{ss:qmevg}--\ref{ss:aoll}.
In Sects.\ \ref{ss:qBRSTt}--\ref{ss:eazje},
however,
we assume that $W$
satisfies the quantum master equation.
Many equations
in those subsections are of a formal nature
due to the singular operator $\Delta$
in the context of local theories.
To obtain a well-defined action of $\Delta$,
a regularization scheme must be used.
See Sect.\ \ref{ss:aoll}.
In Sect.\ \ref{ss:eazje},
the effective action $\Gamma$ is discussed.
Antifields become classical
and acquire a conceptual interpretation:
They are sources for BRST transformations
generated by $\Gamma$.

\subsection{Quantum-BRST Transformation and Its Cohomology}
\label{ss:qBRSTt}

\hspace{\parindent}
Recall that the condition
in Eq.\bref{qme for correlation}
that a functional $X$
produces a gauge-invariant correlation
is
$$
  \left( {X,W} \right) = i \hbar \Delta X
\quad .
$$
Introducing the operator
$\delta_{\hat B}$
\ct{lt85a},
\be
 \delta_{\hat B} X \equiv
  \left( {X,W} \right) - i \hbar \Delta X
\quad ,
\label{qBRST}
\ee
this can be written as
$\delta_{\hat B} X = 0$.
The operator $\delta_{\hat B}$ is
the quantum BRST transformation.
It is the quantum generalization
of the classical BRST transformation $\delta_B$:
As $\hbar$ goes to zero,
$\delta_{\hat B}$
becomes $\delta_B$
in Eq.\bref{def of BRST}.
A functional is said to be
{\it quantum-BRST invariant}
if
\be
\delta_{\hat B} X = 0
\quad .
\label{qBRST invariance}
\ee

Although $\delta_{\hat B}$ is nilpotent
as a consequence of the quantum master equation
\be
  \delta_{\hat B} \delta_{\hat B} X =
  - ( {1 \over 2} \left( {W,W} \right) -
     i\hbar\Delta W,X  ) = 0
\quad ,
\label{nil of qBRST}
\ee
it no longer is a graded derivation since
\be
  \delta_{\hat B}\left( {XY} \right) =
   X \left( {\delta_{\hat B}Y} \right) +
  \left( {-1} \right)^{\epsilon_Y}
   \left( {\delta_{\hat B} X} \right) Y -
  i\hbar \left( {-1} \right)^{\epsilon_Y} \left( {X,Y} \right)
\quad .
\label{qBRST of product}
\ee

Suppose $X$ is the BRST transform
of another functional $Y$, i.e.,
$$
 X = \delta_{\hat B} Y  =
 \left( {Y,W} \right) - i\hbar \Delta Y
\quad .
$$
Then, one refers to $X$ as
{\it quantum-BRST trivial}.
By the nilpotency of $\delta_{\hat B}$,
$X$ is quantum-BRST invariant,
but not in an interesting way.
In fact,
a quantum-BRST trivial operator $X$
produces a zero correlation function,
as can be seen as follows.
The integrand ${\cal I}$
in Eq.\bref{integrand} becomes
$$
  \exp \left( {{i \over {\hbar}}W} \right) X =
    - i {\hbar} \Delta
   \left( {\exp \left( {{i \over {\hbar}}W} \right) Y} \right)
\quad .
$$
Let
$
 \tilde {\cal I} =
   - i \hbar \exp \left( {{i \over {\hbar}}W} \right) Y
$.
Then
$$
  \int {} \left[ \dif \Phi \right]
     \left. { \Delta \tilde {\cal I} } \right|_{\Sigma_\Psi } =
  \int {} \left[ \dif \Phi \right]
   \left. { \left( {-1} \right)^{\epsilon_A+1}
  \left( {{{\partial_r} \over {\partial \Phi^A}}
  {{\partial_r} \over {\partial \Phi_A^*}}
  \tilde {\cal I}} \right)} \right|_{\Sigma_\Psi } =
$$
$$
  \int {} \left[ \dif \Phi \right]
   \left( {-1} \right)^{\epsilon_A+1}
  {{\partial_r} \over {\partial \Phi^A}}
   \left( {
   \left.{
   \left( { { {\partial_r} \over {\partial \Phi_A^*} }
   \tilde {\cal I}} \right) } \right|_{\Sigma_\Psi }
   } \right) +
$$
$$
   \int {} \left[ \dif \Phi \right]  \left( {-1} \right)
  \left. {^{\epsilon_A}\left( {{{\partial_r\partial_r
  \tilde {\cal I}} \over {\partial \Phi_B^*\partial \Phi_A^*}}
  {{\partial_r\partial_r\Psi }
   \over {\partial \Phi^A\partial \Phi^B}}} \right)}
   \right|_{\Sigma_\Psi }
\quad .
$$
The first term produces zero
since it is a surface term
and the second term is zero for symmetry reasons.
Hence, an interesting functional ${\cal O}$
is one that is non-trivial
in the quantum-BRST cohomology,
i.e., ${\cal O}$ is quantum-BRST invariant
but cannot be written
as the quantum-BRST transform of some functional:
\be
  \delta_{\hat B} {\cal O} = 0 \quad {\rm \ but \ }
  {\cal O} \ne \delta_{\hat B} {Y}
\quad ,
\label{observable}
\ee
for any ${Y}$.
An ${\cal O}$ satisfying
Eq.\bref{observable}
is called a {\it quantum observable}.
Observables are considered equivalent
if they differ by a quantum-BRST trivial functional.

Unlike the situation at the classical level,
quantum observables do not form an algebra.
This is
due to the fact that $\delta_{\hat B}$
is no longer a graded derivation,
c.f. Eq.\bref{qBRST of product}.
This is not surprising
since,
in the quantum theory,
the product of two observables
is singular and must be regularized.
The regularization process
may ruin quantum-BRST invariance.
However, if
the space-time supports of
${\cal O}_1$ and ${\cal O}_2$
do not intersect,
then $ \left( { {\cal O}_1 , {\cal O}_2} \right) = 0$.
In such a case,
${\cal O}_1 {\cal O}_2$ is quantum-BRST invariant,
if ${\cal O}_1$ and ${\cal O}_2$ are.

The result obtained above, namely,
\be
 \int {} \left[ \dif \Phi \right]
  \exp \left( {{i \over {\hbar}}W} \right)
  \delta_{\hat B} Y = 0
\quad ,
\label{ward identity}
\ee
for various functionals $Y$
produces identities among correlation functions.
They embody
the Ward identities
associated with the BRST invariance of the theory
\ct{ward50a,takahashi57a,slavnov72a,taylor71a}.

An example of a correlation function
that produces a zero
expectation value is
\be
  \int {} \left[ \dif \Phi \right]
   \exp \left( {{i \over {\hbar}}W_\Psi } \right)
  \left( {{{\partial_l \left( {\left. X
    \right|_{\Sigma_\Psi }} \right)} \over {\partial \Phi^A}} +
    {i \over {\hbar}}{{\partial_l \left( {W_\Psi } \right)}
      \over {\partial \Phi^A}}\left. X
     \right|_{\Sigma_\Psi }} \right) = 0
\quad ,
\label{unSDe}
\ee
obtained by integration by parts
using Eq.\bref{int by parts}.
Up to a normalization factor,
Eq.\bref{unSDe}
is the Schwinger-Dyson equation
\ct{dyson49a,schwinger51a}
\be
  \left\langle {{{\partial_l \left(
   {\left. X \right|_{\Sigma_\Psi }} \right)}
      \over {\partial \Phi^A}} +
    {i \over {\hbar}}{{\partial_l\left( {W_\Psi } \right)}
     \over {\partial \Phi^A}}
   \left. X \right|_{\Sigma_\Psi }} \right\rangle = 0
\quad .
\label{SDe}
\ee
In the above,
$W_\Psi$ denotes the gauge-fixed quantum action
\be
   W_\Psi \equiv \left. W \right|_{\Sigma_\Psi }
\quad ,
\label{gauge-fixed quantum action}
\ee
and $X$ is any quantum-BRST invariant functional,
$
  \delta_{\hat B} X = 0
$.
The derivatives in
Eqs.\bref{unSDe} and \bref{SDe}
involve implicit differentiation
of the gauge-fixing fermion, e.g.,
$$
  {{\partial_l \left( {\left. X \right|_{\Sigma_\Psi }} \right)}
    \over {\partial \Phi^A}} =
   \left. {{{\partial_l X} \over {\partial \Phi^A}}}
     \right|_{\Sigma_\Psi } +
    {{\partial_l \partial \Psi } \over
    {\partial \Phi^A\partial \Phi^B}}
     \left. {{{\partial_l X} \over {\partial \Phi_B^*}}}
     \right|_{\Sigma_\Psi }
\quad ,
$$
and a similar identity holds for
$
  X \to W
$.
The fact that the functional
in Eq.\bref{unSDe}
produces a zero expectation
suggests that it is quantum-BRST trivial.
Indeed, this is the case since
\be
  {{\partial_l \left( {\left. X \right|_{\Sigma_\Psi }} \right)}
    \over {\partial \Phi^A}} + {i \over {\hbar}}
    {{\partial_l \left( {W_\Psi } \right)} \over
    {\partial \Phi^A}}\left. X \right|_{\Sigma_\Psi } =
    \left. {\left( { \delta_{\hat B} \left( {{{
    \left( {-1} \right)^{\epsilon \left( X \right)}} \over {i\hbar}}
   \left( {\Phi_A^*-{{\partial \Psi }
       \over {\partial \Phi^A}}} \right)}
    X \right)} \right)} \right|_{\Sigma_\Psi }
\quad ,
\label{q-BRST triviality of SDe}
\ee
as a short calculation reveals.
One obtains the interesting result
that the Schwinger-Dyson equations
are a consequence of the quantum-BRST symmetry
of the field-antifield formalism
\ct{henneaux90a}.
Reference \ct{ad93a}
argued that antifields originate
as the antighosts of collective fields
that ensure the Schwinger-Dyson equations.
This differs from one viewpoint
that antifields are the sources of BRST transformations.
Equations \bref{ward identity} and
\bref{q-BRST triviality of SDe}
show that the Schwinger-Dyson equations
are certain quantum-BRST Ward identities.

\subsection{Satisfying the Quantum Master Equation}
\label{ss:sqme}

\hspace{\parindent}
The quantum master equation
\be
  {1 \over 2} \left( {W,W} \right) -
     i\hbar \Delta W = 0
\label{qme2}
\ee
is the most important requirement
of the field-antifield formalism
for several reasons:
As demonstrated
in Sect.\ \ref{ss:g},
it guarantees that computations
are independent of the gauge-fixing procedure.
When the quantum action $W$ is
expanded in powers of $\hbar$ via
$
  W = S + O\left( {\hbar } \right)
$,
the $O\left( {\hbar^0} \right)$ term
in Eq.\bref{qme2}
is the classical master equation
\be
  {1 \over 2} \left( {S,S} \right) = 0
\quad ,
\label{cme2}
\ee
which yields the structure equations.
The nilpotency of the quantum BRST operator
depends
on Eq.\bref{qme2}
being satisfied,
as can be seen
from Eq.\bref{nil of qBRST}.
The quantum cohomology
requires the existence of
a nilpotent quantum BRST operator.
It is used
to define the quantum observables of a gauge theory
and to determine when two functionals
are considered equivalent.
In Sect.\ \ref{ss:qmevg},
it is argued that the existence
of gauge anomalies
is related to the violation
of the quantum master equation.
In short,
the field-antifield formalism at the quantum level
depends crucially on Eq.\bref{qme2} being satisfied.

The potential difficulty is due to the operator
$\Delta$
in Eq.\bref{def Delta}
which is singular
when acting on
$S$ or $W$
because usually they are local functionals.
Often terms proportional to delta functions
and derivatives of delta functions are produced.
One therefore needs to regularize $\Delta$.
If one can find a regularization
such that
$
  \left( {\Delta S} \right)_{\rm reg} = 0
$,
while maintaining the classical master equation,
then one can simply let
$
  W = S
$.%
{\footnote{
Strictly speaking,
this statement holds for a finite
or a regularized theory.
Renormalization may require
one to add additional terms to $S$.
}}
Then the quantum master equation is satisfied.
An example of this situation is a theory
without gauge invariances:
Antifields are absent in $S$
since the proper solution is given by $S = S_0$.
Consequently, $\Delta S = 0$.

The quantum master equation has a symmetry
given by quantum BRST transformations,
in which $W \to W - \varepsilon \delta_{\hat B} F$,
where $\eps ( F ) = 1$ and ${\rm gh} [ F ] = -1$
\ct{bv84a,fisch90a,henneaux90a,dejonghe93a,tpbook}.
If Eq.\bref{qme2}
is true,
and
\be
  W = W^{'} +
   \varepsilon \left( { W^{'} , F } \right) +
  i \hbar \varepsilon \Delta F
\quad ,
\label{qme symmetry}
\ee
then $W^{'}$ also satisfies
Eq.\bref{qme2}
to order $\varepsilon^2$.
Eq.\bref{qme symmetry}
has a simple interpretation.
The term $\varepsilon \left( { W^{'} , F } \right)$
is the change in $W$
due to a canonical transformation
(see Eq.\bref{canonical trans on function}).
The extra term
$i \hbar \varepsilon \Delta F$
is due to the non-invariance
of the functional integral measure
under a canonical transformation.
The measure effect can be exponentiated
and leads to the extra term.

A solution to the quantum master equation
ensures that the gauge symmetries survive
the process of quantization.
Although finding a solution
is an important objective,
a solution does not necessarily guarantee
good behavior of the theory.
The theory may still have difficulties
in regard to other issues
such as locality, unitarity
and the presence of Feynman-diagram infinities.
These difficulties may be insurmountable
or
it may be necessary
to add new counterterms
to achieve the desired properties.
There are systems for which the quantum master equation
is satisfied but the theory is non-renormalizable
(see the next subsection).
In short,
the quantum theory is not determined
solely by the quantum master equation.

\subsection{Remarks on Renormalization}
\label{ss:rr}

\hspace{\parindent}
At the one-loop level and beyond,
perturbation theory produces infinities
if Feynman-diagram contributions are not regularized.
If the regularization involves a cutoff,
then standard renormalizability of a field theory
means that the cutoff dependence can be absorbed
into the coefficients
of the terms in the original action
\ct{iz80a,zinnjustin89a}.%
{\footnote{
In dealing with renormalizable theories,
we also assume that a reasonable gauge-fixing procedure
is used, i.e.,
we exclude procedures leading to interactions
that are non-renormalizable
by power counting.
}}
In this manner, infinities are eliminated.
If a regularization procedure can be found
that respects
the structure of the field-antifield formalism
then renormalization is expected to proceed
as in the usual field theory case.
Infinities should be ``absorbable''
into the coefficients of the terms in $S$.
Regularization methods
that seem most convenient for gauge theories,
such as the Pauli-Villars scheme,
are the best candidates for respecting
the field-antifield formalism.

To be concrete,
suppose a gauge theory is renormalizable,
and suppose the regularization of $\Delta$
happens to give
$
  \left( {\Delta S} \right)_{\rm reg} = 0
$.%
{\footnote{
A renormalizable theory
without gauge invariances
is a useful case to keep in mind.}}
Then the $M_n$
in Eq.\bref{hbar expansion of W}
can be chosen to be the counterterms
removing the cutoff dependence of the theory.
The $M_n$ involve terms similar
to the original action $S$.
The quantum master equation
in Eq.\bref{conditions of loop terms}
is then satisfied
because
$ \left( { \Delta M_{n}  } \right)_{\rm reg} = 0 $
and because
$ \left( {M_n,S} \right) = \left( {M_n, M_m} \right) = 0 $.
These equations
hold because of the classical master equation
$\left( {S,S} \right) = 0$,
because
$ \left( {\Delta S} \right)_{\rm reg} = 0 $,
because
the $M_n$ resemble terms in $S$,
and because of analyticity of $S$
in coupling constants.
If
$
  \left( {\Delta S} \right)_{\rm reg} \ne 0
$
and an infinite number of local counterterms
have to be added to the action
to satisfy the quantum master equation,
then the issue of renormalizability is unclear
\ct{anselmi93a}.
This situation arises
for the closed bosonic string field theory
(see Sect.\ \ref{ss:d26cbsft}),
as currently formulated.

The question of whether the quantum master equation is satisfied
and
the question of renormalizability are separate issues
(although somewhat related).
As argued above,
a renormalizable theory without gauge invariances
is guaranteed to satisfy the quantum master equation.
Regularization and renormalization
must be performed,
even if the quantum master equation is satisfied
at all stages of the renormalization process.
When gauge invariances are present,
the interesting issue is whether
the quantum master equation
can be maintained
after renormalization.
A theory can be non-renormalizable
but satisfy the quantum master equation:
A non-renormalizable theory without gauge invariances
is an example.
An infinite number of counterterms must be
added to the action,
but the quantum master equation is satisfied trivially
because no antifields enter.
In the other extreme,
the two-dimensional chiral Schwinger model
is anomalous,
yet the theory still makes sense
and is renormalizable.
It is an example of a renormalizable system
that does not satisfy the quantum master equation.
For more discussion on regularization and renormalization
in the field-antifield formalism,
which are the natural generalizations
of ideas contained in earlier approaches
\ct{zinnjustin75a,lee76a,jl76a},
see refs.\ct{vlt82a,vt82b,tonin92a,%
anselmi93a,dejonghe93a,henneaux93a,lps93a}.

The Zinn-Justin equation
\ct{zinnjustin75a}
has played an important role
in analyzing the renormalizability
of Yang-Mills theories.
The generalization of this equation
within the antibracket formalism
is the subject of the next section.

\subsection{The Effective Action and the Zinn-Justin Equation}
\label{ss:eazje}

\hspace{\parindent}
A useful concept in functional approaches
to field theories
\ct{al73a,iz80a,zinnjustin89a}
is the effective action $\Gamma$.
When used in the classical approximation,
it reproduces computations at the quantum level.
This is accomplished by incorporating
all loop contributions
into effective interactions.
These interactions sum
the one-particle-irreducible diagrams.
Hence, to compute a correlation function
in perturbation theory with $\Gamma$,
one uses tree diagrams only.

The functional $\Gamma$ is obtained
by a Legendre transformation.
One performs the functional integral
over the quantum fields in the presence
of sources.
Classical fields
are obtained by evaluating
the expectation values
of the quantum fields
in the presence of these source terms.
The effective action $\Gamma$
is a functional of the classical fields.

The concept of an effective action
also exists in the antibracket formalism
\ct{bv81a}.
We also denote it by $\Gamma$.
It makes the quantum system
resemble a classical system
by summing loop effects.
Since one wants the analog
of the ``effective antibracket formalism'',
it is necessary to have classical antifields
as well as classical fields.
After this is accomplished,
one can define an antibracket $( \ , \ )_c$
in the space of classical fields and antifields.
Since computations are performed
as though the system is classical,
the classical master equation for $\Gamma$
incorporates the quantum master equation
of the original system.

Consider the functional integral
in the presence of sources $J_A$, i.e.,
\be
  Z \left[ {J,\Phi_c^*} \right] \equiv
   \int {} \left[ \dif\Phi \right] \exp \left[ {{i \over {\hbar}}
   \left( {W\left[ {\Phi ,\Phi_c^* } \right] +
   J_A\Phi^A} \right)} \right]
\quad ,
\label{Z of J}
\ee
where $J_A$ are independent variables
that do not depend on $\Phi$ and $\Phi_c^*$,
and have statistics
$
  \eps \left( {J_A} \right) =
  \eps \left( {\Phi^A} \right) = \eps_A
$.
For reasons that become clear below,
we have replaced the $\Phi^*$
by classical antifields $\Phi_c^*$.
Actually,
to compute perturbatively,
one needs to shift the antifields $\Phi_{cA}^*$ by
${{\partial \Psi } \over {\partial \Phi^A }}$,
where $\Psi$ is an appropriate gauge-fixing fermion.
More precisely,
one performs a canonical transformation
to the gauge-fixed basis
before setting $\Phi_{A}^*$ equal to $\Phi_{cA}^*$.
The net result is that
$
  Z \left[ { J,\Phi_c^* +
  {{\partial \Psi } \over {\partial \Phi }}
            } \right]
$
is the functional integral of interest.
Although a gauge-fixing fermion is present,
we omit
the dependence on $\Psi$
for notational convenience,
in what follows.
Since antifields appear in $S$
in the form
$S = S_\Psi + \Phi_A^* \delta_{B_\Psi} \Phi^A + \ldots $,
they act as sources for the gauge-fixed BRST transformations.

Define a classical field $\Phi_c^A$ by
\be
  \Phi_c^A \left[ {J,\Phi_c^*} \right] \equiv
  {\hbar \over i}{{\partial_l \ln Z
  \left[ {J,\Phi_c^*} \right] } \over {\partial J_A}} =
   \left\langle {\Phi^A} \right \rangle^J
\quad ,
\label{def of classical field}
\ee
where $\left\langle \ \  \right\rangle^J$
denotes an expectation value
in the presence of sources:
\be
  \left\langle X \right\rangle^J \equiv
   { 1 \over {Z \left[ {J,\Phi_c^*} \right] } }
  \int {} \left[ \dif \Phi \right]
  \ X \left[ {\Phi ,\Phi_c^*} \right]
   \exp \left[ {{i \over {\hbar}}
  \left( {W+J_A\Phi^A} \right)} \right]
\quad .
\label{def <>^J}
\ee
Note that $\Phi_c^A$ is a functional of
$J_B$ and $\Phi_{cB}^*$.
In principle,
it is possible to invert this relation
to determine $J_A$ in terms of
$\Phi_c^B$ and $\Phi_{cB}^*$.
We indicate the solution to this inversion
by $J_{cA}$:
$
  J_{cA} = J_A\left[ {\Phi_c,\Phi_c^*} \right]
$.
The effective action
is obtained by a Legendre transformation
\ct{iz80a}:
\be
  \Gamma \left[ {\Phi_c,\Phi_c^*} \right] \equiv
   -i\hbar\ln Z_c -
   J_{cA} \Phi_c^A
\quad ,
\label{def of effective action}
\ee
where $Z_c$ is $Z \left[ { J, \Phi_c^* } \right] $
evaluated at $J_A = J_{cA}$:
\be
  Z_c \equiv Z\left[ {J_c,\Phi_c^*} \right]
\quad .
\label{def of Zc}
\ee
As a result, $Z_c$ is a functional
of classical fields and antifields.
A straightforward calculation of
${{\partial_r \Gamma } \over {\partial \Phi_c^A}}$ gives
$$
  {{\partial_r \Gamma } \over {\partial \Phi_c^A}} =
   - J_{cA} - \left( {-1} \right)^{\eps_A\eps_B}
   {{\partial_r J_{cB}} \over {\partial \Phi_c^A}}
   \Phi_c^B -
   \left. {i\hbar{{\partial_r \ln Z
  \left[ {J,\Phi_c^*} \right] } \over {\partial J_B}}
   {{\partial_r J_{cB} } \over {\partial \Phi_c^A}}} \right|_{J=J_c}
\quad .
$$
Using Eq.\bref{def of classical field},
one finds that the last two terms cancel
so that
\be
  {{\partial_r \Gamma } \over {\partial \Phi_c^A}} =
   - J_{cA}
\quad .
\label{recovery of Jc}
\ee

At this stage,
we have made a transition from fields
to classical fields.
Essentially all quantities are now
functionals of $\Phi_c^A$ and $\Phi_{cA}^*$.
Given two functionals $X$ and $Y$
of $\Phi_c^A$ and $\Phi_{cA}^*$,
define the classical antibracket
$\left( { \ , \ } \right)_c $ by
\be
  \left( {X,Y} \right)_c \equiv
  {{\partial_r X} \over {\partial \Phi_c^A}}
   {{\partial_l Y} \over {\partial \Phi_{cA}^*}} -
   {{\partial_r X} \over {\partial \Phi_{cA}^*}}
   {{\partial_l Y} \over {\partial \Phi_c^A}}
\quad .
\label{def of classical antibracket}
\ee
Since the classical antibracket
is defined in the same manner as the antibracket,
it satisfies the same identities
\bref{antibracket properties} -- \bref{bracket derivation}.
Using Eq.\bref{recovery of Jc},
one obtains
\be
  {1 \over 2}\left( {\Gamma ,\Gamma } \right)_c =
   -J_{cA}{{\partial_l \Gamma } \over {\partial \Phi_{cA}^*}} =
   \left\langle {-i\hbar\Delta W +
   {1 \over 2}\left( {W,W} \right)} \right\rangle_c
\quad ,
\label{anomalous ZJ equation}
\ee
where the final equality is obtained
after some algebra, which makes use of
integration by parts.
In Eq.\bref{anomalous ZJ equation},
$ \left\langle \ \right\rangle_c $
denotes the expectation value in the presence of $J$
but expressed in terms of $\Phi_c$ and $\Phi_c^*$.
More precisely,
if $X$ is a functional of $\Phi$ and $\Phi^*$
then
\be
  \left\langle X \right\rangle_c \equiv
  \left. {\left\langle X \right\rangle^J} \right|_{J=J_c}
\quad ,
\label{def of <>_c}
\ee
where $\left\langle \ \right\rangle^J$ is defined
in Eq.\bref{def <>^J}.
Because of the quantum master equation
\bref{qme2},
Eq.\bref{anomalous ZJ equation} becomes
\be
  \left( {\Gamma ,\Gamma } \right)_c = 0
\quad ,
\label{ZJ equation}
\ee
a result known as the Zinn-Justin equation
\ct{zinnjustin75a}.

Equation \bref{def of <>_c}
allows one to pass from a functional $X$
of the original fields $\Phi$ and $\Phi^*$
to a classical functional
$\left\langle X \right\rangle_c$
of classical fields
$\Phi_c$ and $\Phi_c^*$
by taking the ``classical expectation''
of $X$:
$
  X\left[ {\Phi ,\Phi^*} \right] \to
  \left\langle X \right\rangle_c
$.
We refer to $\left\langle X \right\rangle_c $
as the {\it classical version}
of the quantum functional $X$.
The definition is consistent with
the notation for $\Phi_c^A$ since
\be
    \left\langle \Phi^A \right\rangle_c
  = \Phi_c^A
\quad .
\label{consistent notation for Phi_c}
\ee
The process $X \to \left\langle X \right\rangle_c$
conforms to the idea
that a classical variable
is the expectation value
of the corresponding quantum functional.

Since $\Gamma$ satisfies the Zinn-Justin equation
and plays the role of $S$ in the classical antibracket formalism,
one can construct by analogy
a ``classical-quantum'' BRST transformation $\delta_{B_{cq}}$
having the same properties as $\delta_B$.
Define
\be
  \delta_{B_{cq}} X \equiv
   \left( { X , \Gamma } \right)_c
\quad ,
\label{cq BRST}
\ee
where $X$ is any functional of $\Phi_c$ and $\Phi_c^*$.
The nilpotency $\delta_{B_{cq}}^2 = 0 $ of $\delta_{B_{cq}}$
follows from the Zinn-Justin equation.
Define $X$ to be cq-BRST invariant if $\delta_{B_{cq}} X = 0$.
According to Eq.\bref{ZJ equation},
the effective action $\Gamma$ is cq-BRST invariant since
\be
 \delta_{B_{cq}} \Gamma =
 \left( \Gamma , \Gamma \right)_c = 0
\quad .
\label{ZJ equation2}
\ee
Several of the above-mentioned classical functionals
are also cq-BRST invariant.
A analysis of $ \delta_{B_{cq}}J_{cA} $ reveals that
\be
   \delta_{B_{cq}} J_{cA} \equiv
  \left( {J_{cA}, \Gamma } \right)_c =
  {1 \over 2}\left( {-1} \right)^{\eps_A+1}
  {{\partial_r \left( {\Gamma ,\Gamma } \right)_c}
          \over {\partial \Phi_c^A}}
   = 0
\quad .
\label{cq BRST invariance of Jc}
\ee
Consider
$$
  -i\hbar\delta_{B_{cq}} \ln \left( {Z_c} \right) =
   \left( {\Gamma +J_{cA}\Phi_c^A,\Gamma } \right)_c =
    J_{cA}\left( {\Phi_c^A,\Gamma } \right)_c =
    J_{cA}{{\partial_l \Gamma } \over {\partial \Phi_{cA}^*}} = 0
\quad ,
$$
where the first equality holds
because of
Eqs.\bref{ZJ equation2}
and \bref{cq BRST invariance of Jc},
and
the last equality follows
from Eqs.\bref{qme2} and \bref{anomalous ZJ equation}.
Hence,
\be
  \delta_{B_{cq}}Z_c =
  \left( {Z_c,\Gamma } \right)_c = 0
\quad .
\label{cq BRST invariance of Zc}
\ee

The cq-BRST operator $\delta_{B_{cq}}$
is the effective classical version
of the quantum-BRST operator $\delta_{\hat B} X$.
To understand this statement,
consider
$$
  \delta_{B_{cq}}
  \left( {Z_c\left\langle X \right\rangle_c} \right) =
  \left( {Z_c\left\langle X \right\rangle_c,\Gamma } \right)_c =
  {{\partial_r \left( {Z_c\left\langle X \right\rangle_c} \right)}
     \over {\partial \Phi_c^A}}
   {{\partial_l \Gamma } \over {\partial \Phi_{cA}^*}} -
   {{\partial_r \left( {Z_c\left\langle X \right\rangle_c} \right)}
      \over {\partial \Phi_{cA}^*}}
    {{\partial_l \Gamma } \over {\partial \Phi_c^A}}
$$
$$
  = {{\partial_r \left( {Z \left\langle X \right\rangle^J} \right)}
     \over {\partial J_{cB}}}\left( {J_{cB},\Gamma } \right)_c +
    \left( {-1} \right)^{\eps_A} Z_c
     \left\langle {\left( {{{\partial_r X}
         \over {\partial \Phi_A^*}} +
     {i \over {\hbar}}X{{\partial_r W}
        \over {\partial \Phi_A^*}}} \right)J_A} \right\rangle_c
\quad .
$$
The last step follows because
$  \left\langle { \ } \right\rangle_c $
has dependence on
$\Phi_{c}$ and $\Phi_{c}^*$ through $J_{c}$.
After some algebra which makes use of integration by parts,
one finds that
\be
 \delta_{B_{cq}}
  \left( {Z_c \left\langle X \right\rangle_c} \right)
  = Z_c \left\langle {\delta_{\hat B}X +
    X \left( {\Delta W +
   \frac{i}{2 \hbar}
  \left( {W,W} \right)} \right)} \right\rangle_c +
  (-1)^{\eps_B} Z_c \langle X \Phi_c^B
   \rangle \left( { J_{cB} , \Gamma } \right)_c
\quad .
\label{anomalous cq BRST qBRST identity}
\ee
If
$
  \delta_{B_{cq}}
  \left( { \left\langle X \right\rangle_c } \right)
$
is computed instead,
one obtains the connected part
of Eq.\bref{anomalous cq BRST qBRST identity}
without a $Z_c$ factor.
Since Eq.\bref{cq BRST invariance of Jc}
and the quantum master equation
\bref{qme2} are satisfied,
\be
  \delta_{B_{cq}}
   \left( { \left\langle X \right\rangle_c  } \right) =
   \left\langle { \delta_{\hat B} X} \right\rangle_c
\quad .
\label{cq BRST qBRST identity}
\ee
In other words,
the cq-BRST variation of the classical version of $X$
is the classical version of the quantum-BRST variation of $X$
\ct{anselmi93a}.

The effective action $\Gamma$
in the classical antibracket formalism
plays a role analogous to the proper solution $S$
in the ordinary antibracket formalism.
Properties obeyed by $\Gamma$
are the same as those obeyed by $S$.
Therefore,
one can define a BRST structure
associated with $\Gamma$
\ct{gp93c}.
The BRST structure tensors are encoded in $\Gamma$
and the relations among them
are given by $( \Gamma , \Gamma ) = 0$.
Expanding in a Taylor series in $\Phi_c^*$,
one has
\be
  \Gamma \left[ { \Phi_c ,\Phi_c^* } \right] =
  \Gamma_0 (\Phi_c ) + \Phi_{cA}^* \Gamma^A (\Phi_c )
  + \frac12 \Phi_{cA}^* \Phi_{cB}^* \Gamma^{BA} (\Phi_c )
  + \ldots
\quad .
\label{Gamma antifield expansion}
\ee
Recalling that there
is an undisplayed dependence
on the gauge-fixing fermion $\Psi$,
the above terms have the following interpretation:
$\Gamma_0 (\Phi_c )$
is the one-particle-irreducible
generating functional
for the basic fields including all loop corrections
for the action gauge-fixed using $\Psi$, i.e.,
$S_\Psi$ in Eq. \bref{S sub Psi},
and
$\Gamma^A (\Phi_c )$
is the generator of gauge-fixed cq-BRST transformations.
The gauge-fixed cq-BRST operator $\delta_{B_{cq \Psi} }$,
the analogy of $\delta_{B_{\Psi} }$,
is defined by
\be
   \delta_{B_{cq \Psi}} X \equiv
   \left. {\left( {X, \Gamma } \right)} \right|_{\Sigma_\Psi }
\quad ,
\label{gauge-fixed cq-BRST}
\ee
where $X$ is a functional of the $\Phi_c^A$ only.
In the gauge-fixed basis,
$\Sigma_\Psi$
in Eq.\bref{gauge-fixed cq-BRST}
means that classical antifields
are set to zero.
Thus, one has $\delta_{B_{cq \Psi}} \Phi_c^A = \Gamma^A (\Phi_c )$.
The tensor $\Gamma^{BA} (\Phi_c )$
is related to the on-shell nilpotency
of $\delta_{B_{cq \Psi}}$.
In summary,
the quantum aspects of the classical theory described by $S$
are reproduced by an effective classical theory
governed by $\Gamma$.

\subsection{Quantum Master Equation Violations: Generalities}
\label{ss:qmevg}

\hspace{\parindent}
Suppose Eq.\bref{qme2} is not zero.
Let
\be
  {\cal A} \equiv
  \Delta W + { i \over {2\hbar }} \left( {W,W} \right)
\label{qme anomaly}
\ee
be the violation of the quantum master equation.
A straightforward computation
using
Eqs.\bref{antibracket properties},
\bref{Delta properties} and \bref{qBRST},
but not assuming the validity of the quantum master equation,
reveals that
\be
  \delta_{\hat B}{\cal A} = 0
\quad .
\label{full consistency conditions}
\ee
This equation is consistent with
Eqs.\bref{anomalous ZJ equation},
\bref{def of <>_c} and \bref{cq BRST qBRST identity}
and the Jacobi identity for the antibracket:
One has
$
  \frac12 ( \Gamma,\Gamma )_c =
$
$
  -i \hbar \langle {\cal A} \rangle_c
$
so that
$
0 = \frac{i}{2 \hbar} ( (\Gamma,\Gamma)_c , \Gamma )_c
$
$
 =  (  \langle {\cal A} \rangle_c , \Gamma )_c
$
$
 = \delta_{B_{cq}} \langle {\cal A} \rangle_c
$
$
= \langle \delta_{\hat B} {\cal A} \rangle_c
$.

Suppose that
\be
  i {\cal A} = \delta_{\hat B}\Omega +
  {{\hbar } \over 2}\left( {\Omega ,\Omega } \right) =
   \left( {\Omega ,W} \right) - i\hbar \Delta \Omega +
  {{\hbar } \over 2} \left( {\Omega ,\Omega } \right)
\quad ,
\label{fake anomaly}
\ee
where $\Omega$
is a local functional of the fields and antifields.
The last term may seem surprising
but it is necessary if ${\cal A}$ is to satisfy
Eq.\bref{full consistency conditions}:
When the quantum master equation is violated
as in Eq.\bref{qme anomaly},
the nilpotency of the quantum BRST operator
no longer holds,
as can be seen
from Eq.\bref{nil of qBRST}.
The last term in Eq.\bref{fake anomaly}
is required to compensate for this effect
and ensures
Eq.\bref{full consistency conditions}.
Let
\be
  W^{'} = W + \hbar \Omega
\quad .
\label{remove anomaly}
\ee
Then,
using Eqs.\bref{antibracket properties} and
\bref{Delta properties},
one finds that $W^{'}$ satisfies
the quantum master equation:
$$
  {1 \over 2}\left( {W^{'} , W^{'}} \right) -
    i\hbar \Delta W^{'} = 0
\quad .
$$
Since $W^{'}$ has the same classical limit as $W$, namely,
its order
$  \hbar^0  $ term is $S$,
one can use $W^{'}$ in lieu of $W$ for the quantum action.
Then, since the quantum master equation is satisfied,
a quantum gauge theory can be defined.

When ${\cal A}$ cannot be expressed as
in Eq.\bref{fake anomaly}
for a local functional $\Omega$,
there is an anomaly in the quantum master equation
and an obstruction to maintaining
gauge symmetries at the quantum level.
Since anomalies involve subtleties
and singular expressions,
they are usually not too easy to compute.
The usual approach to this subject uses
a loop expansion:
\be
  {\cal A} =
    \sum\limits_{l=1}^\infty  {}{\cal A}_l\hbar^{l-1} =
   {\cal A}_1 + \hbar {\cal A}_2 + \ldots
\quad .
\label{hbar expansion of anomaly}
\ee
Using Eqs.\bref{hbar expansion of W} and
\bref{hbar expansion of anomaly},
one finds the following expression
for the anomaly at the one-loop level
\be
  {\cal A}_1 \equiv \Delta S + i \left( {M_1,S} \right)
\quad .
\label{one-loop anomaly}
\ee
To this order,
the condition for the absence of an anomaly
in Eq.\bref{fake anomaly}
is that ${\cal A}_1$
is a classical BRST variation, i.\ e.,
\be
  {\cal A}_1 = -i \left( {\Omega_1,S} \right) =
    -i \delta_B \Omega_1
\quad .
\label{one-loop fake anomaly}
\ee
If ${\cal A}_1$ can be expressed as
in Eq.\bref{one-loop fake anomaly}
for some local functional $\Omega_1$,
then,  by setting
\be
  W^{'} = S + \hbar  \Omega_1
\quad ,
\label{one-loop remove anomaly}
\ee
the quantum master equation
in Eq.\bref{qme2}
is satisfied to order $\hbar$.
In words,
Eq.\bref{one-loop fake anomaly}
says that if ${\cal A}_1$ is expressible
as a local BRST variation
then effectively there is no one-loop anomaly.
Since the second term
in Eq.\bref{one-loop anomaly}
is already of this form,
the requirement becomes that $\Delta S$
should be a classical-BRST variation
of a local functional.

To order $\hbar$,
the equation $\delta_{\hat B}{\cal A} = 0$ in
Eq.\bref{full consistency conditions}
is
\be
  \delta_B {\cal A}_1 =
  \left( {{\cal A}_1,S} \right) = 0
\quad .
\label{one-loop consistency conditions}
\ee
In view
of Eqs.\bref{one-loop fake anomaly}
and \bref{one-loop consistency conditions},
the investigation of anomalies
is related to the local BRST cohomology
at ghost number one
\ct{brs74a,baulieu85a,hlw90a,tnp90a,tonin91a}.
Although not obvious,
it turns out that
Eq.\bref{one-loop consistency conditions}
embodies the Wess-Zumino
anomaly consistency equations
\ct{wz71a}.
In field-antifield formalism,
the one-loop master equation anomaly
must be classically BRST invariant.
If anomalies arise beyond the one-loop level,
Eq.\bref{full consistency conditions}
provides the full quantum consistency conditions:
The master equation anomaly must
be quantum-BRST invariant.

Note that $\rm{gh} \left[ {\cal A} \right] = 1$.
Of the fields
in Eq.\bref{field set},
only
${\cal C}^{\alpha_0}_0 \equiv {\cal C}^{\alpha}$
has ghost number one.
Hence, one may write
\be
  i {\cal A} =
    a_\alpha \left( \phi  \right){\cal C}^\alpha + \dots
\quad ,
\label{gh structure of anomaly}
\ee
where the omitted terms involve antifields.
Sometimes these terms are absent
so that
Eq.\bref{gh structure of anomaly}
gives the structure
of the quantum-master-equation anomaly.
Although not obvious,
the coefficients
$a_\alpha \left(  \phi \right)$
are the usual gauge anomalies
\ct{tnp90a}.
In other words,
a quantum-master-equation anomaly and a gauge anomaly
are equivalent.

\subsection{{Canonical Transformations and the Quantum \hfil}
\break
Master Equation}
\label{ss:ctqme}

\hspace{\parindent}
Canonical transformations
preserve the quantum master equation
as long as $W$ is appropriately transformed
\ct{ht92a}.
Consider an infinitesimal canonical transformation
as in Eq.\bref{canonical transformation}
governed by $F$.
Normally a functional $G$ transforms as
$G \to G + \varepsilon (G,F)$.
For $W$, however, one must add an extra term
$i \hbar \varepsilon \Delta F$ to
compensate for ``measure effects''.
Hence, the transformation rule for $W$ is taken to be
\be
  W \to W + \varepsilon ( W, F) +
   i \hbar \varepsilon \Delta F +
   O \left( \varepsilon^2 \right)
  = W - \varepsilon \delta_{\hat B} F +
   O \left( \varepsilon^2 \right)
\quad .
\label{w ct law}
\ee
According
to Eq.\bref{qme symmetry},
Eq.\bref{w ct law}
is a symmetry
of the quantum master equation
\ct{bv84a,henneaux90a,fisch90a,dejonghe93a,tpbook}.
If $W$ is a solution
to the quantum master equation,
changing $W$ as
in Eq.\bref{w ct law}
will not upset the solution.

The same conclusion holds for finite transformations
governed by $F_2$
in Eq.\bref{f2 can trans}.
In this case,
one transforms
from $\{ \Phi , \Phi^* \}$ variables
to $\{ \tilde \Phi , \tilde  \Phi^* \}$ variables
via Eq.\bref{f2 can trans}.
The transformation rule for $W$ is
\ct{bv84a}
\be
\widetilde W [ \tilde \Phi , \tilde  \Phi^*] =
  W [ \Phi , \Phi^* ] - \frac{i\hbar}{2} \ln J
\quad ,
\label{w fct law}
\ee
where the jacobian factor $J$
is the berezinian governing
the change from $\{ \Phi , \Phi^* \}$ variables
to tilde variables.
Then,
$\widetilde W$ satisfies
the quantum master equation exactly in the tilde variables
if $W$ satisfies it in the $\{ \Phi , \Phi^* \}$ variables.
A detailed proof of this result
and a formula for $J$
can be found
in ref.\ct{tnp90a}.
Here, we provide a few key steps.
Define $\widetilde \Delta$
as the analog of $\Delta$
in the transformed variables, i.e.,
$$
  \widetilde \Delta \equiv
   \left( {-1} \right)^{\epsilon_A+1}
   {{\partial_r} \over {\partial \tilde \Phi^A}}
   {{\partial_r} \over {\partial \tilde \Phi_A^*}}
\quad .
$$
If $G$ is an arbitrary functions of
$\tilde \Phi$ and $\tilde  \Phi^*$,
then
\be
  \widetilde \Delta  G \left[ \tilde \Phi , \tilde  \Phi^*  \right] =
   \Delta  G
  \left[ \tilde \Phi \left[  \Phi ,  \Phi^*  \right] ,
   \tilde  \Phi^* \left[  \Phi ,  \Phi^*  \right]  \right]
      - \frac12 \left( G, \ln J \right)
\quad .
\label{fct of Delta}
\ee
When $\Delta$ acts on $G$ on the right-hand side
of Eq.\bref{fct of Delta},
the tilde fields should be regarded as functionals
of $\Phi$ and $\Phi^*$, as indicated.
The chain rule for derivatives is then used.
This produces $\widetilde \Delta  G$ plus an extra term,
which is equal to $\frac12 ( G, \ln J )$ and
needs to be subtracted to obtain the identity
in Eq.\bref{fct of Delta}.
Using Eqs.\bref{w fct law} and \bref{fct of Delta},
one finds,
$$
 i \hbar \widetilde \Delta \widetilde W  -
 \frac12 \left( \widetilde W , \widetilde W \right) =
  i \hbar \left\{ \Delta \widetilde W  -
      \frac12 \left( \widetilde W, \ln J \right) \right\} -
 \frac12 \left( \widetilde W , \widetilde W \right)
$$
$$
  = i \hbar \left\{ {
    \Delta W - \frac{i\hbar}{2} \Delta \ln J -
   \frac12 \left( W - \frac{i\hbar}{2} \ln J , \ln J \right)
                    } \right\}
  - \frac12 \left( { W - \frac{i\hbar}{2} \ln J ,
                     W - \frac{i\hbar}{2} \ln J
                    } \right)
$$
$$
  = i \hbar \Delta W - \frac12 \left( W , W \right) +
  \frac{\hbar^2}{2}
     \left\{ { \Delta \ln J -
           \frac14 \left( \ln J , \ln J \right)
              } \right\}
\quad .
$$
It can be shown
\ct{bv84a,tnp90a}
that
$
  \Delta \ln J = \frac14 \left( \ln J , \ln J \right)
$,
so that the last term is zero.
Hence, if
$W$ satisfies the quantum master equation,
then $\widetilde W$ satisfies the tilde version
of the quantum master equation.
For infinitesimal transformations,
$$
  \tilde \Phi^A = \Phi^A -
    \varepsilon \left( \Phi^A , F \right) +
    O \left( \varepsilon^2 \right)
\quad ,
$$
$$
  \tilde \Phi_A^* = \Phi_A^* -
    \varepsilon \left( \Phi_A^* , F \right) +
    O \left( \varepsilon^2 \right)
\quad ,
$$
\be
  J = 1 - 2 \varepsilon \Delta F +
  O \left( \varepsilon^2 \right)
\quad .
\label{going to infinitesimal}
\ee
Then,
$$
  \widetilde W \left[  \Phi ,  \Phi^*  \right] =
  W \left[  \Phi ,  \Phi^*  \right] +
  \varepsilon \left( W , F \right) +
 i \hbar \Delta F + O \left( \varepsilon^2 \right)
\quad ,
$$
so that
$$
  W \left[  \Phi ,  \Phi^*  \right] \to
  \widetilde W \left[  \Phi ,  \Phi^*  \right] =
  W - \varepsilon \delta_{\hat B} F  +
  O \left( \varepsilon^2 \right)
\quad ,
$$
and one recovers
Eq.\bref{w ct law}.
The identity
$
  \Delta \ln J - \frac14 \left( \ln J , \ln J \right) \sim
 O \left( \varepsilon^2 \right)
$
follows
from Eq.\bref{going to infinitesimal} and
$\Delta^2 = 0$.

One can take advantage of canonical transformations
in analyzing potential anomalies
by going to a basis
for which the computation is simpler.

\subsection{The Anomaly at the One-Loop Level}
\label{ss:aoll}

\hspace{\parindent}
The quantities that appear in the violation
of the quantum master equation in
Sect.\ \ref{ss:sqme}
involve both fields and antifields.
As a consequence,
one must use the action
before any elimination of antifields.
On the other hand,
since propagators are needed
to perform perturbative computations,
a gauge-fixing procedure is required.
Both these requirements can be satisfied
by working
in the gauge-fixed basis described
in Sect.\ \ref{ss:gfb}.
It is achieved by performing
a canonical transformation with
the gauge-fixing fermion $\Psi$
so that
$
  \Phi_A^* \to \Phi_A^* +
  {{\partial \Psi } \over {\partial \Phi^A}}
$.
Throughout the rest of this section,
we assume that an admissible $\Psi$
has been selected and
that the shift to the gauge-fixed basis
has been performed.
According to the result
in Sect.\ \ref{ss:ctqme},
if the quantum master equation
is satisfied and a canonical transformation is performed
to a new basis
then, by appropriately adjusting the action,
the quantum master equation is satisfied
in the new basis.
Hence, the existence or non-existence
of an anomaly is independent
of the choice of basis,
although the form of the anomaly may depend
on this choice.

There are different ways
of obtaining the anomaly.
We mostly follow the approach
of ref.\ct{tnp90a}
and briefly mention other methods
at the end of this subsection.
Reference \ct{tnp90a} obtained general formulas
for the antifield-independent part
of the one-loop anomaly using a Pauli-Villars
regularization scheme.
Since the derivation is somewhat technical,
we present only the final results.
For more details,
see refs.\ct{dtnp89a,tnp90a,vanproeyen91a,%
dstvnp92a,dejonghe93a,tp93a,vp94a,tpbook}.
In particular,
refs.\ct{dejonghe93a,tpbook}
have an extensive discussion
of Pauli-Villars regularization
in the antibracket formalism
to which we refer the reader.

The goal of the next few paragraphs
is to obtain a regularized expression
for $\Delta S$, denoted by
$ \left( { \Delta } S \right) _{\rm reg} $.
The anomaly ${\cal A}_1 $ is essentially
$ \left( { \Delta } S \right) _{\rm reg}$
since the $M_1$ term
in Eq.\bref{one-loop anomaly}
is eliminated as a possible violation
of the quantum master equation
via Eq.\bref{one-loop remove anomaly}
with $\Omega_1 = - M_1$,
i.e., the counterterm
$- \hbar M_1$ is added to the action.
Define
\be
  {K^A}_{B} \equiv
  \lder{}{\Phi^*_A}\rder{}{\Phi^B}
   S \left[ \Phi, \Phi^* \right]
\quad ,
\label{def of K}
\ee
\be
   Q_{AB} \equiv \frac{\partial_l} {\partial\Phi^A}
   \frac{\partial_r}{\partial\Phi^B}
   S \left[ \Phi, \Phi^* \right]
\quad .
\label{def of Q}
\ee
Note that $Q$ involves derivatives with respect to fields
and not antifields.
If expanded about a stationary point,
$Q$ becomes the quadratic form for the fields.
In such an expansion,
the inverse of $Q$ is the propagator.
The properness condition in
the gauge-fixed basis guarantees
that propagators exist.
An operator ${\cal O}$
used to regulate $\Delta S$
is related to $Q$ by
\be
 {{\cal O}^A}_B \equiv
  \left( { T^{-1} } \right)^{AC} Q_{CB}
\quad ,
\label{def of O}
\ee
where $T_{AC}$ is an arbitrary invertible matrix
satisfying
$T_{BA}=(-1)^{\eps_A + \eps_B + \eps_A \eps_B} T_{AB}$.
The inverse of $T$ obeys
$(T^{-1})^{BA}=(-1)^{\eps_A \eps_B} (T^{-1})^{AB}$.
The Grassmann statistics of
$T_{AB}$ and $(T^{-1})^{AB}$
are $\eps_A + \eps_B$ (mod 2).
Eq.\bref{def of O} implies
$Q_{AB} = T_{AC} {{\cal O}^C}_B$.
In the regularization scheme
of ref.\ct{tnp90a},
the matrix $T$ appears in the mass term
for the regulating Pauli-Villars fields.
In that approach,
the violation of the quantum master equation
is shifted from the $\Delta$ term
to the $( S, S )$ term.

The regulated expression for
$\left( \Delta S \right) _{\rm reg}$
is \ct{tnp90a}
\be
   \left( \Delta S \right) _{\rm reg} =
   {\left[ { {F^A}_B
     { \left( { \frac{1}{1 - { {\cal O} \over M } } }
       \right)^B }_{A}
           } \right]}_0
\quad ,
\label{Delta2 S reg}
\ee
where
\be
 {F^A}_C \equiv
  {K^A}_{C} +
  \frac12 (T^{-1})^{AD}
    \left( \delta_B T \right)_{DC}(-1)^{\epsilon_C}
\quad ,
\label{def of F}
\ee
and where $\delta_B T$ denotes
the classical BRST transform of $T$:
$\delta_B T = \left( { T , S } \right)$.
In Eq.\bref{Delta2 S reg},
the sum over $A$ and $B$, leading to the trace,
involves the quadratic form of fields only
and not that of antifields.
The subscript $0$
on the square brackets
in Eq.\bref{Delta2 S reg}
indicates that the term
independent of $M$ is to be extracted.
Here, $M$ denotes a regulator mass.
The one-loop nature
of Eq.\bref{Delta2 S reg}
is evident by the presence
of the propagator factor
$
  - i / \left( { M - {\cal O}} \right)
  \to
 { \left( i M  \right) }^{-1}
     /  \left( { 1 - {\cal O} / M } \right)
$
and the sum over the index $A$
indicating a trace.

When ${\cal O}$ is quadratic in space-time derivatives,
one lets
\be
 {\cal R} = {\cal O}
\quad , \quad \quad
 {\cal M}^2 = M
\quad ,
\label{def2 of R}
\ee
where ${\cal R}$ denotes
the quadratic regulator operator
and ${\cal M}$ denotes the regulating mass or cutoff.
When ${\cal O}$ is linear in space-time derivatives,
it is convenient to multiply
on the right in the trace
in Eq.\bref{Delta2 S reg}
by
$
   1 / \left( 1 + {\cal O} / M \right)
  \ \left( 1 + {\cal O} / M \right)
$
and carry out the multiplication of
$ 1 / \left( 1 - {\cal O} / M \right) $
with $ 1 / \left( 1 + {\cal O} / M \right) $.
Eq.\bref{Delta2 S reg}
can then be manipulated into the form
\ct{tnp90a}
\be
   \left( \Delta S \right)_{\rm reg} =
   {\left[ { {{F^\prime}^A}_B
     { \left( { \frac{1}{ 1 - { {\cal R} \over {\cal M}^2 }  }
              } \right)^B }_{A}
           } \right]}_0
\label{Delta S reg}
\quad ,
\ee
where
\be
 {\cal R} = {\cal O}^2
\quad , \quad \quad
 {\cal M} = M
\quad ,
\label{def1 of R}
\ee
and
\be
 { F^{\prime A} }_C \equiv
  {F^A}_C - \frac{1}{2 M}
       {\left( { \delta_B {\cal O}  } \right)^A}_C
   (-1)^{\epsilon_C}
\label{def of F prime}
\ee
with
$\delta_B {\cal O} = \left( {\cal O} , S \right)$.

Summarizing,
in the quadratic momenta case,
$\left( \Delta S \right) _{\rm reg}$
is given by
Eqs.\bref{Delta2 S reg} and \bref{def of F}.
This is the same as
using Eq.\bref{Delta S reg}
with   ${\cal R}$ and ${\cal M}$ given
in Eq.\bref{def2 of R}
and with $F$
in Eq.\bref{def of F}
replacing $F^\prime$.
In the linear momenta case,
Eqs.\bref{Delta S reg}--\bref{def of F prime} are used.

For the situation in which
${\cal O}$ is quadratic in momenta
or in the case where $\delta_B {\cal O}$
does contribute
in Eq.\bref{Delta S reg},
one can replace
$1 / \left( {  1 - {\cal R} / {\cal M}^2 } \right)$ by
${\rm exp} \left( {  {\cal R} / {\cal M}^2 } \right) $.
This follows by writing
$
 1 / \left( {  1 - {\cal R} / {\cal M}^2 } \right) =
   \int^\infty_0 \exp\left[ { - \lambda
   \left( 1 - {\cal R} / {\cal M}^2 \right) } \right]
   \dif \lambda
$,
and inserting
in Eq.\bref{Delta2 S reg}
or Eq.\bref{Delta S reg}:
$$
 \int_0^\infty  {\dif \lambda } \exp \left( {-\lambda } \right)
  \left[ { {F^A}_B
			{\left( {  \exp
        \left( { {\lambda {\cal R} } \over { {\cal M}^2} } \right)
           }  \right)^B}_A
   } \right]_0 =
$$
$$
  \int_0^\infty  { \dif \lambda }\exp \left( {-\lambda } \right)
   \sum\limits_{n=-p}^\infty  f_n
  \left[ {
   \left( {{ {\lambda } \over {{\cal M}^2}}}
  \right)^n} \right]_0 = f_0 =
$$
$$
  \sum\limits_{n=-p}^\infty
   f_n \left[ {
     \left( {{ {1} \over {{\cal M}^2}}} \right)^n}
    \right]_0 =
 \left[ { {F^A}_B
  { \left( { \exp
            \left( { {\cal R} \over {{\cal M}^2} } \right)
           }  \right)^B}_A
        }  \right]_0
\quad ,
$$
where a Laurent expansion
in ${1 \over {{\cal M}^2} }$
has been performed.
The resulting expression,
\be
   (\Delta S)_{\rm reg} =
   { \left[ {F^A}_{B}
			 { {\left( {
       \exp \left( { {\cal R} \over {{\cal M}^2} } \right)
               } \right)}^B }_{A} \right]
   }_0
\quad ,
\label{Delta S Fujikawa}
\ee
corresponds to the Fujikawa form
\ct{fujikawa80a}
of the regularization
\ct{tnp90a} .
In the limit
${\cal M} \to \infty$,
terms of order $1 / {\cal M}^n$ for $n > 0$ vanish,
whereas terms with $n < 0$ blow up.
The regularization scheme consists
of dropping the terms that blow up.
In the Pauli-Villars regularization,
this is achieved by adding fields
with appropriate statistics, couplings and masses
to cancel all $n < 0$ terms.
As a consequence,
only the $n=0$ term remains.

In the quadratic momentum case,
Eq.\bref{Delta2 S reg}
can be manipulated into the following supertrace form
\ct{gp93c}
\be
   (\Delta S)_{\rm reg}
  =  {\left[-\frac12 {({\cal R}^{-1}\delta {\cal R})^A}_B
   {\left( { \frac{1}{1 - { {\cal R} \over {{\cal M}^2} } }
           } \right)^B}_A
    (-1)^{\epsilon_A} \right]}_0
\quad ,
\label{Delta3 S reg}
\ee
which only involves
the quadratic regulator ${\cal R}$.
Eq.\bref{Delta3 S reg} shows that
if $\delta\gh R = [\gh R, G]$ for some $G$
then the anomaly vanishes,
as a consequence of the cyclicity of the trace
\ct{gp93c}.
Formally,
Eq.\bref{Delta3 S reg} is
\be
   (\Delta S)_{\rm reg} =
   \delta_B {\left[ {
   -\frac12 {    \left( {
    \ln \left( \frac{ -{\cal R} }{ {\cal M}^2 - {\cal R} }
        \right)
                 } \right)^A }_A (-1)^{\epsilon_A}
             } \right] }_0
\quad ,
\label{Delta4 S reg}
\ee
which is tantamount to demonstrating
that $(\Delta S)_{\rm reg}$ satisfies
the one-loop anomaly consistency condition
$\delta_B (\Delta S)_{\rm reg} = 0$
in Eq.\bref{one-loop consistency conditions}
since $(\Delta S)_{\rm reg}$ is a classical BRST variation
and $\delta_B$ is nilpotent.
If the non-BRST-invariant part of the quantity
corresponding to
${ \left[ { \dots } \right] }_0 $
in Eq.\bref{Delta4 S reg}
is local%
{\footnote{
The BRST-invariant part,
which may be non-local,
gives zero contribution to the anomaly
since $\delta_B$ is applied to it.}},
then there is no one-loop anomaly
according
to Eqs.\bref{one-loop fake anomaly}
and \bref{one-loop remove anomaly}.
The expression in square brackets
in Eq.\bref{Delta4 S reg}
turns out to be
the one-loop contribution to the effective action.
Eq.\bref{Delta4 S reg}
says that the one-loop anomaly
is the BRST variation
of this one-loop contribution.

In the approach of
ref.\ct{tnp90a},
Eq.\bref{Delta2 S reg}, \bref{Delta S reg},
or \bref{Delta S Fujikawa}
is evaluated
using standard perturbation theory
about a stationary point.
Antifields are finally set to zero
and ${\cal R}$
is evaluated at $\Sigma$, i.e, on-shell.

Another approach to anomalies,
which retains antifields, is developed
in refs.\ct{vanproeyen91a,gp93c,tp93a,vp94a}.
At one loop,
results agree with the above.
It has the advantage of making it easier
to compute antifield-dependent terms
in the anomaly, if present.
Such terms might arise
if there is an anomalous non-closure
of BRST transformations
or some other difficulty with a BRST-structure equation.
Antifields may be retained or eliminated
at any stage of a computation.
A third approach
is to use the effective action $\Gamma$
in Eq.\bref{anomalous ZJ equation}
\ct{baulieu85a,hlw90a,tonin91a}.
In this method,
antifields must be retained.
Information about anomalies
can be obtained using cohomolgical arguments
based on the Wess-Zumino consistency conditions
\ct{wz71a}.
One must compute the coefficients of candidate terms
using perturbative methods
\ct{dixon76a,bc83a,dtv85a,band86a}.

Eq.\bref{Delta3 S reg} demonstrates
that fields
for which $\delta {\cal R}$ is zero,
do not contribute to the anomaly.
Non-propagating degrees of freedom,
such as gauge-fixed fields and
delta-function-generating Lagrange multipliers $\pi$,
are not expected
to contribute
\ct{tpbook}
because anomalies arise
from loop effects.
In practice,
the evaluation of anomalies is performed perturbatively.
Consequently,
one expands around a stationary point.
In the gauge-fixed basis,
this involves expanding about $\Sigma$,
that is,
around $\Phi^*_A =0$,
if the method retains antifields,%
{\footnote{
In methods for which antifields are eliminated,
one does not expand about $\Phi^*_A =0$
but simply sets $\Phi^*_A =0$.}}
and about $\Phi^A = \Phi_0^A$,
where the $\Phi_0^A$ satisfy
the gauge-fixed equations of motion.
One must be careful to compute
${ F^{\prime A} }_C$
in Eq.\bref{def of F prime}
before expanding about the perturbative saddle point.
In a delta-function implementation
of gauge-fixing,
it is advantageous,
at the beginning of a computation,
to perform a canonical transformation
that shifts the Lagrange multipliers $\pi$
by solutions to equations of motion.
These equations
are generated by the fields
which are being gauge-fixed.
Such a canonical transformation ensures
that gauge-fixed fields and the $\pi$
do not mix on-shell
with the other fields of the system.
This is illustrated
in the first and third sample computations
of Sect.\ \ref{s:sac}.

The choice of $T$
in Eq.\bref{def of O},
which determines the regulator ${\cal O}$,
is at one's disposal.
The requirements on $T$
are that it be invertible
and that it lead to a quadratic regulator
${\cal R}$ that is negative definite
after a Wick rotation to Euclidean space.
Modifying $T$
changes the form of the anomaly.
In particular,
when more than one gauge symmetry is present,
varying $T$
changes the coefficients $a_\alpha$
in Eq.\bref{gh structure of anomaly}.
If some non-zero $a_\alpha$ are made zero
and vice-versa,
the anomaly is shifted
from being associated with one type of gauge symmetry
to another.
This is analogous
to the well known situation
for anomalous chiral gauge theories
in four dimensions:
The anomaly can be moved from the axial vector sector
to the vector sector, if so desired.
See, for example, Sect.\ 4.1 of ref.\ct{tnp90a}.
Although we do not present any examples
of this effect,
it is well illustrated in ref.\ct{tnp90a}.
In performing anomaly calculations,
it is useful to choose $T$ to render
a computation as simple as possible.
For a similar reason,
it is also useful to perform
certain canonical transformations
before commencing a calculation.

\vfill\eject


\section{Sample Anomaly Calculations}
\label{s:sac}

\hspace{\parindent}
In this section,
we present computations of
$\left( {\Delta S} \right)_{\rm reg}$
to see whether the quantum master equation is violated
at the one-loop level.
In general,
the analysis is complicated and lengthy.
For this reason,
we treat only three cases:
the spinless relativistic particle,
the chiral Schwinger model,
and the first-quantized bosonic string.
We use the method of
ref.\ct{tnp90a},
which we have outlined
in Sect.\ \ref{ss:aoll}.
The first step in the procedure is
to transform to the gauge-fixed basis.
One then has the option
of performing additional canonical transformations.
They can be used to partially diagonalize the system,
so that potential contributions to the anomaly
can be calculated separately from various sectors.
The second step
is to compute the matrices
${K^A}_{B}$ and $Q_{AB}$
in Eqs.\bref{def of K} and \bref{def of Q}.
The third step is to select a $T_{AB}$ matrix
so that the operator ${{\cal O}^A}_B$
in Eq.\bref{def of O}
can be obtained.
A judicious choice of $T_{AB}$ can simplify
a computation.
One then obtains ${F^A}_{B}$
from Eq.\bref{def of F}
and ${\cal R}$
from Eq.\bref{def2 of R} or Eq.\bref{def1 of R}.
The final step is to use the anomaly formula
in Eq.\bref{Delta S Fujikawa}.
Standard perturbation theory is performed,
in which one expands about a stationary point
and sets antifields to zero.
For sample computations
using a method that retain antifields
throughout the computation,
see ref.\ct{gp93c,vp94a}.
Other useful results for anomaly calculations
can be found in \ct{tnp90a,dstvnp92a,%
dejonghe93a,gp93c,hull93a,ms93a,%
bh94a,brandt94a,mpsx94a,vp94a}
and references therein.

\subsection{Computation for the Spinless Relativistic Particle}
\label{ss:csrp}

\hspace{\parindent}
In this subsection,
we show that the spinless relativistic particle
of Sect.\ \ref{ss:srp2}
possesses no anomaly.
This example is useful
for illustrating the formalism
of Sect.\ \ref{ss:aoll}
because the computation is relatively simple.

We use
\be
  \Psi =\int {\dif \tau }
    \bar {\cal C} \left( {e - \rho} \right)
\label{rsp psi rho}
\ee
for the gauge-fixing fermion,
where $\rho$ is an arbitrary function of $\tau$.
This allows us to judge
potential dependence on the gauge-fixing procedure
by varying $\rho$.
Next,
a canonical transformation is performed
to the gauge-fixed basis
of Sect.\ \ref{ss:gfb}
using $\Psi$
in Eq.\bref{rsp psi rho}.
One obtains
\be
  S \to \int {\dif \tau }
 \left\{ {
   {{1 \over 2}
   \left( {{{\dot x^2} \over e} - m^2 e } \right) +
   x_\mu^*\dot x^\mu {{ {\cal C} } \over e} +
   \left( {e^* +  \bar {\cal  C} } \right)
    \dot {\cal C}  +
    \bar \pi \left( {
    \bar {\cal C}^* + e - \rho } \right)}
 } \right\}
\quad .
\label{rsp gfb S}
\ee

It is advantageous to perform a canonical transformation
that shifts $\bar \pi$
by the solution
of the equation of motion generated by $e$.
According to the result
in Sect.\ \ref{ss:ctqme},
canonical transformations
do not affect the existence or non-existence
of violations of the quantum master equation.
The variation of $S_\Psi$
with respect to $e$
yields
\be
  \bar \pi - {1 \over 2} {{\dot x^2} \over {e^2}}
     - \frac12 m^2 = 0
\quad .
\label{eos for e}
\ee
The relevant canonical transformation is
$$
  \bar \pi \to \bar \pi +
  {1 \over 2}{{\dot x^2} \over {e^2}} + \frac12 m^2
\quad ,
$$
\be
  e^*\to e^* +
   \bar \pi^*{{\dot x^2} \over {e^3}} \ ,
\quad \quad x_\mu^*\to x_\mu^* +
   {d \over {d\tau }}
  \left( {{{\bar \pi^*\dot x_\mu } \over {e^2}}} \right)
\quad ,
\label{rsp pi ct}
\ee
with the other fields and antifields left unchanged.
The action becomes
$$
  S \to \int { \dif\tau }
   \left\{
  {\left( {{1 \over e} -
  {\rho  \over {2e^2}}} \right)\dot x^2 -
  {1 \over 2}\rho m^2 + \bar{\cal C} \dot{\cal C} +
   \bar \pi \left( {e - \rho } \right) +
         } \right.
$$
\be
  \left. {
    x_\mu^* \dot x^\mu {{{\cal C}} \over e} +
  e^* \dot{\cal C} +
  \bar{\cal C}^*{1 \over 2}
   \left( {{{\dot x^2} \over {e^2}} + m^2} \right) -
   {{ \bar \pi^* {\cal C} \dot x_\mu } \over {e^2}}
   {d \over {d\tau }}
    \left( {{{\dot x^\mu } \over e}} \right)
          } \right\}
\quad .
\label{rsp gfb shifted S}
\ee

Let us first determine the overall structure
of the computation.
{}From Eq.\bref{rsp gfb shifted S},
one finds that the non-zero entries
of the matrix ${K^A}_{B}$ are
\be
  K = \bordermatrix{
                & x^\mu & e & \bar \pi & \bar {\cal C} & {\cal C} \cr
        x_\mu^* & * & * & 0 & 0 & * \cr
            e^* & 0 & 0 & 0 & 0 & * \cr
     \bar \pi^* & * & * & 0 & 0 & * \cr
\bar {\cal C}^* & * & * & 0 & 0 & 0 \cr
     {\cal C}^* & 0 & 0 & 0 & 0 & 0 \cr
   }
\quad ,
\label{rsp non zero K entries}
\ee
where the columns and rows
are labelled by the corresponding
fields and antifields.
We select $T_{AB}$ to be proportional to the identity matrix,
except in the ghost sector for which
\be
  T_{\bar{\cal C}{\cal C}} =
  \bordermatrix{
            & \bar{\cal C} & {\cal C} \cr
              \bar{\cal C} & 0 & -1   \cr
                  {\cal C} & 1 & 0    \cr}
\quad ,
\label{rsp T_ghost}
\ee
and in the $x$ sector for which
\be
  \left( {T_x} \right)_{\mu \nu } = \eta_{\mu \nu }
\quad .
\label{rsp T_x}
\ee
In perturbation theory,
the regulator ${\cal R}$
is evaluated at the stationary point
of the gauge-fixed action.
For $e$,
this corresponds to
\be
  \left. e \right|_\Sigma = \rho
\quad ,
\label{e stationary point}
\ee
where $\Sigma$ indicates
the stationary-point surface in field space.
Using Eq.\bref{rsp gfb shifted S},
a straightforward calculation reveals
that the non-zero entries of
$\left. { {\cal R} } \right|_\Sigma$ are
\be
  \left. { {\cal R} } \right|_\Sigma =
 \bordermatrix{
                & x^\mu & e & \bar \pi & \bar {\cal C} & {\cal C} \cr
        x^\mu & * & 0 & 0 & 0 & 0 \cr
            e & 0 & * & * & 0 & 0 \cr
     \bar \pi & 0 & * & 0 & 0 & 0 \cr
\bar {\cal C} & 0 & 0 & 0 & * & 0 \cr
     {\cal C} & 0 & 0 & 0 & 0 & * \cr
   }
\quad .
\label{rsp non zero regulator entries}
\ee
Eq.\bref{rsp non zero regulator entries}
shows that propagation is diagonal within three sectors:
the $x^\mu$ sector, the $e$-$\bar \pi$ sector
and the ghost sector.
As expected,
the canonical transformation
in Eq.\bref{rsp pi ct}
decouples
$e$ from $x^\mu$:
For the shifted action
in Eq.\bref{rsp gfb shifted S},
one has
\be
  \left. {
     {{\partial _l\partial _rS}
      \over {\partial e\partial x^\mu }}
         } \right|_\Sigma =
   \left. {
     \left( {-2{d \over {d\tau }}
   {1 \over {e^2}}\dot x^\mu +
   2{d \over {d\tau }}{\rho  \over {e^3}}\dot x^\mu } \right)
          } \right|_\Sigma = 0
\quad .
\label{x e decoupling}
\ee

Because a constant $T_{AB}$ matrix has been selected,
the non-zero entries of ${F^A}_B$
in Eq.\bref{def of F} are
the same as
in Eq.\bref{rsp non zero K entries}.
{}From the structure of
Eqs.\bref{rsp non zero K entries}
and \bref{rsp non zero regulator entries},
one sees that the anomaly computation
separates
into contributions from
the $x^\mu$ sector, the $e$-$\bar \pi$ sector
and the ghost sector.
The propagating fields are
$x^\mu$, $\bar {\cal C}$ and ${\cal C}$.
The field $\bar \pi$ serves as a Lagrange multiplier
for setting $e$ equal to $\rho$.
Hence, $e$ and $\bar \pi$ are non-propagating
and should not contribute to the anomaly
according to the analysis
in Sect.\ \ref{ss:aoll}
\ct{tpbook}.
For this particular system,
the contribution is zero
because ${F^A}_B$ in the $e$-$\bar \pi$ sector
is off-diagonal.
It is also clear that
${\cal C}$ and $\bar {\cal C}$
do not contribute to the anomaly
since ${F^A}_B$ is zero for all ghost entries.
One only needs to consider the $x^\mu$ sector.

In what follows
we use a subscript $x$ for quantities
associated with $x^\mu$.
Applying
Eqs.\bref{def of K} and \bref{def of Q}
to Eq.\bref{rsp gfb shifted S},
one arrives at
\be
  {\left( {K_x} \right)^\mu}_\nu =
  {\delta^\mu}_\nu e^{-1}
   {\cal C} {d \over {d\tau }}
\quad ,
\label{rsp K_x}
\ee
and
\be
  \left( {Q_x} \right)_{\mu \nu } =
  \eta_{\mu \nu } Q_x
\quad ,
\label{rsp Q_x}
\ee
where $Q_x$ without $\mu \nu$ subscripts is defined by
\be
  Q_x \equiv -2 {d \over {d\tau }}e^{-1}{d \over {d\tau }} +
  {d \over {d\tau }}\rho e^{-2}{d \over {d\tau }}
\quad .
\label{rsp Q_x noindex}
\ee
Here and below, the derivative
${ {d} \over {d\tau} }$
acts on everything to the right
including $e$, $\rho$ and
the function to which $Q_x$ is applied.
Some contributions
to Eq.\bref{rsp Q_x noindex}
come from the shifts
in Eq.\bref{rsp pi ct}.
We have also dropped terms
proportional to
$\bar \pi^{*}$ and $\bar {\cal C}^*$
because they will not contribute,
when expanding about the stationary point.

Since the ${T_x}$ matrix is
$
  \left( {T_x} \right)_{\mu \nu } = \eta_{\mu \nu }
$,
the regulator matrix
in Eq.\bref{def2 of R}
is
\be
  {\left( {{\cal R}_x} \right)^\mu}_\nu = {\delta^\mu}_\nu Q_x
\quad .
\label{rsp R_x}
\ee
Because $\delta_B {T_x} = 0$,
\be
{\left( {F_x} \right)^\mu}_\nu = {\left( {K_x} \right)^\mu}_\nu
\quad ,
\label{rsp F_x}
\ee
where ${\left( {K_x} \right)^\mu}_\nu$ is given
in Eq.\bref{rsp K_x}.

All the relevant matrices
of Sect.\ \ref{ss:aoll}
for the computation of the anomaly
have been obtained.
At this stage,
one expands about the stationary point
of the gauge-fixed action.
The field $e$ is set equal to the function $\rho$
according to Eq.\bref{e stationary point}.
The regulator matrix becomes
\be
   \left. { {\left( {{\cal R}_x} \right)^\mu}_\nu }
        \right|_\Sigma =
   - {\delta^\mu}_\nu {d \over {d\tau }}
      \rho^{-1}{d \over {d\tau }}
\  ,
\label{R_x at Sigma}
\ee
so that on-shell
$$
  \left( {\Delta S} \right)_{\rm reg} =
$$
\be
    D \int {\dif \tau }
  \left[ {\int\limits_{-\infty }^\infty
   {{{\dif k} \over {2\pi }}}\ \exp \left( {-ik\tau } \right)
   \ \rho^{-1} {\cal C} {d \over {d\tau }}
   \ \exp \left( {{{-{d \over {d\tau }}
   \rho^{-1}{d \over {d\tau }}}
        \over { {\cal M}^2}}} \right)
   \ \exp \left( {ik\tau } \right)} \right]_0
\quad ,
\label{rsp Delta S reg}
\ee
where we have used the form of
$\left( {\Delta S} \right)_{\rm reg}$
in Eq.\bref{Delta S Fujikawa}.
The trace over field indices $A$
produces a factor of $\int {\dif \tau }$
and a factor of $D$
(because the number of $x^\mu$ fields is $D$ and
each contributes equally).
The operator trace is evaluated using
a complete set of momentum-space functions,
thereby generating
the factors $\exp \left( {\pm ik\tau } \right)$.
The calculation
in Eq.\bref{rsp Delta S reg}
is performed in Appendix C,
where it is shown that
the integrand is an odd function of $k$.
Consequently,
\be
  \left( {\Delta S} \right)_{\rm reg} = 0
\quad .
\label{rsp Delta S reg result}
\ee

The calculation
of the $x^\mu$ contribution
is even simpler
using Eq.\bref{Delta3 S reg}.
To compute $\delta_B {\cal R}$,
note that
\be
  \delta_B e = \left( {e,S} \right) =
    \dot {\cal C}
\quad .
\label{rsp BRST var of e}
\ee
Using this equation,
Eq.\bref{rsp Q_x noindex} and Eq.\bref{rsp R_x},
one finds
\be
  {\left( {\delta_B {\cal R}_x} \right)^\mu}_\nu =
   {\delta^\mu}_\nu \left( {2{d \over {d\tau }}
   e^{-2} \dot {\cal C} {d \over {d\tau }} -
    2{d \over {d\tau }}\rho e^{-3}
    \dot {\cal C} {d \over {d\tau }}} \right)
\quad ,
\label{rsp BRST var of R_x}
\ee
so that
\be
  \left. { {\left( {\delta_B {\cal R}_x}
    \right)^\mu}_\nu } \right|_\Sigma = 0
\quad .
\label{rsp BRST var of R_x at Sigma}
\ee
When $ \delta_B {\cal R}_x $ is substituted
into Eq.\bref{Delta3 S reg},
one gets
$$
\left( {\Delta S} \right)_{\rm reg} = 0
\quad ,
$$
in agreement
with Eq.\bref{rsp Delta S reg result}.
The absence of a violation
of the quantum master equation
means that the spinless relativistic particle theory
is gauge-invariant
even at the quantum level.

\subsection{The Abelian Chiral Schwinger Model}
\label{ss:acsm}

\def\Dslash{\slash\mkern-12mu D}
\def\Aslash{\slash\mkern-12mu A}

\hspace{\parindent}
In this subsection,
we analyze the abelian chiral Schwinger model
in two-dimensions.
It is an anomalous gauge theory
and
a particularly simple example
that illustrates the formalism
of Sect.\ \ref{ss:aoll}.
The model contains an abelian gauge field,
i.\ e., ``a photon'' $A_\mu$,
and a charged left-handed fermion.
It is governed by the following classical action
\be
   S_0
  \left[ { A_\mu , \psi , \bar\psi } \right] =
   \int \dif^2 x
     \left[ {
   -\frac{1}{4e^2}
    F^{\mu\nu} F_{\mu\nu} +
      \bar\psi i \Dslash \psi
     } \right]
\quad ,
\label{acsm1}
\ee
where $e$ is the electromagnetic coupling constant.
Although we take $\psi$ to be a Dirac fermion,
we use a covariant derivative
that couples the photon only
to the the right-moving component:
\be
  i\Dslash =
  i \slashit \partial + \Aslash P_{-}
\quad .
\label{acsm2}
\ee
In other words,
$P_{-} \psi$ is charged but $P_{+} \psi$ is neutral.
Here $P_{\pm}$ are the chiral projectors
\be
   P_{-} = \frac12 \left( { 1 - \gamma_5 } \right)
\ , \quad \quad
   P_{+} = \frac12 \left( { 1 + \gamma_5 } \right)
\quad .
\label{acsm3}
\ee
In two-dimensions,
they project onto right- and left-moving states.
Hence, the left-moving fermion $P_{+} \psi$
is a free particle and decouples.

A slash through a vector $V_{\mu}$ represents
$\gamma^{\mu} V_{\mu}$:
$\slashit{V} = \gamma^{\mu} V_{\mu}$.
In two-dimensions the $\gamma^{\mu}$
are $2 \times 2$ matrices
satisfying
$
  \gamma^\mu \gamma^\nu +
   \gamma^\nu \gamma^\mu =
    -2\eta^{\mu \nu }
$,
and $\gamma_5 $ is defined as
$\gamma_5 \equiv \gamma^0\gamma^1$.

The action in Eq.\bref{acsm1} is invariant
under the finite gauge transformations
\be
  A'_{\mu} = A_{\mu} + \partial_\mu\veps
\ , \quad \quad
  \psi' = \exp\{iP_{-}\veps\}\psi
\ , \quad \quad
  \bar\psi' = \bar\psi\exp\{-iP_{+}\veps\}
\quad .
\label{acsm4}
\ee

Although this model
has not been discussed
in previous sections,
it is straightforward to apply
the field-antifield formalism \ct{bm91a,bm94a}.
The proper solution for the gauge sector
corresponds to the solution
for the Yang-Mills example
given in Sect.\ \ref{sss:ymt} for $d=2$
and for a $U(1)$ group.
In addition to the antifield $A^*_\mu$ of the photon,
one has commuting antifields
$\psi^*$ and $\bar\psi^*$
for the fermions.
The proper solution of the master equation is
\be
  S = S_0 + \int\dif^2 x
   \left[ {
  A^*_\mu\partial^\mu \gh C +
     i (\psi^*)^t P_{-}\psi\,\gh C
    -i \bar\psi^* P_{+}^t \bar\psi^t \gh C
     } \right]
\quad ,
\label{acsm5}
\ee
where $\gh C$ is the ghost field
associated with the gauge parameter $\veps$.
The superscript $t$ stands for transpose:
$
  \psi =\left( {
   \matrix{
     \psi^1 \cr
     \psi^2 \cr }
   } \right)
$
and
$
  \bar \psi =\left( { \bar\psi^1 , \bar\psi^2 } \right)
$,
so that
$
  \bar\psi^t =
   \left( { \matrix{
     \bar\psi^1 \cr
     \bar\psi^2 \cr } } \right)
$
and
$
  (\psi^*)^t =
   \left( {  (\psi^1)^* , (\psi^2)^* } \right)
$.

A formal computation
using the expression for $\Delta$
in Eq.\bref{def Delta}
reveals that only the fermion sector
contributes to $\Delta S$.
A more detailed analysis using a regularization procedure
confirms this.%
{\footnote{
The computation made in ref.\ct{dstvnp92a}
for a pure Yang-Mills theories in four dimensions
supports the idea
that gauge fields and ghosts
produce a BRST trivial contribution
to $(\Delta S)_{\rm reg}$.
}}
Therefore,
we focus on the contribution to $\Delta S$
{}from $\psi$ and $\bar\psi$.
Gauge-fixing is not needed because
propagators for the fermions already exist.
For these reasons,
it is not necessary to consider
gauge-fixing auxiliary fields,
nor a non-minimal proper solution.

Using Eqs.\bref{def of K} and \bref{acsm5},
one finds that the $K$ matrix is given by
\be
  K = -i {\cal C}
   \left( { \matrix{
     P_- &    0_2 \cr
     0_2 & -P_+^t \cr}
           } \right)
\quad ,
\label{acsm6}
\ee
where all entries are $2 \times 2$ matrices,
e.\ g.,
$
  0_2 =
  \left( { \matrix{
    0 & 0 \cr
    0 & 0 \cr } } \right)
$,
so that $K$ is a $4 \times 4$ matrix.
In Eq.\bref{acsm6} and throughout this subsection,
we label the rows and columns of matrices
in the order
$\psi^1 , \psi^2 , \bar\psi^1 , \bar\psi^2$.
{}From Eqs.\bref{def of Q} and \bref{acsm1},
the matrix $Q$ for the fermion sector is
\be
  Q =
  \left( {
   \matrix{ 0_2 & -i \tilde{\Dslash} \cr
      i \Dslash & 0_2 \cr }
         } \right)
\quad ,
\label{acsm7}
\ee
where $\tilde{\Dslash}$ is defined by
$
   \int {\dif^2x} \bar \psi i \Dslash \psi =
  -\int {\dif^2x} \psi^t i \tilde{\Dslash} \bar \psi^t
$.
More precisely,
\be
  i \tilde{\Dslash} =
  i \slashit{\partial}^t - \Aslash^t P_+^t
\quad ,
\label{acsm8}
\ee
where, here, the superscript $t$ indicates
the transpose of a matrix in Dirac-index space.

For the matrix $T$, we choose
\be
  T = \left( {
  \matrix{ 0_2 & 1_2 \cr
          -1_2 & 0_2 \cr }
             } \right)
\quad ,
\label{acsm9}
\ee
where
$
  1_2 =
  \left( \matrix{1 & 0 \cr
                 0 & 1 \cr} \right)
$.
Because $\delta_B T = 0$,
the matrix $F$
in Eq.\bref{def of F}
is equal to the matrix $K$
in Eq.\bref{acsm6}.
Using Eq.\bref{def of O}, one finds
\be
  {\cal O} = \left( { \matrix{
    i \Dslash & 0_2 \cr
          0_2 & i\tilde{\Dslash} \cr}
             } \right)
\quad .
\label{acsm10}
\ee
Then from Eq.\bref{def1 of R},
one obtains
\be
  R = \left( {
    \matrix{
 - \Dslash \Dslash &    0_2 \cr
            0_2    & - \tilde{\Dslash} \tilde{\Dslash} \cr}
             } \right)
\quad ,
\label{acsm11}
\ee
for the regulator matrix.

Let us use the following representation
of the gamma matrices:
\be
  \gamma^0 = \sigma^1 =
  \left( \matrix{ 0 & 1 \cr
                  1 & 0 \cr} \right)
\ , \quad \quad
  \gamma^1 = i\sigma^2 =
  \left( \matrix{  0 & 1 \cr
                  -1 & 0 \cr} \right)
\quad ,
\label{acsm12}
\ee
so that
\be
  \gamma^5 =
   \gamma^0\gamma^1 = -\sigma^3 =
   \left( \matrix{-1 & 0 \cr
                   0 & 1 \cr} \right)
\quad ,
\label{acsm13}
\ee
and
\be
  P_- =
   \left( \matrix{ 1 & 0 \cr
                   0 & 0 \cr} \right)
\ , \quad \quad
  P_+ =
   \left( \matrix{ 0 & 0 \cr
                   0 & 1 \cr} \right)
\quad .
\label{acsm14}
\ee
With this representation,
\be
  i \Dslash =
  \left( { \matrix{
           0        & i\partial_+ \cr
  i\partial_- + A_- &      0      \cr} } \right)
\ , \quad \quad
  i \tilde{\Dslash} =
  \left( { \matrix{
      0       & i\partial_- - A_- \cr
  i\partial_+ &         0         \cr} } \right)
\quad ,
\label{acsm15}
\ee
where
\be
  \partial_\pm = \partial_0 \pm \partial_1
\ , \quad \quad
  A_- = A_0 - A_1
\quad .
\label{acsm16}
\ee

The entries in $R$
of Eq.\bref{acsm11}
are
$$
  - \Dslash \Dslash =
   \left( { \matrix{
 i\partial_+ \left( { i\partial_- + A_- } \right) & 0 \cr
    0 &  \left( { i\partial_- + A_- } \right) i\partial_+ \cr
       }  } \right)
\quad ,
$$
\be
  - \tilde{\Dslash} \tilde{\Dslash} =
   \left( { \matrix{
  \left( { i\partial_- - A_- } \right) i\partial_+ & 0 \cr
    0 & i \partial_+ \left( { i\partial_- - A_- } \right) \cr
       }  } \right)
\quad .
\label{acsm17}
\ee
The derivative $\partial_+$ acts to the right,
so that
it differentiates $A_-$
as well as any function
to which $R$ is applied.
Note that $R$ is diagonal.

In general,
the matrix $F'$
in Eq.\bref{def of F prime}
has two contributions.
However, the term proportial to $\delta_B {\cal O}$
does not contribute, upon taking the trace,
because ${\cal O}$
is off-diagonal and $R$ is diagonal
when the above gamma matrices are used.
Hence, one may take
\be
F' = F = K
\quad ,
\label{acsm18}
\ee
where $K$ is given in Eq.\bref{acsm6}.
Summarizing, for the computation of $\Delta S$
in Eq.\bref{Delta S Fujikawa},
one uses $F=K$
in Eq.\bref{acsm6}
and $R$ given
by Eqs.\bref{acsm11} and \bref{acsm17}.
Incorporating the projectors $P_{\pm}$ in $K$,
one obtains
\be
 \Delta S =
    - i  \left[{
     Tr \, {\cal C}
  \left( {
  \exp{ \left( {
     { { R_+ } \over {{\cal M}^2} }
                } \right) } -
  \exp{ \left( {
     { { R_- } \over {{\cal M}^2} }
                } \right) }
         } \right)
   } \right]_0
\ ,
\label{acsm19}
\ee
where there is no trace in Dirac-index space,
only in function space.
In Eq.\bref{acsm19},
\be
 R_\pm =
      \partial^\mu \partial_\mu
    \pm i (\partial_+ A_-) \pm i A_- \partial_+
\quad .
\label{acsm20}
\ee
where
we use a parenthesis around
$(\partial_+ A_-)$
to indicate that $\partial_+$
acts only on $A_-$.

In Appendix C,
the computation
of Eq.\bref{acsm20}
is performed.
One finds
\be
  {\cal A}_1 = \Delta S =
     \frac{i}{4\pi}
 \int \dif^2 x \, {\cal C}
 \left( {
      \epsilon^{\mu \nu} \partial_\mu A_\nu
       - \partial_\mu A^\mu
        } \right)
\quad ,
\label{acsm21}
\ee
where $\epsilon^{10} = 1 = -\epsilon^{01} $.
Note that the anomaly is consistent since
\be
 \delta_B {\cal A}_1 =
      \frac{i}{4\pi}
 \int \dif^2 x \, {\cal C}
 \left( {
      \epsilon^{\mu \nu} \partial_\mu \partial_\nu {\cal C}
      - \partial_\mu \partial^\mu {\cal C}
        } \right)
  = 0
\quad .
\label{acsm22}
\ee
The first term vanishes by the antisymmetry property
of $\epsilon^{\mu \nu}$,
while the second term vanishes by integration by parts
and the anticommuting nature of ${\cal C}$.
A local counterterm $\Omega_1$
cannot be added to the action
to eliminate ${\cal A}_1$
via Eq.\bref{one-loop remove anomaly}.
If one takes
$\Omega_1 = -\frac{1}{8\pi} \int \dif^2 x A^\mu A_\mu$,
then the second term
in Eq.\bref{acsm21}
is eliminated,
but the term
$
   \frac{i}{8\pi} \int \dif^2 x \, {\cal C}
   \epsilon^{\mu \nu} F_{\mu\nu}
$
in ${\cal A}_1$ remains.

\subsection{Anomaly in the Open Bosonic String}
\label{ss:aobs}

\hspace{\parindent}
In this subsection,
we investigate the violation
of the quantum master equation
for the first-quantized open bosonic string
when the dimension of space-time is not $26$.
The bosonic contribution was explicitly computed
in refs.\ct{tnp90a,gp93c}.
We gauge-fix the action
using the fermion $\Psi$
in Eq.\bref{fqbs psi}.
It depends on the conformal factor $\rho$.
It turns out that
when $D \ne 26$,
there is an anomaly.
As a consequence,
the theory depends on $\rho$.
Although the theory is classically gauge-invariant,
one of the four gauge symmetries
is violated by quantum effects.
This anomalous gauge symmetry
cannot be fixed for $D \ne 26$.

Let us apply the one-loop anomaly analysis
given in Sect.\ \ref{ss:aoll}
to the bosonic string.
First, we perform a canonical transformation
with $\Psi$ so that
$
  \Phi_A^* \to \Phi_A^* +
  {{\partial \Psi } \over {\partial \Phi^A}}
$.
This leads to the following shifts in antifield fields
$$
  {e^{*\tau}}_\tau \to {e^{*\tau}}_\tau +
  \bar {\cal C}+\bar {\cal C}_\sigma
\ ,\quad \quad
  {e^{*\sigma}}_\sigma
     \to {e^{*\sigma}}_\sigma +
   \bar {\cal C}-\bar {\cal C}_\sigma \ ,
$$
$$
  {e^{*\tau}}_\sigma\to {e^{*\tau}}_\sigma +
   \bar {\cal C}_{\tau \sigma }+\bar {\cal C}_\tau \ ,
   \quad \quad {e^{*\sigma}}_\tau
    \to {e^{*\sigma}}_\tau + \bar {\cal C}_{\tau \sigma } -
    \bar {\cal C}_\tau \ ,
$$
$$
  \bar {\cal C}^{*\tau } \to \bar {\cal C}^{*\tau } +
    {e_\tau}^{\sigma} - {e_\sigma}^{\tau} \ ,
\quad \quad \bar {\cal C}^{*\sigma } \to
   \bar {\cal C}^{*\sigma }+{e_\tau}^{\tau} -{e_\sigma}^{\sigma} \ ,
$$
\be
  \bar {\cal C}^*\to \bar {\cal C}^* +
   {e_\tau}^{\tau} +{e_\sigma}^{\sigma} -2\rho^{-1/2} \ ,
\quad \quad \bar {\cal C}^{*\tau \sigma } \to
    \bar {\cal C}^{*\tau \sigma } +
      {e_\tau}^{\sigma} +{e_\sigma}^{\tau}
\label{fgbs ct with psi}
\quad .
\ee
Additional terms are produced in the total action
$S_{\rm total} = S + S_{\rm aux}$
of Eqs.\bref{fqbs S} and \bref{aux S for fqbs}
given by
$$
  S \to S +
\int {\dif\tau }\int\limits_0^\pi  {\dif\sigma }
 \left\{ {
    \left( {\bar {\cal C} + \bar {\cal C}_\sigma } \right)
    \left( {{\cal C}^n\partial_n{e_\tau}^{\tau} -
     {e_\tau}^n\partial_n{\cal C}^\tau } \right) +
     \left( {\bar {\cal C}-\bar {\cal C}_\sigma } \right)
     \left( {{\cal C}^n\partial_n{e_\sigma}^{\sigma} -
     {e_\sigma}^n\partial_n{\cal C}^\sigma } \right)
  } \right.
$$
$$
  + \left( {\bar {\cal C}_{\tau \sigma } +
       \bar {\cal C}_\tau } \right)
   \left( {{\cal C}^n\partial_n{e_\tau}^{\sigma} -
   {e_\tau}^n\partial_n{\cal C}^\sigma } \right) +
   \left( {\bar {\cal C}_{\tau \sigma } -
      \bar {\cal C}_\tau } \right)
    \left( {{\cal C}^n\partial_n{e_\sigma}^{\tau} -
    {e_\sigma}^n\partial_n{\cal C}^\tau } \right)
$$
$$
  +\left( {\left( {\bar {\cal C} + \bar {\cal C}_\sigma } \right)
    {e_\tau}^{\tau} + \left( {\bar {\cal C} -
      \bar {\cal C}_\sigma } \right)
    {e_\sigma}^{\sigma} +\left( {\bar {\cal C}_{\tau \sigma } +
     \bar {\cal C}_\tau } \right){e_\tau}^{\sigma} +
    \left( {\bar {\cal C}_{\tau \sigma } -
    \bar {\cal C}_\tau } \right){e_\sigma}^{\tau} } \right){\cal C}
$$
\be
  -\left. {\left( {\left( {\bar {\cal C}+\bar {\cal C}_\sigma } \right)
    {e_\sigma}^{\tau} +\left( {\bar {\cal C}-\bar {\cal C}_\sigma } \right)
   {e_\tau}^{\sigma} +\left( {\bar {\cal C}_{\tau \sigma } +
    \bar {\cal C}_\tau } \right){e_\sigma}^{\sigma} +
    \left( {\bar {\cal C}_{\tau \sigma } -
     \bar {\cal C}_\tau } \right){e_\tau}^{\tau} } \right)
    {\cal C}^{\tau \sigma }} \right\}
\ ,
\label{S with psi ct}
\ee
and
$$
  S_{\rm aux} \to \int {\dif\tau }\int\limits_0^\pi  {\dif\sigma }
  \left\{ {\bar \pi \left( {\bar {\cal C}^* +
     {e_\tau}^{\tau} + {e_\sigma}^{\sigma} -2\rho^{-1/2}} \right) +
  } \right.
$$
\be
 \left.{
    \bar \pi_\tau \left( {\bar {\cal C}^{*\tau }+{e_\tau}^{\sigma} -
      {e_\sigma}^{\tau} } \right) + \bar \pi_\sigma
   \left( {\bar {\cal C}^{*\sigma }+{e_\tau}^{\tau} -
    {e_\sigma}^{\sigma} } \right)+\bar \pi_{\tau \sigma }
    \left( {\bar {\cal C}^{*\tau \sigma }+{e_\tau}^{\sigma} +
    {e_\sigma}^{\tau} } \right)} \right\}
\  .
\label{S aux with psi ct}
\ee

At this stage,
it is desirable to shift fields
by the solutions to the equations of motion
of the ${e_a}^m$,
by using a canonical transformation.
Such a shift guarantees that
the quadratic form $Q_{AB}$
is on-shell diagonal
in the ${e_a}^m$ sector.
This avoids mixing of the $X^\mu$ and ghost sectors
with the $\bar \pi$-${e_a}^m$ sector.
Variations of $S_{\rm total}$ in the gauge-fixed basis
with respect to the ${e_a}^m$
produce the equations for the four $\bar \pi$.
Using a subscript $0$ to denote the solutions,
and ignoring terms proportional
to antifields,
one finds that
$$
   2\left( {\bar \pi } \right)_0 \equiv
     -\left( {{{\partial S} \over {\partial {e_\tau}^{\tau} }} +
   {{\partial S} \over {\partial {e_\sigma}^{\sigma} }}
    - 2\bar \pi } \right) =
$$
$$
  { \left( { {e_\tau}^{\tau} + {e_\sigma}^{\sigma} } \right) }
   e {{\cal L}_X}  -
    {e} \left( {\partial_\tau X^\mu
     D_\tau X_\mu -\partial_\sigma X^\mu D_\sigma X_\mu } \right) +
$$
$$
  2\partial_n\left( {\bar {\cal C}{\cal C}^n} \right) +
    \bar {\cal C}
     \left({ \partial_n{\cal C}^n - 2{\cal C} } \right)
     + \bar {\cal C}_{\tau \sigma }
    \left( {\partial_\tau {\cal C}^\sigma +
    \partial_\sigma {\cal C}^\tau +
     2 {\cal C}^{\tau \sigma } } \right) +
$$
$$
  \bar {\cal C}_\tau \left( {\partial_\tau {\cal C}^\sigma -
   \partial_\sigma {\cal C}^\tau } \right) +
    \bar {\cal C}_\sigma \left( {\partial_\tau {\cal C}^\tau -
    \partial_\sigma {\cal C}^\sigma } \right)
\quad ,
$$
$$
  2\left( {\bar \pi_{\tau \sigma }} \right)_0 \equiv -
    \left( {{{\partial S} \over {\partial {e_\tau}^{\sigma} }} +
   {{\partial S} \over {\partial {e_\sigma}^{\tau} }} -
    2\bar \pi_{\tau \sigma }} \right) =
$$
$$
   -{ \left( { {e_\tau}^{\sigma} + {e_\sigma}^{\tau} } \right) }
    e {{\cal L}_X} -
    {e} \left( {\partial_\sigma X^\mu D_\tau X_\mu -
    \partial_\tau X^\mu D_\sigma X_\mu } \right) +
$$
$$
  2\partial_n
   \left( {\bar {\cal C}_{\tau \sigma }{\cal C}^n} \right) +
   \bar {\cal C}
     \left( {\partial_\tau {\cal C}^\sigma +
    \partial_\sigma {\cal C}^\tau } +
       2 {\cal C}^{\tau \sigma } \right) +
    \bar {\cal C}_{\tau \sigma }
   \left( {  \partial_n{\cal C}^n - 2 {\cal C} } \right)
$$
$$
  -\bar {\cal C}_\tau \left( {\partial_\tau {\cal C}^\tau -
    \partial_\sigma {\cal C}^\sigma } \right) -
   \bar {\cal C}_\sigma \left( {\partial_\tau {\cal C}^\sigma -
   \partial_\sigma {\cal C}^\tau } \right)
\quad ,
$$
$$
  2\left( {\bar \pi_\tau } \right)_0 \equiv
     -\left( {{{\partial S} \over {\partial {e_\tau}^{\sigma} }} -
   {{\partial S} \over {\partial {e_\sigma}^{\tau} }} -
    2\bar \pi_{\tau }} \right) =
$$
$$
  { \left( {{e_\tau}^{\sigma} -{e_\sigma}^{\tau} } \right) }
     e {{\cal L}_X} -
    {e} \left( {\partial_\sigma X^\mu D_\tau X_\mu +
    \partial_\tau X^\mu D_\sigma X_\mu } \right) +
$$
$$
  2\partial_n\left( {\bar {\cal C}_\tau {\cal C}^n} \right) -
    \bar {\cal C}\left( {\partial_\tau {\cal C}^\sigma -
    \partial_\sigma {\cal C}^\tau } \right) -
    \bar {\cal C}_{\tau \sigma }
   \left( {\partial_\tau {\cal C}^\tau -
    \partial_\sigma {\cal C}^\sigma } \right)
$$
$$
  + \bar {\cal C}_\tau
    \left( {\partial_n{\cal C}^n-2{\cal C}} \right) +
    \bar {\cal C}_\sigma
      \left( {\partial_\tau {\cal C}^\sigma +
    \partial_\sigma {\cal C}^\tau -
    2{\cal C}^{\tau \sigma }} \right)
\quad ,
$$
$$
  2\left( {\bar \pi_\sigma } \right)_0 \equiv
     -\left( {{{\partial S} \over {\partial {e_\tau}^{\tau} }} -
    {{\partial S} \over {\partial {e_\sigma}^{\sigma} }} -
     2\bar \pi_\sigma } \right) =
$$
$$
  { \left( { {e_\sigma}^{\sigma} - {e_\tau}^{\tau} } \right) }
      e {{\cal L}_X} -
    e \left( {\partial_\tau X^\mu D_\tau X_\mu +
    \partial_\sigma X^\mu D_\sigma X_\mu } \right) +
$$
$$
  2\partial_n
  \left( {\bar {\cal C}_\sigma {\cal C}^n} \right) +
   \bar {\cal C}\left( {\partial_\tau {\cal C}^\tau -
   \partial_\sigma {\cal C}^\sigma } \right) +
     \bar {\cal C}_{\tau \sigma }
   \left( {\partial_\tau {\cal C}^\sigma -
    \partial_\sigma {\cal C}^\tau } \right)
$$
\be
  + \bar {\cal C}_\tau \left( {\partial_\tau {\cal C}^\sigma +
  \partial_\sigma {\cal C}^\tau -
    2{\cal C}^{\tau \sigma }} \right) +
   \bar {\cal C}_\sigma
   \left( {\partial_n{\cal C}^n-2{\cal C}} \right)
\quad ,
\label{pi_0 defs}
\ee
where the $X^\mu$-lagrangian density ${\cal L}_X$ is defined to be
\be
   {\cal L}_X \equiv
    {e \over {2}}\left( {D_\tau X^\mu D_\tau X_\mu -
    D_\sigma X^\mu D_\sigma X_\mu } \right)
\quad .
\label{L_X def}
\ee
The canonical transformation of interest is given by
$$
  \bar \pi \to \bar \pi +
   \left( {\bar \pi } \right)_0 \ ,
\quad \quad \bar \pi_n\to \bar \pi_n +
   \left( {\bar \pi_n} \right)_0\ ,
\quad \quad \bar \pi_{\tau \sigma }
    \to \bar \pi_{\tau \sigma } +
    \left( {\bar \pi_{\tau \sigma }} \right)_0
\quad ,
$$
\be
  \Phi_A^* \to \Phi_A^* - \bar \pi^*{{\partial_r
  \left( {\bar \pi } \right)_0} \over {\partial \Phi^A}} -
   \bar \pi^{*\tau }{{\partial_r
    \left( {\bar \pi_\tau } \right)_0} \over {\partial \Phi^A}} -
   \bar \pi^{*\sigma }{{\partial_r
    \left( {\bar \pi_\sigma } \right)_0}
          \over {\partial \Phi^A}} -
    \bar \pi^{*\tau \sigma }{{\partial_r
   \left( {\bar \pi_{\tau \sigma }} \right)_0}
        \over {\partial \Phi^A}}
\quad ,
\label{pi e ct}
\ee
where in the first equation,
$n$ stands for $\tau$ or $\sigma$.

The goal of the next few paragraphs
is to obtain
${F^A}_B$ and ${{\cal R}^A}_B$
so that the anomaly in Eq.\bref{Delta S reg}
can be computed.
One must first calculate
$ {K^A}_{B} $ and $ Q_{AB} $
and specify $ T_{AB} $.
Simplifications occur for the following reasons.
In the final gauge-fixed form of the action
in Eq.\bref{fqbs gf action},
the propagating fields are
$X^\mu$, ${\bar {\cal C}}_n$ and ${\cal C}^n$
where $n$ represents
$\tau$ and $\sigma$.\footnote{
This is made clear by setting
$
  {\cal C} = \hat {\cal C} +
  \frac{1}{2} \left( { \partial_\tau {\cal C}^\tau +
  \partial_\sigma {\cal C}^\sigma } \right) -
  2 \rho^{1/2} {\cal C}^n \partial_n  \rho^{-1/2}
$
and
$
  {\cal C}^{\tau\sigma} = \hat {\cal C}^{\tau\sigma} -
  \frac{1}{2} \left( { \partial_\tau {\cal C}^\sigma +
  \partial_\sigma {\cal C}^\tau } \right)
$.
Throughout this subsection,
one should think of
${\cal C}$ and ${\cal C}^{\tau\sigma}$
as standing for these combinations of fields.
In terms of
$\hat {\cal C}$ and $\hat {\cal C}^{\tau\sigma}$,
the gauge-fixed action
in Eq.\bref{fqbs gf action}
is block diagonal
and $\bar {\cal C}$, $\bar {\cal C}^{\tau\sigma}$,
$\hat {\cal C}$ and $\hat {\cal C}^{\tau\sigma}$
are non-propagating fields.
The calculations in this subsection
should be performed in terms of
$\hat {\cal C}$ and $\hat {\cal C}^{\tau\sigma}$.
At the end of the computation,
one sets
$
  \hat {\cal C} = {\cal C} -
  \frac{1}{2} \left( { \partial_\tau {\cal C}^\sigma -
  \partial_\sigma {\cal C}^\tau } \right) +
  2 \rho^{1/2} {\cal C}^n \partial_n  \rho^{-1/2}
$
and
$
  \hat {\cal C}^{\tau\sigma} = {\cal C}^{\tau\sigma} +
  \frac{1}{2} \left( { \partial_\tau {\cal C}^\sigma -
  \partial_\sigma {\cal C}^\tau } \right)
$,
thereby returning to the original fields.}
By the argument
in Sect.\ \ref{ss:aoll},
only these fields contribute;
the ${F^A}_B$ and ${{\cal R}^A}_B$
associated with non-propagating fields
do not enter the calculation
\ct{tpbook}.
Furthermore,
by a judicial choice of $T_{AB}$,
${{\cal R}^A}_B$ can be made diagonal.
Hence, only the diagonal components of
$ {K^A}_{B} $, ${\delta_B {\cal O}^A}_B$ and ${F^A}_B$
for $X^\mu$, ${\bar {\cal C}}_n$ and ${\cal C}^n$
need to be computed.
In what follows,
we use a subscript $X$,
${\bar {\cal C}}$ and ${\cal C}$
to denote respectively
a matrix restricted to
the subspace corresponding to
$X^\mu$, ${\bar {\cal C}}_n$ and ${\cal C}^n$.
The subscript ``ghost'' is used
to denote the combined
${\bar {\cal C}}_n$ and ${\cal C}^n$ subspace.

After the above two canonical transformations
have been performed,
the computation is straightforward.
{}From Eq.\bref{def of K},
one finds that the ${K^A}_{B}$ in the three sectors are
$$
  {\left( {K_X} \right)^\mu}_\nu =
   {\delta^\mu}_\nu {\cal C}^n \partial_n
\quad ,
$$
$$
  {\left( {K_{\bar {\cal C}}} \right)^\tau}_\tau =
    {\left( {K_{\bar {\cal C}}} \right)^\sigma}_\sigma =
     - {\cal C}^n\partial_n
     -{3 \over 2}\left( {\partial_n{\cal C}^n} \right)
      + {\cal C}
\quad ,
$$
$$
  {\left( {K_{{\cal C}}} \right)^\tau}_\tau =
   - {\cal C}^n\partial_n +
    {1 \over 2}\left( {\partial_n{\cal C}^n} \right) +
    {1 \over 2}\left( {\partial_\tau C^\tau -
     \partial_\sigma C^\sigma } \right)
\quad ,
$$
\be
  {\left( {K_{{\cal C}}} \right)^\sigma}_\sigma =
   - {\cal C}^n\partial_n +
   {1 \over 2} \left( {\partial_n{\cal C}^n} \right) -
    {1 \over 2}\left( {\partial_\tau C^\tau -
  \partial_\sigma C^\sigma } \right)
\quad ,
\label{fqbs K comp}
\ee
where the presence of a parenthesis
around a derivative indicates
that it acts only on fields
within the parenthesis.
Throughout this subsection,
the absence of a parenthesis
means that the derivative
acts on everything to the right.
The terms for ${K_X}$ and ${K_{{\cal C}}}$ arise
from differentiating
Eq.\bref{fqbs S},
whereas those for ${K_{\bar {\cal C}}}$ come
from $S_{\rm aux}$ in
Eq.\bref{S aux with psi ct}
after the $\bar \pi_n$ shifts
in Eq.\bref{pi e ct}
have been performed.

The quadratic forms $Q_{AB}$
are
\be
   \left( {Q_X} \right)_{\mu \nu } =
     \eta_{\mu \nu } Q_X
\quad ,
\label{Q_X mu nu }
\ee
where
$$
  Q_X = D_\tau^{\dagger} {e}
   \left( { 1 - e \rho^{-1/2}
  { \left( { {e_\tau}^{\tau} + {e_\sigma}^{\sigma} } \right) }
           } \right) D_\tau -
    D_\sigma^{\dagger} {e}\left( { 1 - e \rho^{-1/2}
     { \left( {{e_\tau}^{\tau} + {e_\sigma}^{\sigma} } \right) }
                        } \right) D_\sigma
$$
\be
  + D_\tau^{\dagger} {{ e \rho^{-1/2}} }\partial_\tau -
    \partial_\tau {{ e \rho^{-1/2}} } D_\tau -
    D_\sigma^{\dagger} {{e \rho^{-1/2}} e}\partial_\sigma +
    \partial_\sigma {{e\rho^{-1/2}} } D_\sigma
\label{Q_X}
\ee
for the $X^\mu$ sector,
where the covariant derivatives
$D_\tau$ and $D_\sigma$
are given in Eq.\bref{gc cov der},
and where
$
  D_\tau^{\dagger} = - \partial_\tau {e_\tau}^{\tau}
    - \partial_\sigma {e_\tau}^{\sigma}
$
and
$
  D_\sigma^{\dagger} = - \partial_\tau {e_\sigma}^{\tau}
    - \partial_\sigma {e_\sigma}^{\sigma}
$.
Since derivatives act to the right,
$\partial_\tau$ and $\partial_\sigma$ in
$D_\tau^{\dagger}$ in first the term in Eq.\bref{Q_X}
act on
$
 {e}\left( { 1- e \rho^{-1/2}
   { \left( { {e_\tau}^{\tau} + {e_\sigma}^{\sigma} } \right) }
            } \right)
$,
on the vielbeins in $D_\tau^{\dagger}$ and $D_\tau$,
and on any function
to which the operator $Q_X$ is applied.
Likewise, for the other derivatives.
The terms in Eq.\bref{Q_X}
come from the original action $S_0$
in Eq.\bref{bosonic string action},
as well as from the $\bar \pi$ shifts
of Eq.\bref{pi e ct}
in $S_{\rm aux}$
of Eq.\bref{S aux with psi ct}.
For the ghost sector,
let
\be
  V = \left( \matrix{
   \bar {\cal C}_\tau \hfill\cr
  \bar {\cal C}_\sigma \hfill\cr
  C^\tau \hfill\cr
  C^\sigma \hfill\cr }
      \right)
\quad .
\label{fqbt def of V}
\ee
Then the
ghost part of the action is
\be
  S_{\rm ghost} =
  {1 \over 2}\int {\dif\tau }\int\limits_0^\pi {\dif\sigma }
    \ V^t Q_{\rm ghost}V
\quad ,
\label{fqbt S ghost}
\ee
where the superscript $t$ on $V^t$ stands for transpose.
The ghost quadratic form is
\be
  Q_{\rm ghost} =
    \left( \matrix{
  0\hfill &0\hfill &{\rho^{-1/2}\partial_\sigma }\hfill
      &{{-\rho^{-1/2}\partial_\tau }}\hfill \cr
  0\hfill &0\hfill &{-\rho^{-1/2}\partial_\tau }\hfill
    &{\rho^{-1/2}\partial_\sigma }\hfill \cr
  {{\partial_\sigma }\rho^{-1/2}}\hfill
     &{{-\partial_\tau \rho^{-1/2}}}\hfill &{0}\hfill &0\hfill \cr
  {-\partial_\tau \rho^{-1/2}}\hfill
   &{{\partial_\sigma }\rho^{-1/2}}\hfill
    &0\hfill &0\hfill \cr} \right)
\quad .
\label{Q ghost}
\ee
The terms in $Q_{\rm ghost}$ originate from
Eq.\bref{S with psi ct} and
from the $\bar \pi$ shifts
of Eq.\bref{pi e ct}
in $S_{\rm aux}$
of Eq.\bref{S aux with psi ct}.
The dependence on ${e_a}^m$ cancels
between the two contributions
leaving only a dependence on $\rho$.

For the matrix $T_{AB}$
of Sect.\ \ref{ss:aoll},
we choose
\be
  \left( {T_X} \right)_{\mu \nu } = e \eta_{\mu \nu }
\quad ,
\label{T_X}
\ee
and
\be
  T_{\rm ghost} = - i
    \left( \matrix{
   0\hfill &{e^{-2}}\hfill &0\hfill &0\hfill \cr
  {-e^{-2}}\hfill &0\hfill &0\hfill &0\hfill \cr
  {0}\hfill &0\hfill &{0}\hfill &{e^{2}}\hfill \cr
  0\hfill &0\hfill &{-e^{2}}\hfill &0\hfill \cr
    } \right)
\quad ,
\label{T_ghost}
\ee
where $T_{\rm ghost}$ acts in the space of antighosts and ghosts
given in Eq.\bref{fqbt def of V}.
In the $X^\mu$ sector
the operator ${\cal O}$,
defined in Eq.\bref{def of O},
is $e^{-1} {\delta^\mu}_\nu {Q_X} $
and is equal to
${\left( {{\cal R}_X} \right)^\mu}_\nu$,
where ${Q_X}$ is given
in Eq.\bref{Q_X}.
In the ghost sector,
${\cal O}$ is
\be
  {\cal O}_{\rm ghost} =
  i\left( \matrix{ 0_2 \hfill &{e^{2}}\hfill
    \rho^{-1/2}\slashit \partial \cr
  {{e^{-2} \slashit \partial }
      \rho^{-1/2}}\hfill & 0_2 \hfill \cr} \right)
\quad ,
\label{O_ghost}
\ee
where each entry is a two by two matrix.
It is somewhat accidental that the Dirac operator
in two-dimensional Minkowski space
\be
  \slashit \partial \equiv \gamma^n \partial_n =
    \gamma^\tau \partial_\tau +
    \gamma^\sigma \partial_\sigma
\quad ,
\label{def of partial slash}
\ee
enters in Eq.\bref{O_ghost},
if the gamma matrices
\be
  \gamma^\tau = \sigma^3 =
  \left( \matrix{1\hfill &0\hfill \cr
  0\hfill &{-1}\hfill \cr} \right)\ ,
\quad \quad
\gamma^\sigma = -i\sigma^2 =
   \left( \matrix{0\hfill &{-1}\hfill \cr
  1\hfill &0\hfill \cr} \right)
\label{def of gamma matrices}
\ee
are used.
They satisfy
\be
  \gamma^n \gamma^m + \gamma^m \gamma^n = -2 \eta^{nm}
\quad .
\label{gamma matrix com rels}
\ee
The ghost-sector regulator matrix
${\cal R}_{\rm ghost}$ is
\be
  {\cal R}_{\rm ghost}={\cal O}^2 =
   \left( \matrix{
  {{\cal R}_{\bar {\cal C}}}\hfill &0_2 \hfill \cr
    0_2 \hfill &{{\cal R}_{{\cal C}}}\hfill \cr} \right)
\quad ,
\label{R ghost}
\ee
where
\be
  {\cal R}_{\bar {\cal C}} = -e^{2}\rho^{-1/2}
   \slashit \partial e^{-2}
   \slashit \partial \rho^{-1/2} \ ,
\quad \quad
    {\cal R}_{{\cal C}} =
  -e^{-2}\slashit \partial e^{2}\rho^{-1} \slashit \partial
\quad .
\label{R C bar and R C}
\ee

To compute ${F^A}_C$
in Eq.\bref{def of F},
one needs the BRST variation of $T_{AB}$.
Using
\be
  \delta_B e =
 {\cal C}^n\partial_ne + e\partial_n{\cal C}^n -
      2e{\cal C}
\label{delta B of e}
\ee
and Eqs.\bref{T_X} and \bref{T_ghost},
it is straightforward to calculate
$
 { \left( T^{-1} \right) }^{AD} \left( \delta_B T \right)_{DC}
$.
The result is combined with ${K^A}_{C}$
in Eq.\bref{fqbs K comp}
to arrive at
$$
  { \left( F_X  \right)^\mu}_\nu = { \delta^\mu }_\nu F_X
\quad ,
$$
where
\be
   F_X = {\cal C}^n\partial_n +
    {1 \over 2}\left( {\partial_n{\cal C}^n} \right) +
  {1 \over 2 }e^{-1}{\cal C}^n
   \left( {\partial_ne} \right) - {\cal C}
\quad ,
\label{fqbt F_X}
\ee
for the $X^\mu$ sector.
For the ghost sector
$$
   {\left( {F_{\bar {\cal C}}} \right)^\tau}_\tau =
   {\left( {F_{\bar {\cal C}}} \right)^\sigma}_\sigma =
    -{\cal C}^n\partial_n-{1 \over 2}
    \left( {\partial_n{\cal C}^n} \right) +
    e^{-1}{\cal C}^n\left( {\partial_ne} \right)-{\cal C}
\quad ,
$$
$$
  {\left( {F_{{\cal C}}} \right)^\tau}_\tau =
     -{\cal C}^n\partial_n-{1 \over 2}
    \left( {\partial_n{\cal C}^n} \right) -
    e^{-1}{\cal C}^n\left( {\partial_ne} \right) +
     2{\cal C}+{1 \over 2}\left( {\partial_\tau {\cal C}^\tau -
    \partial_\sigma {\cal C}^\sigma } \right)
\quad ,
$$
\be
  {\left( {F_{{\cal C}}} \right)^\sigma}_\sigma =
    -{\cal C}^n\partial_n-{1 \over 2}
   \left( {\partial_n{\cal C}^n} \right) -
    e^{-1}{\cal C}^n\left( {\partial_ne} \right) +
   2{\cal C}-{1 \over 2}\left( {\partial_\tau {\cal C}^\tau -
   \partial_\sigma {\cal C}^\sigma } \right)
\quad .
\label{fqbt F_ghost}
\ee
Although $\delta_B {\cal O}_{\rm ghost}$ is non-zero,
it does not contribute because
the regulator matrix ${\cal R}$ is block diagonal
in the $\bar {\cal C}_n$ and ${\cal C}^n$ sectors.
It turns out that,
even in each sector,
only diagonal terms contribute
because of the nature of ${\cal R}$.
Hence,
in Eq.\bref{fqbt F_ghost}
we display only the diagonal part of ${F^A}_{C}$.
Finally,
since
the contribution from ${\cal C}_\tau$
turns out to be equal to
the contribution from ${\cal C}_\sigma$,
the two
$
    \partial_\tau {\cal C}^\tau -
    \partial_\sigma {\cal C}^\sigma
$
terms in Eq.\bref{fqbt F_ghost} for
$ F_{\cal C} $ cancel.
For the rest of this section,
we drop these terms.

The final step is the computation of
$(\Delta S)_{\rm reg}$.
Since $\delta_B {\cal O}$ does not contribute,
we may
use Eq.\bref{Delta S Fujikawa}.
At this stage,
we expand about the classical saddle point,
denoted by $\Sigma$,
corresponding to the solution to the equations of motion.
We set
the $\bar \pi$ equal to zero
and evaluate the ${e_a}^m$
as in Eq.\bref{fqbs gf conditions}.

To determine the regulators and ${F^A}_B$
at $\Sigma$,
note that
\be
   \left. e \right|_\Sigma = \rho
\quad .
\label{e at Sigma}
\ee
It is useful to express the regulators
in terms of symmetric operators $\tilde {\cal R}$.
For $X^\mu$,
\be
  \left. {\left( {{\cal R}_X} \right)} \right|_\Sigma =
   \rho^{-1}\left( {-\partial_\tau \partial_\tau +
    \partial_\sigma \partial_\sigma } \right) =
   \rho^{-1/2}\tilde {\cal R}_X\rho^{1/2}
\quad ,
\label{R_X at Sigma}
\ee
where
\be
  \tilde {\cal R}_X =
  \rho^{-1/2}\partial^n\partial_n\rho^{-1/2}
\quad .
\label{def of R_X tilde}
\ee
For ghosts,
\be
  \left. {{\cal R}_{\bar {\cal C}}} \right|_\Sigma =
   \rho \tilde {\cal R}_{\bar {\cal C}}\rho^{-1}\ ,
\quad \quad
  \left. {{\cal R}_{{\cal C}}} \right|_\Sigma =
\rho^{-1}\tilde {\cal R}_{{\cal C}}\rho
\quad ,
\label{R_ghost at Sigma}
\ee
where
\be
  \tilde {\cal R}_{\bar {\cal C}} =
   -\rho^{1/2} \slashit \partial \rho^{-2}
               \slashit \partial \rho^{1/2} \ ,
\quad \quad
\tilde {\cal R}_{{\cal C}} =
  -\rho^{-1}\slashit \partial \rho \slashit \partial \rho^{-1}
\quad .
\label{def of R_ghost tilde}
\ee

The conjugating factors
in Eqs.\bref{R_X at Sigma} and \bref{R_ghost at Sigma}
can be commuted past the operators ${F^A}_B$
to arrive at the following equivalent form
for $\left( {\Delta S} \right)_{\rm reg}$
$$
   \left( {\Delta S} \right)_{\rm reg} =
    \left[ {
      {\left( {\tilde F_X} \right)^\mu}_\nu
    {\left( {\exp
   \left( {{{\tilde {\cal R}_X} \over {{\cal M}^2}}} \right)}
     \right)^\nu}_\mu +
     } \right.
$$
\be
     \left. {
     {\left( {\tilde F_{\bar {\cal C}}} \right)^n}_n
    {\left( {\exp
      \left( {{{\tilde {\cal R}_{\bar {\cal C}}} \over
          {{\cal M}^2}}} \right)}
    \right)^n}_n +
     {\left( {\tilde F_{{\cal C}}} \right)^n}_n
    {\left( {\exp \left( {{{\tilde {\cal R}_{{\cal C}}}
           \over {{\cal M}^2}}} \right)}
     \right)^n}_n
         } \right]_0
\quad ,
\label{fqbs Delta S reg}
\ee
where
$$
   {\left( {\tilde F_X} \right)^\mu}_\nu =
   {\delta^\mu}_\nu \left( {{\cal C}^n\partial_n +
   {1 \over 2}
  \left( {\partial_n{\cal C}^n} \right)-{\cal C}} \right)
\quad ,
$$
$$
   {\left( {\tilde F_{\bar {\cal C}}} \right)^\tau}_\tau =
   {\left( {\tilde F_{\bar {\cal C}}} \right)^\sigma}_\sigma =
    -{\cal C}^n\partial_n -
  {1 \over 2}\left( {\partial_n{\cal C}^n} \right) -
    {\cal C}
\quad ,
$$
\be
  {\left( {\tilde F_{{\cal C}}} \right)^\tau}_\tau =
  {\left( {\tilde F_{{\cal C}}} \right)^\sigma}_\sigma =
    -{\cal C}^n\partial_n -
   {1 \over 2}\left( {\partial_n{\cal C}^n} \right) +
   2{\cal C}
\quad .
\label{def of F tilde}
\ee

The $T_{AB}$ matrix has been judiciously chosen
so that the violation in the quantum master equation
is proportional to ${\cal C}$.
The coefficient of ${\cal C}^n$
in Eq.\bref{fqbs Delta S reg} vanishes.
To see this,
let $\psi_r$ and $- E_r^2$
be the eigenfunctions and eigenvalues
of any of the $\tilde {\cal R}$ operators:
\be
  \tilde {\cal R} \psi_r = - E_r^2 \psi_r
\quad .
\label{psi_r}
\ee
The $\psi_r$ can be chosen to be real.
The terms
in \bref{def of F tilde}
involving ${\cal C}^n$
enter in the combination
$
 {{\cal C}^n \partial_n +
  {1 \over 2}\left( {\partial_n{\cal C}^n} \right)}
$.
Such a combination gives a zero contribution to
$ \left( {\Delta S} \right)_{\rm reg}$ since
$$
  \left( {{\cal C}^n \partial_n +
  {1 \over 2}\left( {\partial_n{\cal C}^n} \right)} \right)
   \left( {\exp
   \left( {{{\tilde {\cal R}} \over {{\cal M}^2}}} \right)} \right) =
$$
$$
  \int {\dif\tau }\int\limits_0^\pi  {\dif\sigma }\sum\limits_r
   \psi_r\left( {\tau ,\sigma } \right)
   \left( {{\cal C}^n\partial_n +
   {1 \over 2}
  \left( {\partial_n{\cal C}^n} \right)} \right)\psi_r
    \left( {\tau ,\sigma } \right)
   \exp \left( {-{{E_r^2} \over {{\cal M}^2}}} \right) =
$$
$$
  {1 \over 2}
 \int {\dif\tau }\int\limits_0^\pi {\dif\sigma }\sum\limits_r
   \partial_n\left( {\psi_r^2\left( {\tau ,\sigma } \right)
    {\cal C}^n\exp \left( {-{{E_r^2} \over
          {{\cal M}^2}}} \right)} \right)
   \to 0
\quad ,
$$
where, in the last step,
we assume that quantities fall off sufficiently fast
at large $\tau$
and obey appropriate boundary conditions at
$\sigma = 0$ and $\sigma = \pi$.
The reader can also verify
the absence of a ${\cal C}^n$ anomaly directly
by using the methods in Appendix C.

To evaluate the terms in
$\left( {\Delta S} \right)_{\rm reg}$ proportional
to ${\cal C}$,
let
\be
   H_r \equiv -\rho^{-\left( {r+1} \right)/2}
    \slashit \partial \rho^r
    \slashit \partial \rho^{-\left( {r+1} \right)/2}
\quad .
\label{def of H_n}
\ee
The tilde regulators
in Eqs.\bref{def of R_X tilde} and
\bref{def of R_ghost tilde}
are
\be
  \tilde {\cal R}_X = H_0 \ ,
\quad \quad \tilde {\cal R}_{\bar {\cal C}}=H_{-2} \ ,
\quad \quad \tilde {\cal R}_{{\cal C}}=H_1
\quad ,
\label{R tilde in terms on H_n}
\ee
except that, for $X^\mu$,
$\tilde {\cal R}_X$ acts in a $D$-dimensional space,
whereas
$H_0$ acts in a two-dimensional space,
since
$
  - \slashit \partial \slashit \partial =
     I_2 \partial^n \partial_n =
     I_2 \left( { -\partial_\tau \partial_\tau +
             \partial_\sigma \partial_\sigma } \right)
$,
where $I_2$ denotes the two-dimensional unit matrix.
The non-zero terms in
$\left( {\Delta S} \right)_{\rm reg}$
in Eq.\bref{fqbs Delta S reg}
all involve
\be
    {{Tr} \over 2}
   \left[ {\exp \left( {{{H_r} \over
      {{\cal M}^2}}} \right)} \right]_0
   \equiv
  \int {\dif\tau }\int\limits_0^\pi {\dif\sigma } \ \kappa_r
\quad .
\label{def of kappa_n}
\ee
The trace ${Tr}$ is over both $2$ by $2$ gamma space
and function space.
The coefficients $\kappa_r$,
which are computed in Appendix C, are
\be
  \kappa_r = {1 \over {24\pi i}}
   \left( {3r+1} \right)\partial^n\partial_n\ln
    \left( \rho  \right)
\quad .
\label{kappa_n results}
\ee

The contributions to
the violation of the quantum master equation
from the
$X^\mu$, $\bar {\cal C}_n$, and ${\cal C}_n$ sectors
are respectively $ (-1) D \kappa_{0} $,
$ (-1) 2 \kappa_{-2} $, and
$ (+2) 2 \kappa_{1} $,
where the factors in parentheses
are the coefficients of ${\cal C}$
in the $\tilde F$
in Eq.\bref{def of F tilde}.
The total anomaly ${\cal A}$ on-shell is
\ct{polyakov81a,fujikawa82a}
$$
  \left( {\Delta S} \right)_{\rm reg} =
   i{{\left( {D-26} \right)} \over {24\pi }}
   \int {\dif\tau }\int\limits_0^\pi  {\dif\sigma }
   \ {\cal C}\partial^n\partial_n\ln \left( \rho  \right)
$$
\be
  = i{{\left( {D-26} \right)} \over {24\pi }}
   \int {\dif\tau }\int\limits_0^\pi  {\dif\sigma }
   \ {\cal C}
			  \left.
   \partial^n\partial_n\ln \left( e  \right)
     \right|_\Sigma
\quad .
\label{fqbs anomaly}
\ee
It is absent when $D = 26$.
For $D \ne 26 $
no local counter term $\Omega_1$ can be added
to cancel the violation of the quantum master equation
via Eq.\bref{one-loop fake anomaly}
and the theory is anomalous.

\vfill\eject

\section{Brief Discussion of Other Topics}
\label{s:bdot}

\hspace{\parindent}
The following are discussed in this section:
applications to global symmetries,
a geometric interpretation of the field-antifield formalism,
locality, cohomology,
the equivalence
of lagrangian and hamiltonian approaches,
unitarity,
the antibracket formalism in a general coordinate system,
the $D=26$ closed bosonic string field theory, and
the extended formalism for anomalous gauge theories.
One topic not addressed
is the anti-BRST symmetry
\ct{cf76a,ojima80a,ab83a,baulieu85a,%
blt90a,gprr90a,gr90a,hull90a,%
bcg92a,henneaux92a,ikemori93a}.

\subsection{Applications to Global Symmetries}
\label{ss:ags}

\hspace{\parindent}
Certain models
with continuous rigid symmetries
share some of the characteristics of
gauge theories,
such as the closure only on-shell
of the commutator algebra
and the presence of
field-dependent structure constants.
Global supersymmetric theories
without auxiliary fields
and
models employing non-linear realizations
of rigid symmetries
are often
examples of algebras that do not close off-shell.
The antibracket formalism
can be used to assist
in the analysis of such theories
\ct{bbow90a,hlw90a}.
Even though, in the rigid-symmetry case,
the parameters
$\varepsilon^{\alpha}$
in the transformation law
in Eq.\bref{trans gauge}
are not functions of the space-time variable $x$,
there is still the notion of a symmetry structure.
In other words,
the analogs of the structure equations
in Sect.\ \ref{s:ssgt},
such as the Noether identity,
the Jacobi identity, etc.,
still exist.
There are two differences for the globally symmetric case:
(a) everywhere $\varepsilon^{\alpha}$ appears,
it is a constant
and (b)
the compact notation
for Greek indices,
associated with gauge transformations,
involves a discrete sum
but not an integral over space-time.
For Latin indices,
associated with the $\phi^i$,
repeated indices still indicate a space-time integral.
Taking into account the above two differences,
the equations
in Sect.\ \ref{s:ssgt}
hold for the globally symmetric case.

The development of an antibracket-like
formalism proceeds as
in Sect.\ \ref{s:faf}.
Since
the gauge parameters
$\varepsilon^{\alpha}$
are not functions of the space-time variable $x$,
one introduces
{\it constant} ghosts ${\cal C}^\alpha$.
The antifields $\phi_i^*$ for the original fields
$\phi^i$ are space-time functions,
but the antifields for ghosts are constants.
Grassmann statistics and ghost numbers
are assigned as in the gauge-theory case.
In the antibracket and elsewhere,
functional derivatives with respect to ghosts
and antifields of ghosts
are replaced by ordinary partial derivatives.
The proper solution $S$
in Eq.\bref{fsr proper solution}
is a generating functional for the structure tensors.
The structure equations are encoded in
the classical master equation $ ( S , S ) = 0$.
Of course,
since global symmetries
do not affect the rank of the hessian of $S$
at a stationary point,
the concept of properness has little meaning:
If one wants to treat all global symmetries
via an antibracket-like formalism,
one should proceed by mimicking the gauge case.

Since global symmetries
do not upset the development of perturbation theory,
no gauge-fixing procedure
via a fermion $\Psi$ is implemented.
In the quantum theory,
the ghosts are only technical tools.
They should not be considered as quantum fields.
Antifields are still interpreted
as sources for rigid symmetries.
To perform standard perturbation theory,
antifields can be set to zero.
Alternatively,
one can differentiate with respect to antifields
before setting them to zero
to obtain global Ward identities.
A third approach is
to introduce sources $J$ for fields
via Eq.\bref{Z of J},
retain antifields,
and construct the effective action $\Gamma$
as described
in Sect.\ \ref{ss:eazje}.
Anomalous violations of global symmetries
can be analyzed by searching for violations
of the Zinn-Justin equation $( \Gamma , \Gamma )_c = 0$
(see Eq.\bref{anomalous ZJ equation}).
Examples of anomalous global symmetries
are the axial vector currents
of massless four-dimensional QCD.
An application to the $D=4$ supersymmetric
Wess-Zumino model is given
in ref.\ct{hlw90a}.

\subsection{A Geometric Interpretation}
\label{ss:agi}

\hspace{\parindent}
This subsection
discusses a geometric interpretation
of the field-antifield formalism,
as presented by E.\ Witten
\ct{witten90a}.
See also
refs.\ct{khudaverdian91a,henneaux92a,bt93a,schwarz93a,kn93a}.

The geometric intepretation is made clearer
if we first assume
that no fermionic fields are present.
Let ${\cal M}$ denote the manifold
of infinite-dimensional function space.
The classical fields $\Phi^A$ of the theory
are local coordinates for ${\cal M}$.
Then,
${ {\partial} \over {\partial \Phi^A } }$
is a local basis
for the tangent space ${\cal TM}$
of vector fields.
Likewise, $d \Phi^A$ is a basis
for the cotangent space ${\cal T}^{*}{\cal M}$
consisting of differential forms.
There exists a natural quadratic form
on ${\cal TM} \oplus {\cal T}^{*}{\cal M}$
given by
$
 \langle d \Phi^A ,
 { {\partial} \over {\partial \Phi^B } } \rangle
  = \delta_B^A
$,
$\langle d \Phi^A , d \Phi^B \rangle = 0$,
$
 \langle { {\partial} \over {\partial \Phi^A } } ,
   { {\partial} \over {\partial \Phi^B } } \rangle
  = 0
$.
Introduce, in an ad hoc manner,
two quantities $z^A$ and $w_A$ and
associate $z^A$ with $d \Phi^A$
and $w_A$ with ${ {\partial} \over {\partial \Phi^A } }$, i.e.,
\be
 z^A \leftrightarrow d \Phi^A
\ , \quad \quad
 w_A \leftrightarrow { {\partial} \over {\partial \Phi^A } }
\quad .
\label{witten association}
\ee
Consider the Clifford algebra
for $z^A$ and $w_A$
determined by the quadratic form
$\langle \ , \ \rangle $, namely
\be
 \{ z^A , w_B \} = \delta_B^A
\ , \quad \quad
 \{ z^A , z^B \} =0
\ , \quad \quad
 \{ w_A , w_B \} =0
\quad ,
\label{clifford algebra}
\ee
where $\{  \ ,  \  \} $ denotes
the anticommutator:
$  \{ x ,  y  \} = x y + y x $.
A possible representation
of the Clifford algebra
regards the $z^A$ as creation operators
and the $w_A$ as destruction operators.
Then, the most general state
at a point $\Phi$ on the manifold
is created by
$
 \Omega \left( { \Phi , z } \right) =
 \Omega_0 \left( { \Phi  } \right) +
 \Omega_A \left( { \Phi  } \right) z^A +
 \frac12 \Omega_{AB} \left( { \Phi  } \right) z^B z^A +
 \dots
$
acting on a Fock-space vacuum $ | 0 \rangle $,
defined by
$w_A  | 0 \rangle = 0$ for all $A$,
i.\ e.,
it is annihilated by all the $w_A$.
Representing the $w_A$
as ${ {\partial} \over {\partial z^A } }$,
$ | 0 \rangle $ can be taken to be $1$
when considered as a function of the $z^A$.
With the association
$z^A \leftrightarrow d \Phi^A $,
one sees that $\Omega$ is equivalent
to an element of the exterior algebra
of differential forms on ${\cal M}$,
in which differential forms are multiplied by
using the wedge product $\wedge$.
When supplemented with the exterior derivative $d$,
this structure becomes the de Rham complex
\ct{flanders63a,egh80a}.
In summary,
one has an irreducible representation
of the Clifford algebra
in Eq.\bref{clifford algebra}
at each point of the manifold.

The Clifford algebra
in \bref{clifford algebra}
is symmetrical in its treatment
of the elements $z$ and $w$.
Hence,
one can reverse the above viewpoint
and regard the $w_A$ as creation operators
and the $z^A$ as annihilation operators.
In this picture,
let us identify the antifields $\Phi_A^*$
of the antibracket formalism
with the vector-field-like objects $w_A$.
The most general state
at a point $\Phi$
is created by
\be
F \left[ \Phi, \Phi^* \right] =
  F_0 ( \Phi ) + F^A ( \Phi ) \Phi_A^* +
 \frac12	F^{AB} ( \Phi ) \Phi_B^* \Phi_A^* + \dots
\label{dual clifford element}
\ee
acting on a state
$ | 0 \rangle^{\prime} $
that is annihilated by all $z^A$.
In this picture, denoted by $R^{\prime}$ by E.\ Witten,
$z^A = { {\partial}_r \over {\partial \Phi_A^* } } $,
$w_A = \Phi_A^*$,
and $ | 0 \rangle^{\prime} $
can be taken to be $1$
when regarded as a function of the $\Phi_A^*$.
Two elements $F$ and $G$ in the form
of Eq.\bref{dual clifford element}
are multiplied using
$
  \{ \Phi_A^* , \Phi_B^* \} = 0
$.
Exploiting the association
$
  d \Phi^A \to z^A
  = { {\partial}_r \over {\partial \Phi_A^* } }
$
in the $R^{\prime}$ picture,
the exterior derivative
$
 d \equiv
{ {\partial}_r \over { \partial \Phi^A } }  d \Phi^A
$
becomes
$
   { {\partial}_r \over { \partial \Phi^A    } }
   { {\partial}_r \over { \partial \Phi_A^*  } } F
 = - \Delta F
$
when acting on a general functional $F$
of the type
in Eq.\bref{dual clifford element}.
Here, we have used the definition
of $\Delta$ given
in Eq.\bref{def Delta}.
In computing $\Delta F$,
one treats
$\Phi^A$ and $\Phi_B^*$ as independent variables.
Because the exterior derivative is nilpotent,
$\Delta$ satisfies $\Delta^2 = 0$.
In short,
one arrives at a dual picture of the de Rham complex.
It is isomorphic to the standard de Rham complex
but not in a natural way
because there is no preferred manner of associating
the above two Fock-space vacuums
$ | 0 \rangle $ and
$ | 0 \rangle^{\prime} $.
The state
$ | 0 \rangle^{\prime} $
of the $R^{\prime}$ picture
is represented as
$ \left( f_{12 \dots } \right) \prod_A z^A$
in the first picture,
where $f_{12 \dots }$ is arbitrary.
A natural choice {\it does} exist if ${\cal M}$
is endowed with a measure
$
  d \mu =
   \left( {\mu_{12 \dots} } \right)
   d \Phi^1 \wedge d \Phi^2 \wedge \dots
$.
Then, one can take $f=\mu$.

If fermionic fields are present,
${\cal M}$ is a supermanifold.
Then, $\{  \ ,  \  \} $ appears as
a graded commutator:
$
  \{ x ,  y  \} =
  x y - (-1)^{\epsilon_x \epsilon_y} y x
$.
Note that
$
 \epsilon \left( { z^A } \right) =
 \epsilon_A + 1 =
 \epsilon \left( { w_A } \right)
$,
so that $z^A$ and $w_A$ have
the opposite statistics of $\Phi^A$.
For the bosonic case,
$ \epsilon_A = 0$ for all $A$,
and $\{  \ ,  \  \} $
becomes the usual anticommutator.

In the $R^{\prime}$ picture,
the Clifford algebra
in Eq.\bref{clifford algebra}
is satisfied when
\be
w_A = \Phi_A^*
\ , \quad \quad
  z^A =
   (-1)^{\epsilon_A}
  { {\partial}_r \over {\partial \Phi_A^* } }
\quad .
\label{witten identification}
\ee
Then, $-d$ becomes,
when acting on $F$
of Eq.\bref{dual clifford element},
\be
 (-1)^{\epsilon_A + 1}
   { {\partial}_r \over { \partial \Phi^A    } }
   { {\partial}_r \over { \partial \Phi_A^*  } } F
 = \Delta F
\quad .
\label{dual DeRham-like d}
\ee
The nilpotent operator $\Delta$
of the field-antifield formalism
is identified with minus
the exterior derivative.

For elements $F$ and $G$
in Eq.\bref{dual clifford element},
the antibracket $( \ , \ )$ is defined by
$
 \left( { F \left[ \Phi, \Phi^* \right] ,
          G \left[ \Phi, \Phi^* \right] } \right)
$
$
 \equiv
    {{ \partial_r F }
      \over {\partial \Phi^A   }}
    {{ \partial_l G }
      \over {\partial \Phi_A^*  }}
    -
    {{ \partial_r F }
      \over {\partial \Phi_A^*  }}
    {{ \partial_l G }
      \over {\partial \Phi^A    }}
$.
Using $\Delta$,
the antibracket can be expressed as
\be
  \Delta ( FG ) -
  F \Delta ( G ) - (-1)^{\epsilon_G} \Delta ( F ) G =
 (-1)^{\epsilon_G}
  \left( { F [ \Phi, \Phi^* ] , G [ \Phi, \Phi^* ] } \right)
\quad .
\label{DeRham-like antibracket2}
\ee
Eq.\bref{DeRham-like antibracket2}
shows that the antibracket is the obstruction
of $\Delta$ to be a derivation from the right.
Once $\Delta$ has been defined,
one can take the left-hand side of
Eq.\bref{DeRham-like antibracket2}
to be the definition of the antibracket
after multiplying by $ (-1)^{\epsilon_G}$.

Summarizing, one has the following analogy.
If one thinks of function space as a supermanifold,
then antifields are the basis vectors
in the $R^{\prime}$ picture of the de Rham complex.
The operator $\Delta$ is
the analog of the exterior derivative.
The antibracket is the obstruction
to $\Delta$ for it to be a derivation.
Interestingly,
the quantum master equation
\be
  \frac12 ( W , W ) - i \hbar \Delta W = 0
\label{qme in introduction}
\ee
has the same form as the equation of motion
for a Chern-Simons theory
\ct{witten90a}.
The analog of a gauge transformation
in Chern-Simons theory
is a quantum BRST transformation
\ct{lt85a}
in the antibracket formalism.
It is not always too easy
to obtain solutions $W$
to the quantum master equation.
In some sense,
finding an appropriate $W$
is equivalent to obtaining the correct measure,
i.e., specifying $\mu$.

\subsection{Locality }
\label{ss:l}

\hspace{\parindent}
An important but technical aspect
of quantum field theories is locality.
Here,
we study this issue
in the antibracket formalism
\ct{henneaux91a,ht92a,paris92a,gp93b,bbh94a,vp94a}.
In going from the classical action $S_0$
to the proper solution $S$
and to the quantum action $W$,
lagrangian terms are added.
In a theory defined by a local classical action,
the question is whether these terms
are also local.
Local interactions involve fields and
derivatives, up to a finite order, of fields
multiplied at the same space-time point.
Nonlocal terms are likely to lead to difficulties
such as non-renormalizability, non-unitarity
or violations of causality.

The discussion of the gauge structure algebra
in Sect.\ \ref{s:ssgt}
used extensively the consequences of
the regularity condition
in Eq.\bref{consequence of rc}.
An examination of the proof
of Eq.\bref{completesa}
reveals that certain operators need to be inverted
so that nonlocal effects are possible.
Indeed,
it is easy to find a $\lambda^i$
so that the solution to $S_{0,i} \lambda^i = 0$
in Eq.\bref{completesa}
involves a nonlocal operator $T^{ji}$ or a function
$\lambda^{\prime \alpha_0}$ that does not fall off fast
at large space-time distances.
Nonlocality often occurs when the quantity of interest
vanishes because it is an integral of a total derivative.
As an example,
consider $n$ free quantum mechanical particles
governed by the action
$S_0 = \frac12 \int \dif t \ \dot q^i \dot q_i $.
Note that $S_{0,i} = - \ddot q_i $.
Let $\lambda^i = \dot q^i$.
Then,
$
 S_{0,i}\lambda^i = - \frac12 \int \dif t
  {d \over {dt}} \left( { \dot q^i \dot q_i }  \right) \to 0
$.
The solution
in Eq.\bref{completesa}
is $\lambda^{\prime \alpha_0}=0$
and
$$
  T^{ji} \left( {t',t} \right) =
   \left\{ \matrix{ \ \ {{\delta ^{ji}} \over 2} \ ,
   \quad \quad {\rm for \ } \ t>t' \quad , \hfill\cr
  - {{\delta ^{ji}} \over 2} \ ,
    \quad \quad \hbox{\rm for \ } t<t' \quad , \hfill\cr} \right.
$$
since
$
  \lambda^i (t) = \int \dif t'
  \ S_{0,j} \left( {  t' } \right)
  T^{ji} \left( {  t' , t } \right)
$.
The trivial gauge transformation governed by
$ T^{ji} \left( {  t' , t } \right)$
obeys the correct symmetry property
$
  T^{ij} \left( {  t , t' } \right) =
  - T^{ji} \left( {  t' , t } \right)
$.

Hence, an important concept
is local completeness
\ct{henneaux91a}.
Local completeness holds,
when solutions to equations,
such as in Eqs.\bref{completesa},
\bref{consequence of regularity} and
\bref{reduc l 1},
can be satisfied for local functionals
or more precisely,
in non-integrated versions.
The difficulty is that
sometimes these equations are valid
due to total derivatives.

In principle, it is possible
that the gauge structure tensors
involve nonlocal operators.
This issue has been analyzed
in refs.\ct{dhtv91a,henneaux91a,gp93b}.
Given the locality of $S_0$
and that the gauge generators $R^i_{\alpha}$
are local operators,
then the proper solution $S$
of the classical master equation is local.
Reference \ct{henneaux91a}
used cohomological arguments to obtain this result.
The gauge-fixed classical action $S_{\Psi}$
is then guaranteed to be local
if the gauge-fixing fermion $\Psi$ is.
Under these conditions,
the classical BRST operators $\delta_B$
and $\delta_{B_\Psi}$ produce local variations.
The question of quantum locality is more involved.
Since this must be analyzed on a case by case basis,
no general statements about the locality of $W$
can be made.

\subsection{Cohomological Aspects }
\label{ss:ca}

\hspace{\parindent}
In this section,
we introduce the concept of cohomology.
Cohomological methods have been used
to obtain certain general results
\ct{henneaux89a,fisch90a,fh90a,hlw90a,%
bh93a,bbh94a,vp94a,tpbook},
for the field-antifield formalism.
For example,
the existence proof of the proper solution
is based on these methods
\ct{fh90a}.
This section is intended
to assist the reader
in understanding such research.
Because these methods
have been reviewed
in refs.\ct{henneaux90a,ht92a,tpbook},
our discussion is brief.

Consider a series of spaces $F_k$.
The integer $k$ labels different grading levels.
If $\alpha_k \in  F_k$ then $\alpha_k$
is said to have grading $k$.
Let $\delta$ be a nilpotent map from one space
to a successive space
\be
  \ldots \buildrel {} \over \longrightarrow
   F_{k-1}\buildrel \delta  \over \longrightarrow
   F_k\buildrel \delta  \over \longrightarrow
    F_{k+1}\buildrel {} \over \longrightarrow \ldots
\quad .
\label{cohomology map}
\ee
Nilpotency means that
\be
  \delta^2 = \delta \delta = 0
\quad .
\label{nilpotency of delta}
\ee
One can think of $\delta$ as carrying a grading of $1$.
An element $\alpha_k$ in $F_k$
is said to be closed if
$\delta \alpha_k = 0$.
The kernel of $\delta$ for the $k$th space,
${\rm Ker}_k \delta$,
consists of the set of closed elements of $F_k$, i.e.,
\be
  {\rm Ker}_k \delta =
  \left\{ {\alpha_k \mid \alpha_k \in F_k,\delta \alpha_k = 0} \right\}
\quad .
\label{kernel of delta}
\ee
An element $\alpha$
is said to be exact if it is expressible as
$\alpha = \delta \beta$.
The image of $\delta$ in the $k$th space,
${\rm Im}_k \delta$,
are the exact elements of $F_k$, i.e.,
\be
  {\rm Im}_k \delta =
  \left\{ {\alpha_k \mid \alpha_k = \delta \beta_{k-1},\;
   \mbox{for some}\; \beta_{k-1} \in F_{k-1}} \right\}
\quad .
\label{image of delta}
\ee
Due to the nilpotency of $\delta$,
an exact element is automatically closed,
so that
${\rm Im}_k \delta \subset {\rm Ker}_k \delta$.
Consider the equivalence relation $\sim \atop k$
that identifies two elements of $F_k$
if their difference is exact:
\be
  \alpha_k \ { \mathrel{\mathop{k}^{\sim}} } \ \alpha_k^{'}\;
  \mbox{if there is a}\;
  \beta_{k-1} \in F_{k-1}\;
  \mbox{such that}\;
  \alpha_k - \alpha_k^{'} = \delta \beta_{k-1}
\quad .
\label{equivalence relation}
\ee
Define $H_k \left( \delta  \right) $
to be the set of elements of ${\rm Ker}_k \delta$ modulo
the equivalence relation $\sim \atop k$:
\be
  H_k\left( \delta  \right) =
  {\rm Ker}_k\delta / {\rm Im}_k \delta
\quad .
\label{def of H_k}
\ee
The cohomology $ H_k\left( \delta  \right)$
is equivalent to the elements in $F_k$
that are closed but not exact.

A standard example
is the de Rham cohomology
on an $n$-dimensional manifold ${\cal M}$.
The spaces $F_k$ consist of the differential forms
of order $k$ on ${\cal M}$
and $\delta$ is the exterior derivative $d$.
The dimension of $ H_k\left( d  \right)$
is the $k$th Betti number for ${\cal M}$.
In this example,
more structure can be defined.
Differential forms can be multiplied
using the wedge product $\wedge$.
One can add differential forms
so that the formal sum of the $F_k$ spaces
constitutes an algebra.
The exterior derivative respects addition:
$ d ( \alpha + \beta ) = d \alpha + d \beta $,
and it is a graded derivation
from the left of the wedge product:
$
  d \left( { \alpha_j \wedge \beta_k  } \right)
  = d\alpha_j \wedge \beta_k + (-1)^j \alpha_j \wedge d\beta_k
$.
In applications within the antibracket formalism
$\delta$ has these properties,
except, with our conventions,
$\delta$
is a graded derivation from the right:
$
  \delta \left( { \alpha \beta  } \right)
  =  \alpha \delta\beta +
  (-1)^{\epsilon_\beta}
  \left( { \delta\alpha  } \right) \beta
$,
where multiplication is denoted
by juxtaposition of elements.

Cohomological methods can be powerful.
However, they often involve subtle issues
so that one must proceed with strict rigor.
The question of whether a closed element is
expressible as $\delta ( { \rm \ of \ something \ } )$
often involves global issues;
usually, it can be done ``locally''.
Hence, if one is not careful,
one can miscalculate the cohomology.
In regard to the antibracket formalism,
the pitfalls are more severe:
The spaces $F_k$ are almost always infinite dimensional,
and, in quantizing the system,
the multiplication operation becomes singular.
Furthermore,
an ambiguity concerning
the issue of locality
in Sect.\ \ref{ss:l} enters:
one needs to decide whether
local or non-local functionals are permitted.

One cohomology in the antibracket formalism
uses the classical BRST operator $\delta_B$ for $\delta$.
The spaces $F_k$ consist of smooth functionals of fields
with ghost number $k$.
Note that $k$ ranges over all integers,
both positive, negative and zero.
Functionals form an algebra
since they can be added and multiplied.
Furthermore, $\delta_B$ satisfies the correct properties:
it is nilpotent and
is a graded derivation from the right.
The cohomology $H_k\left( \delta_B  \right)$
is the classical space of observables
in the sector with ghost number $k$.

For proving certain results,
two other cohomologies are useful.
The first uses the Koszul-Tate differential $\delta_{kt}$
\ct{koszul50a,borel53a,tate57a}.
Let $G_+$ be the condition of setting all ghost fields to zero:
\be
  G_+ \equiv \left\{ {\Phi^A = 0 \mid
    {\rm gh}\left[ {\Phi^A} \right]\ge 1} \right\}
\quad .
\label{def of G_+}
\ee
Then, the Koszul-Tate differential is defined by
\be
  \delta_{kt} X \equiv \left. {\delta_B X} \right|_{G_+}
\quad ,
\label{def of delta_kt}
\ee
where $\delta_B X = \left( { X, S } \right)$.
When acting on fields (and ghosts),
$\delta_{kt}$
produces zero
\be
   \delta_{kt} \Phi^A
   = \left. { {{\partial_l S} \over {\partial \Phi_A^*}}
            }  \right |_{G_+}=0
\quad ,
\label{delta_kt on fields}
\ee
since $\delta_{kt} \Phi^A$,
having ghost number greater than one,
must be proportional to ghost fields.
Hence, the interesting action
is on antifields:
\be
  \delta_{kt} \Phi_A^* =
  \left( {-1} \right)^{\epsilon_A+1}
   \left. { {{\partial_r S} \over {\partial \Phi^A}}
          } \right|_{G_+}
\quad .
\label{delta_kt on antifields}
\ee

To check the nilpotency of $\delta_{kt}$,
it suffices to compute $\delta_{kt}^2\Phi_A^*$.
One finds
$$
  \delta_{kt}^2\Phi_A^* =
  \left( {-1} \right)^{\epsilon_A + \epsilon_B}
   \left. {\left( {
     \left( { {{\partial_r\partial_rS} \over
        {\partial \Phi_B^*\partial \Phi^A}}
            } \right)
     \left( {{{\partial_rS} \over {\partial \Phi^B}}
     } \right)
    } \right)} \right|_{G_+}
\quad ,
$$
which can be manipulated to give
$$
  \delta_{kt}^2\Phi_A^* =
 \left( {-1} \right)^{\epsilon_B}
  \left. {
    \left( {
      {{\partial_r} \over {\partial \Phi^A}}
         \left( {{{\partial_r S} \over {\partial \Phi_B^*}}
         {{\partial_rS} \over {\partial \Phi^B}}} \right) -
    {{\partial_r S} \over {\partial \Phi_B^*}}
    {{\partial_r\partial_r S}
      \over {\partial \Phi^A \partial \Phi^B}}
    } \right)
   } \right|_{G_+}
\quad ,
$$
which,
using Eqs.\bref{master equation}, \bref{l and r der relation}
and \bref{delta_kt on fields},
produces
$$
  = - {1 \over 2}
   \left. {
     \left( {
    {{\partial_r\left( {S,S} \right)} \over {\partial \Phi^A}}
    } \right)
     } \right|_{G_+} = 0
\quad .
$$
The Koszul-Tate differential
is zero because $S$ satisfies
the classical master equation.
Because $\delta_{kt}$ is constructed
using the BRST operator it is a graded derivation, i.e.,
\be
  \delta_{kt}\left( {XY} \right) =
   X\delta_{kt}\left( Y \right) +
   \left( {-1} \right)^{\epsilon_Y}
    \delta_{kt}\left( X \right)Y
\quad .
\label{graded derivation of delta_kt}
\ee

The action of $\delta_{kt}$
for the antifields with ghosts numbers $-1$, $-2$ and $-3$
is respectively
$$
  \delta_{kt} \phi_i^* =
   \left( {-1} \right)^{\epsilon_i+1}
   S_{0,i}
\quad ,
$$
$$
  \delta_{kt} {\cal C}_\alpha^* =
   \left( {-1} \right)^{\epsilon_\alpha }
   \phi_i^* R_\alpha^i
\quad ,
$$
\be
  \delta_{kt} {\cal C}_{1\alpha_1}^* =
   \left( {-1} \right)^{\epsilon_{\alpha_1}+1}
  \left( {{\cal C}_{\alpha}^* R_{1\alpha_1}^{\alpha} +
  {1 \over 2}\left( {-1} \right)^{\epsilon_i}
   \phi_i^*\phi_j^*V_{1\alpha_1}^{ji}} \right)
\quad ,
\label{action of delta_kt on first few ghosts}
\ee
where the tensors $R$ and $V$ are given
in Sect.\ \ref{s:ssgt}.

An important  result is that
$H_k \left( { \delta_{kt} } \right) = \emptyset $
for $k \le -1 $,
where $\emptyset$ denotes the empty set
\ct{fh90a,henneaux90a}.
Let us formally verify this
for $k=-1$ and $k=-2$.
One can set terms involving ghosts to zero
because they either transform to zero
or are set to zero.
The most general element $\alpha_{-1}$,
with ghost number $-1$
and constructed from antifields, is
$$
  \alpha_{-1} = \phi_i^* \lambda^i
\quad ,
$$
where $\lambda^i$
are functionals of the $\phi$,
with
$
  \epsilon \left( {\lambda^i} \right) =
   \epsilon_i
$
so that
$\epsilon \left( {\alpha_{-1}} \right) = 1$.
Since
$
  \delta_{kt} \alpha_{-1} = -S_{0,i} \lambda^i
$,
$\alpha_{-1}$ is closed if
$
 S_{0,i}\lambda^i = 0
$.
Accordingly,
$\lambda^i$ must be expressible as
in Eq.\bref{completesa},
so that
$$
  \alpha_{-1} =
   \phi_i^*R_\alpha^i\lambda^{'\alpha } +
   \phi_i^*S_{0,j} T^{ji}
\quad .
$$
This is the most general form for a closed
element with ghost number $-1$.
The key question is whether
$\alpha_{-1}$ is exact, i.e.,
$
  \alpha_{-1}
  \lower2pt\hbox{=}\mkern-10mu{^?} \  \delta_{kt} \beta_{-2}
$.
Let
$$
  \beta_{-2} =
   {\cal C}_\alpha^*\lambda^{'\alpha } +
   \left( {-1} \right)^{\epsilon_i+1}
  {1 \over 2}\phi_i^*\phi_j^*T^{ji}
\quad .
$$
Then,
a short computation reveals that
$\delta_{kt} \beta_{-2} = \alpha_{-1}$.
In the $k=-1$ sector,
there are no closed elements that are not exact,
so that the cohomology is trivial
$ H_{-1} \left( { \delta_{kt} } \right) = \emptyset $.

In the $k=-2$ sector,
the most general element is
$$
  \alpha_{-2} =
  {\cal C}_\alpha^* \lambda^\alpha +
  \left( {-1} \right)^{\epsilon_i}
   {1 \over 2} \phi_i^* \phi_j^* M_0^{ji}
\quad ,
$$
where
$
  \epsilon \left( {\lambda^\alpha } \right) =
  \epsilon_\alpha
$
and
$
  \epsilon \left( {M_0^{ji}} \right) =
   \epsilon_i+\epsilon_j
$ (mode 2),
so that
$
 \epsilon \left( {\alpha_{-2}} \right) = 0
$.
The functional $M_0^{ji}$ obeys
$
  M_0^{ji} =
   \left( {-1} \right)^{\epsilon_i\epsilon_j+1} M_0^{ij}
$.
A computation reveals that
$$
  \delta_{kt}\alpha_{-2} =
   \phi_i^* \left( {R_\alpha^i
   \lambda^\alpha -S_{0,j} M_0^{ji}} \right)
\quad ,
$$
so that
\be
 {R_\alpha^i \lambda^\alpha - S_{0,j} M_0^{ji}} = 0
\quad ,
\label{alpha_2 closed condition}
\ee
if $\alpha_{-2}$ is to be closed.
Is
$
  \alpha_{-2}
 \lower2pt\hbox{=}\mkern-10mu{^?} \  \delta_{kt}\beta_{-3}
$.
It takes a little work
to show that $\alpha_{-2}$ is exact.
For $\alpha_{-2}$ to be closed,
$\lambda^\alpha$ must obey
Eq.\bref{completesa}.
Substituting the solution
of Eq.\bref{completesa}
for $\lambda^\alpha$
into Eq.\bref{alpha_2 closed condition} gives
$$
  R_\alpha^i \left( {R_{1\alpha_1}^\alpha
  \lambda^{'\alpha_1} +
   S_{0,j}T_0^{j\alpha }} \right) -
   S_{0,j}M_0^{ji} = 0
\quad ,
$$
or, using Eq.\bref{reduc l 1},
$$
  S_{0,j} \left( {V_{1\alpha_1}^{ji}
   \lambda^{'\alpha_1} +
   \left( {-1} \right)^{\epsilon_j
   \left( {\epsilon_i+\epsilon_\alpha } \right)}
   R_\alpha^iT_0^{j\alpha }-M_0^{ji}} \right) = 0
\quad .
$$
The general solution of this equation is
$$
  V_{1\alpha_1}^{ji}\lambda^{'\alpha_1} +
  \left( {-1} \right)^{\epsilon_j
   \left( {\epsilon_i+\epsilon_\alpha } \right)}
   R_\alpha^iT_0^{j\alpha } -
   \left( {-1} \right)^{\epsilon_i \epsilon_\alpha }
   R_\alpha^jT_0^{i\alpha } - M_0^{ji} =
  S_{0,k} N^{kji}
\quad ,
$$
where $N^{kji}$ must be graded antisymmetric in all indices
and
$
  \epsilon \left( {N^{kji}} \right) =
  \epsilon_i+\epsilon_j + \epsilon_k
$ (mod 2).
In terms of the above functionals and tensors,
let
$$
  \beta_{-3} =
   - {\cal C}_{1\alpha_1}^*
    \lambda^{'\alpha_1} -
    \left( {-1} \right)^{\epsilon_\alpha }
    {\cal C}_\alpha^* \phi_j^*T_0^{j\alpha } +
    {1 \over 6}\left( {-1} \right)^{\epsilon_j}
    \phi_i^* \phi_j^* \phi_k^* N^{kji}
\quad .
$$
Then, a short computation shows that
$
  \alpha_{-2} = \delta_{kt}\beta_{-3}
$.
Since any closed $k=-2$ element
is expressible as an exact form,
$ H_{-2} \left( { \delta_{kt} } \right) = \emptyset $.

Notice that the triviality of the Koszul-Tate cohomology
$H_{-k} \left( { \delta_{kt} } \right) = \emptyset$
for $k > 0$
reflects the consequences of the regularity condition
used in Sect.\ \ref{s:ssgt}.%
{\footnote{
If one does not use a proper solution for $S$,
$H_{-k} \left( { \delta_{kt} } \right)$
can be non-empty for $k > 0$.}}
Although the above discussion has been formal,
a more rigorous analysis can be given.
See refs.\ct{fh90a,henneaux90a}.

Insight into the physical significance
of the Koszul-Tate cohomology is gained
by computing
$H_{0} \left( { \delta_{kt} } \right)$.
A general closed element
of the zero ghost-number sector
is a functional $\alpha_0$ of the fields $\phi$
since
$
  \delta_{kt}\alpha_0\left( \phi  \right)
$
is automatically zero
due to $\delta_{kt} \phi^i = 0$.
Let $\beta_{-1}$ be a general element
of the $-1$ ghost sector, i.e,
$
  \beta_{-1} = - \phi_i^*\lambda^i
$,
where we include a minus sign for convenience,
and where the Grassmann nature of $\lambda^i$ is
$
  \epsilon \left( {\lambda^i} \right) =
   \epsilon_i
$.
Apply the Koszul-Tate differential
to $\beta_{-1}$ to obtain
$
  \delta_{kt}\beta_{-1} =
   S_{0,i}\lambda^i
$.
One concludes that
if
$
\alpha_0 = S_{0,i}\lambda^i
$
then $\alpha_0$ is exact.
Therefore $\alpha_0$ and $\alpha_0{'}$
are not related
under the equivalence relation $\sim \atop 0$
if $\alpha_0$ and $\alpha_0{'}$ differ
on the stationary surface $\Sigma$
where the equations of motion $S_{0,i} = 0$ hold.
Consequently,
$
  H_0\left( {\delta_{kt}} \right)
$
corresponds to the set of distinct functions on $\Sigma$.
More precisely,
\be
 H_0\left( {\delta_{kt}} \right)
  = \left\{ {
    \alpha_0 \left( \phi  \right) \mid
   \alpha_0 \sim \alpha_0^{'} \; ,\;
   \mbox{if}\; \alpha_0 -
   \alpha_0^{'} =
   S_{0,i} \lambda^i\; \mbox{\rm for some}\; \lambda^i
           } \right\}
\quad .
\label{H_0 of delta_kt}
\ee
Suppose that a theory has no gauge invariances.
Then the classical observables correspond to
functionals taking on distinct values on $\Sigma$.
This space is
$ H_0\left( {\delta_{kt}} \right) $.

If a theory has gauge invariances,
then the observables should
be the gauge-invariant elements of
$ H_0\left( {\delta_{kt}} \right) $.
To facilitate the issue of gauge invariance,
one introduces the vertical differential $\delta_{g}$
\ct{fhst89a,henneaux90a}.
An alternative name for $\delta_{g}$
is the ``exterior derivative
along the gauge orbit''.
It is defined as
\be
  \delta_g X =
  \left. { \left( { \delta_B X } \right) } \right|_{G_{-}} =
  \left. { \left( {X,S} \right) } \right|_{G_{-}}
\quad ,
\label{def of delta_g}
\ee
where $G_{-}$ corresponds to the condition
of setting antifields to zero
and going on-shell with respect to the original fields, i.e.,
\be
  G_{-} =
  \left\{ { {\Phi_A^* = 0} , \rder{S}{\phi^i} = 0 } \right\}
\quad .
\label{def of G-}
\ee
Because $\delta_g$ is defined in terms of $\delta_B$,
it is a derivation from the right:
\be
  \delta_g \left( {XY} \right) =
   X \delta_g Y +
   \left( {-1} \right)^{\eps_y}\left( {\delta_g X} \right) Y
\quad .
\label{derivation property of delta_g}
\ee
Antifields can be ignored
in evaluating the vertical differential
because they are either set to zero
or transformed to zero since
\be
  \delta_g \Phi_A^* =
    \left. {-\left( {{{\partial_l S} \over {\partial \Phi^A}}}
    \right)} \right|_{G_{-}} = 0
\quad ,
\label{delta_g on antifields}
\ee
which follows from ghost number considerations
or Eq.\bref{def of G-}.
On fields, one has
\be
  \delta_g \Phi^A =
  \left. { \left( {{{\partial_l S} \over {\partial \Phi_A^*}}
         } \right)} \right|_{G_{-}}
\quad .
\label{delta_g on fields}
\ee
To check the nilpotency of $\delta_g$,
one only needs to check that
$ \delta_g \delta_g \Phi^A = 0 $.
A straightforward calculation gives
$$
  \delta_g^2 \Phi^A =
   \left. {\left( {
   {{\partial_r \partial_l S} \over
          {\partial \Phi^B\partial \Phi_A^*}}
   {{\partial_l S} \over {\partial \Phi_B^*}}
    } \right)} \right|_{G_{-}}
$$
$$
  = \left. {\left( {{{\partial_l} \over {\partial \Phi_A^*}}
    \left( {{{\partial_r S} \over {\partial \Phi^B}}
    {{\partial_l S} \over {\partial \Phi_B^*}}} \right) -
    \left( {-1} \right)^{\left( {\eps_A +1} \right)\eps_B}
   {{\partial_r S} \over {\partial \Phi^B}}
   {{\partial_l\partial_l S} \over {\partial \Phi_A^*\partial \Phi_B^*}}
     } \right)} \right|_{G_{-}}
$$
$$
  = {1 \over 2}\left. {\left( {
   {{\partial_l \left( {S,S} \right)} \over {\partial \Phi_A^*}}
    } \right)} \right|_{G_{-}}=0
\quad ,
$$
where Eqs.\bref{master equation}
and \bref{delta_g on antifields}
have been used.
As a consequence of nilpotency,
a cohomology with respect to $\delta_g$ can be defined.

The physical relevance of $\delta_g$
can be seen by computing $ H_0 \left( {\delta_g} \right)$.
The action of $\delta_g$ on the original fields $\phi^i$ is
\be
  \delta_g \phi^i = R_\alpha^i{\cal C}^\alpha
\quad .
\label{delta_g on phi^i}
\ee
Without loss of generality,
a functional $\alpha_0$ with ghost number $0$
can be taken to be a functional of the $\phi^i$ only.
Such a functional is closed if
$\delta_g \alpha_0 \left( \phi  \right) = 0$.
Using Eq.\bref{delta_g on phi^i},
one finds
$$
  \delta_g \alpha_0 =
   \alpha_{0,i} R_\alpha^i {\cal C}^\alpha =
   0 \ \Rightarrow \alpha_0 {\rm \ is \ gauge \ invariant \ }
\quad .
$$
Since any functional $\beta_{-1}$ with ghost number $-1$
is annihilated by $\delta_g$,
a closed $\alpha_0$ cannot be exact:
$
  \alpha_0 \ne \delta_g \beta_{-1}
$.
The conclusion is that
\be
 H_0\left( {\delta_g} \right) =
 \mbox{\rm  the set of gauge--invariant functionals}
\quad .
\label{H_0 of delta_g}
\ee

With the above insights,
one realizes that observables should
roughly correspond to
$
  H_0 \left( {\delta_{kt}} \right)
   \cap H_0\left( {\delta_g} \right)
$.
However, a difficulty arises.
The intersection
$
  H_0 \left( {\delta_{kt}} \right)
   \cap H_0\left( {\delta_g} \right)
$
does not make sense
unless
$
  \delta_{kt} \delta_g + \delta_g \delta_{kt}
    = 0
$.
The situation
is analogous to the one in quantum mechanics
where one seeks a state that is simultaneously
the eigenvector of two different operators.
Such a state is possible if the two operators commute.
Because of the graded nature
of $\delta_{kt}$ and $\delta_g$,
the analog condition is that
$\delta_{kt}$ and $\delta_g$ anticommute.
The difficulty can be posed as a question:
Should one take the gauge-invariant elements of
$H_0 \left( {\delta_{kt}} \right) $
or should one take the elements of
$H_0\left( {\delta_g} \right)$
modulo the equivalence relation
of Eq.\bref{H_0 of delta_kt}?
If
$
  \delta_{kt} \delta_g + \delta_g \delta_{kt}
    = 0
$,
then the above two procedures yield the same result.
In such a case,
one can define a nilpotent BRST operator
$\delta_B \equiv \delta_{kt} + \delta_g$,
and the observables correspond to the BRST cohomology.
Unfortunately,
$
  \delta_{kt} \delta_g + \delta_g \delta_{kt}
    \ne 0
$
in general.
An inspection of $\delta_B$,
$\delta_{kt}$ and $\delta_g$ reveals that
$\delta_B = \delta_{kt} + \delta_g + {\it extra \ terms}$.
The {\it extra terms} render $\delta_B$ nilpotent,
by compensating
for the failure of the anticommutivity of
$\delta_{kt}$ and $ \delta_g$.
The BRST operator is the natural extension of
$\delta_{kt} + \delta_g$.
The elements of the cohomology of $\delta_B$
are the classical observables
\ct{fh90a,henneaux90a}.
They are the definition
of what one means
by the ``gauge-invariant functionals on $\Sigma$''.

When quantum effects are incorporated,
the quantum BRST transformation $\delta_{\hat B}$
is relevant.
As discussed
in Sect.\ \ref{ss:qBRSTt},
the quantum observables
correspond to the elements
of the cohomology of $\delta_{\hat B}$.

Because canonical transformations
preserve the antibracket,
the cohomology of $\delta_{B}$
is independent of the basis,
as can be seen as follows.
Given a proper solution $S [ \Phi , \Phi^* ]$
in one (untilde) basis,
a solution $\tilde S [ \tilde \Phi , \tilde \Phi^* ]$
in another (tilde) basis
is given by
$
 \widetilde S [ \tilde \Phi , \tilde \Phi^* ]
  \equiv S [ \Phi , \Phi^* ]
$.
Likewise,
given any functional $X [ \Phi , \Phi^* ]$,
one can define a functional $\widetilde X$
of tilde fields using
$
\widetilde X [ \tilde \Phi , \tilde \Phi^* ]
  \equiv X [ \Phi , \Phi^* ]
$.
The tilde antibracket of $ \widetilde X$ and $\widetilde S$,
as a function of tilde fields and antifields,
equals
$(X,S)$ as a function of untilde fields.
Hence, $\widetilde X$ is closed if and only if $X$ is,
and $\widetilde X$ is exact if and only if $X$ is.
Consequently, there is an exact isomorphism
of the cohomologies.

Since the gauge-fixed BRST transformation
$\delta_{B_\Psi}$ is not nilpotent,
one cannot directly define a cohomology
associated with $\delta_{B_\Psi}$.
However, according
to Eq.\bref{nilpot cal of gf BRST},
$\delta_{B_\Psi}^2$ is proportional to the equations
of motion
of the gauge-fixed action $S_{\Psi }$.
Define an equivalence relation,
denoted by $\approx$,
that equates two quantities
if they differ by terms proportional to
the equations of motion for $S_{\Psi }$.
Then, one has $\delta_{B_\Psi}^2 \approx 0$
and a gauge-fixed BRST cohomology
can be defined.
What is the relation between
$H_n \left( { \delta_{B_\Psi} } \right)$
and $H_n \left( { \delta_{B} } \right)$?
The connection is best seen
by going to the gauge-fixed basis
for $\delta_{B}$.
Let
\be
  Y ( \tilde \Phi , \tilde \Phi^*) =
    y ( \tilde \Phi ) +
  y^A ( \tilde \Phi ) \tilde \Phi_A^* + \dots
\label{y-series}
\ee
be the antifield expansion of a functional $Y$
in the gauge-fixed basis
$\tilde \Phi^A$ and $\tilde \Phi_A^*$
of Sect.\ \ref{ss:gfb}.
Eqs.\bref{gauge-fixed BRST} and
\bref{gf BRST trans for antifields}
imply
\be
  \delta_B Y =
  \delta_{B_\Psi} y -
     y^A \lder{S_\Psi}{\tilde \Phi^A}
    + O ( \tilde \Phi^* )
\quad ,
\label{delta relations1}
\ee
so that
\be
  \left.{ (\delta_B Y )  }
   \right|_{ \{ { \tilde \Phi^* = 0 } \} } \approx
  \delta_{B_\Psi} y
\quad ,
\label{delta relations2}
\ee
since the last term
in Eq.\bref{delta relations1}
is proportional to gauge-fixed equations of motion.
Eq.\bref{delta relations2} implies
that if $Y$ is $\delta_B$-closed
then $y$ is $\delta_{B_\Psi}$-closed,
and that if $Y$ is $\delta_B$-exact
then $y$ is $\delta_{B_\Psi}$-exact.
This is not enough to establish any relation
between
$H_n \left( { \delta_{B_\Psi} } \right)$
and $H_n \left( { \delta_{B} } \right)$.
Given a element of $y$
of $H_n \left( { \delta_{B_\Psi} } \right)$,
one must uniquely construct
an element $Y$
of $H_n \left( { \delta_{B_\Psi} } \right)$,
under the condition that
$ \left.{ Y  } \right|_{ \{ { \tilde \Phi^* = 0 } \} } = y $.
In other words,
one must find the higher-order terms
in Eq.\bref{y-series}.
References
\ct{henneaux89a,fh90a,fisch90a}
succeeded in doing this.
For additional discussion,
see refs.\ct{ht92a,tpbook}.
The cohomologies governed by
$\delta_{B_\Psi}$ and $\delta_{B}$
are equivalent.

\subsection{Equivalence with the Hamiltonian BFV Formalism}
\label{ss:ewhBFVf}

\hspace{\parindent}
Gauge theories can also be analyzed
using a hamiltonian formalism.
For the generic theory,
the Batalin-Fradkin-Vilkovisky (BFV) approach
\ct{fv75a,bv77a,ff78a,bf83a,bf86a}
is quite useful.
For the simplest theories,
such as particle models, Yang-Mills theory, and gravitation,
it is not difficult to show that it yields
results equivalent to the field-antifield formulation.
Demonstrating the equivalence in general,
at the classical level
or formally at the quantum level without regularization,
has been the subject of the work
in refs.\ct{bgpr89a,fh89a,siegel89a,siegel89b,%
dfgh91a,ggt91a,ggt92a,nr92a,paris92a,%
dejonghe93a,dejonghe93b}.

A review of the BFV hamiltonian formalism
is given
in ref.\ct{henneaux85a}.
Here, we present only the key ideas.
Let $S_0$ be
the classical action
as determined from a lagrangian $L$
by $S_0 = \int \dif t \ L$.
The hamiltonian $H_{S_0}$
associated with ${S_0}$
is constructed
in the standard manner using%
{\footnote
{In this subsection, we use the convention
that a field index also
represents a spatial position.
An index
appearing twice
represents a sum not only that index
but also an integration over space.
This is the hamiltonian analog of the compact
notation described
in Sect.\ \ref{ss:gt}.
The difference, here, is that time is not included
as part of the integration.  }
  }
\be
  H_{S_0} [ \phi, \pi ] \equiv \dot \phi^i \pi_i - L
\quad ,
\label{H construction}
\ee
where a dot over a field indicates a time derivative,
where the conjugate momentum of $\phi^i$
is $\pi_i \equiv \lder{S_0}{\dot \phi^i}, i=1, \dots , n$,
and where
$H_{S_0}$ is obtained as a function of the $\phi$
and $\pi$
by solving for $\dot \phi$ in terms of the $\pi$ and
$\phi$.
For some systems,
this velocity-momentum inversion process is not possible
due to the presence of primary constraints.
Even in this case, a hamiltonian
$H_{S_0}$ can be uniquely constructed
on the surface of these primary
constraints.
We symbolically represent the procedure
of obtaining a hamiltonian from an action
diagrammatically as
$$
\matrix{
S_0 \cr
\Big\downarrow \cr
H_{S_0} \cr
        }
$$
For a wide class of gauge theories,
$H_{S_0}$ is of the form
\be
H_{S_0} = H_0 \left[ { \varphi , \pi } \right] +
   \lambda^\alpha T_\alpha \left[ { \varphi, \pi } \right]
\quad ,
\label{H_S_0}
\ee
where
the original $n$ variables $\phi^i$
are split into dynamical degrees of freedom
$\varphi^a, a =1, \dots , m \le n$
and Lagrange multipliers $\lambda^\alpha$
for the (secondary) constraints $T_\alpha$.
In Eq.\bref{H_S_0},
$H_0$ and $T_\alpha$ are functions of the $\varphi$
and their momenta only.
The velocities $\dot\lambda^\alpha$ are
usually assumed not to appear
in $S_0$.
This means that
the momenta of the $\lambda^\alpha$,
namely $\pi_\alpha$,
are primary constraints
and do not enter in $H_{S_0}$.
For example,
in a Yang-Mills theory,
the hamiltonian density ${\cal H}_0$ is
${\cal H}_0 = \frac12 E_a^i E^a_i + \frac14 F^a_{ij} F^{ij}_a$,
where $ E_a^{i} = - F_a^{0i} = F_{a0i}$ are the canonical momenta
for the potentials $A^a_i$,
the constraints $T_\alpha$
correspond to Gauss's law:
$
   T_a = - {D_{ia}}^{b} E_{b}^i
$,
and the Lagrange multipliers $\lambda^\alpha$
are $A_0^a$.

For simplicity, assume that the constraints $T_\alpha$
and the hamiltonian $H_0$
are first class, i.e,
\be
   \{ T_\alpha, T_\beta \}_{PB}  =
      T_{\alpha\beta}^\gamma T_\gamma
\ ,\quad \quad
       \{ H_0, T_\alpha \}_{PB}  = V_\alpha^{\beta} T_\beta
\quad ,
\label{first class}
\ee
where $\{ \ , \ \}_{PB}$ denotes the graded Poisson bracket
defined by
\be
 \{ F , G \}_{PB} =
   (-1)^{\eps_i} \rder{F}{\phi^i} \lder{G}{\pi_i} -
      \rder{F}{\pi_i} \lder{G}{\phi^i}
\quad .
\label{Poisson bracket}
\ee
Here, the sum over $i$ is such that all fields
and momenta are included.
If the constraints are second class,
Dirac brackets
\ct{dirac64a}
must be used.
Note that $\{ \pi_i , \phi^j \}_{PB} = - \delta_i^j $.

The BFV program is based on BRST invariance.
One introduces ghosts and their conjugate momenta.
The ghosts needed
correspond to the minimal set,
introduced
in Eq.\bref{field set}.
For the irreducible case,
they are the ${\cal C}^\alpha$.
We use $\agh P_a$
to denote the momentum associated with a ghost
${\cal C}^a$,
where $a$ is a label that enumerate all ghosts.
The Poisson bracket
in Eq.\bref{Poisson bracket}
is then extended to include a sum over ghosts.
With these conventions,
$ \{ \agh P_b , {\cal C}^a \}_{PB} = - \delta_b^a $.
The ghost numbers and statistics of the BFV ghosts
are the same as
in Sect.\ \ref{ss:fa}.
For momenta,
$
  {\rm gh} \left[ { \agh P_a } \right] =
 - {\rm gh} \left[ { {\cal C}^a } \right]
$
and
$
  \eps \left( \agh P_a \right) =
 \eps \left( { {\cal C}^a } \right)
$.
A canonical generator of the BRST transformations
$Q_B$ and an extended hamiltonian $H$
are constructed,
using the requirement that $Q_B$ be nilpotent
and that $H$ be BRST invariant:
\be
 \{ Q_B , Q_B \}_{PB} = 0
\ , \quad \quad
 \{ H , Q_B \}_{PB} = 0
\quad .
\label{Q_B H algebra}
\ee
They can be expanded as a power series in ghost fields,
for which the first few terms are
\be
 H = H_0 +
   {\cal C}^\alpha V_\alpha^\beta \agh P_\beta
      + \dots
\ , \quad \quad
 Q_B  = {\cal C}^\alpha T_\alpha -
   \frac12 (-1)^{\eps_\beta} {\cal C}^\beta {\cal C}^\gamma
   T_{\gamma\beta}^\alpha \agh P_\alpha + \dots
\ \ .
\label{hamiltonian ghost expansion}
\ee
It turns out that the requirement of $\{ Q_B , Q_B\}_{PB} = 0$
reproduces the relations defining the structure of the gauge algebra
at hamiltonian level.
In other words, $Q_B$ plays a role analogous to
the proper solution $S$ of the antibracket formalism.

In this approach,
${\cal O}$ is an observable if it is BRST-invariant, i.e.,
$ \{ {\cal O} , Q_B \}_{PB} = 0$.
Thus, the hamiltonian is an observable.
Two observables ${\cal O}_1$ and ${\cal O}_2$
are considered equivalent if
${\cal O}_2 = {\cal O}_1 + \{ {\cal O}^{'} , Q_B \}_{PB}$,
for some ${\cal O}^{'}$.
A state $\ket\psi$ is called physical
if $Q_B \ket \psi = 0$.
Two states $\ket {\psi_1}$ and $\ket {\psi_2}$ are considered equivalent if
$\ket {\psi_2} = \ket {\psi_1} + Q_B \ket{ \psi'}$,
for some $\ket{\psi'}$.

Given a suitable hamiltonian $H$,
a lagrangian can be constructed via
\be
  \exp \left( {  \frac{i}{\hbar}
      S_H [ \Phi , \dot \Phi ] } \right)
 = \int  \frac{\left[ \dif \Pi \right] }{2 \pi i \hbar}
   \exp \left[ {  { \frac{i}{\hbar}
    \int \dif t \ ( \dot \Phi^i
      \Pi_i - H [ \Phi , \Pi ] ) } } \right]
\quad ,
\label{action construction}
\ee
where $\Phi$ denotes all degrees of freedom and $\Pi$
denotes the corresponding momenta.
We indicate the process of constructing an action $S_H$
from a hamiltonian $H$
by the following diagram
$$
\matrix{
S_H \cr
\Big\uparrow \cr
H \cr
        }
$$

In the BFV formalism,
to obtain a hamiltonian $H_\Psi$,
which is appropriate for insertion in the functional integral,
a fermion $\Psi$ with ghost number minus one
is used.
As in the antibracket formalism,
BRST trivial pairs exist.
Given two fields $\Lambda$ and $\Sigma$,
and their conjugate momenta,
$\agh P_\Lambda$ and $\agh P_\Sigma$,
a term $ \int \dif^{d-1} x \ \Sigma\, \agh P_{\Lambda}$
can be added to $Q_B$ without ruining nilpotency.
The next step in the BFV program
is to introduce
additional fields and their momenta
and add them as trivial pairs to $Q_B$.
These fields are
the analogs of the auxiliary gauge-fixing fields
of Sect.\ \ref{ss:gfaf}.
They include antighosts, extraghosts,
and the Lagrange-multiplier fields
of Eq.\bref{auxiliary fields}.
The fermion $\Psi$
in the hamiltonian formulation
must satisfy conditions similar to those
in Sect.\ \ref{ss:dfgfp}
for the $\Psi$ in the antibracket formalism.
We denote the BRST charge extended by
the inclusion of the additional trivial terms
by $Q^{\rm nm}_B$.
The hamiltonian $H_\Psi$ is given by
\be
  H_\Psi = H - \{ \Psi , Q^{\rm nm}_B \}_{PB}
\quad .
\label{def of H_Psi}
\ee
Let
$
 Z_\Psi = \int [ \dif  \Phi ]
  \exp \left( { \frac{i}{\hbar} S_{H_\Psi} } \right)
$,
where $ S_{H_\Psi}$ is constructed from $H_\Psi$
via Eq.\bref{action construction}.
The Fradkin-Vilkovisky theorem
\ct{fv75a}
states that
$ Z_\Psi$ is independent of $\Psi$.

The equivalence
of the BFV hamiltonian and antibracket methods
is established
if the remaining leg of the following diagram
$$
\matrix{
  & S_0 & \longrightarrow & S_\Psi & \cr
  & \Big\updownarrow &&& \cr
  & H_{S_0} & \longrightarrow & H_\Psi & \cr
        }
$$
can be completed.
In other words,
is $S_{H_\Psi}$,
as constructed from ${H_\Psi}$
via Eq.\bref{action construction},
equivalent to $S_\Psi$
as obtained from the antibracket formalism?
Likewise,
is $H_{S_\Psi}$, as constructed from ${S_\Psi}$
via Eq.\bref{H construction},
equivalent to the BFV hamiltonian ${H_\Psi}$?
Another question
is whether the gauge-fixed BRST charge $Q_{\rm Noether}$,
as constructed from $S_\Psi$ using Noether's theorem,
coincides with the BRST charge
$Q^{\rm nm}_B$ for the BFV formalism.
The affirmative answer to the above questions,
obtained
in refs.\ct{fh89a,dfgh91a},
implies that construction processes in
$$
\matrix{
  & S_0 & \longrightarrow & S_\Psi & \cr
  & \Big\updownarrow & & \Big\updownarrow & \cr
  & H_{S_0} & \longrightarrow & H_\Psi & \cr
        }
$$
commute to give equivalent results.

One can also ask whether
an equivalence occurs before the introduction
of gauge-fixing and $\Psi$:
$$
\matrix{
S \ \cr
\Big\downarrow ? \cr
H \ \cr
        }
$$
Clearly, a straightforward correspondence
cannot exist because $S$ contains antifields.
However, at least for closed irreducible theories,
if certain antifields are set to zero and others
are identified with ghost momenta,
then an equivalence of $H_S$,
as constructed from $S$
via Eq.\bref{H construction},
and the BFV $H$ is achieved
\ct{bgpr89a}.
Similar results have been obtained
in refs.\ct{siegel89a,dfgh91a}.

If sources for the
BRST transformations are included
at the hamiltonian level,
the above correspondence can be made clearer.
Then, the sources in the hamiltonian formulation
can be identified with antifields
in the antibracket formalism.
This method was used
in refs.\ct{ggt91a,dejonghe93a,dejonghe93b}
to establish
the equivalence in the gauge-fixed basis.

An open problem is to extend all of the above analysis
to the quantum case in a rigorous manner.
That situation is more difficult
due to operator ordering problems
and the singular character of field theories.

\subsection{Unitarity}
\label{ss:u}

\hspace{\parindent}
The difficulty in proving unitarity
in covariant approaches
to quantizing gauge theories
is due to the presence of ghosts
and
of unphysical degrees of freedom with negative norms.
One often deals
with indefinite-metric Hilbert spaces.
Unitarity can be spoiled
in theories with
kinetic energy terms of the wrong sign
and/or
non-hermitian interaction terms.
Wrong sign kinetic energy terms
almost always arise in gauge theories
with particles of spin one or higher.
Due to the sign of the metric component $\eta_{00}$,
there are potential difficulties with
the temporal components,
such as $A_0$ in electromagnetism,
$A_0^a$ in Yang-Mills theories,
and $g_{0i}$ in gravity.
Faddeev-Popov and other gauge-fixing ghosts
enter in loops with the wrong sign,
and would lead to a violation of unitarity,
if their contributions were considered in isolation.

Let us summarize how unitarity is established
in certain covariant quantization procedures.
First of all,
one needs to assume that
there are not any non-hermitian
interactions in the original theory and that
the spatial components of tensors have the correct sign
in kinetic energy terms.
In other words,
the theory should be ``naively'' unitary.

The first approach is as follows.
In some theories,
there exists a unitary gauge,
in which it is evident that
the unphysical excitations are not present.
If one can establish the gauge invariance
of the $S$-matrix,
then unitarity can be proven by
going from a covariant gauge to an unitary one
\ct{fs80a}.
Unfortunately, this method is only well developed
for irreducible theories with closed algebras.
For reducible systems,
this approach often encounter difficulties,
although for some specific examples
it has been successfully implemented
\ct{fs88a}.

Another method for checking unitarity
is in perturbation theory
via Feynman diagrams
\ct{thooft71a,tv73a}.
Using the Ward-Takahashi
\ct{ward50a,takahashi57a} or Slavnov-Taylor identities
\ct{taylor71a,slavnov72a},
as well as the Landau-Cutkosky rules
\ct{elop66a},
one tries to show
that contributions from the unphysical polarizations
of the classical fields
are cancelled by contributions from ghost fields
or from other sources.

A third approach proceeds via canonical quantization.
The ``physical sector''
is selected out
by imposing some subsidiary conditions
that
remove negative norm states.
The physical sector
should be stable under time evolution
and
should involve a non-negative metric.
A well-known example of this approach
is the Gupta-Bleuler procedure
\ct{bleuler50a,gupta50a}
for quantizing QED.
All components of the electromagnetic field $A^\mu$
are used;
however only states $\ket \varphi_{\rm phys}$
satisfying
\be
   (\partial_\mu A^\mu)^{+}\ket\varphi_{\rm phys} = 0
\label{gupta}
\ee
are considered,
where $(\partial_\mu A^\mu)^{+}$ denotes
the positive frequency components of $\partial_\mu A^\mu$.
This condition determines
the physical sector ${\cal H}_{\rm phys}$
in the Gupta-Bleuler procedure.
Unfortunately,
when applied to non-abelian Yang-Mills theories,
this method fails to
preserve ${\cal H}_{\rm phys}$ under time evolution.
To quantize covariantly non-abelian gauge theories,
refs.\ct{cf76b,ko78a,ko79a}
proposed
\be
    Q_B \ket\varphi_{\rm phys} = 0
\quad ,
\label{subs cond}
\ee
where $Q_B$ is the hermitian nilpotent BRST operator.
Eq.\bref{subs cond}
is the basis for BRST quantization.
We use $V_{\rm phys}$ to denote the space of states
annihilated by $Q_B$.
In the BRST approach,
the hamiltonian is automatically hermitian
so that the $S$-matrix is a unitarity operator in $V_{\rm phys}$.
However, there is a possible difficulty with $V_{\rm phys}$.
Despite the fact that $Q_B$ commutes with the hamiltonian,
the positive semidefiniteness
of the norm of
$V_{\rm phys}$
is not ensured.
The question of unitarity in
BRST quantization
becomes that of proving
the positive semidefiniteness of $V_{\rm phys}$,
and must be analyzed model by model.

However,
T.\ Kugo and I.\ Ojima \ct{ko78a,ko79a}
(see also \ct{no90a}),
obtained criteria under which unitarity does hold.
They established a connection
with the metric structure of $V_{\rm phys}$
and the multiplets
of the algebra generated by the conserved BRST charge
$Q_B$ and the conserved ghost number charge $Q_C$.%
{\footnote{ We assume there are no anomalies
associated with $Q_B$ and $Q_C$.
For the first-quantized string,
this is actually not the case for $Q_C$,
but BRST quantization is still possible
\ct{hwang83a,ko83a,fgz86a}.
For similar analyses in other models
see refs.\ct{bmp92a,bbrt93a,bln93a,bmp93a}.
}}
With our conventions, $Q_C$ is antihermitian:
$Q_C^{\dagger}=-Q_C$.
These generators satisfy
\be
    [ Q_C , Q_B ] = Q_B
\ , \quad\quad
   [ Q_C , Q_C ] = 0
\ , \quad\quad
    \frac12 \{ Q_B , Q_B \} = Q_B^2 = 0
\quad .
\nonumber
\ee
Three types of multiplets are possible:

\medskip
(a) ``True physical states'': BRST singlets with zero ghost number.
\medskip

(b) Doublets: pairs of BRST singlets related by ghost conjugation.
\medskip

(c) Quartets: pairs of BRST doublets related by ghost conjugation.

\medskip

Roughly speaking, ghost conjugation is the operation
that interchanges
ghosts and antighosts.
Under this operation,
the sign of the ghost number of a state is flipped.
In the next three paragraphs,
we explain the classification of the multiplets.

One can choose states to be eigenfunctions of $Q_c$.
Let $\ket g$ be a state with a non-zero ghost number $g$.
Then $\ket g$ has zero norm
since $\bra g Q_c \ket g = g \bracket{g}{g} = -g \bracket{g}{g}$,
the first equality arising when $Q_c$ acts to the right,
and the second equality arising when $Q_c$ acts to the left.
Non-zero matrix elements occur only when bra and ket states
have opposite ghost numbers.
Under application of $Q_B$,
the ghost number of a state is increased by one.
Such states $\ket s = Q_B\ket{s'}$ also have null norms
since $\bracket{s}{s} = \bra{s'}Q_B Q_B\ket{s'} = 0$.

Due to the nilpotency of $Q_B$,
the representations are either BRST singlets or BRST doublets.
A BRST singlet
$\ket s$ satisfies $Q_B\ket{s} = 0$
and $\ket{s} \ne Q_B\ket{s'}$ for any $\ket{s'}$.
If $Q_B\ket{s'} = \ket{s} \ne 0$,
then $\ket{s'}$ is a member of the BRST doublet
consisting of $\ket{s'}$ and $\ket{s}$.
The upper member of a doublet $\ket{s}$
is annihilated by $Q_B$,
since $Q_B \ket{s} = Q_B Q_B\ket{s'} = 0$,
and it carries one unit of ghost number more than $\ket{s'}$:
$Q_c \ket{s} = Q_c \ket{s'} + 1$.

If $\ket{s}$ is a BRST singlet and carries
ghost number zero, then it is of type (a).
If $\ket{s}$ is a BRST singlet and carries
non-zero ghost number $g$, then it is of type (b).
Under ghost conjugation, another BRST single
with ghost number $-g$ is created,
thus forming the pair.
If $\ket{s}$ and $\ket{s'}$ constitute a BRST doublet,
then ghost conjugation produces another BRST doublet
and a type (c) multiplet is obtained.

For an irreducible gauge theory,
T.\ Kugo and I.\ Ojima in \ct{ko78a,ko79a}
proved that
(i) {\it if type} (a) {\it states have positive definite norm}
and
(ii) {\it if type} (b) {\it states are absent},
then quartets only appear in
$V_{\rm phys}$
through zero norm combinations.
Consequently,
when (i) and (ii) are satisfied,
$V_{\rm phys}$ has a positive
semidefinite norm.
To obtain a unitary theory,
one mods out the null-norm states:
Two states are identified if they differ by a null-norm vector.
Clearly, null-norm states are identified
with the null state.
The modding-out procedure
automatically restricts states
to the zero-ghost number sector,
since states with non-zero-ghost number have zero norms.
Furthermore, because BRST-trivial states
$ Q_B\ket{s'}$ are  null-norm vectors,
all that remains after modding out
are the non-trivial elements
of the $g=0$ BRST cohomology, i.e.,
states with ghost number zero that are annihilated by $Q_B$
and that cannot be expressed as
$ Q_B\ket{s'}$ for any state $\ket{s'}$.
This sector is preserved under time evolution
because $Q_B$ and $Q_C$ commute with the hamiltonian.

In the $g=0$ sector,
it makes sense to identify null-norm states
with the null vector
because they decouple
from matrix elements involving observables,
such as the hamiltonian.
Observables ${\cal O}$ are BRST-invariant operators:
$[ {\cal O} , Q_B ] = 0$.
If $\ket t$ is a BRST-trivial state,
so that $\ket t = Q_B\ket{s'}$,
and if $\ket s$ is any element
of $V_{\rm phys}$, so that $Q_B \ket s = 0$,
then
$
  \bra s {\cal O} \ket t =
  \bra s {\cal O} Q_B\ket{s'} =
  \bra s Q_B {\cal O} \ket{s'} = 0
$.

For reducible systems,
ghosts for ghosts and extraghosts arise,
some of which have zero ghost number.
Hence
a third condition arises for reducible theories:
(iii)
{\it a state of $V_{\rm phys}$
involving ghosts in the $g=0$ sector
must be a member of a quartet multiplet}.
This guarantees that they are null vectors
and do not ruin the
positive semidefiniteness of $V_{\rm phys}$.

The above conditions provide
criteria for establishing the positivity of the norm
and hence unitarity in a covariant formulation.
Reference
\ct{cf76b,ko78a,ko79a}
established unitarity for Yang-Mills theories
by proving (i) and (ii) for this case.

In perturbation theory
and in a Fock space representation,
A.\ Slavnov
in \ct{slavnov89a}
used (i)--(iii)
to obtain simpler criteria.
The most important requirements,
apart from the positivity of the norm of type (a) states,
were that $Q_B$ be nilpotent
and that it have nontrivial action
on all ghost fields or their conjugate momenta.
Under these conditions,
$V_{\rm phys}$ has a positive semidefinite norm.
Then,
S.\ A.\ Frolov and A.\ Slavnov
\ct{fs89a}
using the hamiltonian BFV-BRST formalism
for lagrangians $L$ of the form
in Eqs.\bref{H construction} and \bref{H_S_0},
verified the above-mentioned conditions perturbatively.
The analysis was simplified because
one could use the free BRST charge $Q^{(0)}_B$.
The criteria became that $Q^{(0)}_B Q^{(0)}_B = 0$
and that $Q^{(0)}_B$
have non-trivial action on all ghosts.
Given the validity of perturbation theory,
their result on the unitarity of a gauge theory
holds for the finite reducible case.

S.\ A.\ Frolov and A.\ Slavnov
in ref.\ct{fs90a,slavnov90a},
were able to translate
the above program into a lagrangian approach,
by using an effective action $A_{eff}$.
The action and BRST charge
were perturbatively expanded in a series:
$A_{eff} = A_{eff}^{(0)} + \dots$ and
$Q_B = Q^{(0)}_B + \dots$.
The term $A_{eff}^{(0)}$
was the leading order part
of the general gaussian gauge-fixed action $S_{\Psi}$
of the field-antifield formalism
presented
in Sect.\ \ref{ss:ogfp}.
Requiring nilpotency and BRST invariance of the action
lead to
a series of recursion relations
for the higher order terms
in $A_{eff}$ and $Q_B$.
The action $A_{eff}$, thus obtained,
is constructed using unitarity requirements.
Finally, when certain conditions on the
rank of the
gauge generators are imposed,
the free BRST charge
is seen to act non-trivially on ghosts fields
and unphysical polarizations of the classical fields,
thereby yielding a unitary theory
if the classical gauge-invariant degrees of freedom
have a positive norm.

The problem of unitarity
in the field-antifield formalism was addressed in
\ct{gp92a,paris92a,oppt93a,opt93a}.
A perturbative solution of the proper solution $S$
was obtained in
\ct{gp92a,paris92a,gp93b}
(see also ref.\ct{bh93a}).
Then,
a general gaussian gauge-fixing procedure was performed,
using a fermion $\Psi$
of the type given
in Sects.\ \ref{ss:dfgfp} and \ref{ss:ogfp}.
It was shown that
BRST invariance of the gauge-fixed action
and nilpotency of the gauge-fixed BRST transformation
lead to the same recursion relations
obtained in \ct{fs90a,slavnov90a},
and that
the leading two terms of $S_\Psi$ agree
with $A_{eff}$.
The conclusion
is that the field-antifield formalism
produces an action $S_\Psi$ that coincides
with $A_{eff}$
of ref.\ct{fs90a,slavnov90a}
obtained by unitarity considerations.

The above approaches to unitarity
are formal
in that the difficulties
with field-theoretic infinities are not addressed.
The renormalizability or non-renormalizability
is not used.
To proceed rigorously,
one needs to regulate the theory
with a cutoff,
verify unitarity,
and then make sure that unitarity remains
as the cutoff is removed.
The issue of locality also enters here.
For example,
it may happen that $A_{eff}$ or $S$
contains non-local terms.
This does not necessarily ruin unitarity,
but might signal that the theory is non-renormalizable
or ill-defined.
Studies of unitarity without using perturbation theory
for general systems with finite degrees of freedom,
such as in quantum mechanics,
have been carried out
in ref.\ct{marnelius93a}.

\subsection{The Antibracket Formalism in General Coordinates}
\label{ss:afgc}

\hspace{\parindent}
The antibracket formalism in a general coordinate system
has been developed
in refs.\ct{witten90a,khudaverdian91a,%
bt93a,kn93a,schwarz93a,schwarz93b,hz94a}.
A brief overview is given
in \ct{ht92a}.
This approach sometimes goes by the name
{\it covariant formulation of the field-antifield formalism}.
Possible applications are in mathematics
\ct{ps92a,av93a,lz93a,nersesian93a,horava94a} and
in string field theory
\ct{ps92a,witten92a,lz93a,hz94a,horava94a,sz94a,sz94b}.

Consider a
supermanifold ${\cal M}$ of type $(N,N)$,
meaning that there are
$N$ bosonic and $N$ fermionic coordinates.
Collectively denote these as $z^a$, $a=1,\ldots, 2N$.
In this coordinate system,
a local basis for the cotangent space ${\cal{T^*M}}$
consists of the $1$-forms $d z^a$, $a=1,\ldots, 2N$.
The Grassmann parity
of a differential is the same as that
of the corresponding coordinate:
$\eps(d z^a)=\eps(z^a) = \eps_a$.
Introduce an odd two-form $\zeta$,
$\eps(\zeta)=1$,
which is non-degenerate and closed, i.e.,
$d \zeta = 0$.
In the local basis,
$\zeta$ is expressed as
\be
  \zeta = - \frac12 \zeta_{ab}(z) d z^b \wedge d z^a =
  \frac12  d z^b \wedge \zeta_{ba}(z) d z^a
\ , \quad \quad
  \zeta_{ab}=(-1)^{\eps_a \eps_b + 1} \zeta_{ba}
\quad ,
\label{zeta}
\ee
where $\eps ( \zeta_{ab} ) = \eps_a + \eps_b + 1 $ (mod 2).
Let $\zeta^{ab}$ be the inverse
of the matrix $\zeta_{ab}$.
It obeys
$
 \zeta^{ab} =
  (-1)^{\eps_a + \eps_b + \eps_a \eps_b } \zeta^{ba}
$.
One then defines the antibracket
via Eq.\bref{sym form of antibracket}
but using $\zeta^{ab} (z)$.
Alternatively,
let $X$ be a function on ${\cal M}$.
Then one can define a vector field
$
  {\mathop V \limits^\leftarrow }_X =
  {{\mathop \partial \limits^\leftarrow }
      \over {\partial z^a }}
  \zeta^{ab}\lder{X}{z^b}
$
that acts
from right to left on functions $Y$ of ${\cal M}$:
$
 [Y] {\mathop V \limits^\leftarrow }_X =
 \rder{Y}{z^a}\zeta^{ab}\lder{X}{z^b}
$.
Then the antibracket
can be written as
\be
   (X,Y) = \zeta[ {\mathop V \limits^\leftarrow }_X ,
   {\mathop V \limits^\leftarrow }_Y ] =
   [X] {\mathop V \limits^\leftarrow }_Y
\quad ,
\label{gc def of antibracket}
\ee
where the first equality follows from
$$
dz^b \wedge dz^a [\rder{}{z^c} , \rder{}{z^d} ] =
 \delta_c^a \delta_b^d -
  (-1)^{\eps_a \eps_b} \delta_c^b \delta_a^d
\quad .
$$
In this way,
$\{ {\cal M}, \zeta \}$
becomes an odd symplectic structure.
The antibracket, defined as above,
obeys the properties
in Eqs.\bref{antibracket properties}
and \bref{bracket derivation}.
It turns out that $d \zeta = 0$
is necessary for the Jacobi identity
in Eq.\bref{antibracket properties}.

For ordinary symplectic manifolds,
there exists a natural volume element $d \mu$
obtained by wedging $\zeta$ with itself $N$ times.
Unfortunately, for an old sympletic manifold,
$\zeta \wedge \zeta = 0$.
Hence, a measure must be introduced by hand:
\be
  d \mu(z) = \rho (z) \prod_{a=1}^{2N} d z^a
\quad ,
\label{volume element}
\ee
where $\rho(z)$ is a density.
The divergence of a generic vector field
$
 {\mathop V \limits^\leftarrow } =
{{\mathop \partial \limits^\leftarrow }
      \over {\partial z^a }} V^a
$
is defined in the usual way by
\be
   {\rm div}_\rho {\mathop V \limits^\leftarrow } \equiv
  \frac1{\rho}(-1)^{\eps_a} \lder{(\rho V^a )}{z^a}
\quad .
\label{gc divergence}
\ee
Then,
the laplacian $\Delta_\rho$
acting on a function $X$ is defined
by taking the divergence
of the corresponding vector field via
\be
   \Delta_\rho X\equiv (-1)^{\eps_X}
   \frac12{\rm div}_\rho
  {\mathop V \limits^\leftarrow }_X =
   (-1)^{\eps_X +\eps_a }
   \frac1{2 \rho} \lder{}{z^a}
   \left(\rho\zeta^{ab}\lder{X}{z^b}\right)
\label{gc delta}
\quad .
\ee
Since one would like to use $\Delta_\rho$
as the general coordinate version
of $\Delta$
of Sect.\ \ref{ss:g},
one wants it to be nilpotent.
However, this is not necessarily the case since
$$
    \Delta_\rho \Delta_\rho =
    \frac12 \left[ {
    \Delta_\rho \left(\frac1{\rho}
     (-1)^{\eps_a + \eps_b}
    \lder{}{z^a} \left(\rho\zeta^{ab}\right)\right)
                  } \right]
    \lder{}{z^b}
\quad .
$$
Hence, one requires $\rho$ to satisfy
\be
    \Delta_\rho \left(\frac1{\rho}
     (-1)^{\eps_a + \eps_b}
    \lder{}{z^a} \left(\rho\zeta^{ab}\right)\right)
                   = 0
\quad .
\label{density requirement}
\ee
When Eq.\bref{density requirement}
holds,
$\Delta_\rho$ is formally nilpotent and
a graded derivation of both functional multiplication
and the antibracket, i.e.,
it satisfies
Eq.\bref{Delta properties}
with $\Delta \to \Delta_\rho$.

It turns out that
Eq.\bref{density requirement}
is the necessary and sufficient condition
for the existence of Darboux coordinates locally.
For such coordinates,
$\rho = 1$ and $\zeta^{ab}$
takes the form
in Eq.\bref{sym form of antibracket}.
Then $z^a$ for $a=1,\ldots, N$ can
be identified with fields
and $z^a$ for $a=N+1,\ldots, 2N$ can
be identified with antifields.
Hence,
we employed the Darboux coordinate system
for the antibracket formalism
in Sects.\ \ref{s:faf} -- \ref{s:qea}.
Darboux coordinates suffice
as long as global issues are not important.

Quantization in a general coordinate system proceeds
as in the Darboux case.
Everywhere $\Delta$ appears
in Sects.\ \ref{s:gff} -- \ref{s:qea},
one replaces it by $\Delta_\rho$.
The functional-integral measure
also must be modified.
Integration is restricted
to an $N$-dimensional submanifold ${\cal N}$.
Since little distinction is made
between fermionic and bosonic coordinates
in the covariant formulation,
${\cal N}$ can be an arbitrary $(k, N-k)$ submanifold
as long as $\zeta[ V, V' ] = 0$ on ${\cal{N}}$
for any two tangent vectors $V,\, V'\in {\cal{T N}}$.
One considers a basis
$\{e_1,\ldots, e_N; h^1, \ldots, h^N\}$
for ${\cal{T M}}$,
such that
$\{e_1,\ldots, e_N\}$ is a basis for
${\cal{T N}}$
and $\zeta [ e_i, h^j ] = \delta_i^j$.
Then the volume element on ${\cal N}$ is
\be
  d\mu_{\cal N}(e_1,\ldots, e_N) =
  \left[ d\mu(e_1,\ldots, e_N ; h^1,\ldots,h^N) \right]^{1/2}
\label{gc measure}
\quad .
\ee
Since integration in function space is restricted
to ${\cal N}$,
the above procedure corresponds to a gauge-fixing procedure.
The submanifold ${\cal N}$
can be defined using $N$ linearly independent
constraints $\Psi_A(z)=0$ satisfying
\be
  ( \Psi_A, \Psi_B ) = T^C_{AB}(z) \Psi_C
\quad .
\label{involution}
\ee
The vectors ${\mathop V \limits^\leftarrow }_{\Psi_A}$
are a basis for the tangent space ${\cal{T N}}$ of ${\cal N}$.
Furthermore,
as a consequence
of Eq.\bref{involution},
$
\zeta[ {\mathop V \limits^\leftarrow }_{\Psi_A} ,
   {\mathop V \limits^\leftarrow }_{\Psi_B} ] = 0
$
on ${\cal N}$,
which is a consistency check.
To make contact with
the gauge-fixing procedure
of Sect.\ \ref{s:gff},
one goes to the Darboux coordinate system
and chooses
$
 \Psi_A  = \Phi^*_A - \der{\Psi}{\Phi^A}
$.
One disadvantage of the general coordinate approach,
is that the concept of ghost number becomes obscure.

\subsection{The D=26 Closed Bosonic String Field Theory}
\label{ss:d26cbsft}

\hspace{\parindent}
A review of the current formulation of the closed bosonic string
has been given
in ref.\ct{zwiebach93a}.
Here, we present some of the salient points.

At the first-quantized level,
closed strings possess holomorphic factorization.
This means that, with the possible exception of zero modes,
the integrands of closed-string amplitudes
factorize into two open-string-like integrands,
one for left-moving degrees of freedom
and one for right-moving degrees of freedom.
At the second-quantized level,
there is a similar splitting.
Hence, a closed string field $A$ is a tensor product
of a left string field $A_L$
with a right string field $A_R$,
so that one can write $A$ as
\be
  A = \left( { A_L; A_R } \right)
\quad ,
\label{cs field factorization}
\ee
or as a sum of terms of the form
in Eq.\bref{cs field factorization}.
The field $A_L$ (respectively, $A_R$)
is precisely of the form
of the open string case,
except a subscript $L$ (respectively, $R$)
is appended to all quantities.
One exception is the zero modes
of $X^\mu (\sigma)$, namely
the position and momentum operators.
They are the same for both left and right sectors,
so that
$x_L^\mu = x_R^\mu \equiv x^\mu$
and
$p_L^\mu = p_R^\mu \equiv p^\mu$.
The total string ghost number is the sum
of the left and right string ghost numbers, i.e.,
$
  g\left( A \right) =
  g_L \left( {A_L} \right) + g_R \left( {A_R} \right)
$.

One can attempt to construct closed-string field theory
along the lines
of the open string case described
in Sect.\ \ref{ss:obsft}.
It is easy to see that
not all the open string axioms can
be extended.
When the axioms hold,
Paton-Chan factors
\ct{pc69a}
can be appended to the string field
leading to a non-abelian Yang-Mills gauge group.
However, closed string theories
cannot possess such a non-abelian gauge structure
\ct{schwarz82a}.

Define closed-string integration
$\int\limits_{\rm closed}$
as the product
of left and right open string integrals via
$
  \int\limits_{\rm closed} {} \equiv \int_L {} \int_R {}
$,
i.e.,
for fields in the form
in Eq.\bref{cs field factorization},
one has
\be
  \int\limits_{\rm closed} A = \int_L {A_L} \int_R {A_R}
\quad .
\label{cs integration}
\ee
To generate a non-zero integral
in Eq.\bref{cs integration},
$A$ must have a left-ghost number of $3$
and a right-ghost number of $3$ and
consequently a total ghost number of $6$:
$$
  \int\limits_{\rm closed} A = 0
\ , \quad
  {\rm if} \ g \left( A \right) \ne 6
\quad .
$$

Let $\sigma$ be the first-quantized variable
parametrizing the string.
It varies between $0$ and $2 \pi$
and is periodic.
To define
the closed-string star operation $\circ$,
pick two antipodal points,
e.g. $\sigma =0$ and $\sigma=\pi$.
This divides a string into two halves.
Then, $\circ$ is defined in analogy
to the open string case.
One half of one string overlaps
with one half of the other string
and what remains is the product string.
The first-quantized BRST charge $Q$
is the sum of the right and left parts:
\be
  Q = Q_L + Q_R
\quad ,
\label{cs Q}
\ee
and it carries ghost number one:
$
  g \left( Q \right) = 1
$.
Eq.\bref{cs Q} is equivalent to
\be
  Q A =
  \left( {Q_L A_L; A_R} \right) +
  \left( {A_L; Q_R A_R} \right)
\quad .
\label{cs Q2}
\ee

Even though $\int$, $\circ$ and $Q$
have been defined,
there is a difficulty in obtaining a free action.
Let $C$ denote the closed-string field.
The naive term
$
  \int {C \circ Q C}
$
vanishes because of ghost number considerations.
The total ghost number of the integrand,
which is
$
  2g \left( C \right) + 1
$,
must be equal to $6$.
This constraint cannot be satisfied because
$g \left( C \right)$ is an integer.

To correct the problem,
various schemes can be used
\ct{bfmpp86a,lr86a,kaku88a}.
Let
\be
  \bar c_0^- \equiv \bar c_0^L - \bar c_0^R
\ , \quad \quad
  c_0^- \equiv \frac12 \left( { c_0^L - c_0^R } \right)
\quad .
\label{c minus}
\ee
Impose the following two constraints
on $C$
\be
  \bar c_0^- C = 0
\ , \quad \quad
  \left( { L_0^L - L_0^R } \right)  C = 0
\quad .
\label{c bar minus constraint}
\ee
Anticipating that the free-theory equation
is $Q C = 0$,
one sees that these constraints are consistent since
$
  \left\{ {Q, \bar c_0^-} \right\} = L_0^L - L_0^R
$,
$
  \left[ {Q , L_0^L - L_0^R } \right] = 0
$ and
$
  \left[ {\bar c_0^- , L_0^L - L_0^R } \right] = 0
$.
The condition
$\left( { L_0^L - L_0^R } \right)  C = 0$
is quite natural.
The operator
$L_0^L - L_0^R$ is the generator for rigid rotation
of the first-quantized parameter $\sigma$,
which labels the points along the string.
Since, in the closed-string case,
$\sigma$ is periodic,
there is no preferred choice of an origin.
The condition
$\left( { L_0^L - L_0^R } \right)  C = 0$
does not lead to an equation of motion for $C$
since the $ \frac12 \partial_\mu \partial^\mu $
terms in $L_0^L$ and $L_0^R$ cancel.
{}From Eq.\bref{L 0},
one sees that it implies
that the mass ${\cal M}_L$ of the left-sector must equal
the mass ${\cal M}_R$ of the right-sector,
a well-known constraint of first-quantized
closed-string states.
However, the operation $\circ$
does not preserve the constraint.
One can modify $\circ$ to $\hat \circ$ by averaging
over a rigid rotation that rotates the product string
over angles ranging from $ 0 $ to $2 \pi$.
The new $\hat \circ$ no longer is associative,
as can be checked by drawing some pictures.

The ghost number problem
can be fixed by inserting a factor of $c_0^-$.
Define the quadratic form
\be
  \left\langle {A,B} \right\rangle =
   \int\limits_{\rm closed} {A \hat \circ c_0^- B}
\quad .
\label{cs two form}
\ee
Let the free action be
\be
  S_0^{\left( 2 \right)} =
  {1 \over 2}\left\langle { C, QC } \right\rangle
\quad .
\label{cs free action}
\ee
It is invariant under the gauge transformations
$
  \delta C = Q \Lambda
$,
where $g \left( { \Lambda } \right) =1$
since $g \left( { C } \right) =2$.
There are many ways
of resolving
the difficulty with ghost number
of the free action of closed-string field theory
but they are equivalent to the above.
The tree-level three-point interaction is
\be
  S_0^{\left( 3 \right)} =
  {{2g} \over 3} \int\limits_{\rm closed}
   C \hat \circ C \hat \circ C
\quad .
\label{cs 3-point interaction}
\ee
For on-shell external states,
this interaction correctly produces three-point interactions.

Tree-level gauge invariance
is violated for the theory described by
$ S_0^{\left( 2 \right)} + S_0^{\left( 3 \right)}$
due to the violation of the associativity axiom.
However, by adding higher-order terms
gauge invariance can be restored
\ct{kaku88a,kaku88b,kl88a,kks89a,sz89a,%
ks90a,zwiebach90a}.
The new interactions can be defined by relatively
simple geometrical constraints
\ct{zwiebach90a,zwiebach93a}.
This leads to a tree-level
non-polynomial closed-string field theory.
Unfortunately, the classical theory needs
further modification at the quantum level.
One-loop and higher-loop amplitudes are not produced
by using only tree-level vertices.
It is at this stage where the antibracket formalism
has been of great utility.
Interaction terms proportional to powers of $\hbar$
need to be added
in a manner similar
to Eq.\bref{hbar expansion of W}.
To ensure that the theory
is quantum-mechanically gauge invariant,
the work
in refs.\ct{zwiebach90a,sz93a,zwiebach93a,zwiebach93b,%
hz94a,sz94a}
has relied on the antibracket formalism.
The guiding principle is that
the quantum closed-string field theory
must satisfy
the quantum master equation.

The antibracket is defined
using the quadratic form
in Eq.\bref{cs two form}
\ct{sz93a,zwiebach93a,zwiebach93b}.
As in the open string field theory,
the system is infinitely reducible
so that there are ghosts for ghosts ad infinitum.
The fields can be collected into one object
$\Psi_c$ in a manner similar
to the open string case in Eq.\bref{psi sum}
\be
 \Psi_c \equiv \ldots +
   \stackrel{{\textstyle{s+4}}}{{}^*{\cal C}_s^*}
   + \ldots +
   \stackrel{{\textstyle{4}}}{{}^*{\cal C}_0^*}
 + \stackrel{{\textstyle{3}}}{{}^*C^*}
 + \stackrel{{\textstyle{2}}}{C}
 + \stackrel{{\textstyle{1}}}{{\cal C}_0}
 + \ldots +
   \stackrel{{\textstyle{-s+1}}}{{\cal C}_s}
 + \ldots
\quad ,
\label{cs psi sum}
\ee
where the ghost number is indicated above the field.
The close-string Hodge operation,
denoted by a superscript $*$ in front of a field,
is defined using the quadratic form
in Eq.\bref{cs two form}.
It takes a $p$-form into a $5-p$,
where the order of string form
is the same as the ghost number.

Let
$ \varphi_s $
be a complete set of normalized first-quantized states
for $g \left( {\varphi_s} \right) \le 2$.
Let ${}^* \varphi^s $ denote the corresponding state
transformed by the Hodge star operation.
The ${}^* \varphi^s $
are a normalized complete set of states for ghost numbers
greater than $2$.
With these definitions,
$\Psi_c$
in Eq.\bref{cs psi sum}
can be written
in terms of ordinary particle fields $\psi^s$
via
$
  \Psi_c =
  \sum_s \left(
 {  \varphi_s \psi^s +
    {}^* \varphi^s \psi_s^*
 } \right)$
where $\psi_s^*$ are the antifields of $\psi^s
$.
The quantum master equation is
then the same as the particle case,
namely
$
   {1 \over 2} \left( { W( \Psi_c ), W( \Psi_c ) } \right) =
   i\hbar \Delta W( \Psi_c )
$,
where the antibracket is
$
 \left( {
    X \left( { \Psi_c } \right) ,
    Y \left( { \Psi_c } \right)
         } \right)
  = \frac{\partial_r X} {\partial\psi^s}
     \frac{\partial_l Y} {\partial \psi^*_s}
   - \frac{\partial_r X} {\partial \psi^*_s}
     \frac{\partial_l Y} {\partial\psi^s}
$
and
$
  \Delta =
  \left( {-1} \right)^{\epsilon_s+1}
  {{\partial_r } \over {\partial \psi^s}}
  {{\partial_r } \over {\partial \psi_s^*}}
$.

The solution of the quantum master equation
for the closed-string field theory
is presented
in refs.\ct{zwiebach93a,zwiebach93b}.
This tour-de-force work goes beyond
the goals of our review.
The reader interested in this topic
can consult the above references
for more discussion.

The current formulation of string fields theories
is developed around a particular
space-time background.
Any background is permitted,
as long as it leads
to a nilpotent first-quantized BRST charge.
Such BRST charges
correspond to two-dimensional conformal field theories
with the total central charge
of the Virasoro algebra equal to zero.
Usually, the flat space-time background
in $26$ dimensions is used.
Since string theories contain gravity,
it should be possible to pass
from one background to another.
It is an interesting question
of whether there
is background independence of string field theory
\ct{sen90a,ss92a,witten92a,sz93a,witten93a,sz94a}.
A proof for bosonic string field theories
has been obtained
in refs.\ct{sz93a,sz94a},
for backgrounds infinitesimally close.
The antibracket formalism has played
an important role in the analysis.
The basic idea is that string field theory,
formulated about a particular background ${\cal B}$,
corresponds to a particular solution $S_{\cal B}$
of the classical master equation.
Reference \ct{sz93a}
demonstrated that,
for two nearby backgrounds
${{\cal B}_1}$ and ${{\cal B}_2}$,
$S_{{\cal B}_1}$ and $S_{{\cal B}_2}$
are related by a canonical transformation
of the antibracket.
The conclusion is that string field theory
is background independent,
although not manifestly.
Barring difficulties with singular expressions,
ref.\ct{sz94a}
has extended the result to the quantum case.
For the quantum system, a particular background ${\cal B}$
corresponds to a solution
$W_{\cal B}$ of the quantum master equation.

\subsection{{Extended Antibracket Formalism for \hfil}
\break
Anomalous Gauge Theories}
\label{ss:eafat}

\hspace{\parindent}
In certain cases,
it is possible
to quantize an anomalous gauge theory.
An example is
the first-quantized bosonic string
for $D \ne 26$
discussed in Sect.\ \ref{ss:aobs}.
Polyakov \ct{polyakov81a}
quantized this system
in the presence of a conformal anomaly.
A new degree of freedom,
the Liouville mode, emerged.
Another example is the chiral Schwinger model
in two dimensions.
Despite its anomalous behavior,
it is consistent and unitary
\ct{jr85a}.
In refs.\ct{faddeev84a,fs84a},
additional degrees of freedom were introduced
into four-dimensional anomalous Yang-Mills theories
to cancel the anomalies in the path integral.
This cancellation can be obtained by
adding a Wess-Zumino term
for the extra degrees of freedom.
A careful treatment of
the integration measure in
Faddeev-Popov path-integral quantization
shows how such a Wess-Zumino term can arise naturally
\ct{paris93a}.
For earlier approaches to this subject in the case of
the Schwinger model,
see \ct{bsv86a,ht87a}.
A treatment of anomalous chiral QCD in two dimensions
within the field-antifield formalism was obtained
in refs.\ct{bm91a,bm94a}.
Methods of quantizing anomalous gauge theories
using the antibracket formalism
were developed
in \ct{bm91a,bm93a,gp93a,gp93c,gp93d,jst93a,bm94a}.

Let us describe in general terms
the extended antibracket method
of refs.\ct{gp93a,gp93c,gp93d}.
For simplicity we consider the closed irreducible case.
At the classical level,
the number of dynamical local degrees of freedom
$n_{\rm dof}$ is
the total number of fields $n$
minus the number of gauge invariances $m_0$, i.e.,
$n_{\rm dof} = n - m_0$.
Suppose there are
$r$ anomalous gauge invariances.
Then, due to quantum effects,
$r$ of the $m_0$ gauge degrees of freedom
enter the theory dynamically.
Hence, the true net number of degrees of freedom at quantum level is
$n - m_0 + r$.
Following the ideas
of refs.\ct{wz71a,polyakov81a,zumino83a,faddeev84a,fs84a}
for treating anomalous gauge theories,
one wants to have $r$ extra degrees of freedom.
The proposal is to augment the original set of fields,
$\phi^i$, $i = 1, 2, \dots , n$ with $r$ new fields
$\widehat \phi^{\hat i}$,  ${\hat i} = 1, 2, \dots , r$,
in such a way that
the original gauge structure continues to be maintained
at classical level.
Roughly speaking, the $\widehat \phi$ are fields
parametrizing the anomalous part of the gauge group.
In what follows, a ``hat'' on a quantity indicates
that the quantity is associated
with the extra degrees of freedom
or that the quantity has been generalized
to the extended system.

The key step is to extend the antibracket formalism
to include $\widehat \phi$ variables.
To implement this idea,
one demands the extra fields to transform under the
action of the gauge group:
\be
 \delta \widehat \phi^{\hat i}  =
 \widehat R^{\hat i}_{\alpha}
   \left[ \phi , \widehat \phi \right]
  \varepsilon^{\alpha}
\quad .
\label{extended trans gauge}
\ee
Let $\widehat \phi_{\hat i}^*$
be the antifield of $\widehat \phi^{\hat i}$.
Given Eq.\bref{fsr proper solution},
the classical gauge structure of
the extended theory
should be governed by an action
$
  \widetilde S=S+ \widehat \phi_{\hat i}^*
  \widehat R^{\hat i}_{\alpha} \gh C^{\alpha}
$.
Since field indices range over
$a = 1, 2, \dots , n, n+1, \dots, n+r$, the compact notation
$
 \phi^a \equiv
 [ { \phi^i , \widehat\phi^{\hat i} } ]
$
and
$
  R^a_\alpha \equiv
    [ { R^i_\alpha ,  \widehat R^{\hat i}_\alpha } ]
$
is useful.
To maintain the gauge structure,
the extended generators $R^a_\alpha$
must satisfy the original gauge algebra.
In other words,
when the field content is extended,
Eq.\bref{algebra oberta} must still hold,
where the tensors $T$ and $E$ are
functions of the $\phi^i$ only
and have the same values as in the unextended theory.
This requirement leads
to a set of equations for the generators
$ \widehat R^{\hat i}_{\alpha}$
in Eq.\bref{extended trans gauge}.
Formulas for $\widehat R^{\hat i}_{\alpha}$
in the closed case for which $E=0$
are given in refs.\ct{gp93a,gp93c}.
In the closed case,
it is possible to
solve the antifield independent part of
the original quantum master
equation at one-loop
in a quasilocal way
in the extended theory.
In particular, the Wess-Zumino term
$ M_1 [ \phi,\widehat \phi ]$
can be written as
\be
  M_1 \left[ { \phi,\widehat \phi } \right] =
  -i\int_0^1 ds \gh A_{\hat i}
  \left( { F \left[ { \phi,\widehat \phi s } \right]
         } \right) {\widehat \phi}^{\hat i}
\quad ,
\label{wz}
\ee
where the $\gh A_{\hat i}$
are the antifield-independent part of the anomalies
and $F^i$ is the finite gauge transformation
of the classical fields $\phi^i$ under the anomalous
part of the group.
The BRST variation of $M_1$ gives the original anomaly:
$( M_1 , \widetilde S) = i {\cal A}_\alpha {\cal C}^\alpha$.

However, in the extended antibracket formalism
the action $\widetilde S$ is not proper
\ct{jst93a}.
To have a well defined perturbation expansion,
it is necessary to modify $\widetilde S$
to a new extended action $\widehat S$
that satisfies the classical master equation
\be
  \left( { \widehat S , \widehat S }  \right) = 0
\quad ,
\label{extended cme}
\ee
and is proper, i.e.,
\be
   {\rm rank \ }
    \restric{
       \frac{\partial_l \partial_r \widehat S}
        {\partial z^a \partial z^b} }
  { \rm {on-shell}  }
  = n + m_0 + r
\quad ,
\label{extented def of proper solution}
\ee
where the $z^a$ include
$n$ fields, $r$ extended fields, $m_0$ ghosts,
and their antifields.
One proposal for $\widehat S$ would be
$\widetilde S+\hbar M_1$.
A difficulty is the presence of an order $\hbar$ term.
However, in certain cases
a canonical transformation
which scales $\widehat\phi^{\hat i}$ by $1/\hbar^{\frac12}$
\ct{jst93a}
can be performed
to overcome the problem.
The general structure for $\widehat S$
for anomalous gauge theories
with an anomalous abelian subgroup is
\be
  \widehat S =
   S  -
       \frac{i}2 \widehat\phi^{\hat i}
        \widehat{D}_{ {\hat i} {\hat j} } \widehat\phi^{\hat j}
  + \widehat\phi^*_{\hat i} \widehat{T}^{\hat i}_{{\beta} {\hat j} }
        \widehat\phi^{\hat j} \gh C^{\beta}
\quad ,
\label{extended S}
\ee
where $\widehat{D}$ and $\widehat{T}$ are tensors
which can be found
in ref.\ct{gp93c}.
In particular,
$\widehat{D}_{ {\hat i} {\hat j} }$ is related
to the BRST variation of the anomalies $\gh A_{\hat i}$.
Since $\widehat S$ satisfies the master equation,
a classical BRST symmetry can be defined using
$\delta_B X = \left( { X , \widehat S } \right)$.

The final stage is to find a solution $\widehat W$
to the quantum master equation in the extended space.
Because of the above-mentioned scaling of $\widehat \phi$,
$\widehat W$ is a series in $\hbar^{\frac12}$
\ct{jst93a,vp94a}
rather than $\hbar$:
$
  \widehat W =
  \widehat S + \hbar^{\frac12} \widehat M_{\frac12} + \dots
$.
After regulating the extended theory,
$\left( { \Delta \widehat S } \right)_{\rm reg} $
is computed.
By appropriately adjusting $\widehat S$ and $\widehat M_{\frac12}$,
the antifield-independent part
of the complete one-loop obstruction $\widehat{\cal A}_1$
to the quantum master equation in the extended space
can be cancelled.
In terms of the unscaled extra variables,
this adjustment corresponds to a finite renormalization
of the original expression of the Wess-Zumino term
in Eq.\bref{wz}, $M_1 \rightarrow \widehat M_1$.
Note that
the locality of $\widehat M_{\frac12}$
cannot be guaranteed.
Likewise,
locality of the renormalized Wess-Zumino term
is not assured due to the integral over the variable $s$
in expressions like \bref{wz}.
In some cases,
such as the first-quantized bosonic string
\ct{gp93a,gp93c} or the abelian Schwinger model
\ct{paris93a},
only local action terms are generated.
Then, the anomalous theory makes sense at the quantum level.
However,
for chiral QCD in two dimensions
\ct{gp93a},
the integral over $s$ remains.
Even in these cases,
the violation of locality is in some sense not severe:
The equations of motion are local,
a situation referred to as quasilocal.
When the quantum extended theory is well-defined,
the final stage,
namely gauge-fixing,
proceeds in a manner similar to the non-anomalous case
\ct{gp93a}.

\medskip
\medskip

\noindent
{\large \bf Acknowledgements}

\hspace{\parindent}
We thank
G.\ Barnich, C.\ Batlle, F.\ De Jonghe,
M.\ Henneaux, A.\ K.\ Kosteleck\'y,
J.\ M.\ Pons, J.\ Roca, R.\ Siebelink,
A.\ Slavnov, P.\ Townsend, W.\ Troost,
S.\ Vandoren, A.\ Van Proeyen and F.\ Zamora
for discussions.
This work is supported in part
by the
Comisi\'on Interministerial para la Ciencia
y la Technolog\'ia
(project number AEN-0695),
by a Human Capital and Mobility Grant
(ERB4050PL930544),
by the National Science Foundation
(grant number PHY-9009850),
by a NATO Collaborative Research Grant
(0763/87),
by the Robert A.\ Welch Foundation,
and
by the United States Department of Energy
(grant number DE-FG02-92ER40698).

\vfill\eject

\appendix

\section{Appendix: Right and Left Derivatives}
\label{s:appendixa}

\hspace{\parindent}
In this appendix, we provide more details about
left and right derivatives
\ct{berezin66a,bl75a,leites80a,dewitt84b,berezin87a}.
For any function or functional $X$ of $\phi$,
they are defined as
\be
 {{\partial_l X} \over {\partial \phi }} \equiv
 {{\mathop \partial \limits^\rightarrow }
  \over {\partial \phi }} X
 \  , \quad \quad
 {{\partial_r X} \over {\partial \phi }} \equiv
  X {{\mathop \partial \limits^\leftarrow }
  \over {\partial \phi }}
\quad .
\label{def lr der}
\ee
Left derivatives are the ones usually used.
Right derivatives act from right to left.
The differential $d X(\phi)$ of $X$ is
\be
  d X(\phi) =
  d \phi {{\partial_l X} \over {\partial \phi }} =
 {{\partial_r X} \over {\partial \phi }} d \phi
\quad .
\label{dif of X}
\ee
The formula for the variation $\delta X$ of $X$
with respect to $\phi$
resembles Eq.\bref{dif of X}:
\be
  \delta X(\phi) =
  \delta \phi {{\partial_l X} \over {\partial \phi }} =
 {{\partial_r X} \over {\partial \phi }} \delta \phi
\quad .
\label{variations of X}
\ee

What is the relation between left and right derivatives?
If $\phi$ is commuting
($\epsilon \left( \phi \right) = 0$),
then
$
{{\partial_l X} \over {\partial \phi }} =
{{\partial_r X} \over {\partial \phi }}
$,
so that one only needs to be careful when
$\phi$ is anticommuting
($\epsilon \left( \phi \right) = 1$).
Assume
$\phi$ is anticommuting.
Then $\phi \phi = 0$.
Without loss of generality we may assume that
$X = \phi Y + Z $
where $Y$ and $Z$ have no $\phi$ dependence.
The left and right derivatives of $X$ are then
$
 {{\partial_l X} \over {\partial \phi }} = Y
$
and
$
{{\partial_r X} \over {\partial \phi }} =
 \left( {-1} \right)^{ \epsilon_Y } Y =
\left( {-1} \right)^{\epsilon \left( \phi \right)
  \left( \epsilon_X + 1 \right) } Y
$.
For all cases,
\be
 {{\partial_l X} \over {\partial \phi }} =
 \left( {-1} \right)^{\epsilon \left( \phi \right)
  \left( \epsilon_X + 1 \right) }
 {{\partial_r X} \over {\partial \phi }}
\quad .
\label{l and r der relation}
\ee

As a pedagogical exercise,
let us derive the graded antisymmetry property
of the bracket in Eq.\bref{antibracket properties}.
Start with the definition of
$\left( {Y,X} \right)$
in Eq.\bref{antibracket def}
and interchange the order of derivatives to obtain
$$
  \left( {Y,X} \right) =
  {{\partial_r Y} \over {\partial \Phi^A}}
  {{\partial_l X} \over {\partial \Phi_A^*}} -
  {{\partial_r Y} \over {\partial \Phi_A^*}}
  {{\partial_l X} \over {\partial \Phi^A}}=
$$
$$
  \left( {-1} \right)^{\left( {\epsilon_Y +
  \epsilon_A } \right)
   \left( {\epsilon_X +
   \epsilon_A + 1} \right)}
   {{\partial_l X} \over {\partial \Phi_A^*}}
   {{\partial_r Y} \over {\partial \Phi^A}} -
   \left( {-1} \right)^{\left( {\epsilon_X +
   \epsilon_A } \right)
   \left( {\epsilon_Y +
  \epsilon_A + 1} \right)}
   {{\partial_l X} \over {\partial \Phi^A}}
   {{\partial_r Y} \over {\partial \Phi_A^*}}
\quad .
$$
Then using Eq.\bref{l and r der relation},
one finds that the above is equal to
$$
   \left( {-1} \right)^{\epsilon_X \epsilon_Y +
  \epsilon_X + \epsilon_Y + 1}
  {{\partial_r X} \over {\partial \Phi_A^*}}
  {{\partial_l Y} \over {\partial \Phi^A}} -
   \left( {-1} \right)^{\epsilon_X \epsilon_Y +
   \epsilon_X + \epsilon_Y + 1}
   {{\partial_r X} \over {\partial \Phi^A}}
   {{\partial_l Y} \over {\partial \Phi_A^*}}
$$
$$
   = -\left( {-1} \right)^{
   \left( {\epsilon_X + 1} \right)
   \left( {\epsilon_Y + 1} \right)}
   \left( {X,Y} \right)
\quad .
$$
This is the desired result.

As another exercise,
let us verify the formulas for
$\left( {F,F} \right)$
and
$\left( {B,B} \right)$
in Eq.\bref{() properties}
When $Y=X$,
the second term in Eq.\bref{antibracket def}
can be written as
$$
  {{\partial_r X} \over {\partial \Phi_A^*}}
  {{\partial_l X} \over {\partial \Phi^A}} =
   \left( {-1} \right)^{\left( {
   \epsilon_X + 1} \right)
  \left( {\epsilon_X + 1} \right)}
  {{\partial_r X} \over {\partial \Phi^A}}
  {{\partial_l X} \over {\partial \Phi_A^*}}
\quad ,
$$
using the same manipulations as in the previous paragraph.
When $X=F$ is anticommuting,
the sign factor is plus and the second term
in Eq.\bref{antibracket def}
cancels the first.
When $X=B$ is commuting,
the sign factor is minus
and the two add.

Another useful result concerns
integration by parts.
When $\phi$ is commuting,
one has the standard formula
$
  \int {\dif \phi }{{\partial_r X} \over
    {\partial \phi }} Y =
   -\int {\dif \phi } X
   {{\partial_l Y} \over {\partial \phi }}
$.
In such a case,
$
 {{\partial_r } \over {\partial \phi }} =
 {{\partial_l } \over {\partial \phi }}
$,
so that the left and right subscripts
on $\partial$ are inconsequential.
Suppose $\epsilon \left( \phi \right) = 1$.
Then,
$
\int {\dif \phi }{{\partial_r X} \over {\partial \phi }}Y
$
and
$
\int {\dif \phi }X{{\partial_l Y} \over {\partial \phi }}
$
are both zero
unless both $X$ and $Y$ are linear in $\phi$.
Without loss of generality,
we may assume that
$X = x \phi $
and
$Y = \phi y$
where $x$ and $y$ are independent of $\phi$.
Then,
$
  \int {\dif \phi }{{\partial_r X} \over {\partial \phi }}Y =
  \int {\dif \phi }x\left( {\phi y} \right) =
  \left( {-1} \right)^{\epsilon \left( x \right) }xy
$
and
$
  \int {\dif \phi }X{{\partial_lY} \over {\partial \phi }} =
  \int {\dif \phi }\left( {x\phi } \right)y =
  \left( {-1} \right)^{\epsilon \left( x \right) }xy
$
lead to the same result.
Summarizing, all cases are contained in the formula
\be
  \int {\dif \phi }{{\partial_r X} \over {\partial \phi }}Y =
   \left( {-1} \right)^{\epsilon\left( \phi  \right)+1}
   \int {\dif \phi }X{{\partial_l Y} \over {\partial \phi }}
\quad .
\label{int by parts}
\ee

For second derivatives of $X$, one has
$$
   {{\partial_r \partial_l X} \over
     {\partial \phi \partial \phi '}} =
   {{\partial_l \partial_r X} \over
     {\partial \phi '\partial \phi }}
\quad ,
$$
$$
  {{\partial_l \partial_l X} \over
    {\partial \phi \partial \phi '}} =
  \left( {-1} \right)^{\epsilon \left( \phi  \right)
   \epsilon \left( {\phi '} \right)}
   {{\partial_l \partial_l X} \over
    {\partial \phi '\partial \phi }}
\quad ,
$$
\be
  {{\partial_r \partial_r X} \over
   {\partial \phi \partial \phi '}} =
  \left( {-1} \right)^{\epsilon \left( \phi  \right)
     \epsilon \left( {\phi '} \right)}
   {{\partial_r \partial_r X} \over
   {\partial \phi '\partial \phi }}
\quad .
\label{second der relations}
\ee
In the first equation,
derivatives act from different directions
and hence commute.
In the second and third equations,
one must be careful of the order.

If $X$ is a functional of $Y$
which is a function of $\phi$,
one has the chain rules
$$
  {{\partial_r X \left( {Y\left( \phi  \right)} \right)}
    \over {\partial \phi }}={{\partial_r X}
     \over {\partial Y}}
    {{\partial_r Y} \over {\partial \phi }}
\quad ,
$$
\be
  {{\partial_l X \left( {Y\left( \phi  \right)} \right)}
    \over {\partial \phi }}=
   {{\partial_l Y} \over {\partial \phi }}
   {{\partial_l X} \over {\partial Y}}
\quad .
\label{chain rules}
\ee

\vfill\eject

\section{Appendix: The Regularity Condition}
\label{s:appendixb}

\hspace{\parindent}
If a theory is invariant under the gauge transformations
in Eq.\bref{trans gauge}
then the lagrangian does not depend on all degrees of freedom.
In other words,
$
 S_0 \left[ \phi^{\prime} \right]
 = S_0 \left[ \phi  \right]
$
where $\phi^{\prime}$ stands for finitely transformed fields
produced by any of the infinitesimal variations
in Eq.\bref{trans gauge}.
When this relation is expanded to first order
in the gauge parameters $\varepsilon^\alpha$,
the Noether relations in Eq.\bref{ident noether} are obtained.

Let $\phi_0$ be the stationary point
about which one would like
to perform the perturbative expansion.
Then, in a neighborhood of $\phi_0$
there are other stationary points
given by performing finite gauge transformations on $\phi_0$.
Let $\Sigma$ locally be the surface around $\phi_0$
in $\phi$ configuration space
where the equations of motion vanish.
The regularity condition assumes that
the dimension of $\Sigma$
is maximal and that the quadratic form
generated by expanding the lagrangian
to second order in fields
has a rank $n_{\rm dof}$
on this surface
\ct{bv85a}.
Hence, the number of fields that enter
dynamically in $S_0$ is $n_{\rm dof}$.
The regularity assumption is
important for implementing perturbation theory
since the propagator --
which is the inverse of this quadratic form --
then exists.

Summarizing, the regularity condition is
\be
 {\rm rank \ }
 \restric{{\partial _l S_{0,i}} \over {\partial \phi ^j}}{ \Sigma } =
 \restric{\frac{\partial _l \partial _r S_0}
          {\partial \phi ^i \partial \phi ^j}
          }{ \Sigma }
 = n_{\rm dof}
\quad ,
\label{rc1}
\ee
where $\Sigma$ is the stationary surface defined implicitly by
\be
\restric{S_{0,i}}{ \Sigma } = 0
\quad .
\label{def of sigma}
\ee
In other words,
the on-shell degeneracy
of the hessian
in Eq.\bref{rc1}
is completely due to
the $n - n_{\rm dof}$ independent
null vectors $R^i_\alpha$
associated with gauge transformations
and does not come from some other source
\ct{bv84a,bv85a}:
\be
 \restric{\frac{\partial _l \partial _r S_0}
          {\partial \phi ^i \partial \phi ^j}
          R^i_\alpha}{ \Sigma }=0
\quad .
\label{rc1b}
\ee

An example of a lagrangian that does not satisfy
the regularity condition is ${\cal L} = \phi^{4}$
with no kinetic energy term for $\phi$.
The stationary point $\phi_0 = 0$
has a vanishing quadratic
form even though there is no gauge invariance.
In such a case one can proceed by arbitrarily
adding and subtracting some kinetic energy term
and treating $\phi^4$ minus this kinetic energy term
as a perturbation.
However, throughout this review we assume
that such singular cases do not arise.

In principle, one can separate the degrees of freedom
into propagating degrees of freedom
$\varphi^s$, $s= \ 1,\ 2, \dots , \  n_{\rm dof}$
and gauge degrees of freedom
$\chi^a$, $a= \ 1,\ 2, \dots , \  n - n_{\rm dof}$.
An efficient separation is often difficult
and the $\varphi^s$ and $\chi^a$
are usually quite complicated and nonlocal
functionals of the $\phi^i$.

The regularity conditions are then given by
$$
   S_{0,a} = 0 \ ,  \quad {\rm identically \quad ,}
$$
\be
  {\rm rank }
  \restric{{\partial _l S_{0,s}}
  \over {\partial \phi ^i}}{ \phi = \phi_0 }
  =  n_{\rm dof}
\quad .
\label{rc2}
\ee

The regularity condition assumes that
the equations of motion $S_{0,i}$
constitute a regular representation
of the stationary surface $\Sigma$.
This means that the functions $S_{0,i}$
can be locally split into independent,
$G_s$, and dependent ones, $G_a$,
in such a way that
\begin{enumerate}
\item  $G_a=0$ are a direct consequence of $G_s=0$, and
\item  The rank of the matrix of the gradients $dG_s$
is maximal on $\Sigma$.
\end{enumerate}
In other words,
the regularity condition ensures
that locally the changes of variables
$\phi^i \rightarrow [\varphi^s , \chi^a ]$ or
$\phi^i \rightarrow [G_s , G_a]$
makes sense
\ct{bv84a,bv85a,fh90a}.

When the regularity condition is fulfilled,
it can be shown that any smooth function
that vanishes on the stationary surface $\Sigma$
can be written as
a combination of the equations of motion
\ct{bv85a,fisch90a,fh90a,fhst89a},
i.e.,
\be
   \restric{F(\phi)}{\Sigma} = 0 \Rightarrow
   F(\phi) = \lambda^j S_{0,j}
\quad ,
\label{consequence of rc2}
\ee
where the $\lambda^j$ may be functions of the $\phi^i$.
No restrictions are made on the
$\lambda^j ( \phi )$.
Putting restrictions can lead to
violations of \bref{consequence of rc2}.
An example is
presented in
\ct{vp94a}.
By considering only
local functionals,
ref.\ct{vp94a} found cases
for which
Eq.\bref{consequence of rc2}
could not be satisfied
as a local combination of the
equations of motion.

For more details on regularity conditions
as well as derivations of the above results
consult references
\ct{wh79a,bv84a,bv85a,fhst89a,%
fisch90a,fh90a,henneaux90a}.

\vfill\eject

\section{Appendix: Anomaly Trace Computations}
\label{s:appendixc}

\hspace{\parindent}
In this appendix,
we perform the functional trace calculations
of Sect.\ \ref{s:sac}.
One key idea is to use the Dyson-like expansion
\ct{fh65a,iz80a}
$$
   \exp \left[ {{\cal R}_0 + V} \right] =
    \exp \left[ {{\cal R}_0} \right] +
   \int\limits_0^1 {\dif s}
    \exp \left[ {\left( {1-s} \right){\cal R}_0} \right]
     V \exp \left[ {s{\cal R}_0} \right] +
$$
\be
  \int\limits_0^1 {\dif u}\int\limits_0^u {\dif s}
    \exp \left[ {\left( {1-u} \right){\cal R}_0} \right]
    V \exp \left[ {\left( {u-s} \right){\cal R}_0} \right]
    V \exp \left[ {s{\cal R}_0} \right] + \ldots
\quad .
\label{dyson expansion}
\ee
Typically,
${\cal R}_0$ is independent of the cutoff ${\cal M}$,
and $V$ goes like inverse powers of ${\cal M}$
so that only a few terms
in Eq.\bref{dyson expansion}
need to be kept.

The anomaly equation
Eq.\bref{Delta S reg}
involves a functional trace
\ct{fujikawa80a}.
If one uses momentum-space eigenfunctions
to saturate the sum,
then expressions such as
$$
  \exp \left( {- i k \cdot x} \right)
  \left( {  O \left( {\partial_\mu } \right)  } \right)
   \exp \left( {ik\cdot x} \right)
$$
arise
where $O \left( {\partial_\mu } \right) $
is an arbitrary operator, or a product of operators,
involving the derivative $\partial_\mu$.
By commuting $\exp \left( {ik\cdot x} \right)$
through the expression,
one arrives at
\be
  \exp \left( {- i k \cdot x} \right)
    \left( { O \left( {\partial_\mu } \right) } \right)
   \exp \left( {ik\cdot x} \right) =
  \left( { O \left( {\partial_\mu +ik_\mu }
             \right) } \right) 1
\quad .
\label{momentum conjugation}
\ee
When derivatives in
$O \left( {\partial_\mu +ik_\mu } \right)$
act on the function $1$, they produce zero.

\medskip

For the spinless relativistic particle system,
we begin by taking
Eq.\bref{rsp Delta S reg}
and commuting
$\exp \left( {ik\tau } \right)$
through the expression,
using Eq.\bref{momentum conjugation},
to obtain
$$
  \left( {\Delta S} \right)_{\rm reg} =
$$
$$
   D\int {\dif \tau } \left[ {
   \int\limits_{-\infty }^\infty
  {{ {\dif k} \over {2\pi } }} \rho^{-1}
   {\cal C}
  \left( { {d \over {d\tau } } + ik } \right)
  \ \exp \left( {{ {-\left( { {d \over {d\tau } } + ik } \right)
  \rho^{-1} \left( { {d \over {d\tau }} + ik } \right)} \over
    {\cal M}^2} } \right) 1
                              } \right]_0
\  .
$$
Rescaling $k$ by ${\cal M}$ produces
$$
  \left( {\Delta S} \right)_{\rm reg} =
  D\int {\dif \tau }\int\limits_{-\infty }^\infty
    {{{\dif k} \over { 2 \pi }}}
  \left[ {{\cal M} \rho^{-1} {\cal C}
  \left( {{d \over {d\tau }} +
  i {\cal M} k} \right) \ \times } \right.
$$
\be
  \left. {
     \exp \left( {
       { {k^2} \over \rho } -
  i{ k \over {\cal M} }
        \left( {\rho^{-1} {d \over {d\tau }} +
  {d \over {d\tau } } \rho^{-1}  } \right) -
  {1 \over   {\cal M}^2} {d \over {d\tau }}
   \rho^{-1} {d \over {d\tau } }
                 } \right) 1
  } \right]_0
\quad .
\label{rsp delta S cal 1}
\ee
Eq.\bref{rsp delta S cal 1}
is in the form
of Eq.\bref{dyson expansion}
where $ {\cal R}_0 = { {k^2} / \rho }$.
We use the Dyson-like expansion
in Eq.\bref{dyson expansion}
and pick out the ${\cal M}$-independent term
to arrive at
$$
  \left( {\Delta S} \right)_{\rm reg} =
$$
$$
   -iD\int {\dif \tau }\int\limits_{-\infty }^\infty
     {{{\dif k} \over {2\pi }}}
  \ k \rho^{-1} {\cal C}
 \left\{ {
          {d \over {d\tau }}
   \int\limits_0^1 {\dif s}
  \exp \left( {\left( {1-s} \right){{k^2} \over \rho }} \right)
  \left( {\rho^{-1}{d \over {d\tau }} +
  {d \over {d\tau }}\rho^{-1}} \right)
  \exp \left( {s{{k^2} \over \rho }} \right)
 } \right.
$$
$$
  + \int\limits_0^1 {\dif s}
  \exp \left( {\left( {1-s} \right){{k^2} \over \rho }}
   \right)\left( {{d \over {d\tau }}
    \rho^{-1}{d \over {d\tau }}} \right)
  \exp \left( {s{{k^2} \over \rho }} \right)
$$
$$
  + k^2 \int\limits_0^1
  {\dif u}\int\limits_0^u {\dif s}
  \exp \left( {\left( {1-u} \right){{k^2} \over \rho }} \right)
   \left( {\rho^{-1}{d \over {d\tau }} +
  {d \over {d\tau }}\rho^{-1}} \right) \ \times
$$
\be
 \left. {
  \exp \left( {\left( {u-s} \right)
    {{k^2} \over \rho }} \right)
   \left( {\rho^{-1}{d \over {d\tau }} +
     {d \over {d\tau }}
   \rho^{-1}} \right)
  \exp \left( {s {{k^2} \over \rho }} \right)
  } \right\}
\quad .
\label{rsp delta S cal 2}
\ee
The first term
in Eq.\bref{rsp delta S cal 2}
is zero because it is a total derivative.
To calculate the other two terms
rotate to Euclidean space using
$k \to -ik_E$.
Then the expression
${{k^2} / \rho } \to {{-k_E^{2} } / \rho } $
in the exponents yields
gaussian damping factors,
so that the integrals are convergent.
Even before evaluating the derivatives
${d \over {d\tau }}$,
it is clear that the integrand
is an odd function of $k$
and hence produces a zero integral.

\medskip

For the chiral Schwinger model,
one starts with Eq.\bref{acsm19}.
Using momentum-space wave functions,
the functional trace is
$$
 \Delta S =
    - i \int \dif^2 x \, {\cal C} \times
$$
\be
 \left[{
   \int \frac{\dif^2 k}{ \left( {2\pi} \right)^2}
   \exp{ \left( -ik \cdot x \right) }
  \left( {
  \exp{ \left( {
     { { R_+ } \over {{\cal M}^2} }
                } \right) } -
  \exp{ \left( {
     { { R_- } \over {{\cal M}^2} }
                } \right) }
         } \right)
   \exp{ \left( ik \cdot x \right) }
   } \right]_0
\ ,
\label{acsma0}
\ee
Equation \bref{momentum conjugation}
is used to eliminate the
$\exp { \left( {\pm i k \cdot x} \right) } $
factors.
Then, one scales the momentum $k$ by ${\cal M}$.
The expression for $\Delta S$ becomes
\be
  \Delta S =
   -i\int {\dif^2 x}  \, {\cal C}
   \int {{{\dif^2 k} \over {\left( {2\pi } \right)^2}}
   \left[ {{\cal M}^2
    \left( {\exp \left( {\tilde R_+} \right) -
    \exp \left( {\tilde R_-} \right)} \right)1} \right]_0}
\quad ,
\label{acsma1}
\ee
where
\be
  \tilde R_\pm =
   -k^\mu k_\mu -
  {{-2ik^\mu \partial_\mu \pm A_-k_+} \over {{\cal M}}} +
   {{\partial^\mu \partial_\mu
   \pm i\left( {\partial_+ A_-} \right)
   \pm iA_-\partial_+} \over {{\cal M}^2}}
\quad .
\label{acsma2}
\ee
The parenthesis around
$(\partial_+ A_-)$
indicates that $\partial_+$
acts only on $A_-$.

The notation $\left[ \right]_0$
selects the term in Eq.\bref{acsma1}
independent of ${\cal M}$.
Hence,
one expands
in the factors in the exponentials
proportional to ${\cal M}^{-1}$ and ${\cal M}^{-2}$.
This results in two terms,
$\Delta S_1$ from the leading Taylor series term
in ${\cal M}^{-2}$,
and $\Delta S_2$ from the second order term
in ${\cal M}^{-1}$:
$
  \Delta S = \Delta S_1 + \Delta S_2
$,
where
$$
  \Delta S_1 =
  2\int {\dif^2 x} \, {\cal C}\left( {\partial_+ A_-} \right)
  \int {{{\dif^2 k} \over {\left( {2\pi } \right)^2}}}
  \exp \left( {-k^\mu k_\mu } \right)
\quad ,
$$
\be
  \Delta S_2 =
   i\int {\dif^2 x} \, {\cal C}
  \int {{{\dif^2 k} \over {\left( {2\pi } \right)^2}}}
  \exp \left( {-k^\mu k_\mu } \right)
   2ik^\mu \left( { \partial_\mu A_- } \right) k_+
\quad .
\label{acsma3}
\ee
Integrals are defined by analytic continuation
using Wick rotation, that is,
the $k_0$ integral is performed
by going to Euclidean space.
The line integral contained in
$\int\limits_{-\infty }^\infty  {\dif k_0 }$
can be rotated counterclockwise by $90^{\rm o}$.
Then, one sets
$
k_0 = -ik_0^E
$:
\be
  \int\limits_{-\infty }^\infty  {\dif k_0 } =
  \int\limits_{-i\infty }^{i\infty } {\dif k_0 } =
   -i\int\limits_\infty^{-\infty } {\dif k_0^E=}
    i\int\limits_{-\infty }^\infty  {\dif k_0^E}
\quad .
\label{wick rotation}
\ee
Integrals are then convergent
since
\be
 \exp \left[ {- {k^2} } \right] =
 \exp \left[ {+  k_0^2 - k_1^2  } \right] \to
 \exp \left[ {- {k_0^{E2}} - k_1^2  } \right]
\quad ,
\label{acsma4}
\ee
provides a gaussian damping factor.
The following integration table is obtained
\be
  \int {{{\dif^2 k} \over {\left( {2\pi } \right)^2}}}
   \exp \left( {-k^\mu k_\mu } \right)
  \left\{ \matrix{1\hfill\cr
  k_0^2\hfill\cr
  k_1^2\hfill\cr
  k_0k_1\hfill\cr} \right\} =
  {i \over {8\pi }}
 \left\{ \matrix{2\hfill\cr
  -1\hfill\cr
  1\hfill\cr
  0\hfill\cr} \right\}
\quad .
\label{acsma5}
\ee
Hence,
$$
  \Delta S_1 =
   {{2i} \over {4\pi }}\int {\dif^2 x}
   \, {\cal C}\left( {\partial_+ A_-} \right)
\quad ,
$$
\be
  \Delta S_2 =
   -{i \over {4\pi }}\int {\dif^2 x}
    \, {\cal C}\left( {\partial_+ A_-} \right)
\quad .
\label{acsma6}
\ee
Adding the two contributions,
\be
  \Delta S =
   {i \over {4\pi }}\int {\dif^2 x}
    \, {\cal C}\left( {\partial_+ A_-} \right)
\quad .
\label{acsma7}
\ee
The factor $\partial_+ A_-$
can be written in Lorentz covariant form
using
\be
  \partial_+ A_- =
   \partial_0 A_0-\partial_1 A_1 -
   \partial_0 A_1+\partial_1 A_0 =
   -\partial_\mu A^\mu +
  \varepsilon^{\mu \nu }\partial_\mu A_\nu
\quad .
\label{acsma8}
\ee
Substituting Eq.\bref{acsma8} into Eq.\bref{acsma7}
produces the result in Eq.\bref{acsm21}.

\medskip

For the first-quantized bosonic string theory,
we need to compute $\kappa_r$
of Eq.\bref{def of kappa_n}.
Using momentum-space eigenfunctions,
it can be expressed as
\ct{fujikawa82a,fis90a}
\be
    \kappa_r = {{tr} \over 2}
  \left[ {\int {{{\dif^2k} \over {\left( {2\pi } \right)^2}}}
   \exp \left( {-ik\cdot x} \right)
   \exp \left( {{{H_r} \over {{\cal M}^2}}} \right)
   \exp \left( {ik\cdot x} \right)} \right]_0
\quad ,
\label{kappa_n trace}
\ee
where $tr$ is a trace
over two-by-two gamma-matrix space.
In Eq.\bref{kappa_n trace},
$
  k \cdot x = - k_\tau \tau + k_\sigma \sigma
$.
Using the definition of $H_r$
in Eq.\bref{def of H_n}
and commuting
$\exp \left( {ik\cdot x} \right)$
through the expression
via Eq.\bref{momentum conjugation},
Eq.\bref{kappa_n trace} becomes
$$
  \kappa_r = {{tr} \over 2}
   \left[ {\int {{{\dif^2k} \over {\left( {2\pi } \right)^2}}}
   \exp \left( {-{{\rho^{-\left( {r+1} \right)/2}
   \left( {\slashit \partial +i\slashit k} \right)
   \rho^r\left( {\slashit \partial +i\slashit k} \right)
   \rho^{-\left( {r+1} \right)/2}} \over {{\cal M}^2}}} \right)
    1} \right]_0
$$
$$
  = {{tr} \over 2}
   \left[ {{\cal M}^2\int {{{\dif^2k} \over {\left( {2\pi } \right)^2}}}
   \exp \left( {-{{k^2} \over \rho } - i{A \over {{\cal M}}} +
  {B \over {{\cal M}^2}}} \right)1} \right]_0
\quad ,
$$
where we have performed the rescaling
$
  k \to {\cal M} k
$,
and where
\be
  A \equiv \rho^{-\left( {r+1} \right)/2}
   \left( {\slashit \partial \rho^r \slashit k +
   \slashit k\rho^r\slashit \partial } \right)
    \rho^{-\left( {r+1} \right)/2}
\quad ,
\label{def of A operator}
\ee
\be
  B \equiv -\rho^{-\left( {r+1} \right)/2}
   \slashit \partial \rho^r\slashit \partial
  \rho^{-\left( {r+1} \right)/2}
\quad .
\label{def of B operator}
\ee

For the next step,
we use the Dyson-like expansion
in Eq.\bref{dyson expansion}.
Recalling that
$[ \ ]_0$
indicates that the ${\cal M}$-independent term
is to be selected,
$\kappa_r$ becomes a sum of two terms
$$
  \kappa_r = B {\mbox{-term}} + AA {\mbox{-term}}
\quad ,
$$
where
$$
  AA {\mbox{-term}} =
  - {{tr} \over 2}\int {{{\dif^2k} \over {\left( {2\pi } \right)^2}}}
  \int\limits_0^1 {\dif u}\int\limits_0^u {\dif s} \ \times
$$
$$
  \exp \left[ {-\left( {1-u} \right){{k^2} \over \rho }} \right]
   A \exp \left[ {-\left( {u-s} \right){{k^2} \over \rho }} \right]
   A \exp \left[ {-s{{k^2} \over \rho }} \right]1
\quad ,
$$
and
$$
  B {\mbox{-term}} = {{tr} \over 2}
  \int {{{\dif^2k} \over {\left( {2\pi } \right)^2}}}
   \int\limits_0^1 {\dif s}
   \exp \left[ {-\left( {1-s} \right){{k^2} \over \rho }} \right]
   B \exp \left[ {-s{{k^2} \over \rho }} \right]1
\quad .
$$

In $AA {\mbox{-term}}$,
carry out the differentiations $\slashit \partial$
in both $A$ operators
using the definition of $A$
in Eq.\bref{def of A operator}
to obtain
$$
   AA {\mbox{-term}} =
    -\int {{{\dif^2k} \over {\left( {2\pi } \right)^2}}}
   \int\limits_0^1 {\dif u}\int\limits_0^u {\dif s} \ k^2
   \exp \left[ {- {k^2} } \right]
   \left\{ {{{\partial^n\partial_n\rho } \over \rho }
   \left( { 2 s k^2-1} \right) + } \right.
$$
$$
   \left. {{{\partial^n\rho \partial_n\rho } \over {\rho^2}}
   \left( {-{1 \over 2}\left( {r^2-5} \right) -
     7 s k^2 - u k^2 + 2 s u k^4} \right)} \right\}
\quad ,
$$
where another rescaling
$
  k \to k \rho^{1/2}
$
has been performed.
Next, rotate the $k$ integration from Minkowski space
to Euclidean space.
The line integral contained in
$\int\limits_{-\infty }^\infty  {\dif k_\tau }$
can be rotated counterclockwise by $90^{\rm o}$.
One sets
$
k_\tau = -ik_\tau^E
$
and uses Eq.\bref{wick rotation}
for $k_0 = k_\tau$.
The exponential factor
$
 \exp \left[ {- {k^2} } \right] =
 \exp \left[ {+  k_\tau^2 - k_\sigma^2  } \right]
$
in $AA {\mbox{-term}}$
becomes
$
 \exp \left[ {- {k_\tau^{E2}} - k_\sigma^2  } \right]
$.
One can then do the $s$, $u$ and $k$ integrations
since the latter are now convergent.
The result is
\be
  AA {\mbox{-term}} = {1 \over {4\pi i}}
  \left( {{{\partial^n\partial_n\rho } \over \rho }{1 \over 6} -
   {{\partial^n\rho \partial_n\rho } \over {\rho^2}}
   \left( {{{r^2+1} \over 4}} \right)} \right)
\quad .
\label{result for A-A-term}
\ee

The $B {\mbox{-term}}$ is treated similarly.
One carries out the differentiations $\slashit \partial$
in the $B$ operator and
rescales $k$ by
$ \rho^{1/2}$
to arrive at
$$
   B {\mbox{-term}} = {{tr} \over 2}
  \int {{{\dif^2k} \over {\left( {2\pi } \right)^2}}}
   \int\limits_0^1 {\dif s}\exp \left[ {-k^2} \right]
  \left\{ {{{\partial^n\partial_n\rho } \over \rho }
  \left( {sk^2 - {{r+1} \over 2}} \right) + } \right.
$$
$$
  \left. {{{\partial^n\rho \partial_n\rho } \over {\rho^2}}
   \left( {{{\left( {3-r} \right)
  \left( {r+1} \right)} \over 4} - 3 s k^2
         + s^2 k^4} \right)} \right\}
\quad .
$$
After rotating to Euclidean space,
all integrations can be formed.
The $B {\mbox{-term}}$ is
\be
  B {\mbox{-term}} = {1 \over {4\pi i}}
  \left( {{{\partial^n\partial_n\rho } \over \rho }
  {r \over 2} +
  {{\partial^n\rho \partial_n\rho } \over {\rho^2}}
  \left( {{{r^2} \over 4} -
 {r \over 2} + {1 \over {12}}} \right)} \right)
\quad .
\label{result for B-term}
\ee
The sum of the $AA {\mbox{-}}$ and $B {\mbox{-terms}}$
in Eqs.\bref{result for A-A-term}
and \bref{result for B-term}
is the quoted result for $\kappa_r$
in Eq.\bref{kappa_n results}.

\vfill\eject

\vfill\eject

\end{document}